# Numerical ranges and geometry in quantum information

## *Entanglement, uncertainty relations, phase transitions, and state interconversion*

*Author:*
Konrad Szymański

*Supervisor:*
Prof. Karol Życzkowski

March 2022

UNIWERSYTET JAGIELLOŃSKI
W KRAKOWIE

## *Streszczenie*

Geometria zbiorów pojawiających się podczas analizy problemów informacji kwantowej pozwala dostarczyć wielu informacji na temat tej teorii. W związku z tym warto podejść do mechaniki kwantowej geometrycznie – to spojrzenie zaowocowało wieloma interesującymi wynikami. W tej rozprawie pokazałem wyniki dotyczące *zakresu numerycznego* – zbioru dopuszczalnych wartości oczekiwanych kilku obserwabli. Zastosowałem ten obiekt aby rozwiązać problemy związane z relacjami nieoznaczoności, wykrywaniem splątania czy szacowaniem wielkości przerwy energetycznej. Ponadto, stosując struktury geometryczne zaprezentowałem kryteria osiągalności stanów kwantowych przy użyciu kanałów kwantowych przemiennych z działaniem reprezentacji grupy.

## *Abstract*

Studying the geometry of sets appearing in various problems of quantum information helps in understanding different parts of the theory. It is thus worthwhile to approach quantum mechanics from the angle of geometry – this has already provided a multitude of interesting results. In this thesis I demonstrate results relevant to *numerical ranges* – the sets of simultaneously attainable expectation values of several observables. In particular, I apply this notion in the problems related to uncertainty relations, entanglement detection, and determining bounds for the value of spectral gap. Apart from this, I present geometric structures helping with the question of state interconversion using channels commuting with a particular representation of a group.

They told her "Your music is lovely and pleasant, and all that you show us we cannot but yearn for. But we are the daughters and sons of the Goddess of Cancer, her slaves and creatures. And all that we know is the single imperative *KILL CONSUME MULTIPLY CONQUER*. (...) We wish it were otherwise, but it is not, and your words have no power to move us."

The Goddess of Everything Else only laughed at them, saying, "But I am the Goddess of Everything Else and my powers are devious and subtle. (...) Return to your multiplication, but now having heard me, each meal that you kill and each child that you sire will bind yourself ever the more to my service." She spoke, then dove back in the sea, and a coral reef bloomed where she vanished.

As soon as she spoke it was so, and the animals all joined together. The wolves joined in packs, and in schools joined the fishes; the bees had their beehives, the ants had their anthills, and even the termites built big termite towers; the finches formed flocks and the magpies made murders, the hippos in herds and the swift swarming swallows. And even the humans put down their atlatls and formed little villages, loud with the shouting of children.

Scott Siskind, *The Goddess of Everything Else*

This thesis is dedicated to the people reading this to see if they are mentioned in the dedication.

# Contents

# Introduction 1

The topic of this thesis is the interplay between quantum mechanics, geometry, and algebra. Over a century of scientific activity has been dedicated to connecting these fields. My aim here is to extend the state of knowledge with investigations on the sets of quantum states, their low-dimensional projections, the geometry behind certain quantum resource theories, and algebraic structures related to them.

The subject can be thus summarized as an answer to a basic question: assuming quantum mechanics accurately describes nature, what can be said about the mathematical structures arising from it – and what does it tell us about the physical world?

This thesis serves as an introduction, motivation, and expansion of the content published in articles during my Ph.D. studies. Therefore, I decided to follow a format similar to a textbook: the emphasis is put on basic structures and ideas which influenced certain results already presented in the published papers. Some ideas which expand on the published work are included as well. The topic of my research is primarily the application of geometry in quantum information, and the following articles either further develop the theory or apply its tools in physical scenarios:

[1] Konrad Szymański, Stephan Weis, and Karol Życzkowski. 'Classification of joint numerical ranges of three hermitian matrices of size three'. In: *Linear Algebra and its Applications* 545 (2018),

[2] Jakub Czartowski, Konrad Szymański, Bartłomiej Gardas, Yan Fyodorov, and Karol Życzkowski. 'Separability gap and large-deviation entanglement criterion'. In: *Physical Review A* 100.4 (2019),

[3] Konrad Szymański and Karol Życzkowski. 'Geometric and algebraic origins of additive uncertainty relations'. In: *Journal of Physics A: Mathematical and Theoretical* 53.1 (2019),

[4] Konrad Szymański and Karol Życzkowski. 'Universal witnesses of vanishing energy gap'. In: *Europhysics Letters* 136 (2021),

[5] Timo Simnacher, Jakub Czartowski, Konrad Szymański, and Karol Życzkowski. 'Confident entanglement detection via the separable numerical range'. In: *Physical Review A* 104 (2021).

The thesis is divided in several chapters. I start with a brief sketch of the relevant aspects of quantum mechanics and mathematics, which is followed by the content of my research: the mathematical structures arising from the quantum theory, their properties, and applications. The thesis is finished with a short summary of the work done and a list of potential directions of further research. The appendix contains additional material which some may find useful: stereograms of selected three-dimensional figures appearing in the manuscript and Mathematica code implementing some of the methods presented in this thesis.

## 1.1 Quantum mechanics

Quantum mechanics, one of the most successful physical models, has certain underlying structures. Depending on the approach, one may use a plethora of mathematical tools to aid in understanding: from functional analysis, through category theory, to probability theory and its extensions – an indication of its nondeterministic properties.

In this chapter, the basic elements of quantum mechanics are introduced: quantum states and operations, their properties, and interpretations. Additionally, as a part of this chapter, basic tools of geometry are applied in the analysis of quantum mechanical objects – various approaches are appropriate in different contexts. The geometry of quantum objects connects several fields of mathematics: it is explicitly convex and defined by algebraic constraints. It is therefore appropriate to present the basic concepts of the relevant mathematics: this is accomplished in the following chapter.

## 1.2 Mathematical tools

Origins of geometry as a description of the physical world can be traced back to the ancient cultures of Egypt and Mesopotamia. It was, at this time, a practical tool used in day-to-day life, as well as the language of astronomy. Later on, it was studied as a pure, abstract subject – which certainly benefited the topic, as evidenced by the abundance of modern geometric theorems and concepts. Of particular importance in the context of quantum mechanics is the geometry of convex bodies: convexity arises naturally as a result of the probabilistic nature of quantum mechanics and allows for significant simplification of otherwise involved problems.

Algebra is another idea at the very core of mathematics. The concept of manipulating symbols to reach a certain goal was far from trivial in the early days of science – indeed, *geometric* methods of equation solving were popular at the time – and its invention opened a whole new world of possibilities.

The word *algebra* comes from the 9th century text [6], the first chapter of which describes the elementary problem of solving quadratic equations and related concepts. In this work, *al-ğabr* (آلْجَبْر) is a method of moving subexpressions around in the equation. The idea of equations and manipulating them was limited at this time – they were not seen as interesting objects on their own, the theory was not entirely strict, and was seen as a less consistent part of mathematics. Yet, it was useful, especially in conjunction with geometry. 16th century saw a development of independent field of algebra, led by François Viète, who laid rigorous foundations to it. Soon more results were presented in the fields we recognize today as polynomial theory and theory of matrices.

[6]: Muhammad ibn Musa al-Khwarizmi (circa 820), *Al-Kitab al-mukhtasar fi hisab al-gabr wa'l-muqabala*

Some of the basic ideas of convex geometry and algebra are presented in this thesis. Both of these areas are useful in the analysis of quantum phenomena. In particular, the Joint Numerical Ranges are one of the simplest objects *algebraic geometry* is concerned about, while their convex structure provides an additional effective framework. Matrices – another elementary algebraic structure – also come in handy, since they are the language of quantum mechanics.

## 1.3 Numerical ranges and spectrahedra

Numerical ranges – affine images of the set of quantum states – along with its generalizations are one of the central concepts of this thesis. Here, the theory and applications of numerical ranges are presented: from the basic definitions and properties of relevant objects to entanglement detection, relevance to uncertainty relations, phase transitions, and derivation of bounds for the value of the energy gap. In particularly simple cases, the shapes of numerical ranges can be classified by the features of nonanalytical parts of the boundary: this classification has recently found experimental confirmation [7].

[7]: Xie et al. (2020), 'Observing geometry of quantum states in a three-level system'

## 1.4 State interconversion

In the context of quantum resource theories, one of the most important questions is: when can one state be transformed into another given particular set of allowed operations? Careful analysis reveals that this question has a deeper meaning: the allowed target states are defined by symmetry criteria, which can be translated to the language of convex geometry and algebra.

The case of $U(1)$-covariant operations is studied in detail, with the presentation of the geometric structures appearing during the analysis. Additionally, other exemples of group covariance are presented: the discrete Weyl-Heisenberg group, related to an useful description of quantum mechanics, and $SU(2)$, representing spin rotations.

# Quantum mechanics | 2

Quantum mechanics is the theory of phenomena appearing in the microscopic world, for which the approach of classical physics fails to provide accurate predictions. In full generality, quantum mechanics deals with the behavior of complex systems of partial differential equations, the understanding of which requires the theory of functional analysis. Due to this complexity, quantum mechanics is often simplified in specific cases – as a result, an engineer designing nanoscale electronic devices may be unfamiliar with the jargon and methods used by quantum chemist and vice versa.

The scope of this thesis is related to problems of quantum computation, for which the domain of interest often are finite-dimensional Hilbert spaces. This is already an approximation: even the simplest atom is described by an infinite-dimensional vector space, and the reduction of complexity arises from the selection of the *most important* degrees of freedom (e.g., spin or a discrete subset of spatial states). In this chapter, the theory of finite-dimensional quantum mechanics is presented – even after this simplification, plenty of nontrivial mathematics remains. Some of the consequences of the structure appearing as a result are presented in the following chapters.

## 2.1 Quantum states

The central objects of quantum theory are the *state vectors* or *rays*: they carry the entire information contained within the system of interest and are subject to the evolution [8]. Thus, the physical theory uses the formalism of the complex Hilbert space.

[8]: Sakurai et al. (1995), *Modern Quantum Mechanics*

**Definition 2.1** *Hilbert space is a set $\mathcal{H}$ endowed with the following operations:*

*a) addition $\mathcal{H} \times \mathcal{H} \to \mathcal{H}$,*
*b) scalar multiplication $\mathbb{C} \times \mathcal{H} \to \mathcal{H}$,*
*c) scalar product $\mathcal{H} \times \mathcal{H} \to \mathbb{C}$.*

*The elements of $\mathcal{H}$ are usually denoted with kets $|\psi\rangle$ in the physical context, while the scalar product of $|\alpha\rangle$ and $|\beta\rangle$ is written as $\langle\alpha|\beta\rangle$. With this notation in mind, the axioms of Hilbert space can be written*

*in the following form:*

*a) Vector addition is associative and commutative. There exists a unique identity element of addition, called the zero vector* $0$*, for which* $|\psi\rangle + 0 = |\psi\rangle$*.*
*b) Scalar multiplication is compatible with the multiplicative structure of* $\mathbb{C}$*:* $z(w\,|\psi\rangle) = (zw)\,|\psi\rangle)$ *and* $1\,|\psi\rangle = |\psi\rangle$*.*
*c) Vector and scalar additions are distributive.*
*d) Scalar product is conjugate symmetric (*$\langle\alpha|\beta\rangle = \langle\beta|\alpha\rangle^*$*) and positive definite (*$\langle\psi|\psi\rangle \geq 0$ *with equality only for the zero vector).*
*e) The norm induced by the scalar product* $||\psi\rangle| = \sqrt{\langle\psi|\psi\rangle}$ *is complete.*

Associativity: $(|\alpha\rangle + |\beta\rangle) + |\gamma\rangle = |\alpha\rangle + (|\beta\rangle + |\gamma\rangle)$.
Commutativity: $|\alpha\rangle + |\beta\rangle = |\beta\rangle + |\alpha\rangle$.

Distributivity: $(z + w)(|\alpha\rangle + |\beta\rangle) = z\,|\alpha\rangle + w\,|\alpha\rangle + z\,|\beta\rangle + w\,|\beta\rangle$

In the context of this thesis, the finite-dimensional Hilbert spaces are most important. Such spaces are spanned by a finite collection of vectors: a space is $d$-dimensional if and only if there exist $d$ linearly independent vectors $|1\rangle, \ldots, |d\rangle$ such that every $|\psi\rangle \in \mathcal{H}$ can be written as

In this case, completeness is automatic and can be easily deduced from the rest of the axioms. For infinite-dimensional spaces, completeness is far from being trivial and is indeed a crucial requirement for many of even the most basic properties.

$$|\psi\rangle = \sum_{k=1}^{d} \psi_k\,|k\rangle\,. \tag{2.1}$$

The $d$-dimensional Hilbert space is denoted here by $\mathcal{H}_d$.

Evolution and observable quantities are related to the linear operators on $\mathcal{H}$. Measurement of a physical quantity is described by an Hermitian operator $A$, related to the possible definite outcomes. If an experiment may yield a quantity $a_i$, corresponding to the eigenstate $|a_i\rangle$ of $A$, then the probabilities of measuring $i$-th outcome are $|\langle a_i|\psi\rangle|^2$.

Note that for a Hermitian operator, the eigenstates corresponding to different eigenvalues are orthogonal.

As a result of this probabilistic interpretation, the expectation value of $A$ can be written using the orthogonal eigenbasis $\{|a_i\rangle\}_{i=1}^d$ of $A$:

$$\langle A\rangle_\psi = \sum_{i=1}^{d} a_i |\langle a_i|\psi\rangle|^2, \tag{2.2}$$

which, after rearrangement of the terms yields

Here, $\langle\psi|A|\psi\rangle$ means $\langle\psi|\,(A\,|\psi\rangle)$ – in finite-dimensional case, this reduces to the basic matrix multiplication when a basis is defined, so this notation is justified.

$$\langle A\rangle_\psi = \langle\psi|A|\psi\rangle\,. \tag{2.3}$$

The dynamics of a closed system is also described by linear operators. Every physical system has a corresponding observable (Hermitian operator) describing the energy – its *Hamiltonian* $H$. The Hamiltonian describes the unitary dynamics: if a system is found initially in the state $|\psi\rangle$, then after time $t$, it is in the state

The reduced Planck constant $\hbar$ is often omitted – it can be incorporated in the definition of Hamiltonian $H$ and time $t$, so that it does not appear explicitly in the equations.

$$|\psi(t)\rangle = \exp\left(-\frac{iHt}{\hbar}\right)|\psi\rangle\,, \tag{2.4}$$

$\exp(X) = \sum_{n=0}^{\infty} X^n/n!$

where exp denotes the operator exponent.

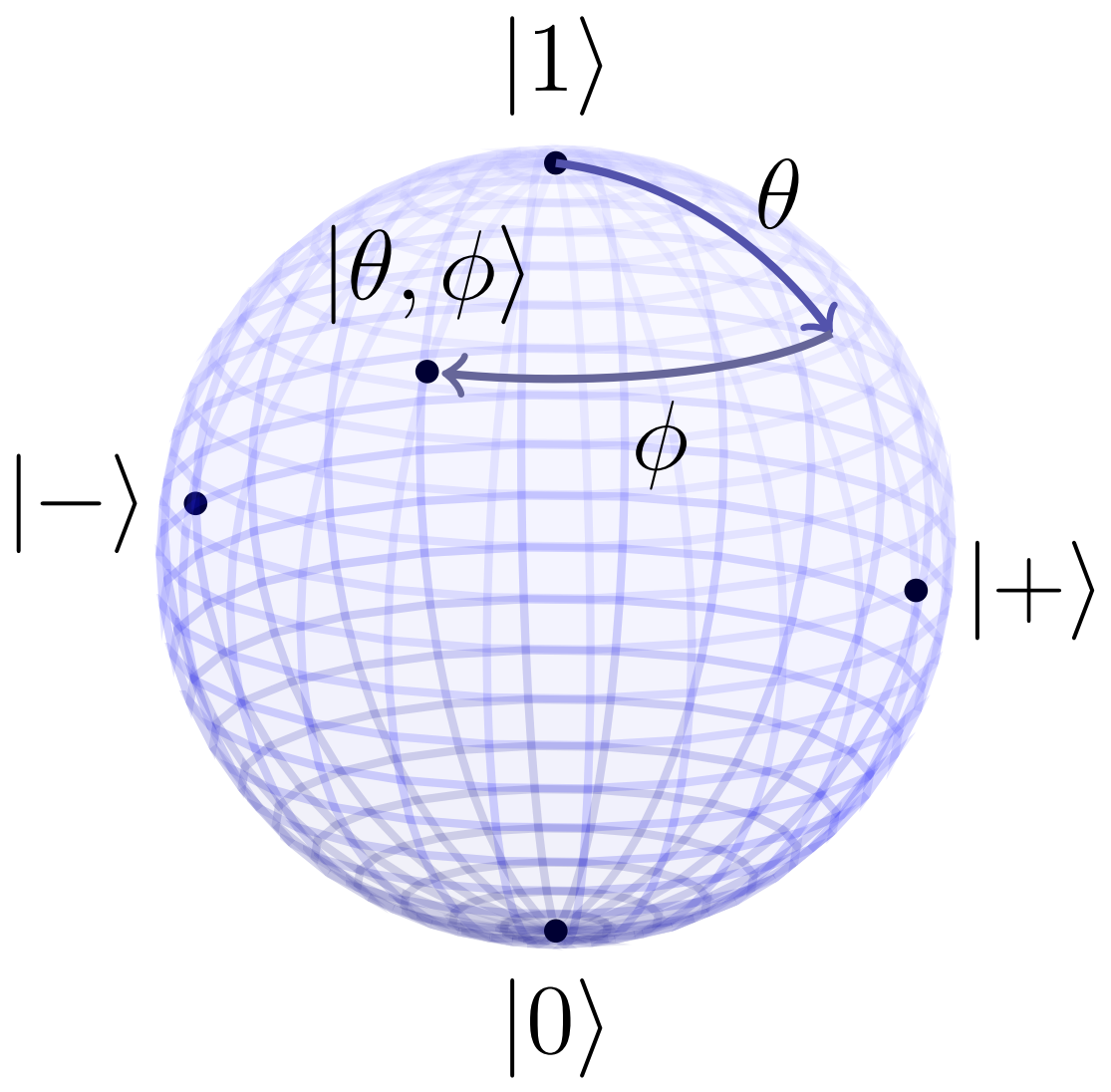

**Figure 2.1:** The Bloch sphere – set of pure states of a qubit, parameterized as to show connection with spherical geometry. The space is spanned by $|0\rangle$ and $|1\rangle$ – note that the orthogonal vectors are antipodal in this picture. Apart from this, the vectors $|\pm\rangle = (|0\rangle \pm |1\rangle)/\sqrt{2}$ and a state described by Eq. (2.6) are shown.

The simplest possible Hilbert space describing a *nontrivial*[1] quantum system is the two-dimensional one: physically, it is a model of a *qubit* system, one of the building blocks of quantum computing. The qubit Hilbert space $\mathcal{H}_2$ is spanned by two orthogonal vectors: $|0\rangle$ and $|1\rangle$. Therefore, one can write any pure state $|\psi\rangle \in \mathcal{H}_2$ as a superposition of these two:

1: *The simplest* is of course one-dimensional, but no dynamics and change is possible in such a scenario.

In line with notation used in classical computation, qubit basis vectors are denoted by $|0\rangle$ and $|1\rangle$ instead of $|1\rangle$ and $|2\rangle$.

$$|\psi\rangle = \psi_0 |0\rangle + \psi_1 |1\rangle . \tag{2.5}$$

The set of single qubit states has two real parameters: $\psi_0$ and $\psi_1$ can be complex, but the physical state is normalized and the overall phase cannot be observed. Traditionally, the parameters are the angles $\theta, \phi$ of trigonometric functions:

$$|\theta, \phi\rangle = \overbrace{e^{i\phi} \cos\left(\frac{\theta}{2}\right)}^{\psi_1} |1\rangle + \overbrace{\sin\left(\frac{\theta}{2}\right)}^{\psi_0} |0\rangle . \tag{2.6}$$

This form of parameterization suggests the geometric interpretation of the set of qubit states: the *Bloch sphere*.

### 2.1.1 Composite systems, ensembles, measurement, and density operators

The formalism of state vectors can be extended to accommodate for the possibility of incomplete information about the system. Such a scenario is not purely theoretical: it comes up naturally during the considerations on multipartite systems and measurement. Let me present this extension with the following example.

The randomness is purely classical, and thus deterministic – here the word 'random' pertains to the lack of knowledge, not instrinsically non-deterministic theory.

**Example 2.1** Consider a device capable of preparing states according to an ensemble $\{(|\psi_i\rangle, p_i)\}$ – a state $|\psi_i\rangle$ is generated upon a push of a button with probability $p_i$. The machine is a black box and its inner operation is not known. The output state can be measured with an arbitrary physical setup.

Several questions come to mind:

a) Can the ensemble $\{(|\psi_i\rangle, p_i)\}$ be reconstructed?
b) If not, what is the maximal amount of information one can gather about the output state of such a device?
c) How to describe the output state in a consistent way?

The answer to the first question is, unfortunately, negative. Consider a qubit system and two devices:

**Device A** emits the state $|0\rangle$ with probability $p_0 = \frac{1}{2}$ and the state $|1\rangle$ with probability $p_1 = \frac{1}{2}$.

**Device B** emits the state $|+\rangle = \frac{1}{\sqrt{2}}(|0\rangle + |1\rangle)$ with probability $p_+ = \frac{1}{2}$ and the state $|-\rangle = \frac{1}{\sqrt{2}}(|0\rangle - |1\rangle)$ with probability $p_- = \frac{1}{2}$.

The only macroscopically accessible quantities are the expectation values of observables – even the measurement probabilities are the expectation values of projection operators. Therefore, if the expectation values of arbitrary operators are equal for the two ensembles, they are indistinguishable.

Straightforward calculation proves the indistinguishability of the states prepared by the device A and B: any Hermitian operator can be written in the form of

$$X = x_0 |0\rangle \langle 0| + x_1 |1\rangle \langle 1| + x_{01} |0\rangle \langle 1| + x_{01}^* |1\rangle \langle 0| . \tag{2.7}$$

The expectation value of $X$ over the first ensemble is

$$\langle X \rangle_{(A)} = \frac{x_0 + x_1}{2}, \tag{2.8}$$

which is equal to the value calculated with respect to the second ensemble:

$$\langle X \rangle_{(B)} = \frac{x_0 + x_1}{2}, \tag{2.9}$$

owing to the fact that the trace of an operator is basis-independent in a finite-dimensional setting.

If one can not reconstruct the original ensemble, what can be inferred about the system? The form of the expression for the expectation value evaluated over an ensemble suggests what information is accessible:

$$\langle X \rangle_{\{(|\psi_i\rangle, p_I)\}} = \sum p_i \langle \psi_i | X | \psi_i \rangle , \tag{2.10}$$

which can be rewritten using basic algebraic operations:[2]

2: Using the cyclic property of trace: $\langle \psi | X | \psi \rangle = \operatorname{Tr} \left( X | \psi \rangle \langle \psi | \right)$.

$$\langle X \rangle_{\{(|\psi_i\rangle, p_I)\}} = \operatorname{Tr} \left[ X \overbrace{\sum p_i |\psi_i\rangle \langle \psi_i|}^{\rho} \right] . \tag{2.11}$$

The operator $\rho$ under the trace is called the *density operator* [9].

[9]: Holevo (2003), *Statistical Structure of Quantum Theory*

Density operators and their properties are one of the central concepts of this thesis. As the ensemble form suggests, the set of all possible density operators can be thought of as a convex hull of the set of projection operators.

**Definition 2.2** *The set of Hermitian square matrices of size d is denoted by* $\mathcal{A}_d$.

**Definition 2.3** *The set of density operators* $\mathcal{M}_d$ *acting on the d-dimensional system* $\mathcal{H}_d$ *is the subset of Hermitian matrices* $\mathcal{A}_d$ *of size d subject to the following constraints: for every* $\rho \in \mathcal{M}_d$,

*a) The trace of* $\rho$ *is equal to 1,* $\operatorname{Tr} \rho = 1$.
*b) The operator* $\rho$ *is positive semidefinite: for all* $|\psi\rangle \in \mathcal{H}_d$, $\langle \psi | \rho | \psi \rangle \geq 0$.

Condition *b)* is often written as $\rho \succcurlyeq 0$.

The idea of density operators also arises in the analysis of multipartite systems. If a description of a particular quantum system naturally splits into two (or more) parts – e.g., it describes two particles, or a particle with spatial and spin degrees of freedom – often the most insightful perspective is to split the Hilbert space in an appropriate way. If the set $\{|\alpha_i\rangle\}$ is a basis of the subsystem $A$ and the basis of $B$ is $\{|\beta_j\rangle\}$, then any pure state $|\psi\rangle$ of a bipartite system can be written using tensor products of the two bases elements:

$$|\psi\rangle = \sum_{i=1}^{n} \psi_{ij} |\alpha_i\rangle \otimes |\beta_i\rangle . \tag{2.13}$$

**Definition 2.4** *Expectation value of an observable described by Hermitian operator X over the state* $\rho$ *is*

$$\langle X \rangle_\rho = \operatorname{Tr} X\rho. \tag{2.12}$$

Evidently, expectation values of observables are linear functionals over the set of density operators.

Supposing only a subsystem $A$ can be accessed,[3] the only measurable observables have the form of $X \otimes \mathbb{1}_B$, where $X$ is Hermitian. Again, this results in the density operator description:

3: This can arise naturally: a single quantum state can span large distances, a measurement device may ignore some degrees of freedom.

$$\langle X \otimes \mathbb{1}_B \rangle_\psi = \operatorname{Tr} X\rho, \tag{2.14}$$

where $\rho$ is the *partial trace* of the total quantum state, defined below.

$$\rho = \operatorname{Tr}_B |\psi\rangle \langle \psi| . \tag{2.15}$$

Similar construction can be performed for the partial trace $\mathrm{Tr}_A$ over the subsystem $A$ and for multipartite systems.

**Definition 2.5** *Let the Hilbert space $\mathcal{H}$ admit a tensor structure: $\mathcal{H} = \mathcal{H}^{(A)} \otimes \mathcal{H}^{(B)}$, such that an operator $M$ acting on $\mathcal{H}$ can be written as a sum of tensor products:*

$$M = \sum_i M_i^{(A)} \otimes M_i^{(B)}. \tag{2.16}$$

*The partial trace of $M$ over the subsystem $B$ is the operator*

$$\mathrm{Tr}_B\, M = \sum_i \left(\mathrm{Tr}\, M_i^{(B)}\right) M_i^{(A)}. \tag{2.17}$$

The set $\mathcal{M}_d$ is an intersection of the cone of positive semidefinite matrices of size $d$ embedded in the space of Hermitian matrices with the hyperplane defined by the unit trace criterion. The description of the set geometry becomes increasingly complicated as the dimension $d$ grows. While $\mathcal{M}_d$ is always convex,[4] the criteria for checking whether a given matrix is positive semidefinite – essentially testing whether all of its eigenvalues are nonnegative – become nontrivial even for $d = 3$.

4: Convex combination of positive semidefinite matrices is always positive semidefinite.

5: Here, $\sigma_x, \sigma_y$ and $\sigma_z$ are the Pauli operators.

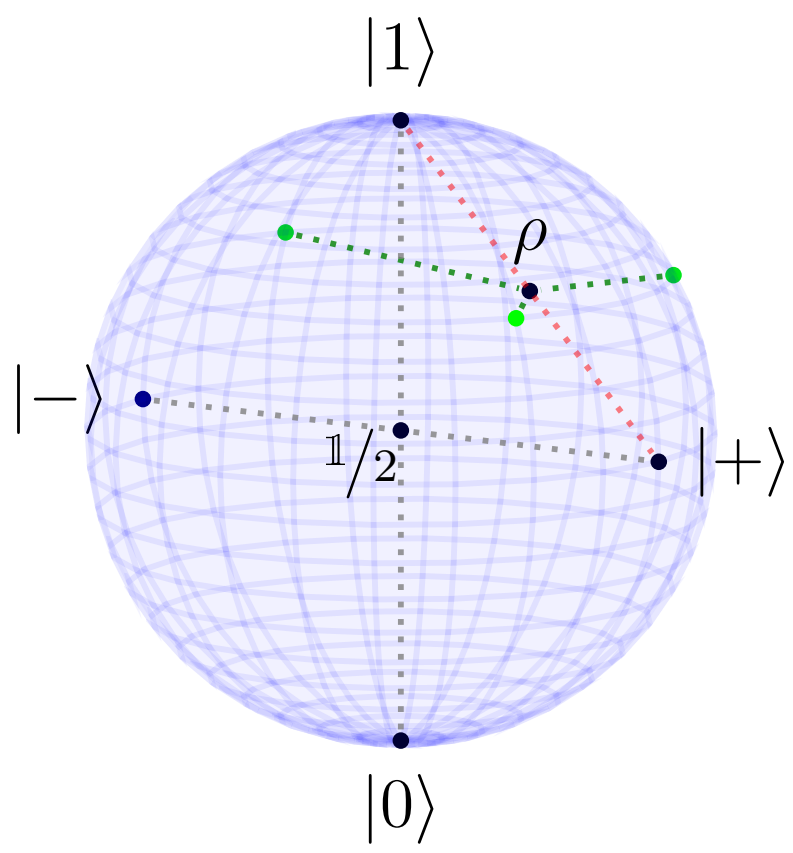

**Figure 2.2:** The set of qubit density operators $\mathcal{M}_2$ – the *Bloch ball*, reminiscent of the Bloch sphere. It is a convex hull of the 1-dimensional projectors – the pure states, which form the boundary of the set. The *maximally mixed state* $\mathbb{1}/2$ lies at the center of $\mathcal{M}_2$.
The decomposition of quantum states into convex sums of pure states is not unique. Owing to the linearity of quantum mechanics, any evolution must respect this ambiguity and can not rely on any preferred decomposition.

### Bloch ball

The $d = 2$ case, however, admits a simple description. Since every unit trace matrix of size 2 can be parameterized as[5]

$$\rho = \frac{\mathbb{1}}{2} + x\sigma_x + y\sigma_y + z\sigma_z, \tag{2.18}$$

and $2 \times 2$ matrices can be diagonalized easily, the positive semidefiniteness criterion becomes

$$x^2 + y^2 + z^2 \leq 1/4. \tag{2.19}$$

The set of qubit density matrices is thus in some sense a *ball* – called the *Bloch ball*. Its structure is compatible with the Bloch sphere description – the boundary is formed by pure states, with mixed states being inside (see Figure 2.2).

### Higher dimensions

The situation changes for higher dimensions: the set of density operators $\mathcal{M}_d$ does not form a ball[6] for $d \geq 3$. There are various ways to analyze its geometry, one of them being the application of the Hilbert-Schmidt geometry.

6: The boundary of $\mathcal{M}_d$ is formed by the positive semidefinite matrices $\rho$ for which $\det \rho = 0$. This polynomial equation does not admit as simple description as Eq. (2.19).

**Definition 2.6** *The Hilbert-Schmidt inner product of two matrices of the same dimensions is defined by*

$$(X, Y)_{\mathrm{HS}} = \operatorname{Tr} XY^{\dagger}. \tag{2.20}$$

*The Hilbert-Schmidt distance is induced by the inner product:*

$$D_{\mathrm{HS}}(X, Y) = \sqrt{(X - Y, X - Y)_{\mathrm{HS}}}. \tag{2.21}$$

The distance can be thought of as a square root of the sum of squares of the absolute values of the matrix elements of the difference, $D_{\mathrm{HS}} = \sqrt{\sum_{ij} |X_{ij} - Y_{ij}|^2}$ (often an additional $1/2$ prefactor is used). It is thus a direct analogue of the Euclidean distance between vectors.

The Hilbert-Schmidt geometry of $\mathcal{M}_d$ is directly related to its parameterization structure. If a complete orthonormal (in the sense of the Hilbert-Schmidt inner product) basis of traceless Hermitian matrices of size $d$ is chosen and denoted by $\{E_i\}_{i=1}^{d^2-1}$, then the distance between the density operators parameterized by

$$\rho(\vec{x}) = \frac{\mathbb{1}}{d} + \sum_{i=1}^{d^2-1} x_i E_i, \tag{2.22}$$

is equal to the Euclidean distance between parameterization vectors:

$$D_{\mathrm{HS}}(\rho(\vec{x}), \rho(\vec{y})) = |\vec{x} - \vec{y}|. \tag{2.23}$$

In this view, the boundary of the set of states defined by the condition $\det \rho(\vec{x})$ becomes a polynomial of variables $\{x_i\}_{i=1}^{d^2-1}$. While the structure of the boundary $\partial\mathcal{M}_d$ is quite elaborate, some of the general properties can be analyzed.

One of the basic questions one can pose is: what is the insphere and outsphere[7] of $\mathcal{M}_d$? Because of the high symmetry of the set of quantum states,[8] the spheres are concentric with the center being the maximally mixed state, $\mathbb{1}/d$. The inradius $r$ and outradius $R$ can be calculated readily: the outradius is the distance between the maximally mixed state and any pure state:

$$R = D_{\mathrm{HS}}\left(\frac{\mathbb{1}}{d}, |\psi\rangle\langle\psi|\right) = \sqrt{\frac{d-1}{d}}, \tag{2.24}$$

while the inscribed sphere of largest radius is tangent to states with $d-1$ equal eigenvalues and one zero eigenvalue:

$$r = D_{\mathrm{HS}}\left(\frac{\mathbb{1}}{d}, \frac{\mathbb{1} - |\psi\rangle\langle\psi|}{d-1}\right) = \sqrt{\frac{1}{d(d-1)}} = \frac{R}{d-1}. \tag{2.25}$$

This can be immediately deduced from the analysis of a section of $\mathcal{M}_d$ with the set of diagonal matrices. Simple analysis reveals that this results in a set equivalent to the probability simplex[9] with the insphere and outsphere of $\mathcal{M}_d$ corresponding to the insphere and outsphere of the simplex (see Figure 2.3).

7: The *insphere* is the largest inscribed sphere and *outsphere* is the smallest circumscribed sphere around $\mathcal{M}_d$.

8: Indeed, the set is unitarily invariant: for every unitary $U$, the relation $U\mathcal{M}_d U^{\dagger} = \mathcal{M}_d$ holds.

9: The diagonal entries are nonnegative and sum up to unity – the same criteria as for the probability simplex.

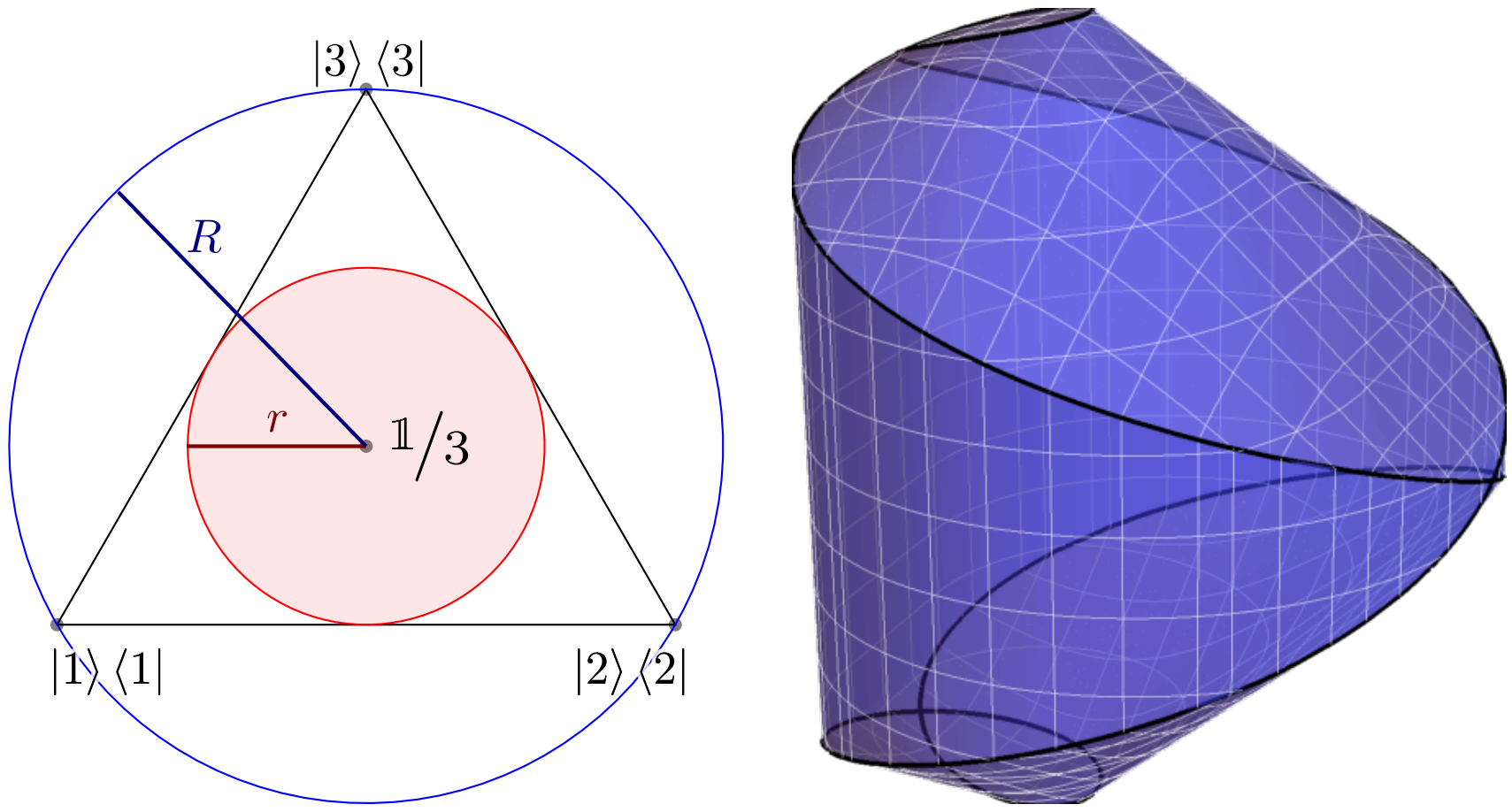

**Figure 2.3:** *Left*: Intersection of the set of quantum states $\mathcal{M}_3$ with the set of diagonal matrices reveals a probability simplex. The insphere and outsphere of $\mathcal{M}_3$ can be deduced from this section.
*Right*: A *model* of the set of states to aid in understanding the higher-dimensional geometry with flat parts and curved surfaces in the boundary.

The set of states contains infinitely many probability simplices embedded inside. Indeed, every orthonormal basis of $\mathcal{H}_d$, $\{|i\rangle\}_{i=1}^d$ defines one: the set $\operatorname{conv}\{|i\rangle\langle i|\}_{i=1}^d$ is always contained in $\mathcal{M}_d$. Furthermore, $\mathcal{M}_d$ *is formed* by the union of all such sets:[10]

$$\mathcal{M}_d = \bigcup_{\text{orth. bases } \{|i\rangle\}} \operatorname{conv}\{|i\rangle\langle i|\}. \tag{2.26}$$

10: Or, alternatively, by the union of all unitary rotations of one simplex: $\mathcal{M}_d = \bigcup_{U\in U(d)} U \operatorname{conv}\{|i\rangle\langle i|\}_{i=1}^d U^\dagger$.

Is $\mathcal{M}_d$ therefore a polyhedron? No – even the qubit case reveals a rounded geometry, which persists in higher dimensions. The boundary of the set of states always contains a family of Bloch balls spanned by every pair of orthogonal pure states $|\alpha\rangle$ and $|\beta\rangle$:[11]

$$\forall_{|\alpha\rangle\perp|\beta\rangle} \left\{ \begin{pmatrix} |\alpha\rangle \\ |\beta\rangle \end{pmatrix}^T \left( \frac{\mathbb{1}}{2} + \vec{r}\cdot\vec{\sigma} \right) \begin{pmatrix} \langle\alpha| \\ \langle\beta| \end{pmatrix} : |\vec{r}| \leq \frac{1}{4} \right\} \subset \partial\mathcal{M}_d. \tag{2.27}$$

11: And, by extension, the boundary $\partial\mathcal{M}_d$ contains appropriately transformed copies of all $\mathcal{M}_k$ for $k < d$.

The notation here means an operator parameterized by $\vec{r} \in \mathbb{R}^3$ with unit trace and eigenvectors lying in the subspace spanned by $|\alpha\rangle$ and $|\beta\rangle$.

A particularly interesting model of how one could imagine the set of states to look like is presented in [10]: a nontrivial union of rotations of the probability simplex (see Figure 2.3). Recently, another three-dimensional model of a qutrit state space appeared in [11], preserving many of the properties of $\mathcal{M}_3$.

[10]: Bengtsson et al. (2013), ‘Geometry of the set of mixed quantum states: An apophatic approach’

[11]: Eltschka et al. (2021), ‘The shape of higher-dimensional state space: Bloch-ball analog for a qutrit’

The analysis based on the flat geometry induced by the Hilbert-Schmidt distance is not the only way to look at the set of quantum states. One of the alternatives is the Bures-Uhlmann geometry [12–14]: every state $\rho \in \mathcal{M}_d$ can be written as a partial trace of a bipartite pure state (called the *purification*),

$$\rho = \sum_{i=1}^d p_i |i\rangle\langle i| = \operatorname{Tr}_B |\psi\rangle\langle\psi|, \tag{2.28}$$

where

$$|\psi\rangle = \sum_{i=1}^d \sqrt{p_i} e^{i\alpha_i} |i\rangle \otimes |i\rangle. \tag{2.29}$$

[12]: Bures (1969), ‘An extension of Kakutani’s theorem on infinite product measures to the tensor product of semifinite $w^*$-algebras’
[13]: Uhlmann (1976), ‘The “transition probability” in the state space of $a^*$-algebra’
[14]: Uhlmann (1993), ‘Density operators as an arena for differential geometry’

One can use the geometry induced by this purification procedure. Every pure bipartite state can be written in the form of

$$|\phi\rangle = \sum_{i,j} \phi_{ij} |i\rangle \otimes |j\rangle , \tag{2.30}$$

and is therefore defined by a matrix of coefficients $\phi_{ij}$. This can be directly reinterpreted as a matrix

$$B_\phi = \sum_{i,j} \phi_{ij} |i\rangle \langle j| , \tag{2.31}$$

with the following relation automatically fulfilled:

$$\mathrm{Tr}_B |\phi\rangle \langle\phi| = B_\phi B_\phi^\dagger . \tag{2.32}$$

Immediately, if $B_\rho$ is a purification of $\rho$ in this manner, so is $BU$ for any unitary $U$. The Bures-Uhlmann metric is precisely the distance between purifications, optimized over all possible unitaries:

$$D^2_{\mathrm{BU}}(\rho, \sigma) = \min_{V,V' \in \mathrm{U(d)}} D^2_{\mathrm{HS}}(B_\rho V, B_\sigma V'). \tag{2.33}$$

This procedure is equivalent to minimizing the fidelity between the purifications.

**Definition 2.7** *The Bures-Uhlmann distance between the density operators $\rho$ and $\sigma$ is*

$$D^2_{\mathrm{BU}}(\rho, \sigma) = \min_{|\phi\rangle, |\psi\rangle} 2\left(1 - |\langle\psi|\phi\rangle|\right) , \tag{2.34}$$

*where the minimization is constrained to bipartite states such that* $\mathrm{Tr}_B |\phi\rangle \langle\phi| = \rho$ *and* $\mathrm{Tr}_B |\psi\rangle \langle\psi| = \sigma$.

The optimization can be done directly with the following result:[12]

$$D^2_{\mathrm{BU}}(\rho, \sigma) = 2\left(1 - \mathrm{Tr}\sqrt{\sqrt{\rho}\,\sigma\sqrt{\rho}}\right) . \tag{2.35}$$

12: Here, $\sqrt{\cdot}$ denotes the matrix square root, properly defined for positive semidefinite matrices.

Unlike the Hilbert-Schmidt metric, the Bures-Uhlmann distance is not flat.

### 2.1.2 Berry phase

Yet another example of nontrivial geometry of quantum states is the existence of geometric phase. This phenomenon naturally appears in the analysis of adiabatic change of system Hamiltonians as an additional prefactor to the overall phase. Here, I present the standard derivation of the basic elements of this theory [15].

[15]: Carollo et al. (2020), 'Geometry of quantum phase transitions'

**Theorem 2.1** *Consider a family of Hamiltonians parameterized by time and the corresponding energy eigenstates:*

$$H(t)\,|n_t\rangle = E_n(t)\,|n_t\rangle\,. \tag{2.36}$$

*If the Hamiltonian $H(t)$ is changing slowly enough, then the initial eigenstate $|\psi(t=0)\rangle = |n_0\rangle$ evolves under the Schrödinger evolution*

$$i\hbar\frac{\mathrm{d}\,|\psi(t)\rangle}{\mathrm{d}\,t} = H(t)\,|\psi(t)\rangle \tag{2.37}$$

*to form a target energy eigenstate multiplied by a phase factor:*

$$|\psi(t)\rangle = \exp\left(-\frac{i}{\hbar}\int_0^t E_n(s)\,\mathrm{d}\,s\right)\times \exp\left(-\int_0^t \langle n_s|\,[{}^{\mathrm{d}}/_{\mathrm{d}\,s}\,|n_s\rangle]\,\mathrm{d}\,s\right)|n_t\rangle \tag{2.38}$$

Here, ${}^{\mathrm{d}}/_{\mathrm{d}\,s}\,|n_s\rangle$ is the derivative of the eigenstate with respect to the parameter appearing in Eq. (2.36), *not* the evolution governed by the Schrödinger equation.

The second phase factor can be written in the form of $e^{i\gamma_n}$, where

$$\gamma_n = i\int_0^t \langle n_s|\left(\frac{\mathrm{d}}{\mathrm{d}\,s}\,|n_s\rangle\right)\mathrm{d}\,s. \tag{2.39}$$

If the Hamiltonian $H(t)$ can be expressed in a parameterized form of $H(\vec{r}(t))$, where $\vec{r} \in \mathbb{R}^k$ is the parameter vector, then the derivatives appearing in this equation can be expanded according to the product rule. Subsequently, the whole expression can be rewritten as a path-dependent integral, instead of a time-dependent one:

$$\gamma_n = i\int_{\vec{a}}^{\vec{b}} \sum_{j=1}^{k} \langle n_{\vec{r}}|\left(\frac{\partial}{\partial r_j}\,|n_{\vec{r}}\rangle\right)\mathrm{d}\,r_j. \tag{2.40}$$

The integrand is called the *Berry connection* $\vec{A}_n(\vec{r})$:

**Definition 2.8** *The Berry connection of a parameterized family of Hamiltonians $H(\vec{r})$ for $\vec{r} \in \mathbb{R}^k$ and $n$-th eigenstate is the expression*

$$\vec{A}_n(\vec{r}) = i\,\langle n_{\vec{r}}|\nabla_{\vec{r}} n_{\vec{r}}\rangle\,. \tag{2.41}$$

*The notation with $\nabla_{\vec{r}}$ is to be interpreted the following way: the $j$-th coordinate of $\vec{A}_n$ reads*

$$\left(\vec{A}_n(\vec{r})\right)_j = \langle n_{\vec{r}}|\left(\frac{\partial}{\partial r_j}\,|n_{\vec{r}}\rangle\right). \tag{2.42}$$

An interesting corollary of this reasoning is the fact that the Berry phase of 1D quantum mechanical systems, $H = {}^{p^2}/_{2m} + V(x)$, is *always* zero – the eigenstates always can be chosen to be purely real, so their derivatives are real, and since $\langle n(\vec{r})|\nabla_{\vec{r}} n(\vec{r})\rangle$ is always imaginary, it must vanish. This generalizes to every system with time reversal symmetry.

Berry connection explicitly depends on the chosen gauge (the phases of the eigenstates can be chosen arbitrarily for different $\vec{r}$) – it is therefore not a directly observable quantity. The only case when the integral does not depend on the gauge is the cyclic integral over the closed path Γ:

$$\oint_\Gamma \overbrace{i \langle n_{\vec{r}} | \nabla_{\vec{r}} n_{\vec{r}} \rangle}^{\vec{A}_n(\vec{r})} \cdot \mathrm{d}\,\vec{r}. \tag{2.43}$$

This value – the phase accumulated by an eigenstate during cyclic and adiabatic evolution – is called the *geometric phase* or *Berry phase* [16]. It is tempting to use the generalized Stokes theorem and rewrite this expression using a surface integral. Indeed, this defines a *Berry curvature*:

[16]: Berry (1984), 'Quantal phase factors accompanying adiabatic changes'

**Definition 2.9** *Berry curvature* $\Omega_n(\vec{r})$ *is an antisymmetric tensor of rank two fulfilling*

$$\oint_\Gamma \overbrace{i \langle n_{\vec{r}} | \nabla_{\vec{r}} n_{\vec{r}} \rangle}^{\vec{A}_n(\vec{r})} \cdot \mathrm{d}\,\vec{r} = \int_\Sigma \Omega_n(\vec{r}) \cdot \mathrm{d}\,\sigma, \tag{2.44}$$

*where* $\partial\Sigma = \Gamma$ *and the integration on the right side is carried over the surface* $\Sigma$.

The effect of the Berry phase is visible even in the simplest systems. Below I present the standard example of the qubit interacting with an external magnetic field. If the magnetic field at time $t = 0$ has the same value as at $t = T$, the phase change can be measured – it turns out to be proportional to the spherical area enclosed by the magnetic field vector.

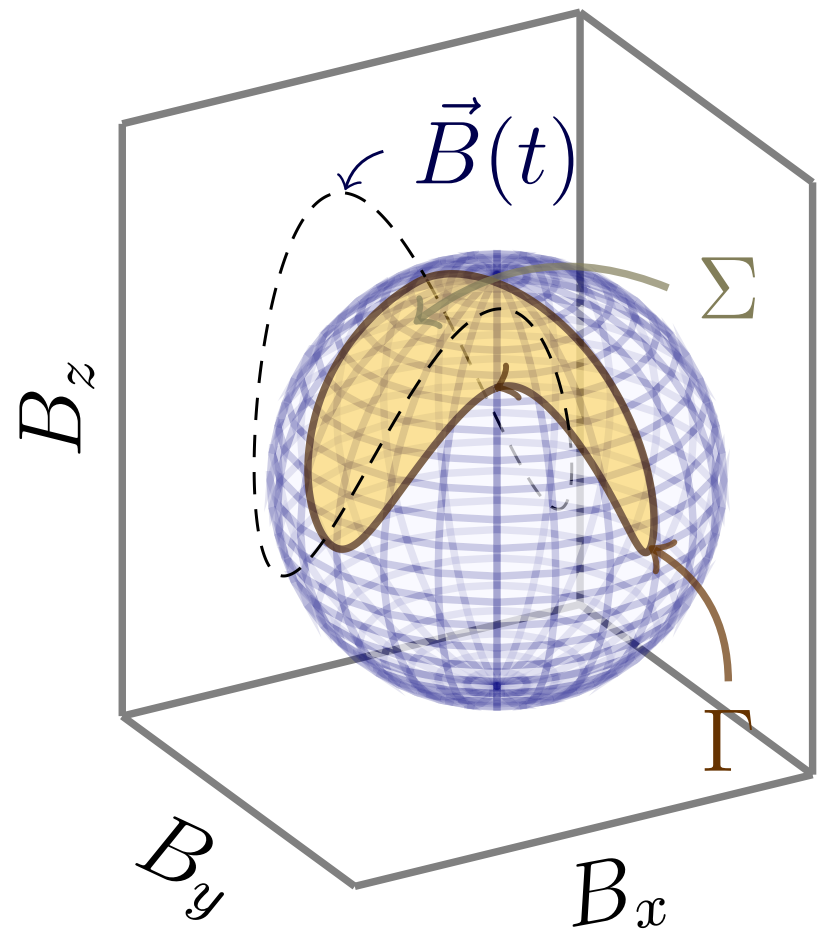

**Figure 2.4:** Cyclic and adiabatic evolution of a spin-1/2 Hamiltonian (Eq. (2.45)) parameterized by the magnetic field $\vec{B}$ induces a Berry phase on the eigenstates. The geometric phase is proportional to the area Σ of the curve Γ traced on the unit sphere by the projection $\vec{B}(t)/|\vec{B}(t)|$ of the magnetic field $\vec{B}(t)$ during its evolution.

**Example 2.2** Consider a particle with total spin $s = 1/2$ interacting with the magnetic field $\vec{B}$. The Hamiltonian is

$$H_{\vec{B}} = \vec{\sigma} \cdot \vec{B}, \tag{2.45}$$

where the elements of $\vec{\sigma}$ are Pauli matrices and the dot product is to be understood as the sum $\sum_{i=1}^{3} \sigma_i B_i$. For any $\vec{B}$ there are two eigenstates $|\pm_{\vec{B}}\rangle$ of Hamiltonian $H_{\vec{B}}$ defined by

$$H_{\vec{B}} \, |\pm_{\vec{B}}\rangle = \pm|\vec{B}| \, |\pm_{\vec{B}}\rangle \,. \tag{2.46}$$

The eigenstates can be multiplied by arbitrary phases, potentially depending on $\vec{B}$. One possible choice is the following:

$$|+_{\vec{B}}\rangle = \begin{pmatrix} \cos\left(\frac{\theta}{2}\right) e^{-i\phi} \\ \sin\left(\frac{\theta}{2}\right) \end{pmatrix}, \quad |-_{\vec{B}}\rangle = \begin{pmatrix} \sin\left(\frac{\theta}{2}\right) e^{-i\phi} \\ -\cos\left(\frac{\theta}{2}\right) \end{pmatrix}, \tag{2.47}$$

where the angles come from the spherical parameterization of the magnetic field,

$$\vec{B} = |B| \begin{pmatrix} \sin\theta\cos\phi \\ \sin\theta\sin\phi \\ \cos\theta \end{pmatrix}. \tag{2.48}$$

With such choice of the state phase, the Berry connection of the states depends only on the angles $\theta$ and $\phi$:

$$\vec{A}_{\pm}(\vec{B}) = i \langle \pm_{\vec{B}} | \nabla_{\vec{B}} (\pm_{\vec{B}}) \rangle = \frac{1 \pm \cos\theta}{2} \hat{\phi}. \tag{2.49}$$

Here, $\vec{A}_{\pm}$ is described in the spherical coordinates and $\hat{\phi}$ denotes the azimuthal unit vector of this coordinate system.

If the magnetic field $\vec{B}$ evolves according to the trajectory $\vec{B}(t)$ such that $\vec{B}(0) = \vec{B}(T)$, the states accumulate phase according to the Theorem 2.1. Therefore,

The first phase factor depends on the average energy during the evolution and is unrelated to the geometric phase effects. I omitted $\hbar$ for clarity.

$$|\pm(T)\rangle = \exp\left(\mp i \int_0^T |\vec{B}(t)| \,\mathrm{d}\, t\right) \exp\left(\pm \frac{i}{2} \Sigma[\vec{B}]\right) |\pm(0)\rangle , \tag{2.50}$$

where $\Sigma[\vec{B}]$ is the spherical area enclosed by the vector $\vec{B}/|\vec{B}|$ during its evolution. Note that a different choice of phases of $|\pm_{\vec{B}}\rangle$ leads to a different connection, but the resulting *phase change* is always the same.

### 2.1.3 Measurements, inference and quantum de Finetti theorem

The fundamental method of gaining knowledge about a physical system, not only in the context of quantum mechanics, is through measurement. Measurement, whether direct or through a complicated inference scheme, allows for probing specific properties of a quantum system, potentially even reconstructing the quantum state in full, but also unavoidably disturbs it – mechanism of this disturbance is still a matter of heated debate.[13]

13: Measurement results are unambiguously described by the expectation values – but *how does the measurement occurs* and what is the quantum state afterwards is a much more involved subject.

For some, quantum mechanics is entirely an operational theory, without any *ontic* meaning – it is *just a model* of reality, just as a calculation done by an engineer is, and questions about the mechanism of measurement are pointless, unless the answer will give rise to a more accurate description. Others treat state vectors and their unitary dynamics as a basic element, on which the universe *truly* operates. There is also room for plenty of other interpretations in between: with objective intermittent collapse or with classical particles interacting in a specific way to reconstruct quantum mechanical predictions.

Here, I am mostly interested in the mathematical description of measurement effects, which does not depend any particular interpretation. Indeed, most of these, unless extended in some way, yield the same predictions, and are therefore experimentally undistinguishable.[14]

14: This is not to criticize proponents of different interpretations of quantum mechanics – many of those are intellectually absorbing in their own right, regardless whether they are 'true' or not.

### Basic effects of measurement

The most general way to describe all kinds of interactions upon which a *classical measurement result* is observed, is the usage of Kraus operators and Positive Operator-Valued Measures (POVM) [9, 17]. The slightly cryptic name denotes a special set of operators:

[9]: Holevo (2003), *Statistical Structure of Quantum Theory*
[17]: Davies (1976), *Quantum Theory of Open Systems*

**Definition 2.10** *POVM is a set $\{E_i\}$ of positive semidefinite operators on $\mathcal{H}$ such that $\sum_i E_i = \mathbb{1}$. The effect of measurement of state $\rho$ is a result – index of an operator $i$ and the change of a state. Each result comes with probability*

$$P_i = \operatorname{Tr} \rho E_i, \tag{2.51}$$

*and the state change is described by a Kraus operator $K_i$ such that $E_i = K_i^\dagger K_i$. Such a decomposition is not unique and the exact form is determined by the details of a physical system. Upon measuring the $i$-th outcome, the state reads*

$$\rho' = \frac{K_i \rho K_i^\dagger}{P_i}. \tag{2.52}$$

Using multiple measurements on copies of the original state $\rho$, one may infer its form through a process called *quantum tomography*. The basic idea is straightforward: since the expectation value is a linear function over the set of quantum states $\mathcal{M}_d$, the knowledge of the value $\langle X \rangle$ for one observable $X$ ensures that a quantum state $\rho$ lies on a known hyperplane intersecting the set of density operators. If the expectation values of sufficiently many observables are known, the intersection of hyperplanes is a single point – the sought quantum state. This procedure can be rephrased in the language of POVMs as well. A particularly elegant approach to the problem of quantum tomography is the application of symmetric measurements – the *symmetric, informationally complete positive operator-valued measures* (SIC-POVM) [18] . Such measurements are interesting from a mathematical perspective, since they can are associated with various problems superficially unrelated to quantum mechanics. It is conjectured that SIC-POVMs exist for arbitrary dimensions.

**Definition 2.11** *A SIC-POVM on $\mathcal{M}_d$ is the set of $d^2$ positive semidefinite operators $\{E_i\}_{i=1}^{d^2}$ such that*

$$\operatorname{Tr} E_i E_j = \begin{cases} 1 & i = j, \\ 1/d+1 & i \neq j. \end{cases} \tag{2.53}$$

[18]: Appleby et al. (2021), 'SIC-POVMs from Stark Units'

There is an additional subtlety to the problem of quantum tomography, mirroring the Bayesian/frequentist divide of classical probability theory.

Just as probability can be interpreted as a state of knowledge from the Bayesian perspective or as an objective property of a system defined as the relative frequency of an event as the number of experiments is taken to infinity, there are two possible views on what *is* a quantum state $\rho$ of a repeated experiment [19]:

[19]: Caves et al. (2002), 'Unknown quantum states: the quantum de Finetti representation'

a) $\rho$ is merely a description of one's knowledge about reality,
b) $\rho$ is a description of the objective state of reality.

Just as in the classical case, the two views on $\rho$ eventually coincide as the quantum tomography procedure is repeated infinitely many times. A problem arises if the knowledge of $\rho$ is imperfect, reminiscent of the question: *does 'probability of probabilities' make sense?* If a device repeatedly prepares a quantum state for measurement, then in the *objective* view the state of $N$ samples is the tensor product of $N$ identical, yet unknown, quantum states $\rho$:

$$\sigma^{(N)} = \rho^{\otimes N}, \tag{2.54}$$

and a proper experiment allows for the determination of $\rho$.[15] The Bayesian view is different: *a priori* state of $N$ samples $\sigma^{(N)}$ could be arbitrary – e.g., correlated or entangled with the $N+1$ sample. Under certain assumptions this problem does not appear [19]:

15: In a way, the process of taking $N$ samples might be interpreted as a single experiment with the aim to infer the total state.

**Theorem 2.2** (Quantum de Finetti theorem) *Consider a sequence of quantum states $\left(\sigma^{(N)}\right)$ such that $\sigma^{(N)} \in \mathcal{M}_{d^N}$. The sequence is called exchangeable if the following two conditions are met.*

*a) The state is symmetric under permutations of the subsystems: for any permutation $\pi$ acting on basis vectors through the swap operator* $\mathrm{SWAP}_\pi$,

$$\mathrm{SWAP}_\pi \left| i_1 \otimes \ldots \otimes i_N \right\rangle = \left| i_{\pi(1)} \otimes \ldots \otimes i_{\pi(N)} \right\rangle, \tag{2.55}$$

*the state $\sigma^{(N)}$ is invariant under permutation for any $N$ and $\pi$:*

$$\sigma^{(N)} = \mathrm{SWAP}_\pi \, \sigma^{(N)} \, \mathrm{SWAP}_\pi^\dagger . \tag{2.56}$$

*b) The state $\sigma^{(N)}$ is extendible: it is a marginal of $\sigma^{(N+M)}$ for any $M$:*

$$\sigma^{(N)} = \mathrm{Tr}_M \, \sigma^{(N+M)}. \tag{2.57}$$

*If a sequence $\left(\sigma^{(N)}\right)$ is exchangeable, then there exists a probability measure $P$ on $\mathcal{M}_d$ such that for every $N$*

$$\sigma^{(N)} = \int_{\mathcal{M}_d} P(\rho) \rho^{\otimes N} \, \mathrm{d}\, \rho. \tag{2.58}$$

*Furthermore, the probability measure $P(\rho)$ is uniquely determined by $\left(\sigma^{(N)}\right)$.*

As a consequence, it is valid in the Bayesian view to consider a sequence of experiments as a probabilistic mixture of tensor copies of states. Experiment outcomes simply update the Bayesian probability $P(\rho)$.

The quantum de Finetti theorem is a nontrivial consequence of the structure of the set of states $\mathscr{M}$ and its marginals. If any of the assumptions are dropped, the theorem does not hold. For instance, the state $|\mathrm{GHZ}\rangle = \frac{1}{\sqrt{2}}\left(|0\rangle^{\otimes 3} + |1\rangle^{\otimes 3}\right)$ is symmetric, but not extendible, so it can not describe a Bayesian state of three consecutive samples of a qubit experiment. Quantum mechanics restricted to real symmetric density matrices does not allow for the analogue of the quantum de Finetti theorem either: the sequence

$$\sigma^{(N)} = \frac{1}{2}\left(|+i\rangle\langle+i|^{\otimes N} + |-i\rangle\langle-i|^{\otimes N}\right), \quad |\pm i\rangle = \frac{|0\rangle \pm i\,|1\rangle}{\sqrt{2}}, \tag{2.59}$$

is exchangeable and real symmetric[16] for any $N$, but does not admit any decomposition as a mixture of tensor products of real symmetric states – it is already written as the (unique) mixture of states involving complex numbers. In the standard (complex) quantum mechanics, it simply corresponds to an experimental setup with the promise to output *always* $|+i\rangle$ or *always* $|-i\rangle$.

16: In the basis formed by tensor products of $|0\rangle$ and $|1\rangle$.

## 2.2 Quantum channels, dual states, and complete positivity

**Definition 2.12** *A quantum channel is an affine linear map* $\mathscr{E} : \mathscr{A}_d \longrightarrow \mathscr{A}_{d'}$ *which:*

a) *Is positive and trace-preserving: for every* $\rho \in \mathscr{M}_d$*, the resulting operator*
$$\rho' = \mathscr{E}(\rho) \tag{2.60}$$
*has an unit trace and is positive semidefinite. Therefore,* $\mathscr{E}$ *maps quantum states to quantum states:* $\mathscr{E}(\mathscr{M}_d) \subset \mathscr{M}_{d'}$.

b) *Every extension with identity of arbitrary dimension has the same property:* $\mathscr{E} \otimes \mathrm{id}$ *maps quantum states to quantum states (*$\mathscr{E}$ *is completely positive).*

Here, $\mathscr{A}_d$ denotes the set of Hermitian matrices of size $d$ and id is the identity map: $\mathrm{id}(\rho) = \rho$.

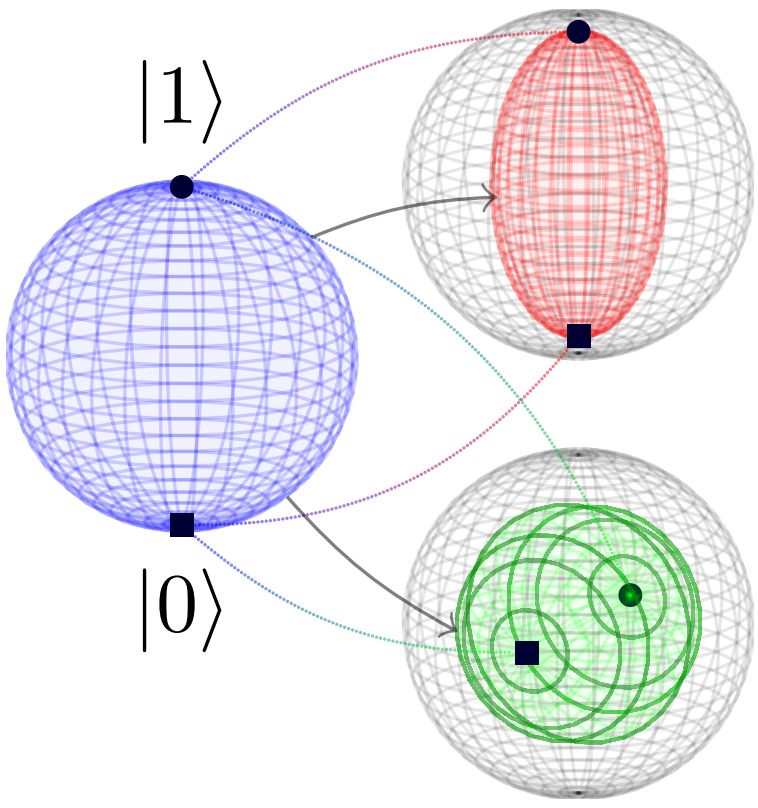

**Figure 2.5:** *Quantum channel* maps quantum states into quantum states. Depicted are several possible quantum channels from qubit to qubit; in this case complete positivity can be deduced from the geometric properties of the resulting ellipsoid.

This definition exactly captures the essence of physically realizable operations – the part involving extensions is truly crucial. For instance, the transpose map $\rho \mapsto \rho^T$, clearly preserving eigenvalues, maps quantum states to quantum states, but its extension does not. It is an indication of the fact that the transpose operation $\rho \mapsto \rho^T$ is not realizable in the physical world.

An insightful view on quantum channels can be constructed by means of the Choi-Jamiołkowski isomorphism, presenting parallels between quantum states and quantum channels [20, 21].

[20]: Choi (1975), 'Completely positive linear maps on complex matrices'
[21]: Jamiołkowski (1972), 'Linear transformations which preserve trace and positive semidefiniteness of operators'

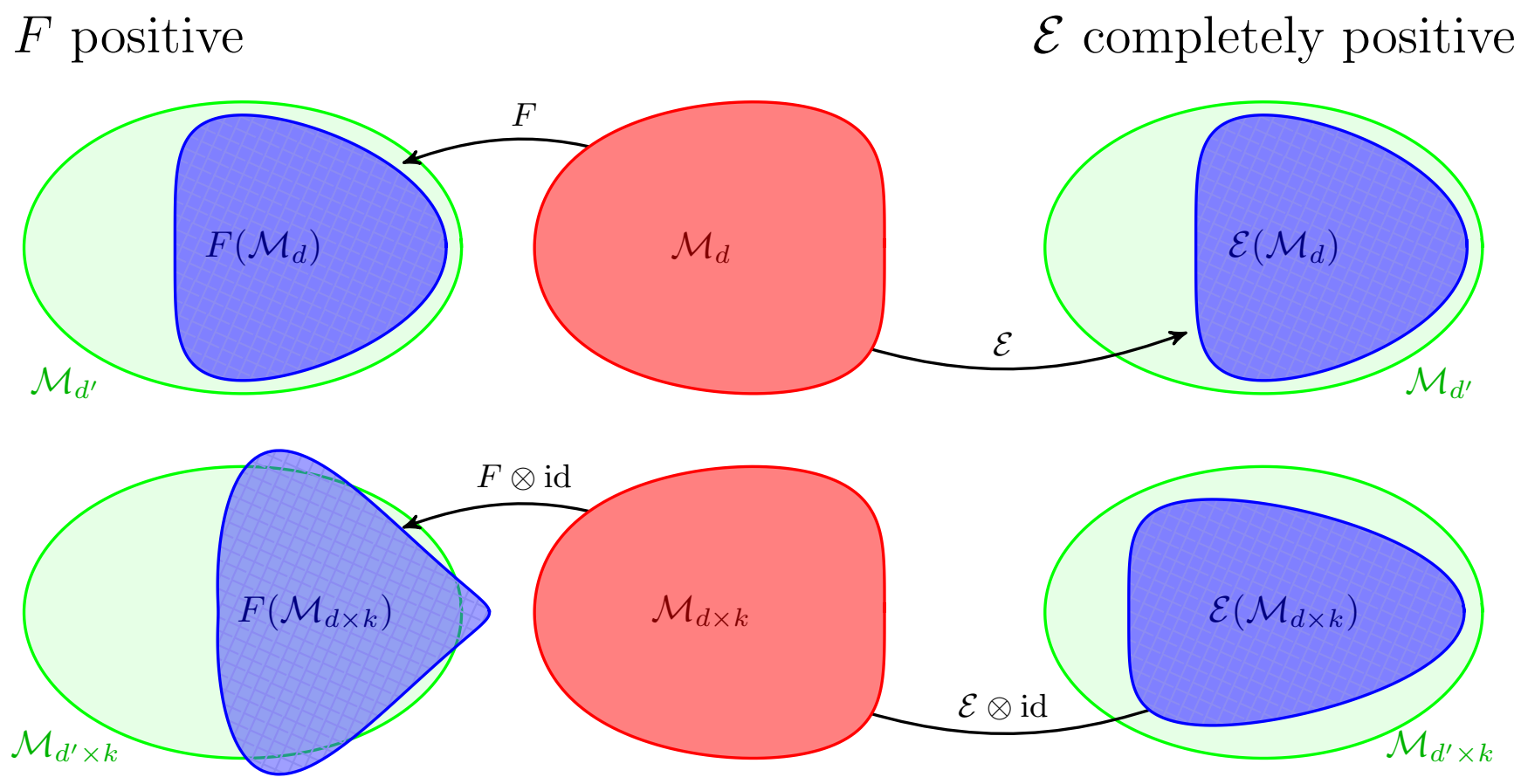

**Figure 2.6:** *Quantum channel* $\mathscr{E}$ is a positive trace-preserving map which stays positive after extension by $k$-dimensional identity id – the image of set of states $\mathscr{M}_{d\times k}$ under $\mathscr{E} \otimes \mathrm{id}$ is a subset of the set of states $\mathscr{M}_{d'\times k}$. Positive, but not completely positive map $F$, upon extension by identity transform some states of composite systems into operators which do not correspond to any legitimate state.

To explain the meaning of this isomorphism, let us start with an initial state $\sigma$ of a tripartite system, parameterized by the density operator $\rho \in \mathscr{M}_d$ of system A:

$$\sigma = \rho \otimes |\omega\rangle\langle\omega| \,, \tag{2.61}$$

where $|\omega\rangle$ is the following maximally entangled bipartite state of the subsystem BC,

Here, $|a, b\rangle$ is used in place of $|a\rangle \otimes |b\rangle$ for readability purposes.

$$|\omega\rangle = \frac{1}{\sqrt{d}} \sum_{i=1}^{d} |i, i\rangle \,. \tag{2.62}$$

By projecting[17] the state of the first two subsystems onto $|\omega\rangle$ and tracing these subsystems out, the state $\rho$ is obtained in the last subsystem – this is a fairly standard example of quantum teleportation [22]:

17: Here, *projection* means multiplication by $\Pi_\omega = d^2 |\omega\rangle\langle\omega| \otimes \mathbb{1}_C$ – the scaling is important. Multiplication by this operator followed by the partial trace corresponds in the physical world to the postselection of the results of an experiment involving measuring $|\omega\rangle\langle\omega|$ in the subsystem $AB$.

$$\overbrace{\sum_k \langle k, k|}^{AB} \Big( \underbrace{\rho}_{A} \otimes \underbrace{\sum_{i,j} |i, i\rangle\langle j, j|\Big) \overbrace{\sum_l |l, l\rangle}^{AB}}_{BC} = \underbrace{\rho}_{C} \,. \tag{2.63}$$

[22]: Dür et al. (2005), 'Standard forms of noisy quantum operations via depolarization'

Now, one could act on the last subsystem with a quantum channel $\mathscr{E}$, yielding $\mathscr{E}(\rho)$. These two steps – projection and applying $\mathscr{E}$ – do commute, since they act on disjoint subsystems. Thus, the end result should be the same, regardless of the order of operations. If the operations are reversed, the map $\mathscr{E}$ is applied first to the subsystem $C$, and the total state reads

$$\sigma' = (\mathrm{id}_{AB} \otimes \mathscr{E})\,(\rho \otimes |\omega\rangle\langle\omega|). \tag{2.64}$$

Let me denote the quantum state of the last two subsystems by $\Phi_{\mathscr{E}}$ – it is known as the Choi-Jamiołkowski [20, 21] dual state of the

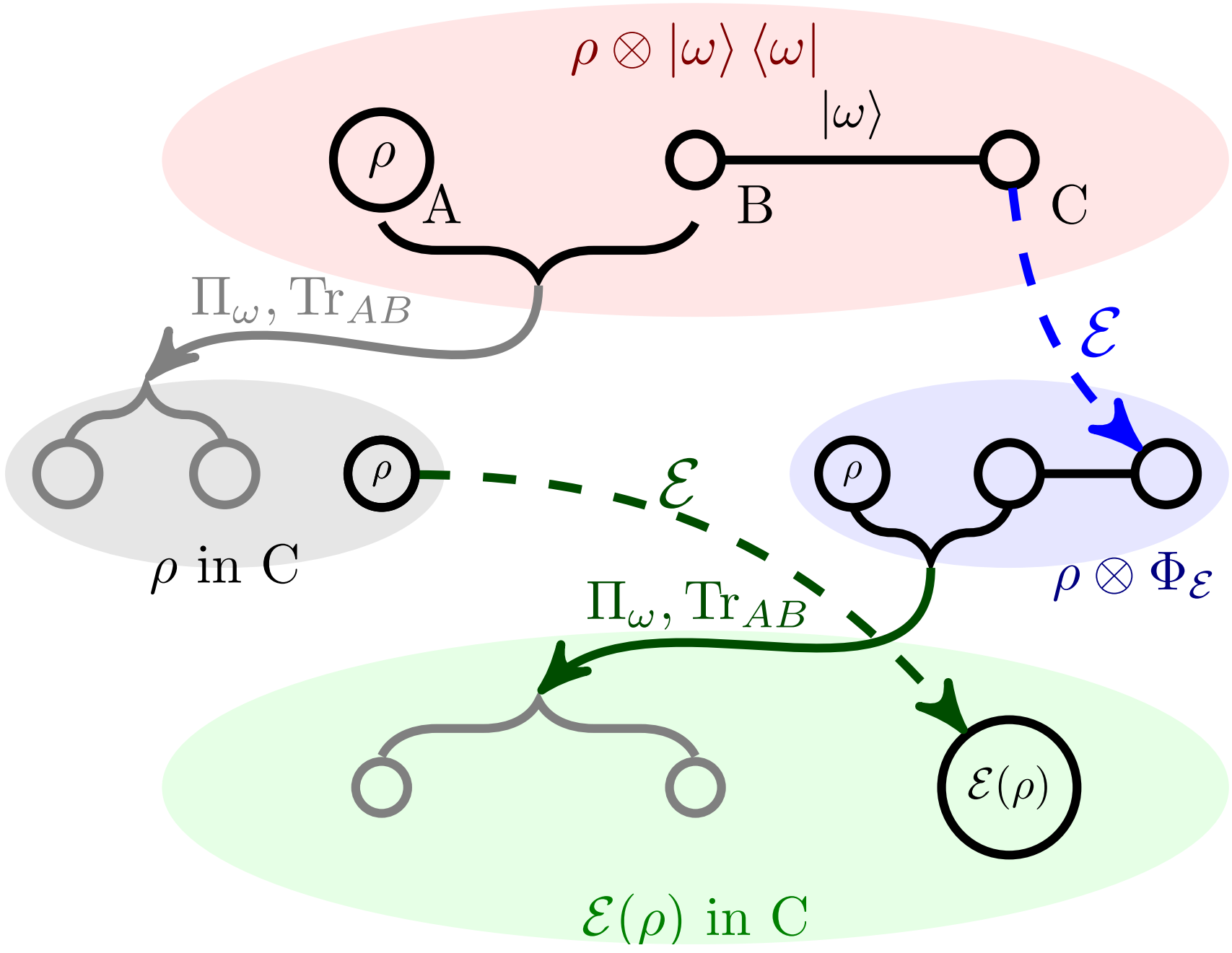

**Figure 2.7:** The Choi-Jamiolkowski isomorphism can be interpreted in the terms of a quantum teleportation protocol: projection onto the maximally entangled state $|\omega\rangle$ on the subsystem BC and action of a quantum channel $\mathscr{E}$ do commute, and in the middle of the procedure a state $\Phi_{\mathscr{E}}$ dual to the channel $\mathscr{E}$ is formed.

channel $\mathscr{E}$. It reads simply

$$\Phi_{\mathscr{E}} = (\mathrm{id} \otimes \mathscr{E})\,(|\omega\rangle\langle\omega|). \tag{2.65}$$

The Choi-Jamiołkowski dual $\Phi_{\mathscr{E}}$ is positive semidefinite if and only if $\mathscr{E}$ is completely positive.

Now, by projecting the two subsystems AB onto $|\omega\rangle$ and tracing them out, the final state is again $\mathscr{E}(\rho)$. Thus, the action of a channel $\mathscr{E}$ can be simulated by a quantum teleportation scheme [22] amounting to

$$\mathscr{E}(\rho) = \mathrm{Tr}_A[(\rho^T \otimes \mathbb{1}_B)\Phi_{\mathscr{E}}]. \tag{2.66}$$

[22]: Dür et al. (2005), 'Standard forms of noisy quantum operations via depolarization'

The transposed state $\rho^T$ appears naturally during the calculations – to obtain an analogue of Eq. (2.66) without transpose, an alternative definition of the operator dual to a channel may be devised [23]. However, the dual is no longer a quantum state.

[23]: Jiang et al. (2013), 'Channel-state duality'

## 2.3 Weyl-Heisenberg group and Wigner representation

Of all nonstandard representations and interpretations of quantum mechanics, the one provided by the stabilizer formalism is perhaps one of the most useful ones. It immediately provides a simple way to simulate classically a large portion of quantum operations, delineating the difference between the quantum and the classical world, and enables quantum error correction. Furthermore, its modification serves as an interesting toy theory adjacent to quantum mechanics, the mathematics behind the formalism is used in the construction of a particularly symmetric and elegant scheme for the reconstruction of a quantum state from obervable values (SIC-POVM). It is interesting from the viewpoint of geometry as well, serving as a generalization of the cyclic majorization discussed

The interpretation served via this construction represents quantum states as *quasiprobability distributions* – the quasiprobabilities are real and sum up to unity, but may in general be negative. However, for a large portion of states – even the ones with a large amount of entanglement – the quasiprobabilities are nonnegative, which allows a classical interpretation of the underlying dynamics.

[24]: Koukoulekidis et al. (2021), 'Constraints on magic state protocols from the statistical mechanics of Wigner negativity'

in one of the following chapters. Here, I present the construction of the basic elements of the theory appearing in [24].

Consider a $p$-dimensional quantum system, where $p$ is an odd prime. In this system, an orthonormal basis $\{|n\rangle : n \in \mathbb{Z}_p\}$ is fixed and the operators $X$ and $Z$ are defined by

The $p = 2$ case has significant difficulties. While this dimension is explicitly excluded, the symbols $X$ and $Z$ serve as a reminder of qubits. In this scenario, $X = \sigma_x$ and $Z = \sigma_z$.
$X$ and $Z$ are thus unitary, instead of Hermitian, generalizations of Pauli matrices.

$$X|n\rangle = |n + 1 \,(\mathrm{mod}\, p)\rangle\,, \quad Z|n\rangle = \omega^n |n\rangle\,, \tag{2.67}$$

where $\omega = \exp\left(i\frac{2\pi}{p}\right)$. Using these operators, one can consider a discrete subgroup of $SU(p)$, composed of all operators which can be written as a product of $X$, $Z$, and $\omega$. It can be easily proved that $ZX = \omega XZ$ and

$$X^p = Z^p = \mathbb{1}_p. \tag{2.68}$$

Thus every element of such a group can be written as an operator proportional to $X^x Z^q$ for some $x$ and $q$. This can be extended to nonprime dimensions in a manner consistent with the factorization.

**Definition 2.13** *Let p be an odd prime number. The Weyl-Heisenberg group is composed of elements parameterized by pairs $(x, q) \in \mathbb{Z}_p^2$ called the displacement operators and defined by*

$$D_{x,q} = (-\kappa)^{xq} X^x Z^q, \tag{2.69}$$

*where*

$$\kappa = \exp\left(i\frac{\pi}{p}\right). \tag{2.70}$$

*For nonprime odd dimension $d$ with the prime factorization of $d = p_1 \cdot p_2 \cdot \ldots \cdot p_n$, the displacement operators are tensor products of the corresponding operators for each of the factors:*

Here, $\vec{x} = (x_1, \ldots, x_n)$ and analogously for $\vec{q}$.

$$D_{\vec{x},\vec{q}} = D^{(p_1)}_{x_1,q_1} \otimes \ldots \otimes D^{(p_n)}_{x_n,q_n}. \tag{2.71}$$

Using the displacement operators, the Wigner distribution of a finite-dimensional quantum state may be defined.

**Definition 2.14** *Consider a state $\rho \in \mathcal{M}_d$ such that $d$ is odd and has prime factorization $d = p_1 \cdot \ldots \cdot p_n$. Let $\mathfrak{X} = \mathbb{Z}_{p_1} \times \ldots \times \mathbb{Z}_{p_n}$. The (discrete) Wigner distribution of a state $\rho$ is the function $W : \mathfrak{X}^2 \longrightarrow \mathbb{R}$:*

$$W_\rho(\vec{x}, \vec{q}) = \frac{1}{d} \operatorname{Tr} \rho A_{\vec{x},\vec{q}}, \tag{2.72}$$

*where the phase point operator $A_{x,q}$ is defined using the displacement operators:*

$$A_{\vec{x},\vec{q}} = \frac{1}{d} D_{\vec{x},\vec{q}} \left( \sum_{\vec{x}',\vec{q}' \in \mathfrak{X}} D_{\vec{x}',\vec{q}'} \right) D^\dagger_{\vec{x},\vec{q}}. \tag{2.73}$$

Wigner distribution is always real, but not necessarily positive – see Figure 2.8. However, the marginal distributions – the Wigner function projected onto $\vec{x}$ or $\vec{q}$ – are always probability distributions equal to the probability distributions of observing eigenstates of $Z$ and $X$.
The object defined here is the *discrete Wigner function*, suitable for finite-dimensional systems. In other fields of quantum theory various different continuous Wigner functions are used.

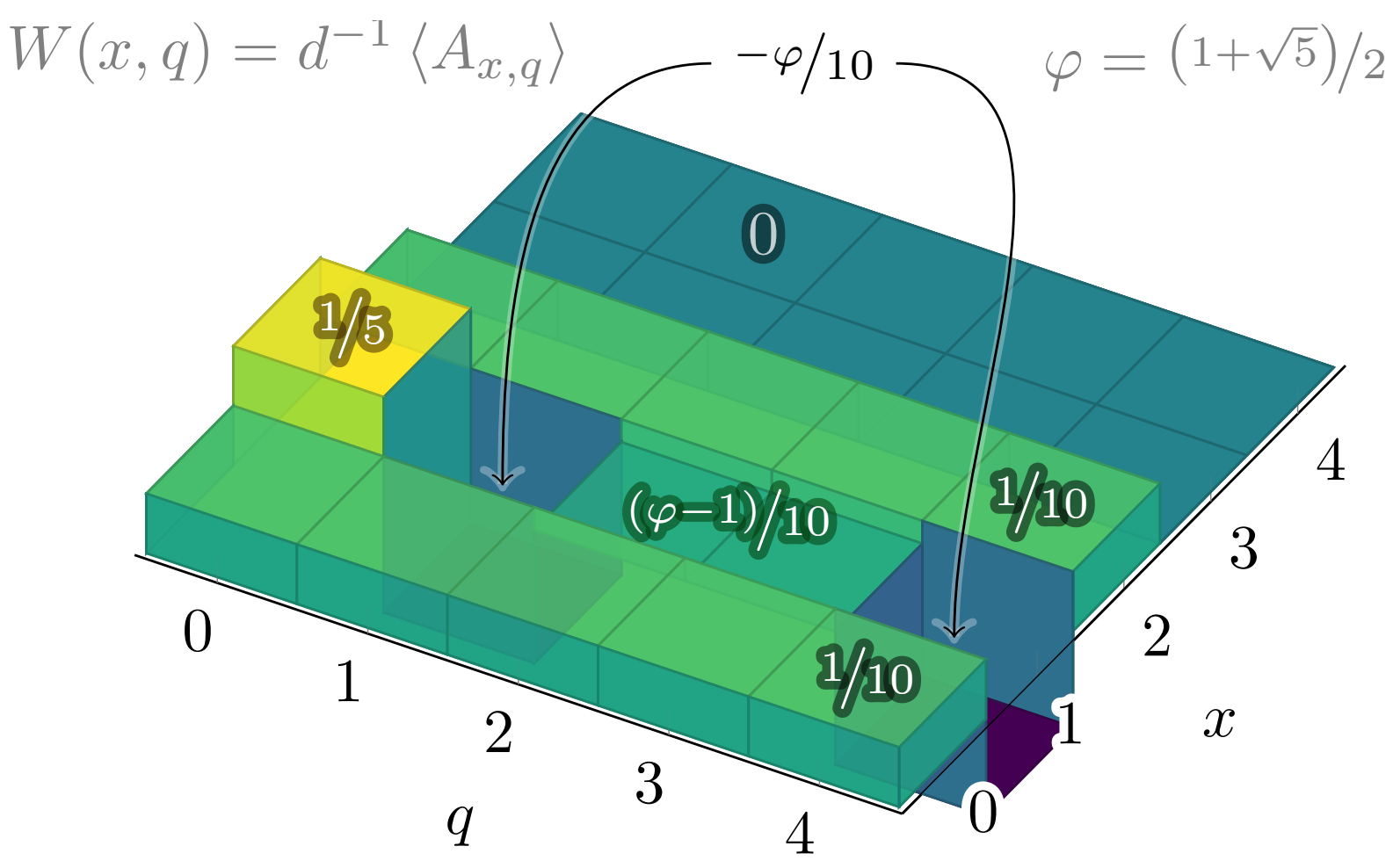

**Figure 2.8:** Wigner function $W(x,q)$ of a state $|\psi\rangle = (|0\rangle + |1\rangle)/\sqrt{2}$ of a system of dimension 5. Here, $\varphi = \left(1+\sqrt{5}\right)/2$ denotes the golden ratio.

By construction, the phase point operators $A_{\vec{x},\vec{q}}$ are Hermitian, pairwise orthogonal, and the following relation holds:

$$\sum_{\vec{x},\vec{q}\in\mathfrak{X}} A_{\vec{x},\vec{q}} = d\mathbb{1}. \tag{2.74}$$

Thus, the Wigner distribution may in some cases be interpreted as a probability distribution completely defining the state $\rho$ through the identity

$$\rho = \sum_{\vec{x},\vec{q}\in\mathfrak{X}} W_\rho(\vec{x},\vec{q})A_{\vec{x},\vec{q}}. \tag{2.75}$$

It is not only the quantum states which are linked to the Wigner distributions. Indeed, quantum channels are representable in this formalism as well. For a channel $\mathscr{E} : \mathscr{H}_n \to \mathscr{H}_m$, let the following transition quasiprobability matrix:

$$W_{\mathscr{E}}(\vec{x},\vec{q}|\vec{x}',\vec{q}') = W_{\Phi(\mathscr{E})}(\vec{x}' \oplus \vec{x}, (-\vec{q}') \oplus \vec{q}), \tag{2.76}$$

where $\Phi(\mathscr{E})$ is the Choi-Jamiołkowski state dual to $\mathscr{E}$. Elementary calculation shows that the action of a quantum channel can be described by means of this matrix [24]:

[24]: Koukoulekidis et al. (2021), ‘Constraints on magic state protocols from the statistical mechanics of Wigner negativity’

**Lemma 2.3** (Koukoulekidis-Jennings) *Wigner function of the state $\mathscr{E}(\rho)$ splits into the sum of Wigner function elements of $\rho$ multiplied by the transition quasiprobability described by Eq.* (2.76)*:*

$$W_{\mathscr{E}(\rho)}(\vec{x},\vec{q}) = \sum_{\vec{x}',\vec{q}'\in\mathfrak{X}} W_{\mathscr{E}}(\vec{x},\vec{q}|\vec{x}',\vec{q}')W_\rho(\vec{x}',\vec{q}'). \tag{2.77}$$

As noted before, this construction fails for qubits: the naive generalization of phase point operators and Wigner distributions does not provide an accurate description of quantum states. Qubit extensions with similar properties can be considered, although the construction is different: for reference see [25].

[25]: Raussendorf et al. (2020), ‘Phase-space-simulation method for quantum computation with magic states on qubits’

## 2.4 Important subsets of the set of density matrices $\mathcal{M}$

It is not only the entire set $\mathcal{M}_d$ of all possible $d$-dimensional quantum states which is of interest – its convex subsets are often considered in a wide array of problems in quantum mechanics. Convexity arises naturally in some fields: it is not uncommon for some property of states $p(\rho)$ to hold under convex combination: that is, if $p(\rho_0)$ and $p(\rho_1)$, then also $p(\rho_t)$, where $t \in [0, 1]$ and

$$\rho_t = t\rho_1 + (1-t)\rho_0. \tag{2.78}$$

Some of the examples of interesting convex subsets of $\mathcal{M}_d$ discussed in the following sections are:

**Separable states** $\mathcal{M}^{\mathrm{SEP}}$ – i.e., those which are not entangled. Clearly, any mixture of such two states can not produce entanglement.

Separability and entanglement are properties of states of composite systems – therefore, the set of separable states $\mathcal{M}^{\mathrm{SEP}}_{d_1\times\ldots\times d_n}$ is a subset of $\mathcal{M}_d$ (with $d = d_1 \times \ldots \times d_n$) – the set of states of a multipartite system with local dimensions $d_1, \ldots, d_n$.

**Positive partial transpose states** $\mathcal{M}^{\mathrm{PPT}}$ for which the partial transpose yields a valid density operator. Closely related to $\mathcal{M}^{\mathrm{SEP}}$.

**Free states** of some resource theory $\mathcal{M}^{\mathrm{free}}$ – if one can obtain $\rho$ and $\sigma$ without much effort, mixing these two is allowed as well, as it is equivalent to forgetting some classical information.

**Translationally invariant states** $\mathcal{M}^{\Delta}$ of lattice systems (e.g., spin chains). This set contains the eigenstates of translationally invariant Hamiltonians.

**Reduced states** $\mathcal{M}^{\mathrm{red}}$ of translationally invariant states $\mathcal{M}^{\Delta}$. The set $\mathcal{M}^{\mathrm{red}}$ is composed of partial traces of states from $\mathcal{M}^{\Delta}$, and thus $\mathcal{M}^{\mathrm{red}}$ encodes information about the eigenstates of translationally invariant lattice systems in a finite-dimensional form.

### 2.4.1 Entanglement: separable and positive partial transpose states

One of the most important behaviors of quantum theory, distinguishing it from classical counterparts, is the presence of *entanglement* – a phenomenon linking correlations, nonlocality, and foundations of quantum mechanics. It is thus productive to characterize entanglement – for instance, to have a method determining whether a state is entangled or not.

Is is certainly known what an entangled state is not: if a bipartite state of the form

$$\rho_A \otimes \rho_B \tag{2.79}$$

is observed, then every property, value of an observable quantity etc. can be measured locally in $A$ and $B$, and any knowledge of the other part of the state does not help with the determination of properties on 'our' side. This object, a *product state*, is not entangled.

One can go further and allow classical correlations to appear: if an ensemble of product states, $\{(p_i, \rho_A^{(i)} \otimes \rho_B^{(i)})\}$ is prepared, the state is called *separable*:[18]

$$\sum_i p_i \rho_A^{(i)} \otimes \rho_B^{(i)}, \tag{2.80}$$

as it does not possess any quality characteristic of entanglement. Correlations do appear, but are classical in nature and can be described by local hidden variables, as suggested by the ensemble form. If a state $\rho$ can not be described in this form, the correlations between the subsystems may be unexplainable by any classical model [26]. The counterintuitive properties of such states earned them their own name.

18: This construction works for multipartite systems as well:

**Definition 2.15** *The set of separable states of a $n$-partite system with local dimensions $d_1, \dots, d_n$ is denoted by $\mathcal{M}^{\text{SEP}}_{d_1\times\ldots\times d_n}$.*

[26]: Werner (1989), 'Quantum states with Einstein-Podolsky-Rosen correlations admitting a hidden-variable model'

**Definition 2.16** *A state $\rho$ is said to be entangled if and only if it is not separable.*

*Positive partial transpose* (PPT) states are a further relaxation of the restrictions imposed on states: it is the set of states such that one of the methods of entanglement detection fails. For a state of the form

$$\rho = \sum_i c_i \chi_i \otimes \xi_i, \tag{2.81}$$

the partial transpose reads

$$\rho^{T_A} = \sum c_i \chi_i^T \otimes \xi_i. \tag{2.82}$$

No restrictions are put on the variables here – $c_i$ may be arbitrary complex numbers, $\chi_i, \xi_i$ are arbitrary operators.

**Definition 2.17** *A bipartite state $\rho$ has a positive partial transpose (is PPT) if and only if $\rho^{T_A} \succcurlyeq 0$. The set of all PPT states of a bipartite system with local dimensions $m$ and $n$ is denoted by $\mathcal{M}^{\text{PPT}}_{m\times n}$.*

Observe that if a state is separable and therefore described by Eq. (2.80), then its partial transpose,

$$\sum_i p_i \left(\rho_A^{(i)}\right)^T \otimes \rho_B^{(i)}, \tag{2.83}$$

is also a valid quantum state and a convex mixture of product states. Therefore, if a state $\rho$ does not possess the positive partial transpose property, then $\rho$ is entangled.

This stems from a characterization of quantum channels: for a given map $\mathcal{E} : \mathcal{M}_d \to \mathcal{M}_d$ to be realizable in the physical world, it must transform valid states into valid states, as well as any extension must do so. Transpose does not possess this property: indeed, if $\rho$

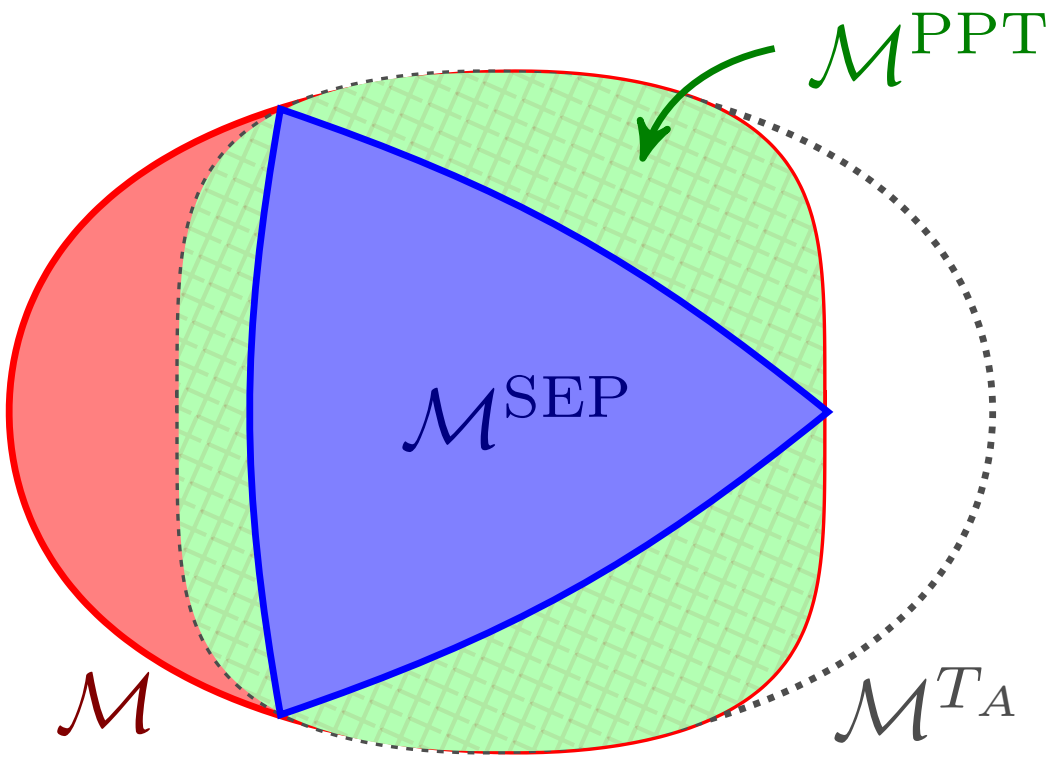

**Figure 2.9:** The set of separable states $\mathcal{M}^{\mathrm{SEP}}$ is the convex hull of the set of product states, while the set of positive partial transpose states $\mathcal{M}^{\mathrm{PPT}}$ is the intersection of $\mathcal{M}$ with its reflection through partial transpose map, $\mathcal{M}^{T_A}$. Separable states always have positive partial transpose, but the converse is true only for bipartite qubit-qubit and qubit-qutrit cases.

Here, $T$ is the transpose map: $T(\rho) = \rho^T$.

is a valid state, so is $\rho^T$, but the extension $T \otimes \mathrm{id}$ fails on a subset of entangled states.

In some scenarios, the mere presence of entaglement is of little utility in itself. More important is the question whether given a state $\rho \in \mathcal{M}_d$, can one transform a collection of $n$ copies of this state, $\rho^{\otimes n} \in \mathcal{M}_{d^n}$, into a single highly entangled state $\sigma \in \mathcal{M}_d$ using local operations and classical communication between subsystems? Positive partial transpose criterion offers a partial solution to this problem: if a state $\rho$ is PPT, this task can not be performed and the states $\rho$ is *nondistillable* [27]. There exist some results in the converse direction as well [28].

This protocol is referred to as *entanglement distillation*.

[27]: Chen et al. (2016), 'Distillability of non-positive-partial-transpose bipartite quantum states of rank four'

[28]: Verch et al. (2005), 'Distillability and positivity of partial transposes in general quantum field systems'

### 2.4.2 Phase transitions and quantum information: marginal states

An interesting perspective on the problem of phase transitions links it to the properties of certain subsets of density matrices describing systems of finite size. Consider a translationallly invariant 1D chain Hamiltonian in the general form:

$$H_{(h)} = \sum_{n \in \mathbb{Z}} h_{n,n+1}, \tag{2.84}$$

where $h$ is the interaction operator between the neighboring sites. A standard way to some of the properties of the ground state like one- or two-site expectation values would involve diagonalizing the total Hamiltonian $H_{(h)}$.

Another approach to this problem is to characterize the set of reductions of translationally invariant states. Consider a set of states $\mathcal{M}_\infty$ of an infinite number of qudits. Now, the pure translationally invariant states are the eigenstates of the operator $\Delta$ defined by its action on the product states

$$\Delta \,|\ldots, \underbrace{k_n}_{\text{site } n}, \underbrace{k_{n+1}}_{\text{site } n+1}, \ldots\rangle = |\ldots, \underbrace{k_{n-1}}_{\text{site } n}, \underbrace{k_n}_{\text{site } n+1}, \ldots\rangle\,. \tag{2.85}$$

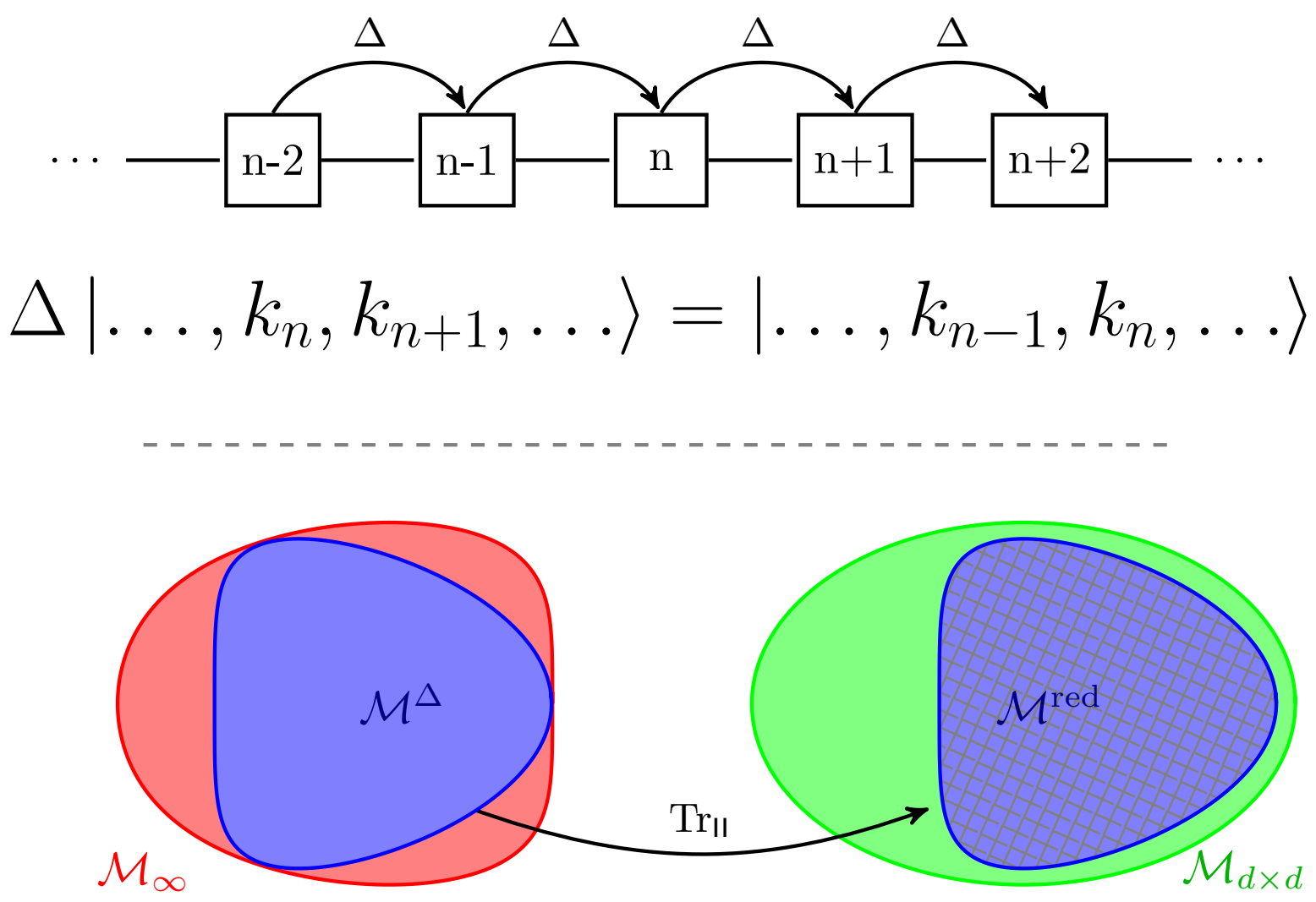

**Figure 2.10:** *Top*: The shift operator $\Delta$ acting on a chain system translates the quantum state by a single step on the chain. The set of translationally invariant states is denoted by $\mathcal{M}^{\Delta}$. *Bottom*: Marginal states $\mathcal{M}^{\mathrm{red}} = \mathrm{Tr}_{\mathrm{II}}\,\mathcal{M}^{\Delta}$ are the partial traces of the set of translationally invariant states $\mathcal{M}^{\Delta}$ into the set of bipartite states of adjacent sites.

The shift operator $\Delta$ is unitary. Since it commutes with every translationally invariant Hamiltonian of the form described by Eq. (2.84), a common eigenbasis for both of them is possible. Therefore, the set of translationally invariant states, defined by

$$\mathcal{M}^{\Delta} = \{\rho \in \mathcal{M}_{\infty} : \Delta\rho\Delta^{\dagger} = \rho\}, \tag{2.86}$$

is worthy of consideration, since $\mathcal{M}^{\Delta}$ contains all information characterizing the ground states of translationally invariant Hamiltonians. The set of reductions of $\mathcal{M}^{\Delta}$ has been analyzed previously [29]: with the notation of $\mathrm{Tr}_{\mathrm{II}}$ being the partial trace *into* two adjacent sites defined by its action on product states,

[29]: Verstraete et al. (2006), 'Matrix product states represent ground states faithfully'

$$\mathrm{Tr}_{\mathrm{II}} \bigotimes_{n\in\mathbb{Z}} \rho_n = \rho_0 \otimes \rho_1, \tag{2.87}$$

the set of reductions is

Note that here it does not matter *which* pair of adjacent sites are taken – the states are *translationally invariant*.

$$\mathcal{M}^{\mathrm{red}} = \mathrm{Tr}_{\mathrm{II}}\,\mathcal{M}^{\Delta} = \{\mathrm{Tr}_{\mathrm{II}}\,\rho : \rho \in \mathcal{M}_{\infty}, \Delta\rho\Delta^{\dagger} = \rho\}. \tag{2.88}$$

It is a subset of the set of bipartite density operators $\mathcal{M}_{d\times d}$. One of its properties is immediately apparent: since for any state $\rho \in \mathcal{M}^{\Delta}$ and $h$ describing the interaction between adjacent sites as in Eq. (2.84), the following is true:

$$\langle h_{n,n+1}\rangle_{\rho} = \langle h\rangle_{\mathrm{Tr}_{\mathrm{II}}\,\rho}\,, \tag{2.89}$$

there exists a direct connection between the ground states of local Hamiltonians and the set of reductions of translationally invariant states.

**Lemma 2.4** *The partial trace $\rho_h$ of the ground state $|\psi(h)\rangle$ of $H_{(h)}$ is contained in the set $\mathcal{M}^{red}$. The state $\rho_h$ minimizes the expectation value of $h$ in this set:*

$$\rho_h = \underset{\sigma\in\mathcal{M}^{\Delta}}{\arg\min}\,\langle h\rangle_{\sigma}\,. \tag{2.90}$$

[30]: Hudson et al. (1976), 'Locally normal symmetric states and an analogue of de Finetti's theorem'
[31]: Chen et al. (2017), 'Joint product numerical range and geometry of reduced density matrices'

Characterization of this set is, however, famously difficult. Similar approaches in related scenarios have yielded some results: in $\infty$ dimensions, rather than 1D, it turns out that the set of reductions of translationally invariant states exactly coincides with the set of separable states [30, 31].

## 2.5 Concluding remarks

The set of mixed quantum states $\mathscr{M}_d$ of a $d$-dimensional system possesses intricate geometric properties useful in the analysis of the quantum theory. The concepts introduced in this chapter will be applied in the presentation of the original results of this thesis contained in chapters 4 and 5.

# Mathematical tools: convex geometry and algebra | 3

This chapter presents some of the mathematical tools used in the analysis of the set of quantum states: convex geometry and algebraic structures. While the basic notions (like the convex hull or a polynomial equation) may be used intuitively, it is beneficial to present some of the more involved ideas explicitly. The material introduced here will be used in the following chapters: the convexity of the set of quantum states and its affine images has nontrivial consequences; its structure, defined by polynomials, reveals useful information.

## 3.1 Convex geometry

Convex structure of sets appearing in quantum mechanics is frequently employed in various scenarios – one of the basic questions one may ask about a set $X$ are the extrema of a function $f : X \to \mathbb{R}$ defined over $X$, and the convexity of $X$ (along with assumptions about the structure of the function $f$) considerably reduces the complexity of this problem. The basic criterion of convexity is an expression of the following request: *every segment with endpoints in $X$ should be contained in $X$* [32].

[32]: Lay (2007), *Convex Sets and Their Applications*

**Definition 3.1** *A set $X$ is convex if and only if every convex combination of two points from $X$ is contained in $X$, that is, for every $x, y \in X$ and $0 \leq \alpha \leq 1$*

$$\alpha x + (1-\alpha) y \in X. \tag{3.1}$$

**Lemma 3.1** *A strictly convex function $f : X \to \mathbb{R}$ defined over convex and compact $X$ attains its maxima on the boundary $\partial X$ of the set $X$; additionally, every minimum of $f$ is also its global minimum.*

*Proof.* If a point $x \in X$ can be decomposed as a convex combination of the form $x = \sum_{i=1}^{n} p_i x^{(i)}$, then $f(x) \leq \max_i f\left(x^{(i)}\right)$ – therefore, $x$ can correspond to a maximum only if no nontrivial decomposition exists, which may happen only for $x \in \partial X$. Similarly, assume that $x \in X$ is a minimum of $f$ in some neighbourhood $O \subset X$ and $y \in X$ exists such that $f(y) < f(x)$. Then for $x_\alpha = \alpha x + (1-\alpha) y$ such that $x_\alpha \in O$ and $0 < \alpha < 1$, $f(x_\alpha) > f(x)$ following from the assumption of local minimum. However, $f(x_\alpha) < f(x)$ by reasoning analogous to the case of maxima. No such $y$ may exist and $x$ is a global minimum. □

Sufficiently regular convex sets admit a *dual* description characterized by the linear inequalities fulfilled by the set. There are multiple related concepts originating in this idea: supporting points, hyperplanes and half-spaces, functions, and polar sets. The entire reasoning starts with half-spaces:

**Definition 3.2** *A supporting half-space of a set $X$ is a closed half-space $H$ satisfying the following criteria:*

*a) $X$ is a subset of $H$.*
*b) Boundary $\partial X$ of $X$ has a nonempty intersection with the boundary $\partial H$ (a hyperplane) of $H$.*

The nomenclature associated with the boundary structure goes further:

**Definition 3.3** *Let $H$ denote a supporting half-space of a convex compact set $X \subset \mathbb{R}^d$. Then, the half-space is defined by the vector $\vec{y}$ and the associated linear inequality:*

$$\forall_{\vec{x} \in H} \vec{x} \cdot \vec{y} \leq h_X(\vec{y}). \tag{3.2}$$

*a) The boundary $\partial H = \{\vec{x} \in \mathbb{R}^d : \vec{x} \cdot \vec{y} = h_X(\vec{y})\}$ is called the support hyperplane whenever $\vec{y} \neq \vec{0}$.*
*b) The function $h_X(\vec{y})$ is called the support function. It admits the form of*

$$h_X(\vec{y}) = \max_{\vec{x} \in X} \vec{x} \cdot \vec{y}. \tag{3.3}$$

*c) The intersection $\partial H \cap \partial X$ is an exposed face of $X$. It is the set of maximizers of the above expression.*

Exposed faces and support functions are defined even for $\vec{y} = \vec{0}$: the entire $X$ is its own exposed face and $h_X(\vec{0}) = 0$.

Further analysis of the structure induced by supporting objects suggests the definition of "dual" (called *polar* instead) set: the vectors $\vec{y}$ for which the support function is bounded by unity.

The notation varies across sources, but the most accepted variant defines the *dual* set $X^*$ to be precisely the negative of the *polar*, $X^* = -X^\circ$.

**Definition 3.4** *Polar set $X^\circ$ of a convex compact $X \subset \mathbb{R}^n$ is defined by the support function $h_X$,*

$$X^\circ = \{\vec{y} \in \mathbb{R}^n : h_X(\vec{y}) \leq 1\}. \tag{3.4}$$

Polar sets often offer a perfect description of the originals in the sense that $X^\circ$ contains the information needed to reconstruct $X$. The conditions are not too stringent and are easily attained:

**Lemma 3.2** *Let $X \subset \mathbb{R}^n$ be compact and convex with $\vec{0} \in \operatorname{int} X$. In this case, $\vec{0} \in X^\circ$ and $(X^\circ)^\circ = X$.*

Here, $\operatorname{int} X$ is the *interior* of $X$ – the largest open set contained in $X$.
The reason behind the requirement $\vec{0} \in \operatorname{int} X$ is the conservation of information. For any convex set $X$, the following holds: $X^\circ = (X \cup \{\vec{0}\})^\circ$, so if a set $X$ is separated from $\vec{0}$, its polar does not determine it uniquely.

*Proof.* Since $X$ is compact, there exists a ball $B(\vec{0}, R) \supset X$. The polar of this ball is $B(\vec{0}, \frac{1}{R})$, which is contained in $X^\circ$ – thus, $\vec{0} \in X^\circ$. The proof of the second statement splits into two cases:

**$X \subset (X^\circ)^\circ$:** For every $\vec{x} \in X$ and $\vec{y} \in X^\circ$, $\vec{x} \cdot \vec{y} \leq 1$, therefore the point $\vec{x}$ is also contained in $(X^\circ)^\circ$.

**$X \supset (X^\circ)^\circ$:** If $\vec{z}$ is not contained in $X$, then there exists a vector $\vec{y}$ such that $\vec{x} \cdot \vec{y} \leq 1$ for all $\vec{x} \in X$ and $\vec{z} \cdot \vec{y} > 1$. Therefore, $\vec{y} \in X^\circ$ and $\vec{z} \notin (X^\circ)^\circ$.

□

Some of the information contained in $X$ is easily accessible through its polar set: the nonanalytical parts of the boundary of $X$ are immediately visible in $X^\circ$, support functions can be directly deduced from the norms of the points on the boundary of the polar, points on $\partial X$ correspond to the normal vectors of $X^\circ$.

**Lemma 3.3** *Let $X$ be a convex and compact set with $\vec{0} \in \operatorname{int} X$. Additionally, let the vectors $\vec{x} \in \partial X$ and $\vec{y} \in \partial X^\circ$ satisfy $\vec{x} \cdot \vec{y} = 1$. In such case,*

*a) The vector $\vec{y}$ is proportional to an unit length normal vector of $X$ at $\vec{x}$.*
*b) The vector $\vec{x}$ is proportional to an unit length normal vector of $X^\circ$ at $\vec{y}$.*

*Proof.* A vector $\vec{n}$ is normal to the boundary of a convex set $X$ at $\vec{x}$ if and only if

$$\forall_{\vec{x}' \in X} \vec{x}' \cdot \vec{n} \leq \vec{x} \cdot \vec{n}. \tag{3.5}$$

It is possible for a set to have multiple normals at a single point, e.g., at corners.

The vector $\vec{y} \in \partial X^\circ$ obeying $\vec{x} \cdot \vec{y} = 1$ is therefore normal to $\partial X$ at $\vec{x}$ by definition, and the unit normal is given by

$$\frac{\vec{y}}{|\vec{y}|}. \tag{3.6}$$

The proof of the second part follows from Lemma 3.2. □

One of the observations – the projection-intersection duality – will come in handy later:

**Theorem 3.4** *Let $X \subset \mathbb{R}^{n+m}$ be a convex and compact set such that $\vec{0} \in \operatorname{int} X$ and let $X^\circ$ denote its polar. If $\pi : \mathbb{R}^{n+m} \longrightarrow \mathbb{R}^n$ is an orthogonal affine projection such that*

$$\pi(\vec{x}' \oplus \vec{x}'') = \vec{x}', \tag{3.7}$$

*then the polar set $\pi(X)^\circ$ of $\pi(X)$ is the intersection $S$ of $X^\circ$ with a hyperplane selected by $\pi$:*

$$\pi(X)^\circ = \overbrace{\left\{\vec{y}' \in \mathbb{R}^n : \vec{y}' \oplus \vec{0} \in X^\circ\right\}}^{S}. \tag{3.8}$$

Here, $\oplus$ denotes simple sum – in the context of Euclidean vectors in $\mathbb{R}^d$, it can be thought of as a simple concatenation, $(a_1, \ldots, a_n) \oplus (b_1, \ldots, b_m) = (a_1, \ldots, a_n, b_1, \ldots, b_m)$.
Here, the symbol $\vec{0}$ is the zero vector of $\mathbb{R}^m$,

*Proof.* If $X^\circ$ is polar to $X$, then every $\vec{y} \in X^\circ$ such that $\vec{y} = \vec{y}' \oplus \vec{y}''$ fulfills $\vec{y} \cdot \vec{x} \leq 1$ for every $\vec{x} \in X$. Since

$$\pi(X) = \left\{\vec{x}' \in \mathbb{R}^n : \exists \vec{x}'' \in \mathbb{R}^m,\ \ \vec{x}' \oplus \vec{x}'' \in X\right\}, \tag{3.9}$$

the following holds:

$$\pi(X)^\circ = \left\{\vec{v} \in \mathbb{R}^n : \forall \vec{x}' \oplus \vec{x}'' \in X,\ \ \vec{x}' \cdot \vec{v} \leq 1\right\}. \tag{3.10}$$

This is the same exact set as $S$: $\vec{y}' \in S$ if and only if $\vec{y}' \oplus \vec{0} \in X^\circ$, which is equivalent to

$$S = \left\{\vec{y}' \in \mathbb{R}^n : \forall \vec{x}' \oplus \vec{x}'' \in X,\ \ \vec{x}' \cdot \vec{y}' \leq 1\right\}. \tag{3.11}$$

□

## 3.2 Algebra of polynomials

In many cases, geometric methods can be supplemented by the application of algebra. In the topics presented in this thesis, the algebraic structure of polynomials is often beneficial: the dual geometry of numerical ranges (discussed in the following chapter) is described by a generalization of characteristic polynomials, which helps with classification. Additionally, polynomials naturally implement convolution as multiplication – therefore, they are convenient in the analysis of a certain case of group majorization, in which convolution plays a crucial role.

In this section, a basic introduction to the theory of polynomials will be presented. The theorems and approaches to problems are well known (the proofs can be found in [33]) – the connection with questions arising in physics is novel.

[33]: Cox et al. (2013), *Ideals, varieties, and algorithms: an introduction to computational algebraic geometry and commutative algebra*

### 3.2.1 Basic properties of polynomials

The fields $K$ appearing in the definitions of polynomial rings are taken to be infinite and algebraically complete, usually $\mathbb{C}$. Lifting of any of these restrictions leads to a significantly more involved theory. This sometimes introduces undesirable results: often one would like to work with polynomials over $\mathbb{R}$ (not an algebraically complete field), and the remnants of using $\mathbb{C}$ instead are visible in the final result.

Let me start with the most basic element appearing in the theory of polynomials: a *monomial*.

**Definition 3.5** *A monomial in variables $x_1, \ldots, x_k$ is an expression defined by powers $d_1, \ldots, d_k \in \mathbb{N}$:*

$$x_1^{d_1} \cdot x_2^{d_2} \cdot \ldots \cdot x_k^{d_k}. \tag{3.12}$$

*The set of monomials admits a multiplication structure:*

$$\left(x_1^{d_1} \cdot x_2^{d_2} \cdot \ldots \cdot x_k^{d_k}\right)\left(x_1^{d'_1} \cdot x_2^{d'_2} \cdot \ldots \cdot x_k^{d'_k}\right) = x_1^{d_1+d'_1} \cdot x_2^{d_2+d'_2} \cdot \ldots \cdot x_k^{d_k+d'_k}. \tag{3.13}$$

*Polynomials are finite sums of monomials multiplied by coefficients in K. The set of polynomials in variables $x_1, \ldots, x_k$ over the field $K$ forms a ring denoted by*

$$K[x_1, \ldots, x_k]. \tag{3.14}$$

In particular, the set of polynomials is a vector space over $K$ with monomials as basis vectors.

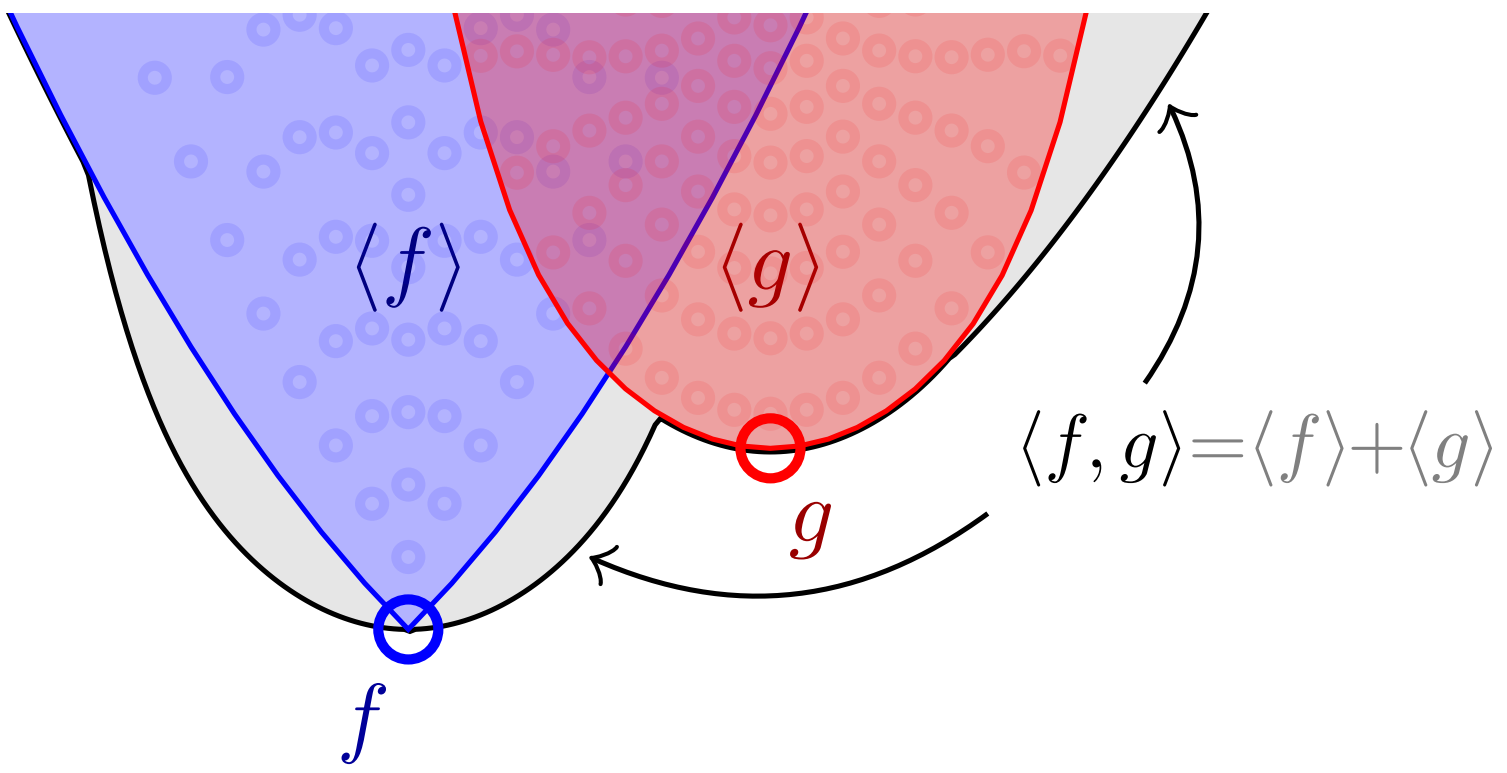

**Figure 3.1:** Polynomial ideal $I$ is a set closed under multiplication by any polynomial and addition of two elements from $I$. According to the Hilbert basis theorem, every ideal is generated by a finite set of polynomials. Pictured is a graphical representation of three ideals: $\langle f \rangle$ generated by $f$, $\langle g \rangle$ generated by $g$ and $\langle f, g \rangle$ generated by $f$ and $g$. The ideal $\langle f, g \rangle$ can be thought of as a Minkowski sum of $\langle f \rangle$ and $\langle g \rangle$ in the space of polynomials.
The *generators* are generally not unique. Certain set of generators called Gröbner basis aids in calculations involving polynomials.

Sets of polynomials are associated with geometric structures – subsets of *affine spaces*, which in this context is equivalent to powers $K^n$ of the field $K$. The connection between polynomials and peculiar subsets of this space becomes apparent upon the realization that a polynomial $f = \sum c_{d_1,\ldots,d_k} x^{d_1} \ldots x^{d_k}$ – up to this point an abstract expression – defines a *polynomial function*:

$$f(x_1, \ldots, x_k) = \sum c_{d_1,\ldots,d_k} x^{d_1} \ldots x^{d_k}. \tag{3.15}$$

In the cases relevant to this thesis, the polynomial functions are isomorphic to polynomials. Indeed, in all infinite fields there is a bijection between them, and the fields used here are $\mathbb{C}$ and $\mathbb{R}$. The connection between subsets of $K^k$ and sets of polynomials is the notion of *affine variety*.

**Definition 3.6** *An affine variety $V$ of polynomials $f_1, \ldots, f_n$, is the maximal subset of $K^n$ of solutions to the polynomial equations $f_i(\vec{x}) = 0$:*

$$V(f_1, \ldots, f_n) = \{\vec{x} \in K^n : \forall_i f_i(\vec{x}) = 0\}. \tag{3.16}$$

The set of all affine varieties over $K^n$ is exactly the class of closed sets under *Zariski topology*.

**Definition 3.7** (Zariski topology) *Closed sets are defined to be exactly the algebraic sets, while open sets are their complements.*

### 3.2.2 Ideals and Gröbner bases

The geometric structure of affine varieties is closely connected to the notion of an *ideal*: a subset of all polynomials closed under specific operations. The utility of ideals in the analysis of the geometry

**Definition 3.8** *Ideal is a subset $I$ of a ring $R$ such that*

*a) for all $x, y \in I$, $x + y \in I$,*
*b) for every $r \in R$ and $x \in I$, $rx \in I$.*

of affine varieties comes from the fact that the set of polynomials vanishing over a variety $V$ exactly fits in this definition: if $f(\vec{x}) = 0$ and $g(\vec{x}) = 0$ for $\vec{x} \in V$, then so does $(f+g)(\vec{x})$ and $(fh)(\vec{x})$ for arbitrary polynomial $h$. Sets of polynomials therefore define ideals, but in a manner that is not unique – often there exists more than one basis of an ideal $I$.

**Definition 3.9** *An ideal generated by polynomials $f_1, \dots, f_k$ in variables $x_1, \dots, x_k$ is defined by*

$$\langle f_1, \dots, f_k \rangle = \left\{ \sum_{i=1}^{k} f_i h_i : h_i \in K[x_1, \dots, x_k] \right\}. \tag{3.17}$$

*The set $\{f_1, \dots, f_k\}$ is called the basis of the ideal.*

**Theorem 3.5** (Hilbert basis theorem) *Every polynomial ideal $I$ has a finite generating set: there exist $f_1, \dots, f_k$ such that $I = \langle f_1, \dots, f_k \rangle$.*

**Theorem 3.6** (Nullstellensatz) *Let $V$ denote the algebraic variety defined by polynomials $f_1, \dots, f_k$. Then, if $g(x) = 0$ for all $x \in V$, $g$ is a power of some polynomial in $I(V)$.*

A finite basis always exists for any ideal (Theorem 3.5). However, some are more useful than others, and often a basis with certain properties (with given order of the leading terms or with variables separated) might come in handy. A Gröbner basis is the solution to this problem. It requires an additional structure – a particular order on the set of monomials.

**Definition 3.10** *A monomial ordering is a linear well ordering on the monomials of $K[x_1, \dots, x_k]$ which is preserved under multiplication (if $p$, $q$, and $r$ are monomials and $p \preccurlyeq q$, then $pr \preccurlyeq qr$ and $p \preccurlyeq pr$).*

Linear (also called total) ordering is reflexive ($a \preccurlyeq a$), transitive ($a \preccurlyeq b \wedge b \preccurlyeq c \implies a \preccurlyeq c$), antisymmetric ($a \preccurlyeq b \wedge b \preccurlyeq a \implies a = b$) and total ($a \preccurlyeq b \vee b \preccurlyeq a$). Well ordering means that every nonempty subset has a least element.

**Definition 3.11** *The* leading term *$LT(f)$ of the polynomial $f$ with respect to the monomial order $\preccurlyeq$ is the monomial appearing in $f$ with the highest order according to $\preccurlyeq$.*

Gröbner basis $G$ of an ideal $I$ is exactly the basis compatible with the monomial ordering. Before stating the definition, let me show the intuition behind the introduction of this kind of basis: one would like to reliably determine whether a given polynomial $f$ is an element of the ideal $I$ or not, and easily provide the decomposition in terms of basis elements. Monomial order and Gröbner basis accomplishes exactly this: any polynomial $f \in I$ can be reduced by some basis element $g \in G$ – that is, brought to the form of

$$f = gh + f', \tag{3.18}$$

such that the total order of the remainder $f'$ is lower than of the original, $LT(f') \preccurlyeq LT(f)$. The polynomial $f'$ can then be reduced by other element $g'$ of the Gröbner basis $G$, and so on – due to $\preccurlyeq$ being a well order, the process eventually stops. The decomposition of $f$ in terms of basis elements can be then read from the steps taken during successive reductions.

**Definition 3.12** *Gröbner basis of a polynomial ideal $I$ with respect to the monomial ordering $\preccurlyeq$ is a finite set $G = \{g_1, \dots, g_k\}$ such that*

*a) $G$ generates $I$:*

$$I = \langle g_1, \dots, g_k \rangle, \tag{3.19}$$

*b) Leading term $LT(f)$ of every element $f \in I$ is divisible by the leading term $LT(g)$ of some element $g \in G$. Therefore, the result of reduction (partial division) of $f$ by $g$ has a lower leading term order.*

The choice of monomial ordering greatly affects the resulting basis. This can be put to use in the problems related to elimination of variables.

**Theorem 3.7** (Gröbner elimination theorem) *Consider an ideal $I$ of $K[x_1, \dots, x_k, y_1, \dots, y_l]$. Let $\preccurlyeq_X$ and $\preccurlyeq_Y$ denote any monomial ordering on $K_X := K[x_1, \dots, x_k]$ and $K_Y := K[y_1, \dots, y_l]$, respectively. Then, let the composite ordering be defined by*

$$f_X h_Y \preccurlyeq f'_X h'_Y \iff \begin{cases} h_Y \preccurlyeq_Y h'_Y & \text{if } f_X = f'_X, \\ f_X \preccurlyeq_X f'_X & \text{otherwise}, \end{cases} \tag{3.20}$$

This is akin to a lexicographic ordering.

*where symbols with $X$ and $Y$ subscripts are monomials in $K_X$ and $K_Y$, respectively.*

*In such case, the Gröbner basis $G$ of $I$ with respect to $\preccurlyeq$ eliminates the $X$ variables: the $X$-independent basis elements generate the $X$-independent part of the ideal $I$:*

$$\langle G \cap K_Y \rangle = I \cap K_Y. \tag{3.21}$$

*Furthermore, $G \cap K_Y$ is a Gröbner basis of $I \cap K_Y$.*

This observation is of great help in some problems: if a particular algebraic set is defined by a system of polynomial equations with auxiliary variables, one can get rid of them easily: if the task is to solve

$$\begin{cases} f_1(x_1, \dots, x_k, y_1, \dots, y_l) &= 0, \\ \qquad \dots & \\ f_k(x_1, \dots, x_k, y_1, \dots, y_l) &= 0, \end{cases} \tag{3.22}$$

and one is interested in the solution of $Y$ variables only, then the $X$-independent part of Gröbner basis of $\langle f_1, \dots, f_k \rangle$ with respect to the composite ordering presented in Theorem 3.7 exactly describes this set.

Let me illustrate this with the following example, presenting the algebraic approach to the problems of geometry.

**Example 3.1** Let $S$ be a semialgebraic set – the intersection of the unit disk and a closed half-plane (see Figure 3.2):

$$S = \left\{ \vec{x} \in \mathbb{R}^2 : |\vec{x}| \leq 1 \wedge x_1 + x_2 \geq -1 \right\}. \tag{3.23}$$

The task is to determine its polar, $S^\circ$. It is the set

$$S^\circ = \{ \vec{y} \in \mathbb{R}^2 : \sup_{\vec{x} \in S} \vec{x} \cdot \vec{y} \leq 1 \}. \tag{3.24}$$

The boundary of $S^\circ$ is composed of $\vec{y}$ such that they saturate the defining inequality. The vector $\vec{y} \in \partial S^\circ$ corresponds to $\vec{x} \in S$, and $\vec{y}$ is proportional to the normal vector of $\partial S$ at $\vec{x}$.

The boundary $\partial S$ is part of the set of solutions of the following polynomial equation in $\mathbb{R}[x_1, x_2]$:

$$\overbrace{(x_1^2 + x_2^2 - 1)(x_1 + x_2 + 1)}^{f} = 0. \tag{3.25}$$

The normal vector can be straightforwardly extracted from this equation: it is proportional to $\vec{\nabla} f$. There are therefore 4 equations in 5 variables $\vec{x}, \vec{y}, \lambda$ which link together the sets $S$ and $S^\circ$:

$$\begin{cases} 0 = & f, \\ 0 = & \vec{\nabla} f - \lambda \vec{y}, \\ 0 = & \vec{x} \cdot \vec{y} - 1. \end{cases} \tag{3.26}$$

The Gröbner basis of the ideal $\langle f, \vec{\nabla} f - \lambda \vec{y}, \vec{x} \cdot \vec{y} - 1 \rangle$ calculated with respect to the elimination ordering is the following singleton:

$$\{ (y_1^2 + y_2^2 - 1)(y_1 + 1)^2 (y_2 + 1)^2 \} \tag{3.27}$$

The set of real solutions of this equation is the unit radius circle centered at $(0, 0)$ and the point $(-1, -1)$ – which exactly spans the polar set $S^\circ$ through the convex hull.

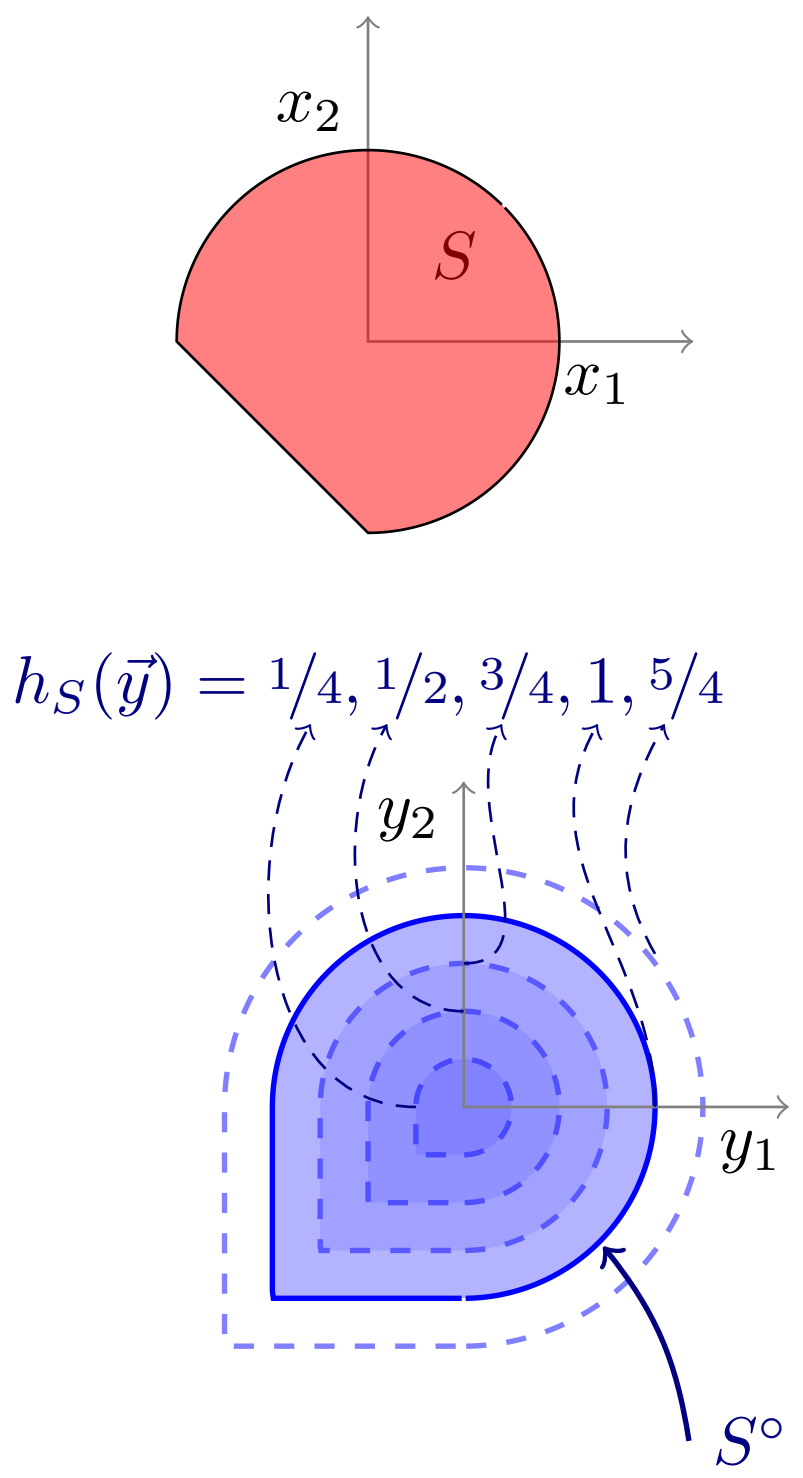

**Figure 3.2:** *Top*: The convex set $S$ (red) defined in Example 3.1.
*Bottom*: Polar $S^\circ$ (blue) of $S$. The polar $S^\circ$ can be interpreted as a sublevel set of the support function $h_S$.

Note that the raw result contains points inside $S^\circ$ (that is: not on the boundary). This is a result of using a polynomial description of $\partial S$ (Eq. (3.25)), which set of solutions contains points not present in $S$. Therefore, some care must be taken when interpreting the results of such calculations.

## 3.3 Concluding remarks

The ideas presented in this chapter will be useful in the main body of this thesis – the next two chapters. Most of the notions regarding convex geometry and polynomials appearing here are standard results [32, 33], they are contained in this chapter for completeness. The Theorem 3.4 is – to my knowledge – novel, albeit straightforward.

[32]: Lay (2007), *Convex Sets and Their Applications*
[33]: Cox et al. (2013), *Ideals, varieties, and algorithms: an introduction to computational algebraic geometry and commutative algebra*

# Numerical ranges and spectrahedra 4

The convex set $\mathcal{M}_d$ of quantum states of dimension $d$ becomes less and less comprehensible as the system size $d$ grows, as a result of the increasing complexity of the polynomial description of its boundary. A rational idea is to omit some of the information needed to describe the convex set $\mathcal{M}_d$ in its entirety and leave only the part needed in the problem under consideration.

There are at least two ways to formalize this concept: one can consider low-dimensional projections of the set $\mathcal{M}_d$ or cuts through it – the affine subsets. The two approaches are closely connected, as will be shown in this chapter: both stem from the linear parameterization of a general density operator $\rho \in \mathcal{M}_d$ with a complete basis of Hermitian traceless matrices $\{X_i\}_{i=1}^{d^2-1}$:

$$\rho(x_1, \ldots, x_{d^2-1}) = \frac{\mathbb{1}}{d} + \sum_{i=1}^{d^2-1} x_i X_i. \tag{4.1}$$

The matrices $\{X_i\}_{i=1}^{d^2-1}$ must be traceless to ensure that $\rho$ has always unit trace. For $\rho$ to be positive semidefinite, constraints on $(x_i)_{i=1}^{d^2-1}$ must be placed – nontrivial geometry of $\mathcal{M}_d$ manifests itself this way.

One of the possible simplifications – forcing a subset of the parameters to be zero – is the basis of the idea behind the *spectrahedral geometry*. The other puts no additional restrictions beyond $\rho$ being a density operator, but some of the parameters are ignored – and results in the *numerical range*.

First mentions about the numerical range come from the early 20th century [34, 35]– the image of *Rayleigh quotient* map for a given square, not necessarily Hermitian, matrix $M$ of size $d$,

[34]: Toeplitz (1918), 'Das algebraische Analogon zu einem Satze von Fejér'
[35]: Hausdorff (1919), 'Der Wertvorrat einer Bilinearform'

$$\mathcal{H}_d \setminus \{0\} \ni v \mapsto \frac{\langle v, Mv\rangle}{\langle v, v\rangle} \in \mathbb{C}, \tag{4.2}$$

has been considered in relation to the operator norms and algebraic properties of $M$. The *numerical range*, originally only of mathematical interest, has found use in quantum information science.

**Definition 4.1** *The joint numerical range of k Hermitian operators* $X_1, \ldots, X_k$ *of size d is the subset of* $\mathbb{R}^k$ *defined by*

$$W(X_1, \ldots, X_k) = \{(\langle X_1 \rangle_\rho, \ldots, \langle X_k \rangle_\rho) : \rho \in \mathcal{M}_d\}, \tag{4.3}$$

*where the expression* $\langle X \rangle_\rho = \operatorname{Tr} \rho X$ *is the quantum mechanical expectation value.*

The joint numerical range[1] captures the important information about a variety of problems in a low-dimensional form – hence, it is useful in various scenarios. Much of the power behind the use of numerical range relies on the connections between algebraic properties of the operators in the definition and geometry of the resulting set. The basic properties are the following:

1: What is called here the *joint numerical range* has multiple names in the literature: *joint algebraic numerical range* or simply *numerical range* are not uncommon. The symbol $W$ originates from German word *Wertevorrat*, or the *field of values* of Rayleigh quotient. In some contexts, the *numerical range* denotes the object described by an equation analogous to Eq. (4.3), but with pure states only: $\rho = |\psi\rangle\langle\psi|$. With this definition, the resulting set is not always convex.

**Theorem 4.1** *The joint numerical range* $W(X_1, \ldots, X_k)$ *of k Hermitian operators of size d has the following properties:*

I refer to the numerical range as $W$ in the list.

a) *W is convex and compact.*
b) *If* $k = 2$, *the numerical range* $W(X, Y)$ *interpreted as a subset of the complex plane contains the spectrum of* $X + iY$.
c) *If* $k = 1$, *the numerical range* $W(X)$ *is the closed segment between the smallest and largest eigenvalue of X.*
d) *The numerical range W has full dimensionality (equivalently,* $\operatorname{vol} W > 0$*) if and only if the operators shifted to make them traceless,*

$$\tilde{X}_i = X_i - \frac{\operatorname{Tr} X_i}{d} \mathbb{1}_d, \tag{4.4}$$

*are linearly independent. The dimension of W is equal to the dimension of* $\operatorname{span}\{\tilde{X}_1, \ldots, \tilde{X}_k\}$.
e) *For* $d = 2$ *and* $k = 2$, *the numerical range is an ellipse which degenerates to an interval if the defining operators commute. If* $d = 2$ *and* $k = 3$, *the numerical range is an ellipsoid, provided it is of full dimension.*
f) *Provided full dimensionality, the antipodal points of the numerical range correspond to orthogonal states: for* $\vec{n} \in \mathbb{R}^k$ *and*

$$\rho_+ = \arg\max_{\rho \in \mathcal{M}_d} \langle \vec{n} \cdot \vec{X} \rangle_\rho, \quad \rho_- = \arg\min_{\rho \in \mathcal{M}_d} \langle \vec{n} \cdot \vec{X} \rangle_\rho, \tag{4.5}$$

*the states have zero overlap:* $\operatorname{Tr} \rho_+ \rho_- = 0$.
g) *If there exists an orthonormal basis* $\{|1\rangle, \ldots, |d\rangle\}$ *such that* $X_i |n\rangle = \lambda(n)_i |n\rangle$, *then the numerical range is the convex hull of the joint eigenvalues:*

$$W(X_1, \ldots, X_k) = \operatorname{conv}\{\vec{\lambda}(1), \ldots, \vec{\lambda}(d)\}. \tag{4.6}$$

Existence of such a basis is equivalent to the matrices being simultaneously diagonalizable.

h) *The support function of W is the largest eigenvalue of a certain combination of the defining operators:*

$$h_W(\vec{y}) = \lambda_{\max}\left(\sum_{i=1}^{k} X_i y_i\right). \tag{4.7}$$

*Proof.* a), b), and c) are classic results of the theory of numerical ranges – proofs can be found in [36].

[36]: Psarrakos et al. (2002), *Numerical range: (in) a Matrix Nutshell*

d) Shifting the defining operators by identity only translates the numerical range, and therefore does not change its dimension. Let me assume therefore that the operators $\tilde{X}_1, \ldots, \tilde{X}_k$ are traceless. The numerical range $W$ is an image of the $(d^2-1)$-dimensional set $\mathcal{M}_d$ under the linear map

$$\rho \mapsto (\mathrm{Tr}\,\rho\tilde{X}_1, \ldots, \mathrm{Tr}\,\rho\tilde{X}_k). \tag{4.8}$$

Dimension of the joint numerical range is therefore the rank of this linear map, which is equal to the dimension of the linear subspace spanned by the linear forms[2] $\{\mathrm{Tr}\cdot\tilde{X}_1, \ldots, \mathrm{Tr}\cdot\tilde{X}_k\}$. This in turn is equal to the dimension of $\mathrm{span}\{\tilde{X}_1, \ldots, \tilde{X}_k\}$ itself. If the operators are linearly independent, then the dimension of the numerical range is equal to the number of operators $k$.

2: Here, $\mathrm{Tr}\cdot\tilde{X}$ denotes the function $\rho \mapsto \mathrm{Tr}\,\rho\tilde{X}$.

e) Follows from reasoning analogous to the proof of d) and the fact that $\mathcal{M}_2$ is affinely equivalent to a three-dimensional ball.

f) The state $\rho_+$ maximizing the expectation value of $\vec{n}\cdot\vec{X}$ is a convex combination of pure states from the eigenspace to the largest eigenvalue of $X$ – and similar reasoning applies to $\rho_-$. Therefore, the overlap $\mathrm{Tr}\,\rho_+\rho_-$ may be nonzero only if the largest and smallest eigenvalue eigenspaces are not ortogonal, and in this situation the operator $\vec{n}\cdot\vec{X}$ is proportional to the identity, which is not possible if the dimension of the numerical range $W$ is full as a result of d).

g) The set of quantum states $\mathcal{M}_d$ is the convex hull of pure states, therefore the joint numerical range $W$ is the convex hull of the expectation values over pure states. In this situation, it is straightforward to prove that the latter has exactly the form of Eq. (4.6) and is already convex.

h) Follows from the proof of f) and property c).

□

The first physical application and the most fundamental result comes from the following observation: the states corresponding to the boundary of the numerical range minimize certain expectation values, and are therefore the ground states of related observables (see Figure 4.1):

**Theorem 4.2** *Point $\vec{p}$ on the boundary $\partial W$ of the numerical range $W(X_1, \ldots, X_k)$ with inward-pointing normal $\vec{n}$ is an image of the ground state of the combined operator $\vec{n}\cdot\vec{X} = \sum_{i=1}^{k} n_i X_i$.*

This theorem is an immediate result of the theory of numerical ranges and similar reasoning can be found in the works of Felix Hausdorff [35] and Otto Toeplitz [34].

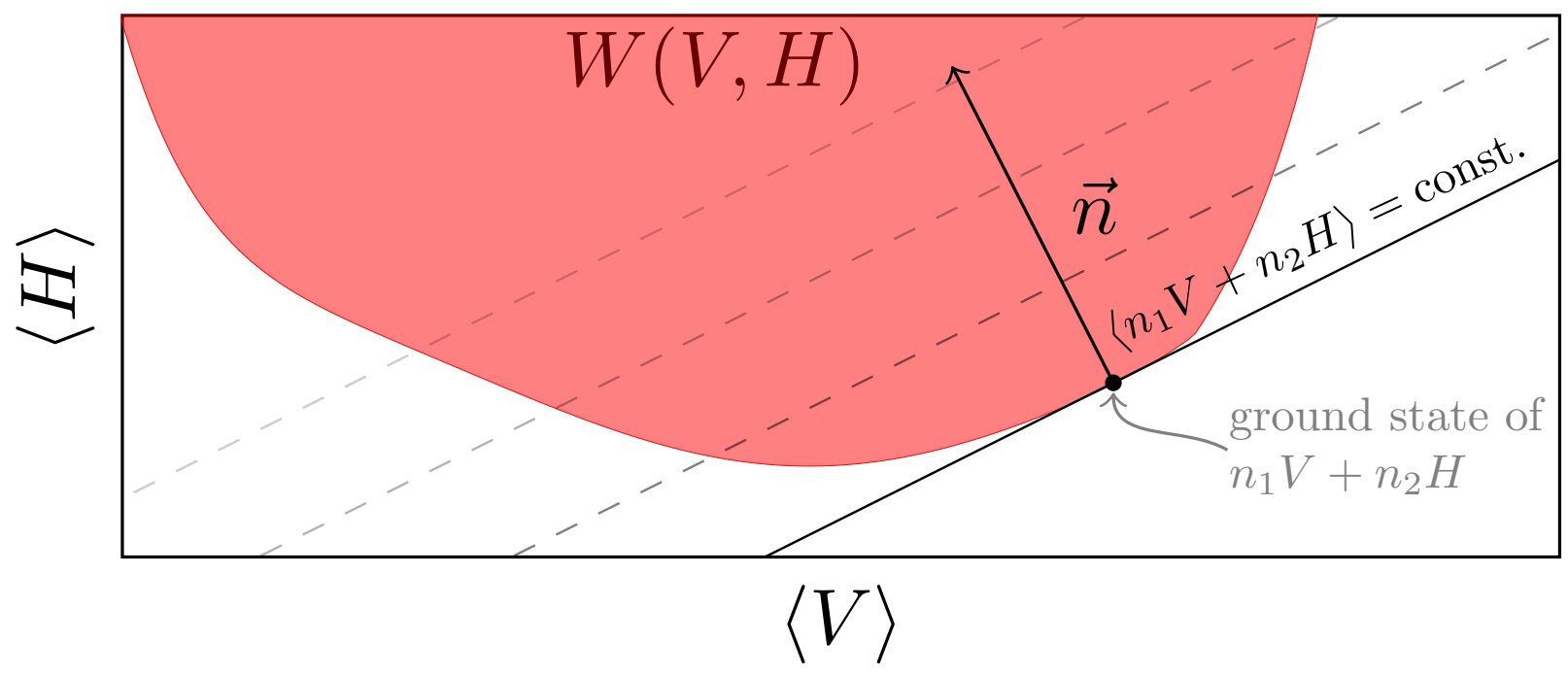

**Figure 4.1:** The boundary of the joint numerical range $W(V, H)$ is formed by the ground states of combinations of $H$ and $V$. This result generalizes to arbitary number of operators.

*Proof.* Joint numerical range $W(X_1, \ldots, X_k)$ is a convex set – therefore, if a point $\vec{p}$ has an inward-pointing normal $\vec{n}$, it is a minimizer of a certain linear form over $W$:

$$\vec{p} = \arg\min_{\vec{x} \in W} \vec{n} \cdot \vec{x}. \tag{4.9}$$

Since $\vec{x} = (\langle X_1 \rangle_\rho, \ldots, \langle X_k \rangle_\rho)$ for some state $\rho$, $\vec{p}$ is an image of a state minimizing $\langle \sum_{i=1}^k n_i X_i \rangle$. This is precisely the definition of the ground state of $\sum_{i=1}^k n_i X_i$. □

The most interesting behavior occurs if the ground state of $\vec{n} \cdot \vec{X}$ changes rapidly as the parameter vector $\vec{n}$ is varied – that is, at the point of a first-order phase transition. In such cases, the support points generically suddenly jump, leading to flat parts of the boundary. The properties of the flat parts of the boundary $\partial W$ may serve as a basis for the classification of numerical ranges, which is discussed in the next section.

The formalism of numerical ranges might at this point look like a reiteration of already known methods – it does yield the same information as other approaches. However, it does so in a concise way, allowing for straightforward visualisation and a different viewpoint on the problems. Let me present one of the unexpected applications of numerical ranges: the distinguishability of unitary operations discussed in [37].

[37]: Gawron et al. (2010), 'Restricted numerical range: a versatile tool in the theory of quantum information'

**Example 4.1** The problem is to determine when two unitaries $U$ and $V$ can be distinguished using single-shot experiment. This may look like a formidable task at first glance, however there is a simple geometrical approach to this problem. If there exists a vector $|\psi\rangle$ such that $U|\psi\rangle$ is orthogonal to $V|\psi\rangle$, one can distinguish between the outcomes of an experiment with an unknown quantum channel – simply apply the unknown unitary of $|\psi\rangle$ and measure the projector onto $U|\psi\rangle$ on the output state. Then, a positive result is *equivalent* to $U$ being implemented instead of $V$.

How to determine such a state? Consider the operator $U^\dagger V$ which decomposes into Hermitian and anti-Hermitian parts:

$$U^\dagger V = X + iY, \;\; \Re \langle U^\dagger V \rangle = \langle X \rangle, \;\; \Im \langle U^\dagger V \rangle = \langle Y \rangle . \quad (4.10)$$

The operator $U^\dagger V$ is unitary, and thus normal – therefore, $X$ commutes with $Y$ and they do share the eigenvectors. Basic properties of numerical ranges then ascertain that $W(X, Y)$ is a polytope, the convex hull of the joint eigenvalues. Existence of $|\psi\rangle$ discriminating between $U$ and $V$ is then equivalent to the point $(0, 0)$ being contained in $W(X, Y)$, since this determines that

$$\langle \psi | U^\dagger V | \psi \rangle = 0, \quad (4.11)$$

or, in other words: that $U |\psi\rangle$ is orthogonal to $V |\psi\rangle$.

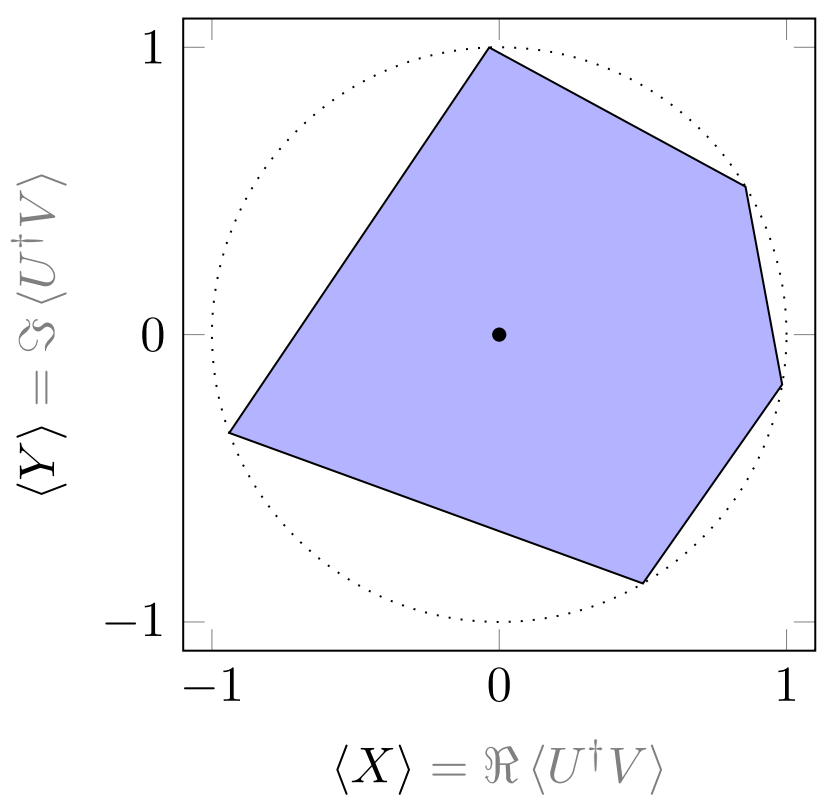

**Figure 4.2:** An unitary $U$ is one-shot distinguishable from the unitary $V$ if the numerical range of $X + iY = U^\dagger V$ contains the origin – $\vec{0} \in W(X, Y)$. In such a case, there exists a state $|\psi\rangle$ such that $U |\psi\rangle$ is orthogonal to $V |\psi\rangle$, which can be therefore perfectly distinguished.

## 4.1 Spectrahedra

The geometry of the numerical ranges can be interpreted using the dual description offered by affine subsets of the cone of positive semidefinite operators:

**Definition 4.2** *A spectrahedron centered at the operator $X_0 \in \mathcal{A}_d$ and spanned by $k$ operators $X_1, \ldots, X_k \in \mathcal{A}_d$ is the set of parameters corresponding to positive semidefinite operators:*

$$\mathrm{Spec}(X_0; X_1, \ldots, X_k) = \{\vec{x} \in \mathbb{R}^k : X_0 + \sum_{i=1}^{k} X_i x_i \succcurlyeq 0\}. \quad (4.12)$$

*The set of parameters is isomorphic to the set of operators itself:*

$$\{X \in X_0 + \mathrm{span}\{X_1, \ldots, X_k\} : X \succcurlyeq 0\}. \quad (4.13)$$

*Therefore, the notation* Spec *may refer to the set of operators, when it is unambiguous to do so (e.g., in expressions like $X \in$ Spec, the symbol $X$ refers to a matrix).*

The symbol $\mathcal{A}_d$ here is the set of Hermitian matrices of size $d$.

A well-known result [38] is the duality between numerical ranges and the respective spectrahedra.

[38]: Ramana (1997), 'An exact duality theory for semidefinite programming and its complexity implications'

**Theorem 4.3** *Let $(X_i)_{i=1}^k$ be a sequence of Hermitian operators of size $d$. The numerical range $W(X_1, \ldots, X_k)$ is polar to* $\mathrm{Spec}(\mathbb{1}; -X_1, \ldots, -X_k)$.

This theorem has been discovered independently many times, even in fields unrelated to numerical ranges – this is the case with [38].

*Proof.* The point $\vec{y}$ is contained in the polar $W^\circ$ to $W(X_1, \ldots, X_k)$ if and only if for every $\vec{x} \in W$, $\vec{y} \cdot \vec{x} \leq 1$. The vectors $\vec{x}$ are expectation values, so this can be rewitten as

$$\forall \rho : \sum_{i=1}^{k} y_i \langle X_i \rangle_\rho \leq 1. \quad (4.14)$$

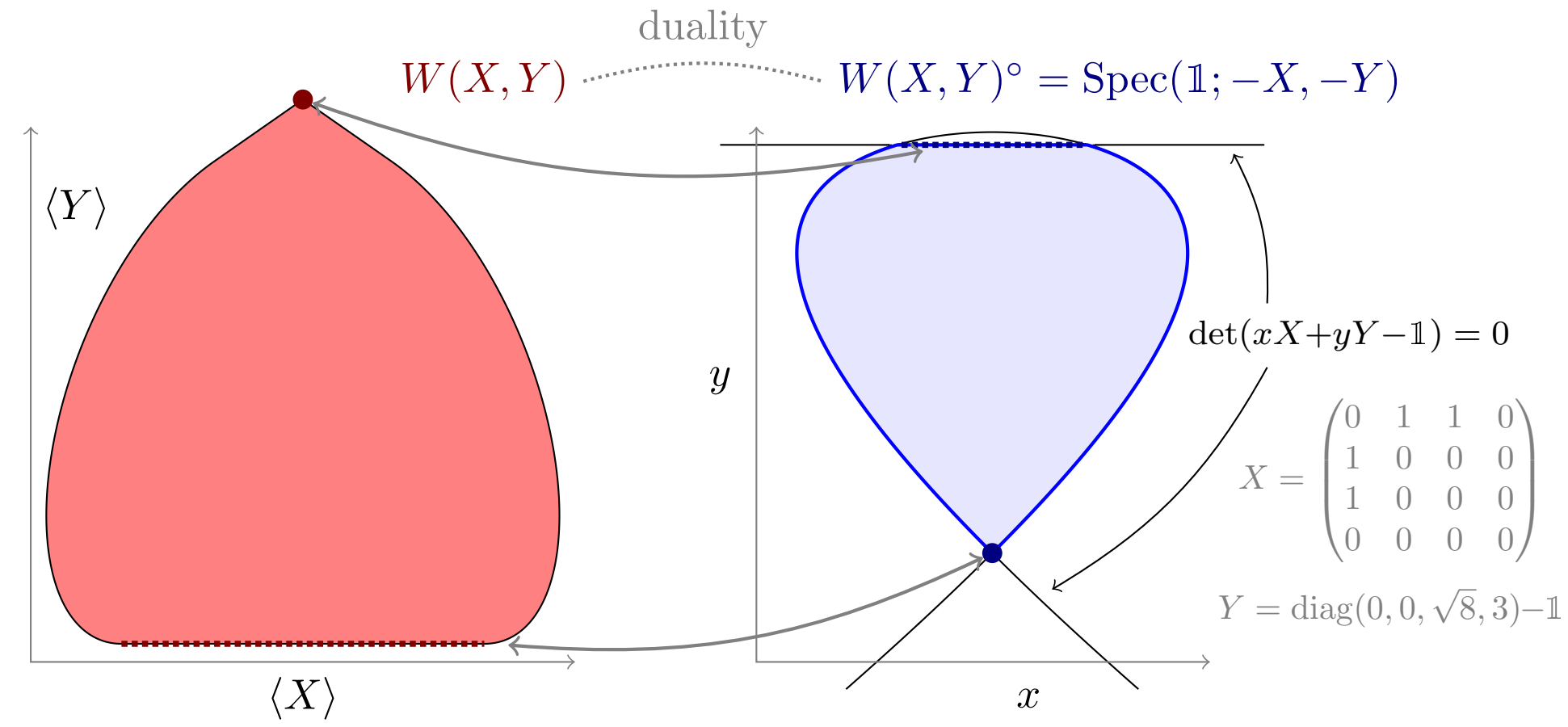

**Figure 4.3:** Joint numerical ranges are dual (polar) to spectrahedra – sets of parameters defining a positive semidefinite operators. Points at the boundary of a spectrahedron satisfy certain polynomial equation – an analogue of characteristic polynomial. Nonanalytical parts of the boundary of the numerical range are transformed to dual counterparts in the boundary of the spectrahedron: cusps map to segments and vice versa.

This is equivalent to saying that the largest eigenvalue of $\sum_{i=1}^{k} y_i X_i$ is smaller than one – that is,

$$\mathbb{1} - \sum_{i=1}^{d} y_i X_i \succcurlyeq 0. \tag{4.15}$$

Therefore, $\vec{y} \in W^\circ$ if and only if it is contained in the spectrahedron $\mathrm{Spec}\,(\mathbb{1}, -X_1, \ldots, -X_k)$. ☐

Spectrahedral geometry can be easily described using polynomial equations. In some cases, notably two Hermitian operators, an explicit construction of its polar – the joint numerical range – can be given. In any case, the boundary of the joint numerical range is a semialgebraic set. The construction starts with the result of Theorem 4.3: spectrahedra and numerical ranges are connected through polar duality. The boundary of a spectrahedron is composed of matrices of incomplete rank, since only then an eigenvalue can change sign, and therefore admits a polynomial description:

$$\vec{y} \in W^\circ \implies \overbrace{\det\left(\mathbb{1} - \vec{y} \cdot \vec{X}\right)}^{f} = 0. \tag{4.16}$$

Lemma 3.3 (page 31) additionally implies that the normal vector of polar set at $\vec{y}$ is proportional to the maximizer in the original set. Combining these results, one arrives at the set of polynomial equations linking the vectors at the boundary of the spectrahedron $\vec{y} \in \partial\,\mathrm{Spec}(\mathbb{1}, -X_1, \ldots, -X_k)$, the points at the boundary of the numerical range $\vec{x} \in \partial W(X_1, \ldots, X_k)$:

$$\begin{aligned} \overbrace{\det\left(\mathbb{1} - \vec{y} \cdot \vec{X}\right)}^{f} &= 0, \\ \nabla_{\vec{y}} f - \lambda \vec{x} &= \vec{0}, \\ \vec{y} \cdot \vec{x} - 1 &= 0. \end{aligned} \tag{4.17}$$

The equations can be simplified using the Gröbner basis approach – the goal is to eliminate $\vec{y}$ and leave $\vec{x}$ only and provide a polynomial description of the numerical range boundary $W$. For $k = 2$ operators, this approach does not fail – a result first proved by Kippenhahn [39]. The exact statement of the theorem uses projective algebraic geometry to avoid some of the ambiguities[3] of Eq. (4.17):

[39]: Kippenhahn (1951), 'Über den Wertevorrat einer Matrix'

3: While the set of solutions to Eq. (4.17) is guaranteed to contain the important points of $\partial W$, it may also include other unphysical solutions. Slightly more strict, although less intuitive approach, solves this problem.

**Theorem 4.4** (Kippenhahn)
*The joint numerical range $W(X_1, X_2)$ of two operators $X_1, X_2$ of size $d$ is the convex hull of the solutions $\vec{x} \in \mathbb{R}^2$ of the polynomial equations in $\lambda, \vec{x} \in \mathbb{R}^2, \vec{y} \in \mathbb{R}^3$:*

$$\overbrace{\det(\mathbb{1}_d y_0 + y_1 X_1 + y_2 X_2)}^{f} = 0, \tag{4.18}$$
$$(1, x_1, x_2) - \lambda \vec{\nabla}_{\vec{y}} f = 0.$$

In addition, an explicit formula for the boundary for the $k = 2$ case is known [40]:

[40]: Fiedler (1981), 'Geometry of the Numerical Range of Matrices'

**Theorem 4.5** (Fiedler) *If the boundary $\partial W(X, Y)$ of the numerical range $W(X, Y)$ is an irreducible algebraic curve, then it is defined by the polynomial equation*

$$\det \begin{pmatrix} \langle X, X\rangle & \langle X, Y\rangle & \langle X, \mathbb{1}_d\rangle & x\mathbb{1}_n \\ \langle X, Y\rangle & \langle Y, Y\rangle & \langle Y, \mathbb{1}_d\rangle & y\mathbb{1}_n \\ \langle X, \mathbb{1}_d\rangle & \langle Y, \mathbb{1}_d\rangle & \mathbb{1}_n & \mathbb{1}_n \\ x\mathbb{1}_n & y\mathbb{1}_n & \mathbb{1}_n & 0 \end{pmatrix} = 0, \tag{4.19}$$

*where $n = 1/2 \times d(d-1)$ and $\langle \cdot, \cdot \rangle$ denotes the second mixed compound defined as*

$$\langle X, Y\rangle_{(i,r),(j,s)} = \frac{1}{2}\left(X_{ir}Y_{js} + X_{js}Y_{ir} - X_{is}Y_{jr} - X_{jr}Y_{is}\right). \tag{4.20}$$

For $k \geq 3$ operators, the direct analogue of Kippenhahn's theorem fails [41] – the surface defined analogously to Eq. 4.18 may contain parts not present in the boundary of the numerical range (e.g., infinite lines). This phenomenon arises due to the implicit complex structure of the field over which the polynomials are defined [42]. Regularization procedure for this case exists, but requires tools outside the typical approach involving Gröbner bases [43]. In the typical usage, this does not pose a problem, as the spurious parts of the resulting variety can be easily identified.

[41]: Chien et al. (2010), 'Joint numerical range and its generating hypersurface'

[42]: Schwonnek et al. (2020), 'The Wigner distribution of n arbitrary observables'

[43]: Plaumann et al. (2021), 'Kippenhahn's Theorem for joint numerical ranges and quantum states'

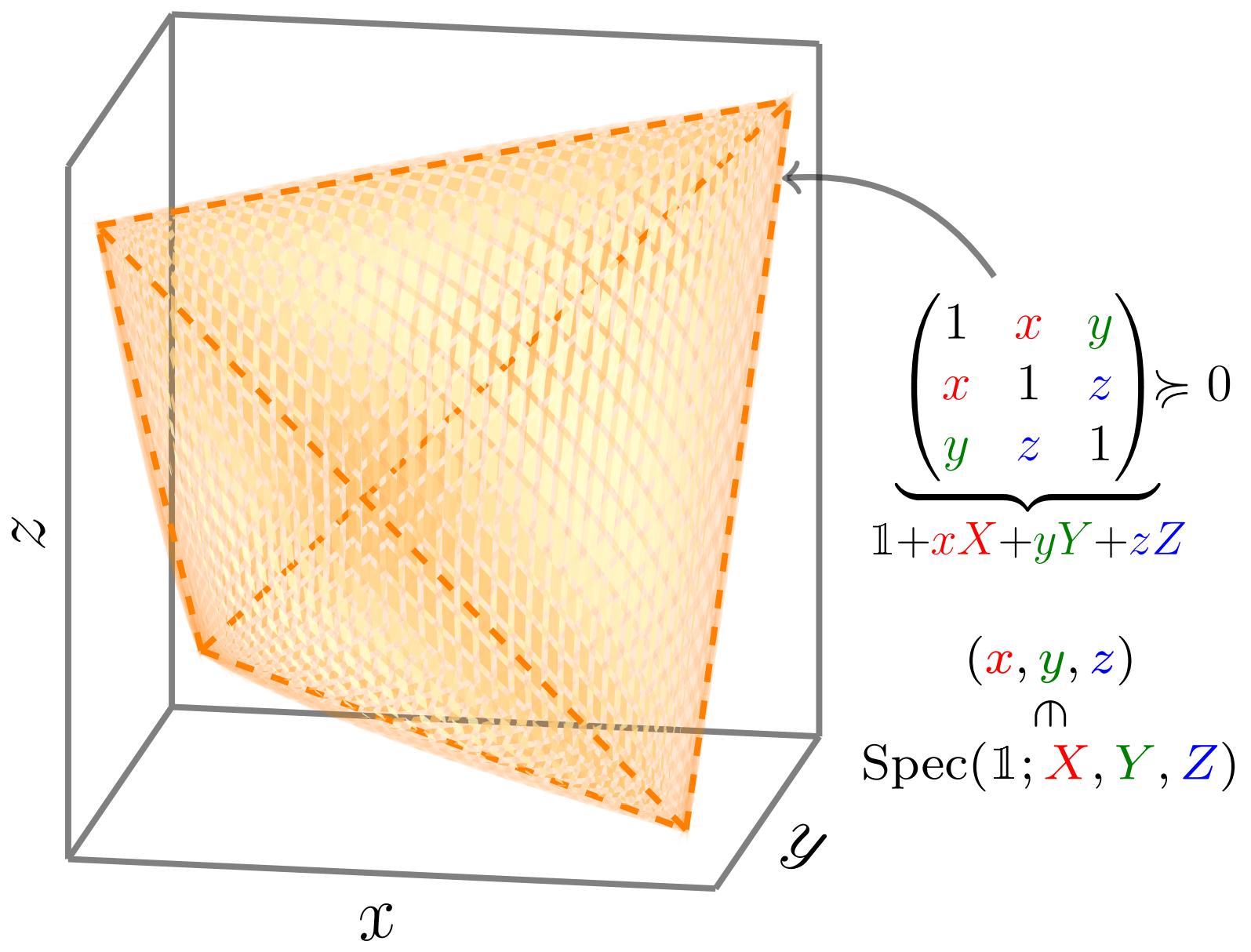

**Figure 4.4:** The *elliptope*, a well-known example of a spectrahedron. Elliptope resembles inflated tetrahedron and is defined by a linear matrix inequality presented on the picture. As a result of the Theorem 4.3, this set is polar to the numerical range plotted for $e = 4$ and $s = 0$ of Figure 4.5. The stereographic (3D) version of this picture is available in Figure A.1 (Appendix A, page 102).

### 4.1.1 Semidefinite optimization

Spectrahedra play an important role in the class of optimization problems named *semidefinite optimization* [44]. The basic statement of the problem has the following form: given matrices $C$ and $\{A_i\}_{i=1}^n$ (elements of $\mathcal{A}_d$, the set of Hermitian matrices of size $d$) and a vector $\vec{b} \in \mathbb{R}^n$,

[44]: Michałek et al. (2021), *Invitation to Nonlinear Algebra*

- minimize the following expression over $X \in \mathcal{A}_d$:

$$\mathrm{Tr}\, X^\dagger C, \tag{4.21}$$

- subject to

$$\mathrm{Tr}\, X^\dagger A_i = b_i, \text{ for } i = 1, \dots, n, \text{ and } X \succcurlyeq 0. \tag{4.22}$$

**Definition 4.3** *A matrix X consistent with the constraints of Eq.* (4.22) *is called feasible.*

The constraints naturally define a spectrahedron: the set of matrices fulfilling Eq. (4.22) is exactly an affine subset of the set of positive semidefinite matrices. If there exists any matrix $X_0$ consistent with the constraints, then the spectrahedron is generated by the linearly independent matrices $\{B_i\}_{i=1}^{d^2-n}$ orthogonal to all of $A_i$:

That is, for any pair of $A_i$ and $B_j$, $\mathrm{Tr}\, A_i^\dagger B_j = 0$.

$$X \text{ is feasible if and only if } X \in \mathrm{Spec}(X_0; B_1, \dots B_{d^2-n}). \tag{4.23}$$

There is an additional structure to this, directly presenting and utilizing the spectrahedral geometry: the semidefinite optimization problem has a dual formulation. For a given vector $\vec{b} \in \mathbb{R}^n$ and Hermitian matrices $C, A_1, \dots, A_n$, the optimization problem can be written in the form of

$$\min_{X \succcurlyeq 0} \max_{\vec{y} \in \mathbb{R}^n} \left[ \mathrm{Tr}(X^\dagger C) + \sum_{i=1}^{n} y_i \left( b_i - \mathrm{Tr}(X^\dagger A_i) \right) \right]. \tag{4.24}$$

A global optimum of this expression – if it exists – can only be found on the subspace enforced by the constraints (Eq. (4.22)) – any other $X$ causes the sum to be unbounded during the maximization over $\vec{y}$.[4] The max-min inequality[5] then can be used to ascertain that the following expression is not larger than Eq. (4.24):

$$\max_{\vec{y}\in\mathbb{R}^n}\min_{X\succcurlyeq 0}\left[\vec{b}\cdot\vec{y}+\operatorname{Tr}\left(X^\dagger\left[C-\sum_{i=1}^{n}y_iA_i\right]\right)\right]. \tag{4.25}$$

Now, the minimization over any positive semidefinite $X$ ensures that the matrix $\left(C-\sum_{i=1}^{n}y_iA_i\right)$ is positive semidefinite.[6] The dual formulation has thus the following form: given $\vec{b}\in\mathbb{R}^n$ and Hermitian matrices $C$ and $\{A_i\}_{i=1}^{n}$ of size $d$,

- maximize
$$\vec{b}\cdot\vec{y}, \tag{4.26}$$
- subject to
$$C-\sum_{i=1}^{n}y_iA_i\succcurlyeq 0. \tag{4.27}$$

The constraints explicitly define a spectrahedron. The dual problem – maximization of $\vec{b}\cdot\vec{y}$ – always has a solution which bounds the solution to the primal problem – minimization of $\operatorname{Tr} X^\dagger C$ – from below. The solutions are equal – the bound is tight – when certain conditions are met [38].

**Theorem 4.7** (Slater) *The primal and dual semidefinite optimization problems have equal solutions if any of the following conditions are met:*

*a) There exists a feasible $X$ of the primal problem which is strictly positive definite, i.e., a matrix $X$ satisfying the constraints has all eigenvalues strictly positive.*
*b) There exists a vector $\vec{y}\in\mathbb{R}^n$ such that the Hermitian matrix $C-\sum_{i=1}^{n}y_iA_i$ is strictly positive definite.*

Semidefinite optimization appears in multiple contexts: correlation matrices appearing in statistics are positive semidefinite and thus one can optimize statistical quantities easily; basic problems of computer science can often be translated (or approximated using) to the language of matrix inequalities; it also forms a basis for entanglement detection [45].

The approach of semidefinite programming approach has its limitations, however. It is straightforward to *minimize the largest eigenvalue* over a spectrahedron – and similarly, maximize the smallest eigenvalue – but the *minimization of the smallest eigenvalue* is much more involved. Many of the problems discussed in the

4: This is essentially a Lagrange multiplier method.

5: The original inequality is stated in suprema and infima, but in this case the extrema are attained.

**Lemma 4.6** (Max-min inequality) *For any $f: A\times B\to\mathbb{R}$,*

$$\sup_{a\in A}\inf_{b\in B} f(a,b)\le\inf_{b\in B}\sup_{a\in A} f(a,b).$$

6: If it is not, the minimization becomes unbounded.

[38]: Ramana (1997), 'An exact duality theory for semidefinite programming and its complexity implications'

[45]: Brandao et al. (2004), 'Robust semidefinite programming approach to the separability problem'

**Example 4.2** Minimization of the largest eigenvalue over a spectrahedron $S=\operatorname{Spec}(C;A_1,\dots,A_n)$ is equivalent to the following semidefinite optimization problem:

- minimize $t$,
- subject to $t\mathbb{1}-X\succcurlyeq 0$, where $X\in S$.

later part of this chapter are explicitly the kind of questions that semidefinite optimization can not provide an answer for.[7] Other methods are therefore needed – and those based on joint numerical ranges do provide solutions to these problems.

7: Minimization of the sum of variances or determination of the smallest separable expectation value explicitly call for the *minimization of the smallest eigenvalue*.

## 4.2 Classification(s)

In this section I will provide a classification of joint numerical ranges of three operators of size three, originally appearing in [1]. The reasoning presented here is a significant simplification of the original proof. Let me start with the following, most general statement about flat parts of the boundary of a general numerical range.

[1]: Szymański et al. (2018), 'Classification of joint numerical ranges of three hermitian matrices of size three'

Faces are introduced in Definition 3.3. Nontrivial faces $F$ of $W$ are those faces which are neither singletons nor the whole set $W$.

**Theorem 4.8** *Let $\{X_i\}_{i=1}^{k}$ act on $d$-dimensional system. For every nontrivial face $F$ on the boundary of the numerical range $W(X_1, \ldots, X_k)$, there exist operators $\{Y_i\}_{i=1}^{k-1}$ acting on at most $(d-1)$-dimensional system such that $F$ is an affine image of $W(Y_1, \ldots, Y_{k-1})$.*

*Proof.* A nontrivial face $F$ is a set of maximizers of a linear functional over the numerical range:

$$F = \underset{\langle \vec{X} \rangle \in W}{\arg\max}\, \vec{n} \cdot \langle \vec{X} \rangle \,. \tag{4.28}$$

An image of a quantum state $\rho$ can belong to $F$ if and only if every decomposition of $\rho$ into the convex hull of projectors,

Here, all $p_i > 0$ and $\sum_{i=1}^{n} p_i = 1$.

$$\rho = \sum_{i=1}^{n} p_i \left|\psi_i\right\rangle \left\langle\psi_i\right| , \tag{4.29}$$

contains state vectors $|\psi_i\rangle$ which all are eigenvectors to the largest eigenvalue $\lambda_{\max}$ of $\vec{n} \cdot \vec{X}$. This can be easily proven by assuming the opposite – in such case, the expectation value $\langle \vec{n} \cdot \vec{X} \rangle_\rho$ is lower than $\lambda_{\max}$, so the image of $\rho$ can not lie in $F$.

The set of maximizer states,

$$M = \underset{\rho \in \mathcal{M}_d}{\arg\max}\, \langle \vec{n} \cdot \vec{X} \rangle_\rho \,, \tag{4.30}$$

is therefore affinely equivalent to the entire set of quantum states $\mathcal{M}_{d'}$ of dimension $d'$ equal to the degeneracy of $\lambda_{\max}$.This follows from the construction of the set of states as the convex hull of projectors – every state $\rho$ described by Eq. (4.29) has exactly the required form. Therefore, the face $F$ is formed by the joint numerical range of operators restricted to $\mathcal{M}_{d'}$ and has dimension

The following result is a direct application of Theorem 4.8 for the case of three observables of size three:

**Lemma 4.9** *The flat parts in the numerical range of three operators acting on a qutrit are segments and ellipses. Every pair of flat parts share a common point.*

*Proof.* In this case, a flat part can only arise as an affine image of a qubit numerical range – hence, the flat parts are segments and ellipses. Since two two-dimensional subspaces in a three-dimensional space do share a common point, all flat parts must intersect. □

of at most $k-1$ (since it lies on a hyperplane). By selection of correct combinations of operators $X_1, \ldots, X_k$, the affine equivalency of $F$ with lower-dimensional joint numerical range can be proved. □

These observations come in handy during the classification of all possible joint numerical ranges of three Hermitian operators of size three.

**Theorem 4.10** *Let $\vec{X} = X_1, X_2, X_3$ denote a triple of Hermitian matrices of size three, linearly independent with identity which do not share a common eigenvector. The joint numerical range $W(X_1, X_2, X_3)$ has the following properties:*

*a) $W(X_1, X_2, X_3)$ is convex and has a nonzero volume.*
*b) Let $\mathcal{F}$ denote the nontrivial exposed faces of $W(X_1, X_2, X_3)$ contained in the $\partial W(X_1, X_2, X_3)$ – in other words, the flat parts of the boundary. Elements of $\mathcal{F}$ are segments and ellipses. Pairs of exposed faces share common points: for $F, F' \in \mathcal{F}$ such that $F \neq F'$, the intersection $F \cap F'$ is a single point.*
*c) If $F \in \mathcal{F}$ intersects a segment $S \in \mathcal{F}$, it does so at one of the endpoints of $S$.*
*d) At most one segment is contained in $\mathcal{F}$.*
*e) There are at most 4 ellipses on the boundary. If a segment is present, at most two ellipses belong to $\mathcal{F}$.*

Linear independence with identity here means that no combination $\sum_{i=1}^{3} n_i X_i$ is proportional to $\mathbb{1}_3$.

*Proof.* a) If $W(X_1, X_2, X_3)$ has zero volume, it lies on a surface contained in $\mathbb{R}^3$. If the normal vector of this surface is $\vec{n}$, then $\vec{X} \cdot \vec{n}$ is proportional to the identity matrix, which contradicts the assumptions.

b) This is a result of Lemma 4.9.

c) Suppose the opposite occurs: there exists a segment $S \in \mathcal{F}$ and other exposed face $F \in \mathcal{F}$ such that $S \cap F$ is not one of the endpoints of $S$. Let the point of intersection be named $\vec{p}$. $S$ is an image of two-dimensional subspace, hence there exist two distinct pure vectors such that

$$\vec{p} = \langle \alpha | \vec{X} | \alpha \rangle = \langle \beta | \vec{X} | \beta \rangle . \tag{4.31}$$

The point $\vec{p}$ is contained in the other exposed face $F$ corresponding to the normal vector $\vec{n}$. Pick any $\vec{q} \in F$ distinct from $\vec{p}$, with its pure preimage being $|\gamma\rangle$. The three states $|\alpha\rangle, |\beta\rangle, |\gamma\rangle$ are linearly independent eigenvectors to the extremal eigenvalue of $\vec{n} \cdot \vec{X}$, which is therefore proportional to identity – the eigenvectors span the entire space. This contradicts the assumptions: the numerical range is flat, which violates the first property.

Here, $\vec{X} \cdot \vec{n} = \sum_{i=1}^{3} n_i X_i$. Similarly, $\langle \psi | \vec{X} | \psi \rangle$ denotes the vector of expectation values $\left( \langle X_i \rangle_\psi \right)_{i=1}^{3} \in \mathbb{R}^3$.

This reasoning does not apply to endpoints: their pure preimages are unique.

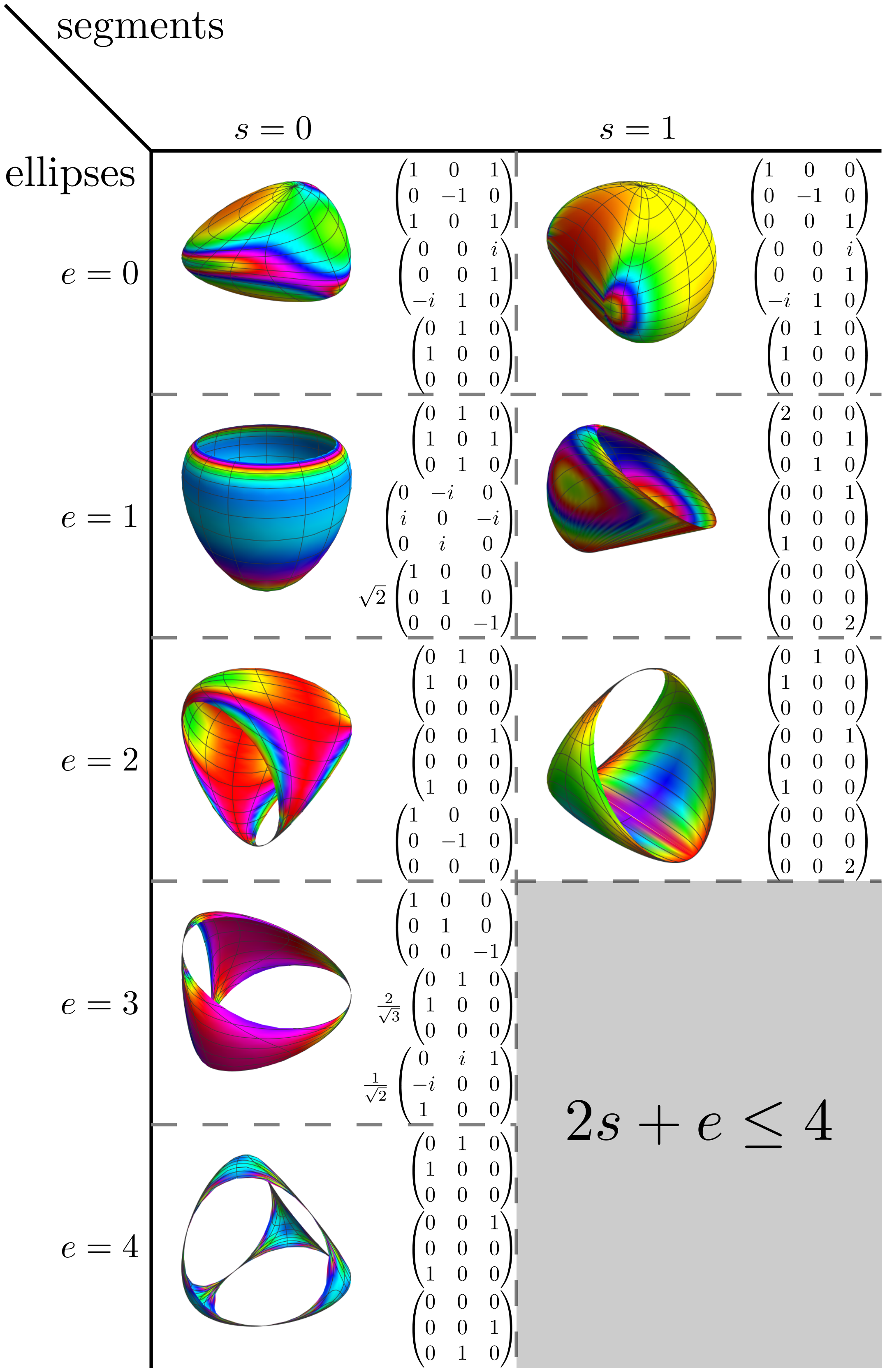

**Figure 4.5:** Nondegenerate joint numerical ranges of three Hermitian matrices of size three belong to one of eight classes, characterized by the number $e$ of ellipses and $s$ of segments in the boundary $\partial W$. Examples of the numerical ranges $W(X_1, X_2, X_3)$ belonging to each possible class are presented along with the triples of matrices $X_i$. The numerical ranges are all convex, but only their boundaries (excluding ellipses) are shown and colored according to the boundary curvature. For $s = 1$ segment maximum of $e = 2$ ellipses is possible. This classification was obtained in an article coauthored by the author of this thesis [1].

d) Suppose two such segments exist: $A, B \in \mathcal{F}$, the point of intersection corresponding to the state $|\psi\rangle$, and the other endpoints to $|\alpha\rangle$ and $|\beta\rangle$ (see Figure 4.6).
Since $A \neq B$, the segments have different normal vectors $\vec{n}_A, \vec{n}_B$, respectively. The two segments lie on a surface with normal $\vec{u}$.[8] This means that the expectation value is constant

$$\langle \vec{u} \cdot \vec{X} \rangle_\phi = \text{const.} \tag{4.32}$$

for the superpositions of the following form:

$$|\phi\rangle \in \text{span}\{|\psi\rangle, |\alpha\rangle\} \cup \text{span}\{|\psi\rangle, |\beta\rangle\}, \langle\phi|\phi\rangle = 1. \tag{4.33}$$

This fact along with the polarization identity leads to

$$\langle\alpha|\vec{u} \cdot \vec{X}|\psi\rangle = \langle\beta|\vec{u} \cdot \vec{X}|\psi\rangle = 0. \tag{4.34}$$

Now, since the matrices are of size three, $\vec{u} \cdot \vec{X} |\psi\rangle$ can be written as a superposition of $|\psi\rangle, |\alpha\rangle$, and $|\beta\rangle$. While in general $\langle\alpha|\beta\rangle \neq 0$ even though $\langle\alpha|\psi\rangle = \langle\beta|\psi\rangle = 0$, this equation still implies that

$$\vec{u} \cdot \vec{X} |\psi\rangle \propto |\psi\rangle. \tag{4.35}$$

Therefore, the state $|\psi\rangle$ is an eigenstate of three linearly independent operators: $\vec{n}_A \cdot \vec{X}, \vec{n}_B \cdot \vec{X}, \vec{u} \cdot \vec{X}$. It is therefore an eigenstate of all three $X_1, X_2, X_3$, which violates the assumptions.

e) The connection graph of the exposed faces $\mathcal{F}$ is full: every two elements share a common point. A convex polytope composed of linear extensions of the exposed faces has the same connection graph; however, all such graphs are dual to nets of convex polyhedra, and therefore planar. Since the largest full planar graph has four vertices [46], the set of faces $\mathcal{F}$ contains at most four exposed faces: $e \leq 4$. If a segment is present ($s = 1$), as a result of the property *c)* it may connect to at most two other exposed faces ($e \leq 2$).

□

8: The normal $\vec{u}$ does not lie in the span of $\vec{n}_A$ and $\vec{n}_B$. If it did, $A$ and $B$ would be parts of a bigger exposed face.

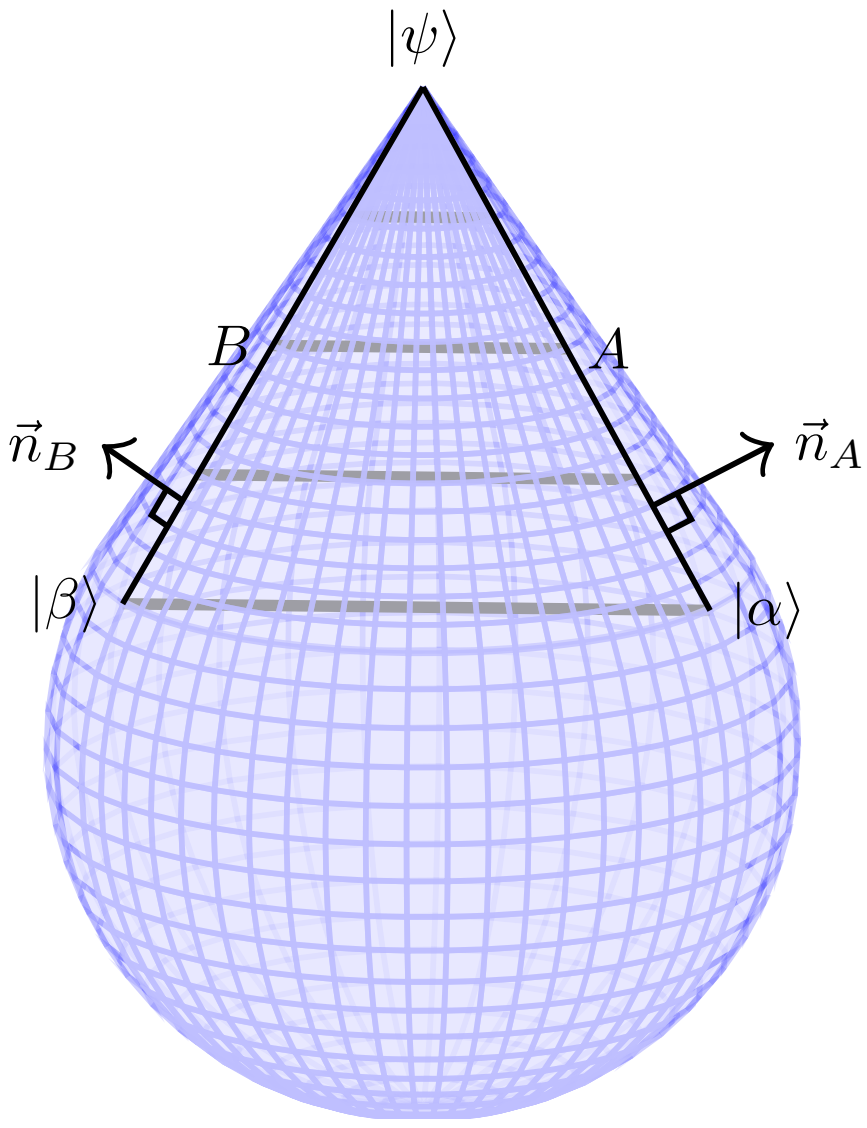

**Figure 4.6:** Graphical representation of the proof of the property d) of Theorem 4.10: if two segments $A$ and $B$ are present, they share a common point being an image of the state $|\psi\rangle$. $A$ and $B$ lie on a single surface (sketched with gray lines) with normal $\vec{u}$, which proves that $|\psi\rangle$ is an eigenstate of $\vec{u} \cdot \vec{X}$.

Other shapes are possible if the assumptions are not met: when $\{X_1, X_2, X_3, \mathbb{1}_3\}$ are linearly dependent, the resulting shape is flat and affinely equivalent to the known case of two qutrit operators [47]. If there exists a common eigenvector of all three matrices, they are unitarely equivalent to the three matrices with block diagonal form

$$X_i = \left(\begin{array}{c|cc} x_i & 0 & 0 \\ \hline 0 & \multicolumn{2}{c}{\multirow{2}{*}{$Y_i$}} \\ 0 & & \end{array}\right). \tag{4.36}$$

[46]: Kuratowski (1930), 'Sur le problème des courbes gauches en topologie'

[47]: Keeler et al. (1997), 'The numerical range of 3×3 matrices'

Thus, the numerical range decomposes into a convex hull of two sets:

$$W(X_1, X_2, X_3) = \text{conv}\left(\{\vec{x}\} \cup W(Y_1, Y_2, Y_3)\right). \tag{4.37}$$

The second part, $W(Y_1, Y_2, Y_3)$, is a numerical range of three *qubit* operators, and thus can only form an ellipsoid, a flat ellipse embedded in the 3D space, a segment, or a point (Theorem 4.1, property e)). In several cases, when $\vec{x} \notin W(Y_1, Y_2, Y_3)$, there exist multiple segments in the boundary of $W(X_1, X_2, X_3)$, in a manner consistent with the reasoning presented in the above proof. One of such cases is presented in Figure 4.6 – the numerical range is a convex hull of a point and a ball.

### 4.2.1 Experimental analysis of $\mathscr{M}_3$ and numerical ranges

The quantum-mechanical nature of the numerical ranges allows for an experimental realization of their geometric structure. Recently, the classification presented in Theorem 4.10 and originally described in [1] was confirmed experimentally using a photonic scheme [7].

[1]: Szymański et al. (2018), 'Classification of joint numerical ranges of three hermitian matrices of size three'

[7]: Xie et al. (2020), 'Observing geometry of quantum states in a three-level system'

Empirical detection of the geometry of quantum states relies on a single photon source. The photon, upon creation, undergoes an arbitrary unitary evolution to three target modes, encoding the qutrit state. The measurement is performed by further unitary operations, which realize the active transformation of the basis to the eigenbasis of an arbitrary Hermitian operator with subsequent photon detection.

## 4.3 Uncertainty relations

A particular type of uncertainty relations is of use in finite-dimensional systems: the one involving the sum of variances. If $|\psi\rangle$ is an eigenstate of $X$, then the variance $\Delta^2 X_\psi$ vanishes – and eigenstates are always attainable in finite dimension. Thus, state-independent uncertainty relations involving the *product* of variances are always trivial.[9]

9: For any $X$ and $Y$, $\Delta^2 X \Delta^2 Y \geq 0$ and the bound is tight.

This problem does not arise in the case of the sum of variances – therefore, such an approach has been employed in finite-dimensional problems [48]. The goal is therefore to provide a lower bound $C$ for the sum of two variances of arbitrary operators $X, Y$:

[48]: He et al. (2011), 'Planar quantum squeezing and atom interferometry'

$$\Delta^2 X + \Delta^2 Y \geq C(X, Y). \tag{4.38}$$

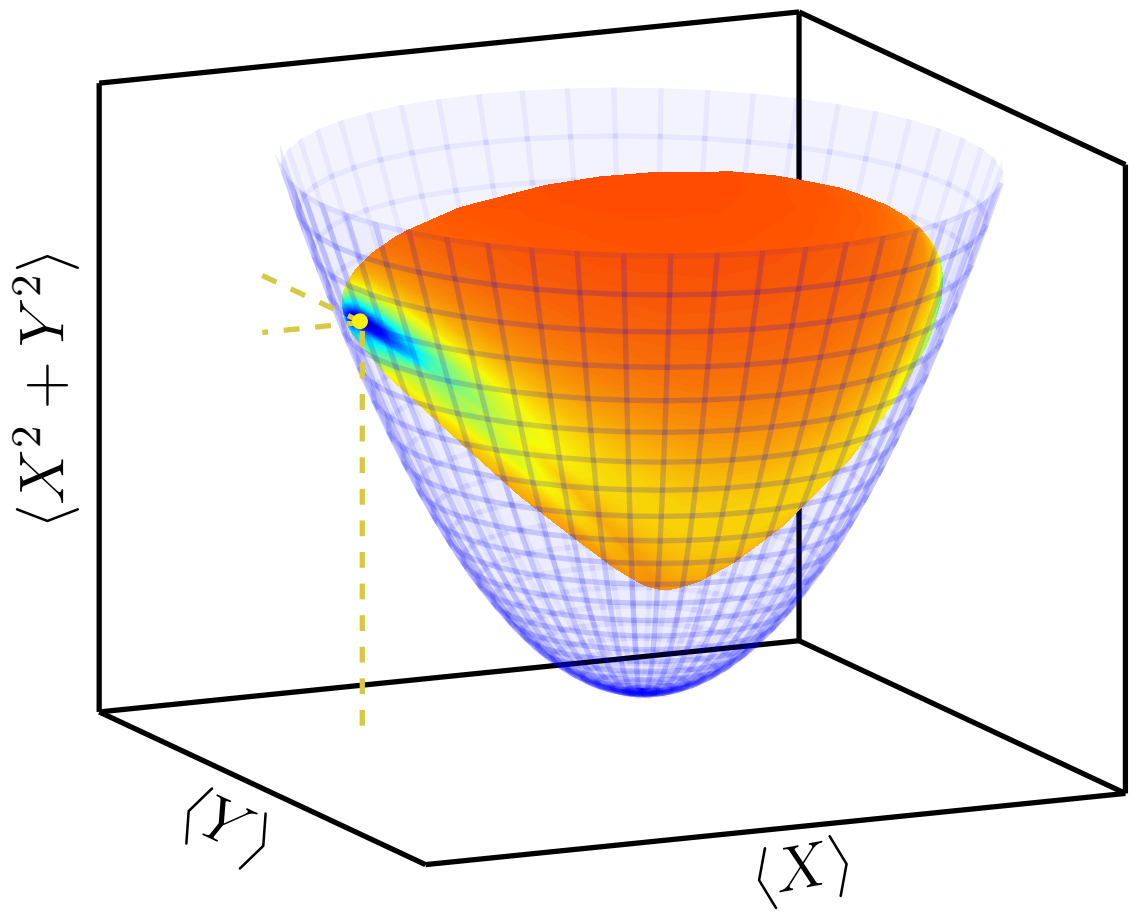

**Figure 4.7:** Bounds for the sum of variances of operators $X$ and $Y$ can be extracted from the joint numerical range $W(X, Y, X^2 + Y^2)$.

There are multiple approaches to this problem: it can be presented as the minimization of the expectation value of a certain observable over separable states [49], the algebraic properties of the operators can be used to provide a bound [50], or in some cases a simplified model of the physical system in question can be solved [51]. Here, I am interested in providing an approach based on the geometry of numerical ranges and associated algebraic structures. The crucial observation is the following: the sum of variances can be expanded to yield the following identity:

[49]: Giorda et al. (2019), 'State-independent uncertainty relations from eigenvalue minimization'
[50]: Maccone et al. (2014), 'Stronger Uncertainty Relations for All Incompatible Observables'
[51]: Dammeier et al. (2015), 'Uncertainty relations for angular momentum'

$$\Delta^2 X + \Delta^2 Y = \langle X^2 + Y^2 \rangle - \langle X \rangle^2 - \langle Y \rangle^2 . \tag{4.39}$$

As demonstrated in [3], the problem can be reduced to the optimization of a convex function defined on the numerical range $W(X, Y, X^2 + Y^2)$.

[3]: Szymański et al. (2019), 'Geometric and algebraic origins of additive uncertainty relations'

**Theorem 4.11** *Finding* $\min_\rho \Delta^2 X_\rho + \Delta^2 Y_\rho$ *is equivalent to the determination of a paraboloid of revolution taken from a certain family tangent to* $W(X, Y, X^2 + Y^2)$.

*Proof.* The following equation is equivalent to $\Delta^2 X + \Delta^2 Y = \alpha$:

$$\langle X^2 + Y^2 \rangle - \langle X \rangle^2 - \langle Y \rangle^2 = \alpha . \tag{4.40}$$

Therefore, the level sets of $\Delta^2 X + \Delta^2 Y$ are paraboloids of revolution in the space of expectation values parameterized by the parameter $\alpha$. The smallest $\alpha$ for which there exists a state $\rho$ such that the sum of variances $\Delta^2 X_\rho + \Delta^2 Y_\rho$ is equal to $\alpha$ corresponds to a paraboloid tangent to the numerical range $W(X, Y, X^2 + Y^2)$. □

Since numerical ranges can be efficiently approximated, this offers a simple way to numerically estimate the bounds for state-independent uncertainty relations involving sums of variances with arbitrary accuracy [52].

[52]: Schwonnek et al. (2017), 'State-independent uncertainty relations and entanglement detection in noisy systems'

In some cases, tight analytical bounds can be provided through the polynomial description of the boundary of the numerical ranges – see Eq. (4.17). Straightforward implementation does work: the system of polynomial equations defines the boundary of $W(X, Y, X^2 + Y^2)$, as well as the paraboloid of constant sum of variances, and enforces that both surfaces are tangent. However, in some cases the solution is attained faster using a family of linear approximators of variance. The following lemma encapsulates this idea:

**Lemma 4.12** *For every quantum state $\rho$ and $x, y \in \mathbb{R}$,*

$$\Delta^2 X_\rho + \Delta^2 Y_\rho \geq \langle X^2 + Y^2 \rangle_\rho - 2x \langle X \rangle_\rho - 2y \langle Y \rangle_\rho + x^2 + y^2, \quad (4.41)$$

*with equality if and only if $\langle X \rangle_\rho = x, \langle Y \rangle_\rho = y$.*

*Proof.* For any $x$, $\langle (X - x\mathbb{1})^2 \rangle \geq \Delta^2 X$. The equality is attained if and only if $x = \langle X \rangle$. The lemma follows from expanding this expression for both variances. □

Such approximations naturally correspond to linear functions over $W(X, Y, X^2 + Y^2)$. The minima of linear functionals over numerical ranges correspond to the ground states of combinations of operators (Theorem 4.2), which results in the following observation [3].

[3]: Szymański et al. (2019), 'Geometric and algebraic origins of additive uncertainty relations'

**Theorem 4.13** *Finding* $\min_\rho \Delta^2 X_\rho + \Delta^2 Y_\rho$ *is equivalent to the minimization of the smallest eigenvalue of* $(X - x\mathbb{1})^2 + (Y - y\mathbb{1})^2$ *over real $x$ and $y$.*

The problem of finding a bound for the sum of variances can therefore be stated in a purely algebraic form: as a system of three polynomials equations in three variables: the eigenvalue $\lambda$ and parameters $x$ and $y$:

$$\begin{cases} \overbrace{\det\left[(X - x\mathbb{1})^2 + (Y - y\mathbb{1})^2 - \lambda\mathbb{1}\right]}^{f} = 0, \\ \partial_x f = 0, \\ \partial_y f = 0. \end{cases} \quad (4.42)$$

As always is the case with systems of polynomial equations, care must be taken as to take into account only the physical solutions, corresponding to only the real roots of the polynomials.

The two approaches can be used to provide tight and analytical bounds for a physically important case: the sum of variances of two spin components.

| $j$ | $\min \Delta^2 J_X + \Delta^2 J_Y$ | order | $j$ | $\min \Delta^2 J_X + \Delta^2 J_Y$ | order |
|---|---|---|---|---|---|
| 1/2 | 1/4 = 0.25 | 1 | 1 | 7/16 = 0.4375 | 1 |
| 3/2 | ≈ 0.6009 | 3 | 2 | ≈ 0.7496 | 3 |
| 5/2 | ≈ 0.8877 | 7 | 3 | ≈ 1.018 | 6 |
| 7/2 | ≈ 1.142 | 13 | 4 | ≈ 1.260 | 10 |
| 9/2 | ≈ 1.374 | 21 | 5 | ≈ 1.484 | 15 |
| 11/2 | ≈ 1.591 | 31 | 6 | ≈ 1.695 | 21 |
| 13/2 | ≈ 1.796 | 43 | 7 | ≈ 1.894 | 28 |
| 15/2 | ≈ 1.991 | 57 | 8 | ≈ 2.085 | 36 |
| 17/2 | ≈ 2.178 | 73 | 9 | ≈ 2.268 | 45 |
| 19/2 | ≈ 2.358 | 91 | 10 | ≈ 2.445 | 55 |

**Table 4.1:** Tight bounds for the uncertainty relation for sum of variances $\Delta^2 J_X + \Delta^2 J_Y$ can be found using the analytical description of the joint numerical range of $W(J_X, J_Y, J_X^2 + J_Y^2)$. The bounds as function of the total angular momentum $j$ are described as roots of polynomials of varying order.

**Example 4.3** (Planar squeezing) An important instance of the problem of additive uncertainty relations arises in atomic optics, where *squeezed states* are used in interferometry and metrology [48] . In this context, the sensitivity of the measurement is determined by the sum of variances of two spin components. The spin matrices are indexed with the total spin $j$ taking integer and half-integer values. With fixed total spin $j$, the matrices are of size $(2j+1)$, and can be explicitly written out in the eigenbasis $\{|m\rangle\}_{m=-j}^{j}$ of one of them:

[48]: He et al. (2011), 'Planar quantum squeezing and atom interferometry'

$$
\begin{aligned}
\langle n|J_X|m\rangle &= \frac{\delta_{n,m+1} + \delta_{n,m-1}}{2}\sqrt{j(j+1) - nm}, \\
\langle n|J_X|m\rangle &= \frac{\delta_{n,m+1} - \delta_{n,m-1}}{2i}\sqrt{j(j+1) - nm}, \\
\langle n|J_Z|m\rangle &= \delta_{m,n} m.
\end{aligned}
\tag{4.43}
$$

The question is: given $j$, what is the minimum sum of variances of two of these operators? The exact choice of which one to use does not matter – the matrices are related to a highly symmetric $SU(2)$ group – and the goal is to provide a tight bound $c_j$ such that

$$\Delta^2 J_X + \Delta^2 J_Y \geq C_j. \tag{4.44}$$

The approach presented in the Theorems 4.11 and 4.13 is highly efficient in this case. Both can be used to provide a polynomial description of the tight bound $c_j$ in the following form: given the total spin $j$, a polynomial $h_j(C_J)$ with integer coefficients and of the smallest possible order can be determined such that

$$h_j(C_j) = 0. \tag{4.45}$$

The results of this approach are presented in Table 4.1. The detailed discussion can be found in [3].

[3]: Szymański et al. (2019), 'Geometric and algebraic origins of additive uncertainty relations'

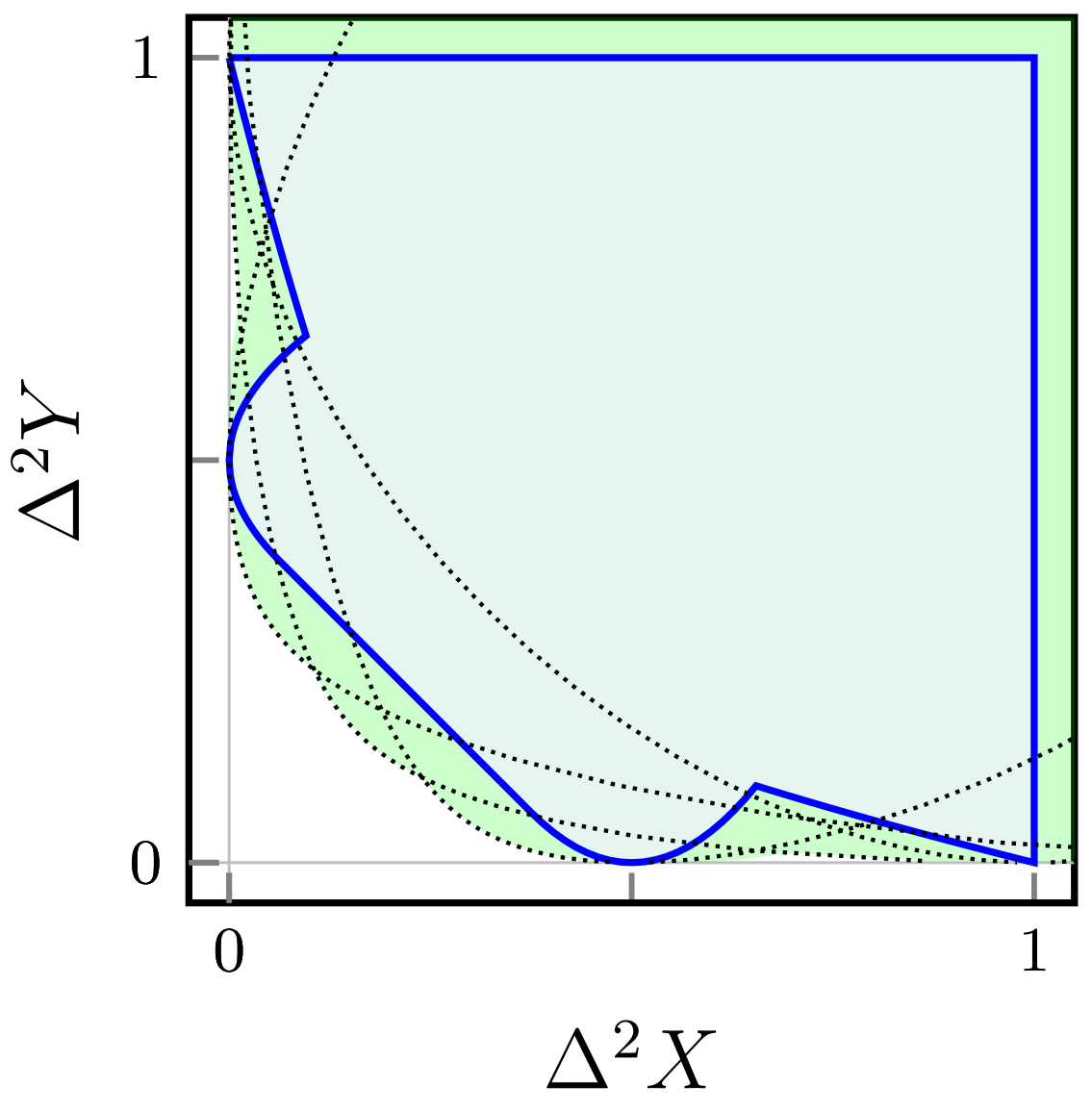

**Figure 4.8:** The uncertainty range – a set of simultaneously attainable pairs of variances – is shown in blue. Here, the observables are spin operators for total angular momentum $j = 1$: $X = J_X$ and $Y = J_Y$. The uncertainty range is covered by the set union of the numerical ranges (shown in green with dotted boundaries) of linear variance approximators – operators of form described by Eq. (4.49). This approach allows for determination of approximate and analytical uncertainty relations.

### 4.3.1 Uncertainty range

The main idea of numerical range is the visualization of important variables and omission of the insignificant ones. This can be extended in various ways: the numerical ranges are images of linear maps over the set of states, but other functionals are possible [51]. In this section, I provide a method of systematic approximation of one of these extensions: the image of map measuring the variances of observables.

[51]: Dammeier et al. (2015), 'Uncertainty relations for angular momentum'

**Definition 4.4** *The uncertainty range is the set of simultaneously attainable variances of a given set of observables:*

$$V(X_1, \dots, X_k) = \left\{ (\Delta^2_\rho X_1, \dots \Delta^2_\rho X_k) : \rho \in \mathcal{M}_d \right\}. \quad (4.46)$$

For $k = 2$ operators – the simplest nontrivial case – this set can be viewed as an image of a map

$$W(X, X^2, Y, Y^2) \ni (x, x', y, y') \mapsto (x' - x^2, y' - y^2) \in V(X, Y). \quad (4.47)$$

By analyzing its structure, the entire family of additive uncertainty relations of the following form can be found:

$$\alpha \Delta^2 X + (1 - \alpha) \Delta^2 Y \geq c(\alpha). \quad (4.48)$$

The structure is, however, highly complicated: being an image of a nonlinear map, it may be – and often is – nonconvex and the tools used previously do not work straightaway. This problem can be avoided by finding a set of linear approximations to the variance, which together allow for a partial analysis of the geometry of the uncertainty range.

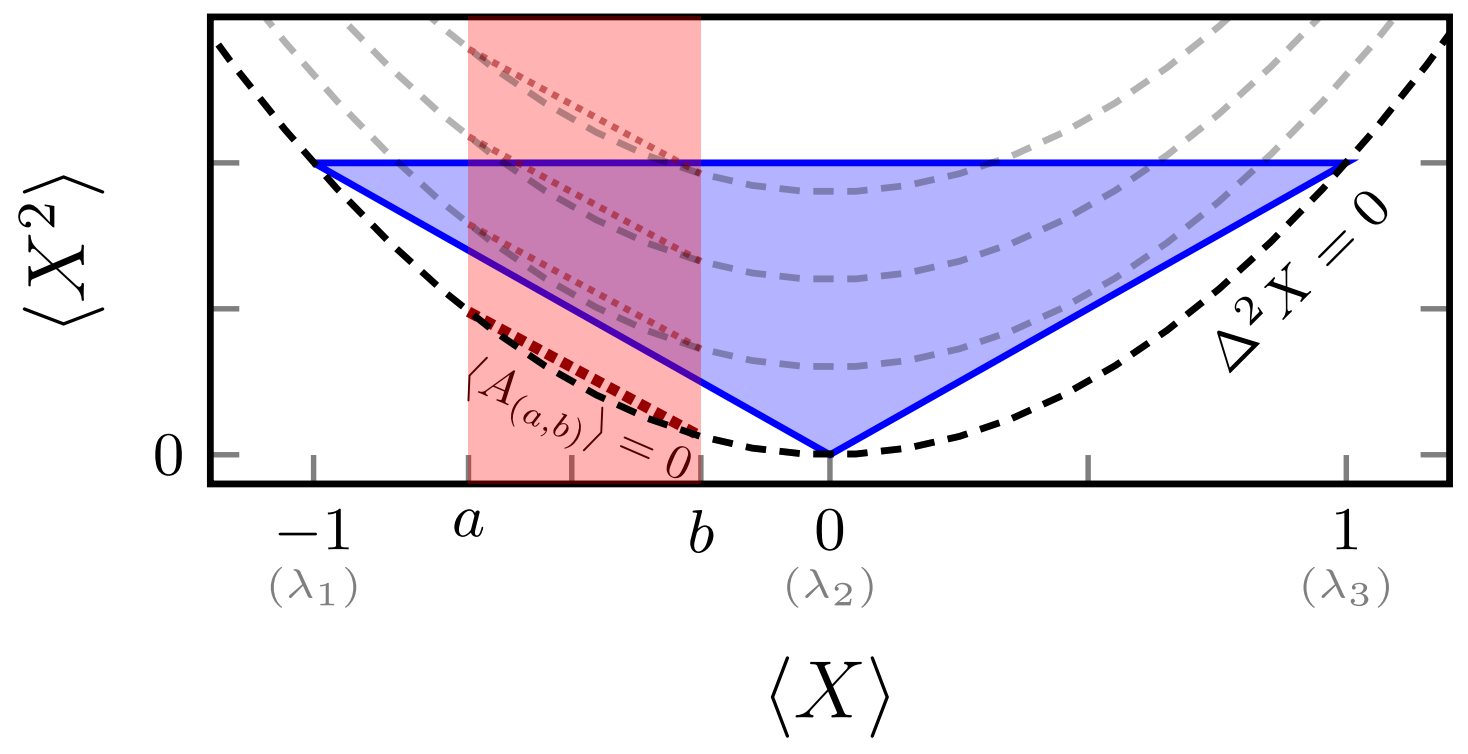

**Figure 4.9:** Linear approximations to variance $\Delta^2 X$ can be determined with the aid of the numerical range $W(X, X^2)$. For given real $a$ and $b$, the operator $A_{(a,b)}$ defined in Eq. (4.49) provides a lower bound for variance: if $a \leq \langle X \rangle \leq b$, then $\langle A_{(a,b)} \rangle \leq \Delta^2 X$. The variance can be bounded from below with a finite number of such linear approximants.

**Definition 4.5** *The sector approximant of variance of an observable $X$ is the Hermitian operator $A_{(a,b)}$ which expectation value provides a lower bound for $\Delta^2 X$ when $a \leq \langle X \rangle \leq b$ for arbitrary $a, b$. The operator reads*

$$A_{(a,b)} = X^2 - (a+b)X + ab. \tag{4.49}$$

By stitching together the entire range of $\langle X \rangle$ with different sector approximants, one can provide a controlled approximation of variance:

**Theorem 4.14** *Let $\Lambda(X)$ denote the set of eigenvalues of $X$. For an increasing sequence $(x_i)_{i=1}^N$ containing $\Lambda(X)$, the variance of an arbitrary state can be bounded from below by*

$$\Delta^2 X \leq \min_i \langle X^2 - (x_{i+1} + x_i)X + x_{i+1}x_i \rangle . \tag{4.50}$$

By providing sector decompositions of two operators one can approximate the uncertainty range:

**Theorem 4.15** *If the sequences $(x_i)_{i=1}^N$ and $(y_i)_{i=1}^M$ provide approximations to variance in the sense of Theorem 4.14, the sum of variances is bounded by*

$$\Delta^2 X + \Delta^2 Y \geq \overbrace{\min_{i,j} \lambda_{\min}(X_i + Y_j)}^{c} . \tag{4.51}$$

*The approximation error is bounded:*

$$\Delta^2 X + \Delta^2 Y - c \leq \overbrace{\left( \frac{\max_i (x_{i+1} - x_i)}{2} \right)^2}^{\delta_X} + \overbrace{\left( \frac{\max_j (y_{j+1} - y_j)}{2} \right)^2}^{\delta_Y} . \tag{4.52}$$

**Definition 4.6** *The Minkowski sum of two sets* $A, B \subset \mathbb{R}^k$ *is the following set:*

$$A \oplus B = \{\vec{a} + \vec{b} : \vec{a} \in A, \vec{b} \in B\}.$$

*Furthermore, the uncertainty range* $V(X, Y)$ *is related to the union of numerical ranges* $W(X_i, Y_j)$*:*

$$V(X, Y) \subset \left( \bigcup_{i,j} W(X_i, Y_j) \right) \oplus ([0, \delta_X] \times [0, \delta_Y]), \qquad (4.53)$$

*where* $\oplus$ *denotes the Minkowski sum of two sets and* $\times$ *is the Cartesian product.*

## 4.4 Phase transitions

The possibility of applying numerical ranges in the topic of phase transition was hinted in Theorem 4.2. The applications go further than the basic properties of ground states: by using the geometry of numerical ranges it is possible to show in a straightforward manner multiple facts concerning phase transitions. Here I present two of them: unreachability of the derivative state and estimation of the value of the energy gap.

### 4.4.1 Unreachability of the derivative state

**Theorem 4.16** *Consider a parameterized Hamiltonian* $H(\lambda) = H + \lambda V$ *and its* $\lambda$*-dependent ground state* $|\psi_\lambda\rangle$ *(assumed nondegenerated and in gauge such that* $\langle\psi_\lambda| \left({}^{\mathrm{d}}/_{\mathrm{d}\lambda} |\psi_\lambda\rangle\right) = 0$*). If* $|\sigma_\lambda\rangle$ *denotes the normalized derivative of* $|\psi_\lambda\rangle$ *with respect to* $\lambda$*:*

$$|\sigma_\lambda\rangle \propto \frac{\mathrm{d}\,|\psi_\lambda\rangle}{\mathrm{d}\,\lambda}, \quad \langle\sigma_\lambda|\sigma_\lambda\rangle = 1, \qquad (4.54)$$

*then* $|\sigma_\lambda\rangle$ *is never a ground state* $H(\kappa)$ *for any pair of* $\lambda$ *and* $\kappa$*.*

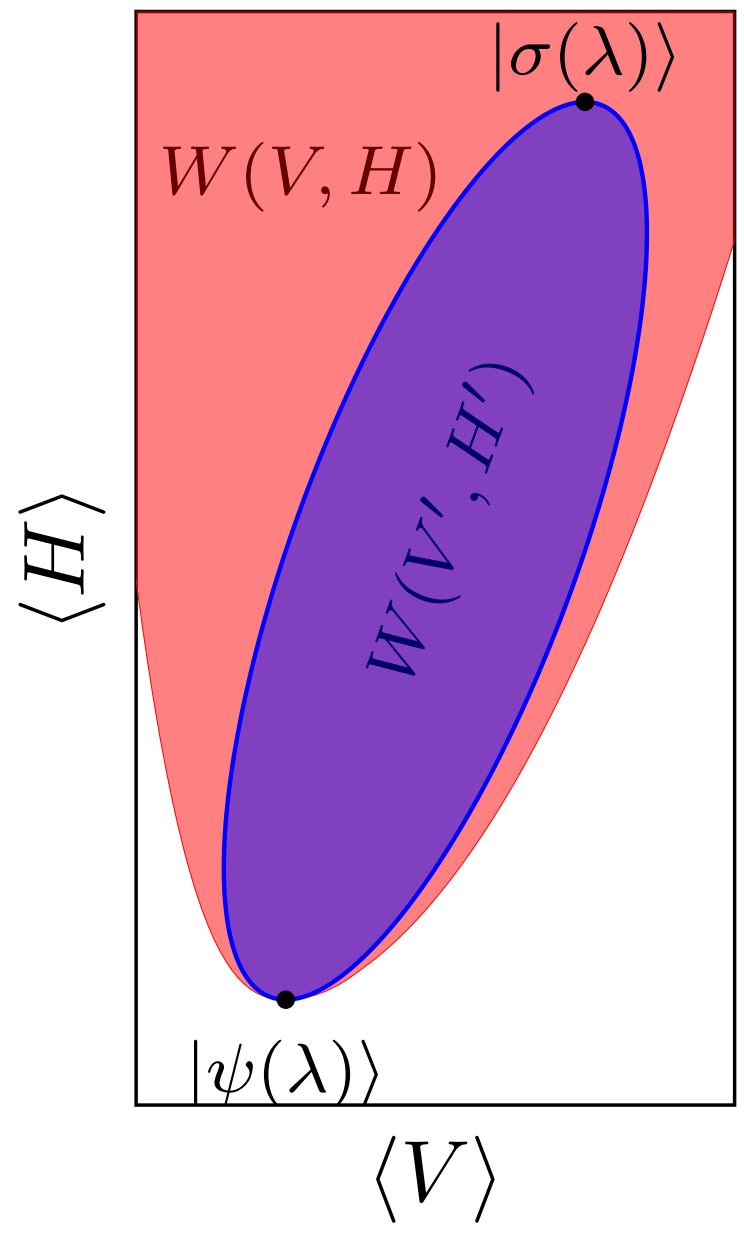

**Figure 4.10:** Derivative of the ground state of $H + \lambda V$ with respect to the parameter $\lambda$ is never a ground state of $H + \kappa V$ for any pair of $\lambda$ and $\kappa$. Geometrical structure of the restricted numerical range $W(V', H')$ allows for a proof of this fact.

*Proof.* Consider the numerical range $W(V, H)$ and the numerical range of the operators restricted to the subspace spanned by $|\sigma_\lambda\rangle$ and $|\psi_\lambda\rangle$. The vectors can be chosen without loss of generality to be orthogonal, therefore the matrix elements read

$$H' = \begin{pmatrix} \langle H\rangle_\psi & \langle\psi|H|\sigma\rangle \\ \langle\sigma|H|\psi\rangle & \langle H\rangle_\sigma \end{pmatrix}, \quad V' = \begin{pmatrix} \langle V\rangle_\psi & \langle\psi|V|\sigma\rangle \\ \langle\sigma|V|\psi\rangle & \langle V\rangle_\sigma \end{pmatrix}. \quad (4.55)$$

Since $|\psi_\lambda\rangle$ is an eigenvector of $H$, then $\langle\sigma|H|\psi\rangle = 0$ and the matrix $H'$ is diagonal. The off-diagonal elements of $V'$ are nonzero – this follows from the perturbation theory – and the restricted numerical range $W(V', H')$ is thus always a nondegenerate ellipse.[10] However, since $|\sigma_\lambda\rangle$ is orthogonal to $|\psi_\lambda\rangle$, the images of these states are antipodal points on the ellipse. The image of $|\sigma_\lambda\rangle$ therefore cannot lie in the lower part of the boundary of $W(V, H)$ – the bulk part of

10: Alternatively: the derivative denotes the direction in which the ground state is changing; thus the numerical range $W(V', H')$ is tangent to $W(V, H)$ and therefore nondegenerate.

the ellipse of $W(V', H')$ ensures this. Thus, $|\sigma_\lambda\rangle$ is never a ground state of any Hamiltonian $H + \kappa V$ for any $\kappa$ and $\lambda$.[11] □

11: But $|\sigma_\lambda\rangle$ might be a ground state of $-H + \kappa V$ – note the minus sign.

### 4.4.2 Bounds for the energy gap

The connection between the set of ground states and the boundary of the joint numerical range yields another result: the geometry of the numerical range does in certain cases provide information about the value of the energy gap [4]. This quantity, also called *spectral gap*, is of central importance in the optimal speed of adiabatic quantum evolution (affecting adiabatic quantum computation), its dynamics gives an important indicator of quantum phase transitions, and is connected to the properties of the ground states. However, energy gap is difficult to estimate – in some cases the value of the gap is not a computable number [53] – and there are multiple possible approaches to this problem.

[4]: Szymański et al. (2021), 'Universal witnesses of vanishing energy gap'

[53]: Cubitt et al. (2015), 'Undecidability of the spectral gap'

Here, I present a method to provide a bound for the value of the energy gap based on the geometry of the numerical range of suitably chosen observables. For a given Hamiltonian $H$, the goal is to estimate the energy gap $\delta(H)$. If one can find an auxiliary perturbation Hamiltonian $V$ possessing certain properties, then the upper bound for $\delta(H)$ can be extracted from the boundary of the numerical range $W(V, H)$.

**Definition 4.7** *The* energy gap *or* spectral gap $\delta(H)$ *of a Hamiltonian* $H$ *is the difference between energies of the ground state and the first excited state, excluding the degeneracy. If the system is defined by a sequence of Hamiltonians* $H_{(N)}$ *(as a function of system size* $N$*), then the energy gap is*

$$\delta(H) = \limsup_{n\to\infty} \delta\left(H_{(N)}\right). \quad (4.56)$$

The method stems from a straightforward observation relating the commutation properties of operators with the geometry of the numerical range:

**Lemma 4.17** *If a cusp in the boundary of* $W(X, Y)$ *is an image of a pure state* $|\psi\rangle$*, then* $|\psi\rangle$ *is a common eigenvector of* $X$ *and* $Y$*. Therefore, both operators and their joint numerical range decompose:*

$$\begin{aligned} X &= X_0 \oplus X_\perp, \\ Y &= Y_0 \oplus Y_\perp, \\ W(X, Y) &= \operatorname{conv}\left(W(X_0, Y_0) \cup W(X_\perp, Y_\perp)\right), \end{aligned} \quad (4.57)$$

*where* $X_0$ *and* $Y_0$ *act on a one-dimensional subspace spanned by* $|\psi\rangle$*.*

*Proof.* A cusp has multiple independent normal vectors. Therefore, by virtue of Theorem 4.2, the vector $|\psi\rangle$ is a ground state of two independent linear combinations of $X$ and $Y$, hence it is an eigenvector of both of them. The theorem follows by writing any density operator $\rho$ in the block form consistent with this observation and identifying diagonal blocks which contribute to the expectation values $\langle X\rangle_\rho$ and $\langle Y\rangle_\rho$. The expectation values turn out to be convex mixtures of points from $W(X_0, Y_0)$ and $W(X_\perp, Y_\perp)$. $\square$

The cusp geometry is affected by the image of the subspace orthogonal to the state $|\psi\rangle$. In particular, if the operators $X$ and $Y$ are finite-dimensional, only a cusp of the form (a) of Figure 4.11 can appear.

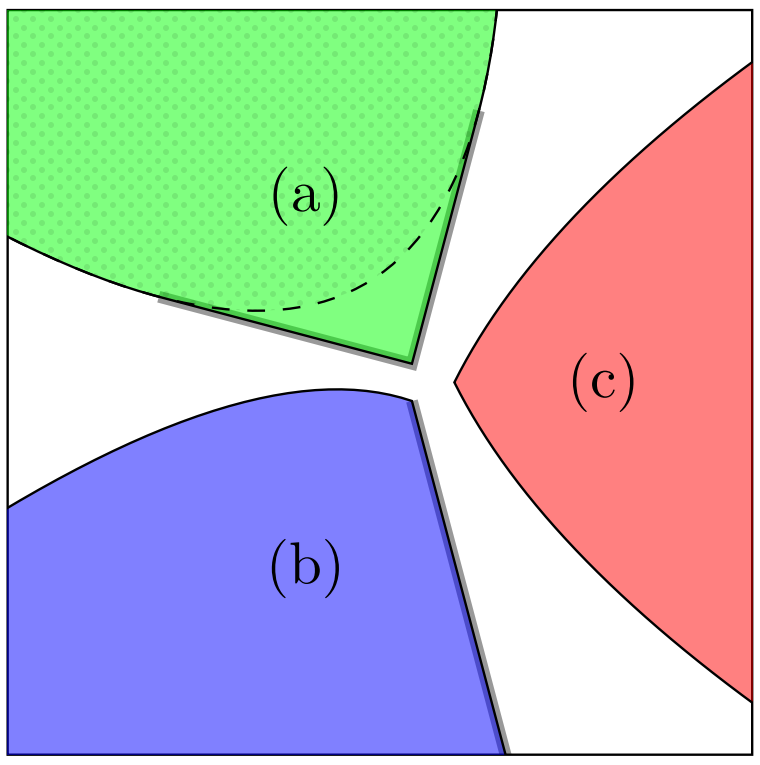

**Figure 4.11:** Different kinds of cusps of 2D convex sets. The flat parts of the boundary (segments) are highlighted with thick lines. The numerical range of two finite-dimensional Hermitian operators may contain only cusps similar to (a), in which case the cusp itself is isolated from the rest of the boundary. If the operators are infinite-dimensional (e.g., describe an infinite spin chain), then the cases (b) and (c) are possible and provide a signature of the vanishing energy gap.

If the cusp is an image of a ground state of one of the operators, then the cusp geometry provides a bound for the value of the energy gap. While the language of geometry might be unwieldy to use in normal contexts, it can be translated to a statement about dynamics of the expectation values as a parameter of a Hamiltonian is varied.

**Theorem 4.18** *Assume that for a Hamiltonian H there exists a Hermitian operator V such that:*

*a) The ground state of $H + \lambda V$, denoted by $|\psi_\lambda\rangle$, is constant (equal to $|\psi_{\lambda=0}\rangle$) in $\lambda \in [0, \lambda^*]$.*
*b) At $\lambda = \lambda^*$, a discontinuity of $\langle H\rangle_\lambda = \langle\psi_\lambda|H|\psi_\lambda\rangle$ appears, of size*

$$\varepsilon = \left(\lim_{\lambda\to\lambda^*_+} \langle H\rangle_\lambda\right) - \langle H\rangle_{\lambda=0}\,. \tag{4.58}$$

*In such case, the energy gap of H is bounded from above:*

$$0 \le \delta(H) \le \varepsilon. \tag{4.59}$$

*If the bound vanishes – the expectation value $\langle H\rangle_\lambda$ is continuous at $\lambda = \lambda^*$ – then the Hamiltonian H is gapless.*

*Proof.* The situation can be described by the numerical range $W(V, H)$ – see Figure 4.12. Ground state of $H + \lambda V$ being constant in a range of parameters is equivalent to the cusp being present at the bottom of the numerical range. Then, the discontinuous change of expectation values corresponds to the sudden change of the support point as a normal vector is varied.

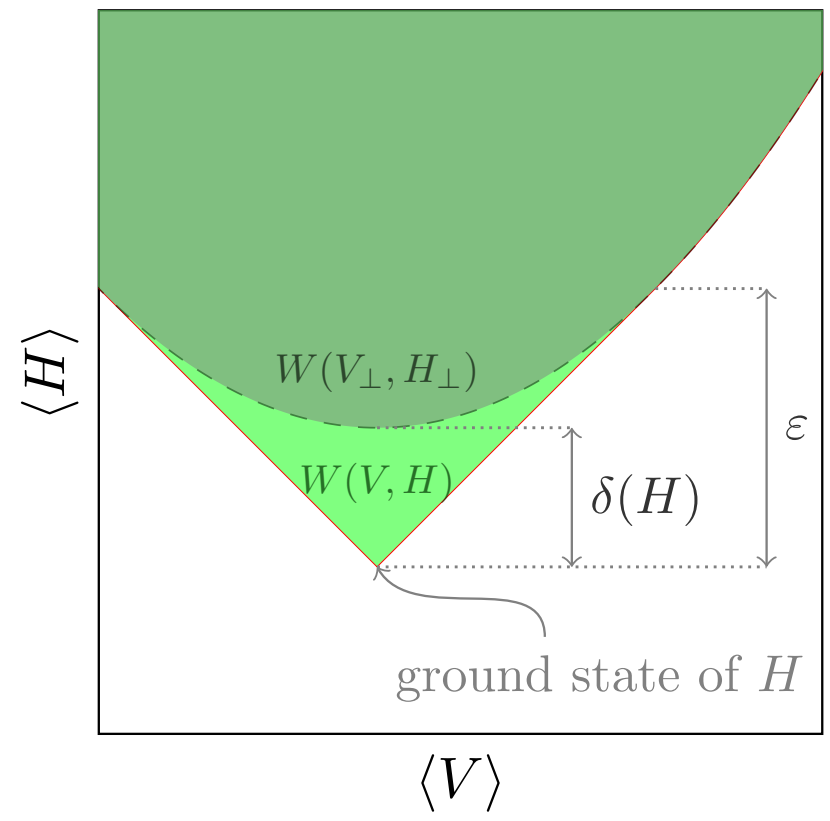

**Figure 4.12:** Graphical representation of Theorem 4.18 and a sketch of its proof. Discontinous change of the expectation value corresponds to a segment being present in the boundary of the numerical range $W(V, H)$ – a fact which can be used to bound the value of the spectral gap $\delta(H)$.

By virtue of Lemma 4.17 the operators admit a joint block diagonal form and the numerical range splits:

$$\begin{gathered} H = H_0 \oplus H_\perp, \quad V = V_0 \oplus V_\perp, \\ W(V, H) = \operatorname{conv}\left(W(V_0, H_0) \cup W(V_\perp, H_\perp)\right), \end{gathered} \tag{4.60}$$

where the operators with the subscript 0 act on the ground state of $H$.

The ground state $|\psi_\lambda\rangle$ of a Hamiltonian $H + \lambda V$ always lies in the boundary of the numerical range $W(V, H)$. If a nonzero energy gap is present, the boundary does contain segments. The existence of the segments implies that a support point of the numerical range changes discontinously around some normal vectors – this corresponds to the sudden change of the ground state as the parameter $\lambda$ is varied. Indeed, the support point changes from the image of the ground state $|\psi_{\lambda=0}\rangle$ to a point of $W(V_\perp, H_\perp)$. This can be used to estimate the energy gap: the size of the jump discontinuity of $\langle H \rangle_\lambda$ bounds the energy gap $\delta(H)$ from above.

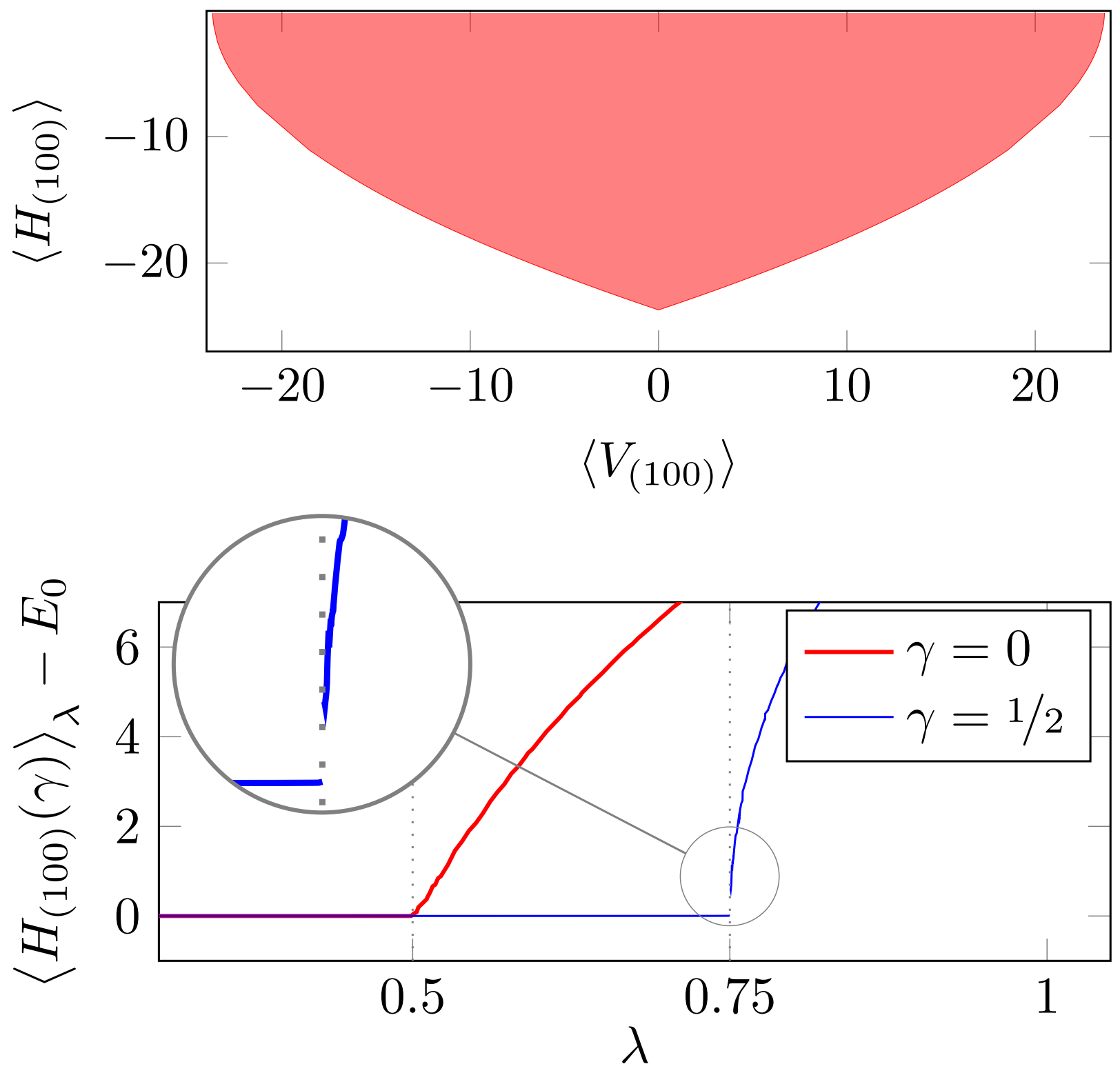

**Figure 4.13:** *Top*: Numerical range of a finite approximation of the $XY$ model Hamiltonian $H$ and a gaplessness witness term $V$ for $N = 100$ sites. The energy gap is numerically indistinguishable from 0 and the numerical range is consistent with this observation: the kind of cusp visible closely approximates case (c) of Figure 4.11, which may only arise in the gapless case.
*Bottom*: Size of the jump discontinuity of the expectation value of $H$ over a ground state of $H + \lambda V$ as a function of $\lambda$ provides an upper bound to the value of the energy gap. For $\gamma = 0$, the system is numerically gapless and no discontinuity is present. For $\gamma = 1/2$, the discontinuity becomes visible – a sign of nonzero energy gap.

If the expectation value $\langle H \rangle_\lambda$ is continuous as a function of $\lambda$, the numerical range $W(V_\perp, H_\perp)$ is infinitesimally close to the image of the ground state $|\psi_{\lambda=0}\rangle$. In other words, there exist states orthogonal to $|\psi_{\lambda=0}\rangle$ arbitrarily close to it in energy and the Hamiltonian $H$ is gapless. □

**Example 4.4** *(Main result of [4])* The $XY$ model is a spin chain described by the Hamiltonian parameterized by the asymmetry parameter $\gamma$ and the system size $N$:

[4]: Szymański et al. (2021), 'Universal witnesses of vanishing energy gap'

$$H_{(N)}(\gamma) = \sum_{n=1}^{N-1} \frac{1+\gamma}{2} \sigma_x^{(n)} \sigma_x^{(n+1)} + \frac{1-\gamma}{2} \sigma_y^{(n)} \sigma_y^{(n+1)}, \tag{4.61}$$

where $\vec{\sigma}$ are local Pauli matrices labeled by the site number $n$. The $XY$ model is gapless for $\gamma = 0$ as $N \longrightarrow \infty$ [54].

[54]: Lieb et al. (1961), 'Two soluble models of an antiferromagnetic chain'

Theorem 4.18 can be used to provide bounds for the value of the energy gap of the $XY$ model for finite $N$. In practice, the upper bound $\varepsilon$ for the spectral gap of $H_{(N)}$ quickly becomes numerically indistinguishable from 0.

Three-body interaction operator $V_{(N)}$ acting on an $N$-site chain can be used to detect the vanishing energy gap of $H_{(N)}$ for $\gamma = 0$,

[55]: Lado (2021), *DMRGPY: Python library to spin and fermionic Hamiltonians with DMRG*

$$V_{(N)} = \sum_{n=2}^{N-1} \sigma_x^{(n-1)} \sigma_z^{(n)} \sigma_y^{(n+1)} - \sigma_y^{(n-1)} \sigma_z^{(n)} \sigma_x^{(n+1)}. \qquad (4.62)$$

In the calculations, the interactions are tapered towards the ends of the spin chain to eliminate edge effects – see [4] for details.

The system can be simulated numerically for finite $N$ and the ground states of $H_{(N)}(\gamma) + \lambda V_{(N)}$ are found using the matrix product states method [55]. Calculations for $N = 100$ indicate that the numerical range $W(V_{(100)}, H_{(100)})$ for $\gamma = 0$ contains a characteristic cusp which numerically closely approximates the gapless case even though the system size is finite – see Figure 4.13. Behavior of the expectation values puts a numerically tight bound on the value of the energy gap of $H_{(100)}(\gamma = 0)$ (the upper bound is indistinguishable from zero within the accuracy limit). In the case of $\gamma = 1/2$, the visible discontinuity of the expectation values is consistent with the fact that the system is gapped.

## 4.5 Separable numerical range and entanglement detection

In the case of systems consisting of at least two distinct and independently observable parts, additional structure arises in the set of quantum states. Properties like entanglement or Schmidt rank induce certain interesting subsets of the entire set $\mathcal{M}_d$ of states of dimension $d$, and these subsets can be analyzed using techniques closely related to numerical ranges. One of the most basic ideas of this research direction is the separable numerical range.

$\mathcal{M}^{\mathrm{SEP}}$ is described in Definition 2.15 on page 25.

**Definition 4.8** *Let the Hermitian operators $X_1, \ldots, X_k$ of size $d$ act on a $n$-partite system with local dimensions $d_1, \ldots, d_n$, such that $d = d_1 \times \ldots \times d_n$. The separable numerical range of $X_1, \ldots, X_k$ is the range of simultaneously attainable expectation values over the set of separable states $\mathcal{M}^{\mathrm{SEP}}_{d_1 \times \ldots \times d_n}$:*

$$W_{\mathrm{SEP}}(X_1, \ldots, X_k) = \left\{ (\langle X_1 \rangle_\rho, \ldots, \langle X_k \rangle_\rho) \in \mathbb{R}^k : \rho \in \mathcal{M}^{\mathrm{SEP}} \right\}. \qquad (4.63)$$

Geometric features of joint and separable numerical ranges can be used for entanglement detection. One of the early applications of geometry in this context is the theory of entanglement witnesses. First, let me define the class of block positive operators.

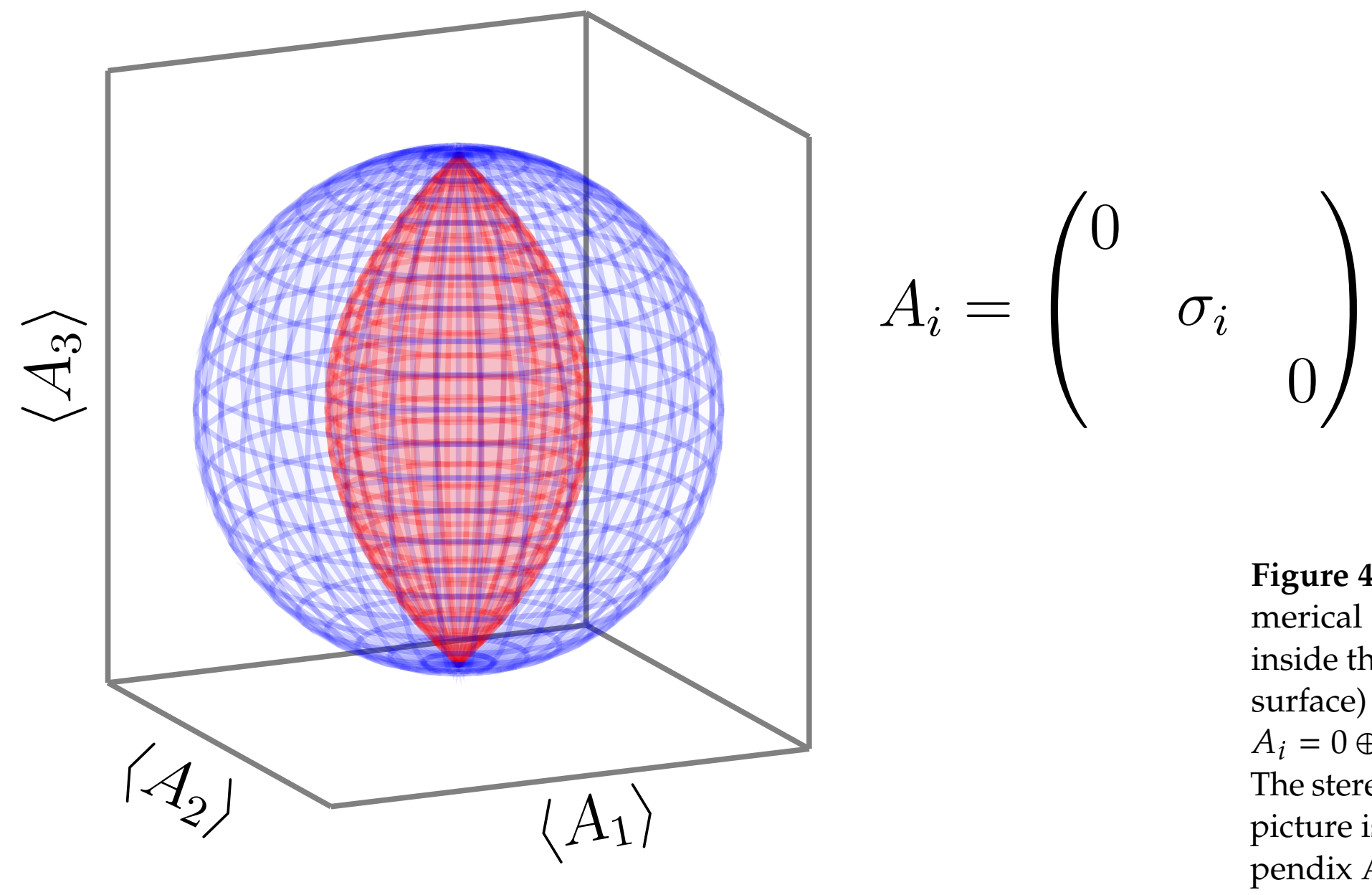

**Figure 4.14:** Exemplary separable numerical range (red surface) shown inside the joint numerical range (blue surface) of three two-qubit operators $A_i = 0 \oplus \sigma_i \oplus 0$ for $i = 1, 2, 3$. The stereographic (3D) version of this picture is available in Figure A.2 (Appendix A, page 102).

**Definition 4.9** *A Hermitian operator $X$ acting on a bipartite Hilbert space $\mathcal{H}_A \otimes \mathcal{H}_B$ is block positive if and only if its expectation value over every product state is nonnegative:*

$$\forall_{|\alpha\rangle \in \mathcal{H}_A, |\beta\rangle \in \mathcal{H}_B} \langle \alpha \otimes \beta | X | \alpha \otimes \beta \rangle \geq 0. \tag{4.64}$$

Equivalently, $W_{\text{SEP}}(X) = [a, b]$ with $a \geq 0$.

With this in mind, the entanglement of state $\rho$ can be witnessed with *some* block positive operator $X$ – for entangled states there always exists a block positive $X$ such that $\langle X \rangle_\rho < 0$. Existence is guaranteed for any entangled state: $\mathcal{M}^{\text{SEP}}$ is a convex set, and by the Hahn-Banach separation theorem [56] there always exists a linear functional

$$f : \mathcal{M}_d \to \mathbb{R}, \tag{4.65}$$

such that $f(\omega) \geq 0$ for $\omega \in \mathcal{M}^{\text{SEP}}$ and $f(\rho) < 0$. Since linear functionals over $\mathcal{M}_d$ are isomorphic to the expectation values of operators, the entanglement witness exists for any entangled state.

[56]: Edwards (1965), *Functional Analysis, Theory and Applications, Holt, Rinehart and Winston, New York*

There exist extensions of the idea of entanglement witnesses, taking into account the expectation values of other observables. Separable numerical ranges can be viewed as a complete generalization of such approaches – the bounds implied by the set $W_{\text{SEP}}(X_1, \ldots, X_k)$ take into account every information that can be inferred frpm the expectation values of the defining observables.

Since the set of separable states $\mathcal{M}^{\text{SEP}}$ is convex, so is $W_{\text{SEP}}$, being its image under a linear map. This suggests possible way to approximate the separable numerical range in a controlled way: convex sets are defined by their boundaries. Every point on the boundary of a convex set is a maximizer of a certain linear functional – thus, to sample a point from $\partial W_{\text{SEP}}$ it is sufficient to find a separable state that maximizes a certain expectation value.

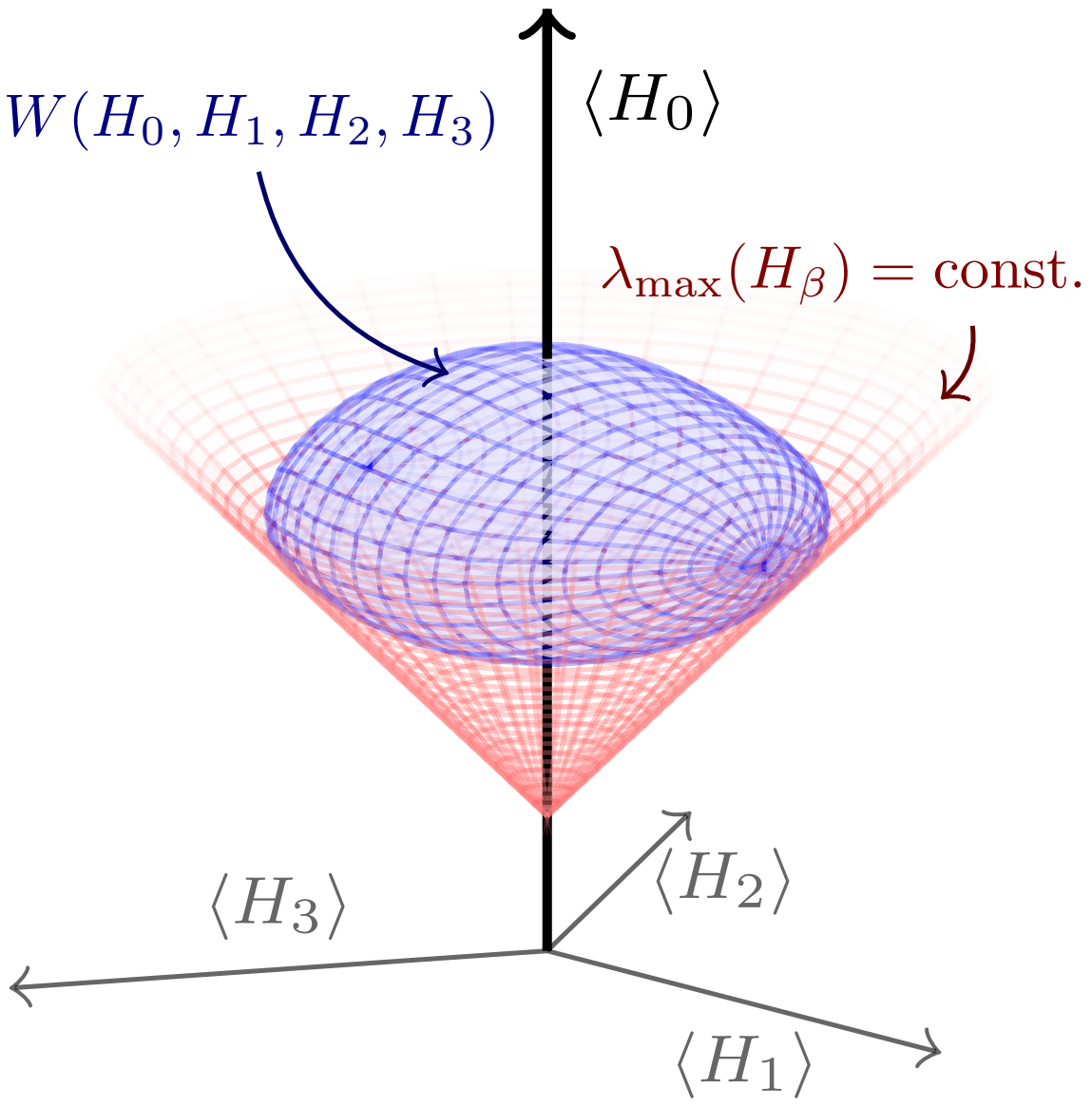

**Figure 4.15:** Visualisation of the algorithm finding $\lambda^{\otimes}_{\max}(H)$, the largest expectation value of an observable $H$ among separable states for qubit-qudit bipartite systems. The observable $H$ is decomposed into $\{H_i\}_{i=0}^{3}$ acting on qudit – numerical range $W$ (blue) of these operators can be used to determine $\lambda^{\otimes}_{\max}(H)$. The process involves finding the hypercone (red) tangent to $W$, and is always a two-dimensional problem regardless on the dimensionality of qudit.

### 4.5.1 Approximation procedure

As a result of this observation, the problem of approximating a separable numerical range $W_{\text{SEP}}(X_1, \ldots, X_k)$ can be reduced to the following one: given a normal vector $\vec{n} \in \mathbb{R}^k$, find the maximum expectation value $\lambda^{\otimes}_{\max}(H)$ of $H = \vec{n} \cdot \vec{X}$ among the separable states. Due to the linear nature of the problem and the convex structure of separable states,[12] it suffices to optimize over the pure product states:

Here, $X_i$ and $H$ act on a bipartite space $\mathcal{H}_A \otimes \mathcal{H}_B$.

12: The set $\mathcal{M}^{\text{SEP}}$ is the convex hull of pure product states.

$$\lambda^{\otimes}_{\max}(H) := \max_{|\alpha\rangle \in \mathcal{H}_A, |\beta\rangle \in \mathcal{H}_B} \langle \alpha \otimes \beta | \, H \, | \alpha \otimes \beta \rangle \, . \tag{4.66}$$

It is easy to see that:

$$\lambda^{\otimes}_{\max}(H) = \max_{|\alpha\rangle, |\beta\rangle} \langle \alpha | \, H_\beta \, | \alpha \rangle \, , \tag{4.67}$$

This is related to the fact that the separable numerical range is a convex hull of the sum of the joint numerical ranges of the reduced operators: $W_{\text{SEP}}(\vec{X}) = \text{conv} \bigcup_{\rho \in \mathcal{M}_B} W(\vec{X}_\rho)$.

where $H_\beta = \text{Tr}_B \left[ H(\mathbb{1} \otimes |\beta\rangle \langle \beta|) \right]$ is an operator acting on the subsystem A. If $|\alpha\rangle$ describes a qubit state, the complexity of the problem is reduced significantly, since $H_\beta$ is a matrix of size 2. With $|\beta\rangle$ fixed, the maximum expectation value of $H_\beta$ is equal to its largest eigenvalue,

$$\max_{|\alpha\rangle} \langle \alpha | \, H_\beta \, | \alpha \rangle = \overbrace{\frac{\langle H_0 \rangle_\beta + \sqrt{\langle H_1 \rangle_\beta^2 + \langle H_2 \rangle_\beta^2 + \langle H_3 \rangle_\beta^2}}{2}}^{\lambda_{\max}(H_\beta)} , \tag{4.68}$$

where the operators are defined by

Here, $\sigma_i$ are Pauli matrices.

$$H_i = \text{Tr}_A \left[ H(\sigma_i \otimes \mathbb{1}) \right] . \tag{4.69}$$

This line of thought culminates in the following theorem:

The numerical implementation of this procedure is the following: the four-dimensional numerical range is approximated as two convex polyhedra: the superset, formed by sampling normal vectors from the 3-sphere and intersecting the defining half-spaces, and subset, which is the convex hull of points found on the boundary. Then, the upper bound to $h$ is found by evaluating the function on the vertices of the superset polyhedron, and the lower bound is given by the same procedure for the subset polyhedron. This can be further simplified by noticing that the only relevant normal vectors form a two-dimensional subset of the 3-sphere – see Figure 4.15.

**Theorem 4.19** *Let an operator $H$ act on a bipartite qubit-qudit system. Maximization of $\langle \alpha \otimes \beta | H | \alpha \otimes \beta \rangle$ is equivalent to the maximization of the following (convex) function $h$ over the 4-dimensional numerical range $W(H_0, H_1, H_2, H_3)$:*

$$h(\langle H_0 \rangle, \langle H_1 \rangle, \langle H_2 \rangle, \langle H_3 \rangle) = \frac{\langle H_0 \rangle_\beta + \sqrt{\langle H_1 \rangle_\beta^2 + \langle H_2 \rangle_\beta^2 + \langle H_3 \rangle_\beta^2}}{2}, \tag{4.70}$$

*where the operators $\{H_i\}_{i=0}^3$ act on the qudit subsystem and are defined by Eq.* (4.69).

This can be easily solved numerical with arbitrary accuracy; it can be further simplified if the operator $H_0$ is proportional to the identity, since the maximization problem depends on $W(H_1, H_2, H_3)$ only.

The method is not limited to qubit-qudit optimization. To see this, assume that the $d$-dimensional part also possesses a separable structure[13] of dimensions $2 \times d' = d$ – then, the optimization of the function defined in Eq. (4.68) over $W(H_0, H_1, H_2, H_3)$ can be substituted with optimization of the same convex function over the separable numerical range $W_{\mathrm{SEP}}(H_0, H_1, H_2, H_3)$. The final result is then the maximum product expectation value over tripartite product states. [14]

13: That is, the quantum system is composed of two qubits and a $d'$-dimensional qudit.

14: The maximization is done over $\mathcal{M}_{2\times 2\times d'}^{\mathrm{SEP}}$.

The methods presented here were put to use in the analysis of entanglement witnesses constructed from generic observables taken from Gaussian Orthogonal Ensemble [2, 5].

[2]: Czartowski et al. (2019), 'Separability gap and large-deviation entanglement criterion'
[5]: Simnacher et al. (2021), 'Confident entanglement detection via the separable numerical range'

### 4.5.2 Calculation of separable numerical range is NP-hard

As with other problems related to entanglement and its detection, the determination of the separable numerical range is a NP-hard problem. This section presents the reasoning shown originally in [57].

[57]: Friedland et al. (2018), 'Nuclear norm of higher-order tensors'

**Definition 4.10** *An undirected graph is a tuple $G = (V, E)$ of the set of vertices $V$ and the set of edges – unordered 2-tuples of elements in $V$.*
*A graph is called complete if the set of edges is complete, i.e., $E = V \times V$.*
*A subgraph of graph $G = (V, E)$ restricted to $V' \subset V$ is the graph $G' = (V', E')$, where*

$$E' = \{(a, b) \in E : a, b \in V'\}. \tag{4.71}$$

A complete subgraph of a graph $G$ is called a *clique* of $G$. Determination of a *maximum clique* is a widely known NP-hard problem. As it turns out, finding the maximum clique is equivalent to the maximization of the expectation value of a certain bipartite Hermitian operator over the product states.

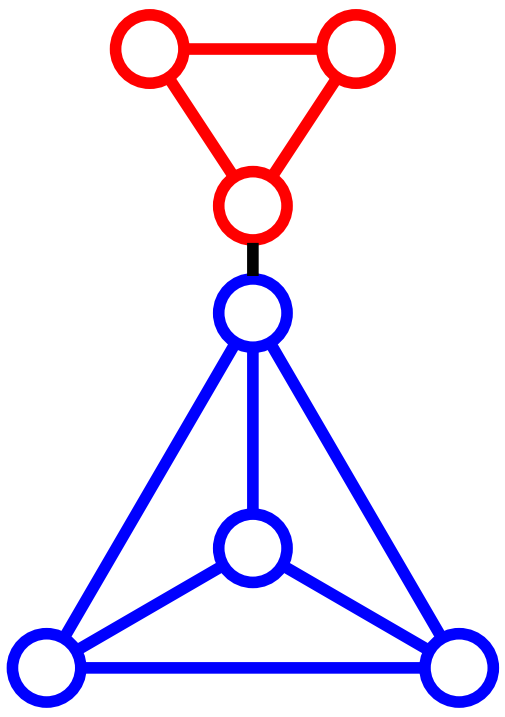

**Figure 4.16:** An exemplary graph with 7 vertices and 10 edges. Cliques of a graph include 4-clique (in blue), 3-clique (in red), the four 3-cliques of 4-clique, cliques defined by the edges (10 of them) and singletons formed by the vertices (7 of them).

For a graph $G = (V, E)$ with $n = |V|$ vertices, let me define the square matrix $A_G$ associated with the graph,

$$A_G = \sum_{(a,b)\in E} F_n^{(a,b)}, \tag{4.72}$$

where

$$\left(F_n^{(a,b)}\right)_{(i,j);(k,l)} = \begin{cases} \frac{1}{2} & (i,j,k,l) \in \left\{ \begin{matrix} (a,b,a,b), & (b,a,b,a), \\ (a,b,b,a), & (b,a,a,b) \end{matrix} \right\}. \\ 0 & \text{otherwise} \end{cases} \tag{4.73}$$

**Theorem 4.20** (Friedland-Lim) *The matrix $A_G$ defined by Eq.* (4.72) *is Hermitian, positive semidefinite, and of size $n^4$. With the separable structure of $\mathcal{M}^{\mathrm{SEP}}_{n^2\times n^2}$, the maximum expectation value of $A_G$ over product states is equal to*

$$\max_{|\alpha\rangle,|\beta\rangle} \langle \alpha\otimes\beta|\, A_G \,|\alpha\otimes\beta\rangle = \frac{\kappa(G)-1}{\kappa(G)}, \tag{4.74}$$

*where $\kappa(G)$ is the size of a maximum clique of $G$.*

Since the determination of a maximum expectation value is the central core of calculating separable numerical ranges – its boundary is formed by the points maximizing some combination of operators – the problem of approximating $W_{\mathrm{SEP}}$ is at least as hard, and therefore in general NP-hard.[15]

15: This is a significant problem for large system sizes, not so much for low dimensions.

However, there is a light of hope: algorithms determining the clique number are much faster than any direct numerical maximization of expectation values over the set of product states. Perhaps it is possible to rewrite them as graph-theoretic problems and use the more efficient methods?

### 4.5.3 Numerical range restricted to the states with positive partial transpose

The set of states with positive partial transpose $\mathcal{M}^{\mathrm{PPT}}$ coincides with $\mathcal{M}^{\mathrm{SEP}}$ for the two simplest cases (qubit-qubit and qubit-qutrit systems) and forms its nontrivial superset for the higher dimensional bipartite systems. It is thus worthwhile to consider the restricted numerical ranges for which the set of allowed states is $\mathcal{M}^{\mathrm{PPT}}$:

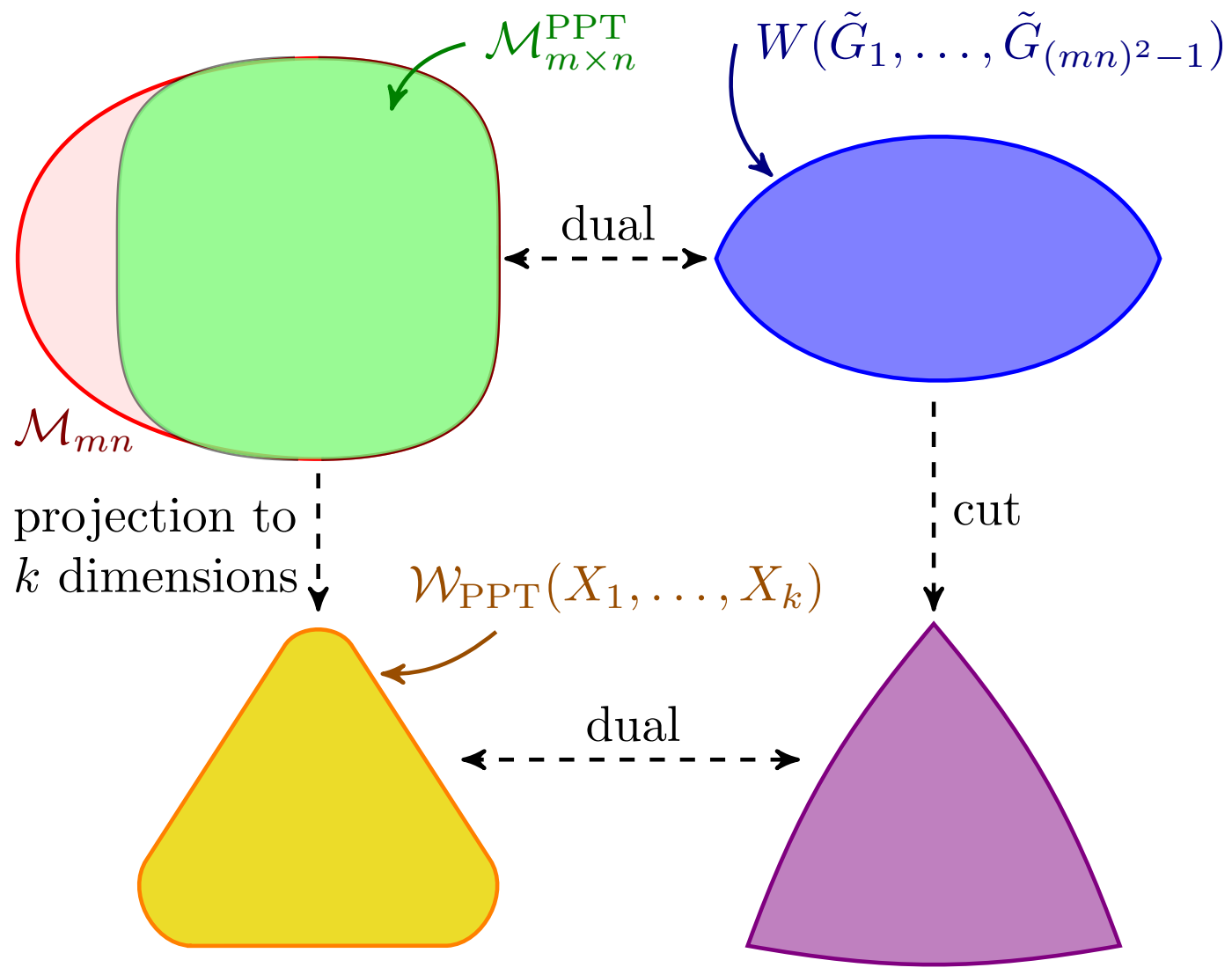

**Figure 4.17:** The structure behind the set $\mathcal{M}^{\mathrm{PPT}}_{m\times n}$ of states of a bipartite $m\times n$-dimensional quantum system with positive partial transpose, PPT numerical range $W_{\mathrm{PPT}}(X_1,\dots,X_k)$, joint numerical range dual to $\mathcal{M}^{\mathrm{PPT}}$ and its intersections.

**Definition 4.11** *The PPT numerical range of $k$ Hermitian operators $X_1,\dots,X_k$ of size $nm$ is*

$$W_{\mathrm{PPT}}(X_1,\dots,X_k) = \left\{\left(\langle X_1\rangle_\rho,\dots,\langle X_k\rangle_\rho\right) : \rho\in\mathcal{M}^{\mathrm{PPT}}_{m\times n}\right\}. \tag{4.75}$$

Since $\mathcal{M}^{\mathrm{PPT}}$ is convex, so is $W_{\mathrm{PPT}}$. PPT numerical range is thus defined by its extremal points and determination of the boundary can be reduced to a family of maximization problems.

Similarly to the case of $W_{\mathrm{SEP}}$, an operator $H=\sum_i n_i X_i$ is constructed and its expectation value is maximized over $\mathcal{M}^{\mathrm{PPT}}$. The result provides a point on $\partial W_{\mathrm{PPT}}$ with a normal vector equal to $\vec{n}$.

Maximization itself can be performed using known semidefinite optimization techniques. To present the PPT maximization problem in this way, one can introduce a linear parameterization of density operators in $\mathcal{M}_{mn}$ describing a bipartite system of local dimensions $m$ and $n$:

$$\rho(\vec{x}) = \frac{\mathbb{1}}{mn} + \sum_{i=1}^{(mn)^2-1} G_i x_i, \tag{4.76}$$

where the matrices $\{G_i\}_{i=1}^{(mn)^2-1}$ are Hermitian, traceless, and linearly independent. Then the PPT optimization problem can be expressed as

- maximize $\operatorname{Tr}\rho(\vec{x})H$ as a function of $\vec{x}$,
- subject to $\begin{pmatrix}\rho(\vec{x}) & 0\\ 0 & \rho(\vec{x})^{T_A}\end{pmatrix} \succcurlyeq 0$.

Such a formulation offers an interesting view on $\mathcal{M}^{\mathrm{PPT}}$: since there is a direct correspondence between $\rho\in\mathcal{M}^{\mathrm{PPT}}_{m\times n}$ and $\begin{pmatrix}\rho(\vec{x}) & 0\\ 0 & \rho(\vec{x})^{T_A}\end{pmatrix}\in\mathcal{M}_{2mn}$, the set $\mathcal{M}^{\mathrm{PPT}}_{m\times n}$ can be considered to be equivalent to a *spectrahedron* in $\mathcal{M}_{2mn}$. Spectrahedra are dual to numerical ranges, and therefore the set $\mathcal{M}^{\mathrm{PPT}}_{m\times n}$ is dual to a numerical range of $(mn)^2-1$ operators of size $2mn$; a fact outlined in the following theorem.

**Theorem 4.21** *Let the set of bipartite states $\mathcal{M}_{mn}$ with local dimension m and n be described as a spectrahedron spanned by the set of orthonormal operators $\{G_i\}_{i=1}^{(mn)^2-1}$:*

$$\mathcal{M}_{mn} = \mathrm{Spec}\left(\frac{\mathbb{1}}{mn}, G_1, \ldots, G_{(mn)^2-1}\right). \tag{4.77}$$

*In such a case, the set of PPT states $\mathcal{M}_{m\times n}^{\mathrm{PPT}}$ of this system is dual (polar) to the joint numerical range $W(\tilde{G}_1, \ldots, \tilde{G}_{(mn)^2-1})$, where*

$$\tilde{G}_i = -(mn)G_i \oplus G_i^{T_A}. \tag{4.78}$$

Here, $\oplus$ denotes the simple sum – composition of diagonal blocks.

*Furthermore, since PPT numerical range is an affine image of $\mathcal{M}_{m\times n}^{\mathrm{PPT}}$, the dual of PPT numerical range is the intersection of $W(\tilde{G}_1, \ldots, \tilde{G}_{(mn)^2-1})$ with a certain hyperplane.*

*Proof.* Duality of the set of positive partial transpose states $\mathcal{M}_{m\times n}^{\mathrm{PPT}}$ and the numerical range $W(\tilde{G}_1, \ldots, \tilde{G}_{(mn)^2-1})$ is a direct application of Theorem 4.3. With this ascertained, the duality of the PPT numerical range $W_{PPT}(X_1, \ldots, X_k)$ and a certain affine subset of $W(\tilde{G}_1, \ldots, \tilde{G}_{(mn)^2-1})$ is a result of Theorem 3.4. ☐

### 4.5.4 Restricted Schmidt rank numerical range

**Definition 4.12** *The Schmidt rank of a pure bipartite state $|\psi\rangle$ is the smallest number r of pure product states needed to compose $|\psi\rangle$:*

$$|\psi\rangle = \sum_{i=1}^{r} \sqrt{p_i}\, |\alpha_i\rangle \otimes |\beta_i\rangle\,, \tag{4.79}$$

*Such states are also called r-entangled.*

**Definition 4.13** *The convex hull of pure states defined by Definition 4.12 is the set of mixed Schmidt rank r states (also called r-entangled states), denoted by the symbol $\mathcal{M}^{r-\mathrm{ent.}}$.*

One of the most important[16] cases is the Schmidt rank equal to 2 (2-entangled states, $\mathcal{M}^{2-\mathrm{ent.}}$). Problems related to 2-entangled states include entanglement distillation, which can be reduced to the determination of the maximum expectation value over the set $\mathcal{M}^{2-\mathrm{ent.}}$. This can be achieved with the help of the separable numerical range.

16: Apart from $r = 1$ which corresponds to product states.

If the goal is to maximize the expectation value of $H$ over $\mathcal{M}_{m\times n}^{2-\mathrm{ent.}}$, one can rephrase the problem using a separable structure on the extended state space. If two orthogonal states $|\psi\rangle = |\alpha_1\rangle \otimes |\beta_1\rangle$ and $|\phi\rangle = |\alpha_2\rangle \otimes |\beta_2\rangle$ are fixed, the maximization of $\langle H\rangle$ over the subspace spanned by these vectors is fairly simple: the matrix elements of the restricted operator[17] are

The 2-entangled states here are a subset of $\mathcal{M}_{mn}$ for a bipartite system with local dimensions $m$ and $n$.
The notation is as follows: $|\alpha_i\rangle \in \mathcal{H}_m$ and $|\beta_i\rangle \in \mathcal{H}_n$.

17: This reduces the complexity a lot: the original operator $H$ is of size $mn$, and size of $H'$ is two.

$$H'_{ij} = \begin{pmatrix} \langle H\rangle_\psi & \langle\psi|H|\phi\rangle \\ \langle\phi|H|\psi\rangle & \langle H\rangle_\phi \end{pmatrix}_{ij}. \tag{4.80}$$

The largest eigenvalue of $H'$ corresponds to the maximum expectation value over the subspace spanned by $|\psi\rangle$ and $|\phi\rangle$. Direct calculation yields

$$\lambda_{+} = \frac{\langle H\rangle_{\psi} + \langle H\rangle_{\phi} + \sqrt{4|\langle\psi|H|\phi\rangle|^2 + (\langle H\rangle_{\psi} - \langle H\rangle_{\phi})^2}}{2}. \quad (4.81)$$

This form form suggests a method of maximization $\langle H\rangle$ over $\mathcal{M}^{2-\text{ent.}}$ globally, with the aid of a particular separable numerical range.

**Theorem 4.22** *Maximization of $\langle H\rangle$ over $\mathcal{M}^{2-ent.}$ is equivalent to finding the maximum (with constraints) of the following expression:*

$$\chi = \frac{\langle H_{+}\rangle_{\psi} + \sqrt{\langle H\otimes H\cdot \text{SWAP}\rangle_{\psi} + \langle H_{-}\rangle_{\psi}^{2}}}{2}. \quad (4.82)$$

*Here, $H_{\pm} = H \otimes \mathbb{1}_{mn} \pm \mathbb{1}_{mn} \otimes H$ and* SWAP *acts in the following way:*

$$\text{SWAP}\,|\alpha_1 \otimes \beta_1 \otimes \alpha_2 \otimes \beta_2\rangle = |\alpha_2 \otimes \beta_2 \otimes \alpha_1 \otimes \beta_1\rangle\,. \quad (4.83)$$

*The constraints are the following: the optimization of $\chi$ is to be done over the subset of*

$$W_{\text{SEP}}(H_{+}, H_{-}, H\otimes H\cdot \text{SWAP}, \text{SWAP}), \quad (4.84)$$

*for which $\langle \text{SWAP}\rangle_{\psi} = 0$.*

Here, the operators act on an extended space composed of two copies of the original bipartite system (four parts in total). The set of separable states used in $W_{\text{SEP}}$ is $\mathcal{M}^{\text{SEP}}_{m\times n\times m\times n}$. Therefore, the SWAP operator used here exchanges the two copies of the original bipartite system: $|\alpha_i\rangle \in \mathcal{H}_m$ and $|\beta_i\rangle \in \mathcal{H}_n$.

*Proof.* The condition $\langle \text{SWAP}\rangle_{\psi} = 0$ is equivalent to the two substates being orthogonal.[18] With this constraint, Eq. (4.82) defining $\chi$ can be recognized as equal to Eq. (4.81), which is a maximum expectation value of $H$ over a subspace of pure 2-entangled states spanned by two orthogonal pure product states. Clearly, every pure 2-entangled state lies on an appropriate subspace of this form – therefore, the maximum of Eq. (4.82) is equal to the maximum expectation value of $H$ over $\mathcal{M}^{2-\text{ent.}}$. □

18: Indeed, $\langle \text{SWAP}\rangle \geq 0$ over separable states – the subset to optimize over is thus a three-dimensional face of the analyzed four-dimensional separable numerical range $W_{\text{SEP}}$ defined in Eq. (4.84).

A hierarchy of problems can be recognized in this structure: determination of the maximal expectation value over $\mathcal{M}^{2-\text{ent.}}$ reduces to a problem involving separable numerical ranges, which in turn can be approximated with joint numerical ranges (Theorem 4.19). Finally, the joint numerical range is determined by the ground states of a combination of the defining observables (Theorem 4.2).

## 4.6 Concluding remarks

Sets of simultaneously attanaible expectation values – the *numerical ranges* – find numerous applications throughout quantum mechanics. In this chapter, I sketched the theory of these objects and presented my own results in this subject. Numerical ranges inherently shift the focus to only on a subset of variables defining a quantum state – this approach allows for significant simplification of various problems, and the applications presented in this chapter are based exactly on this idea. The most important results here are new perspectives on uncertainty relations, relation between the value of the energy gap and the geometry of numerical ranges, and entanglement detection.

# State interconversion 5

## 5.1 Introduction

Among multiple approaches to quantum information, the formalism of resource theories seems one of the most powerful ones: many different phenomena can be analyzed with a single underlying set of principles, including quantum thermodynamics, entanglement, quantum computations and symmetry of quantum states under arbitrary groups. In this chapter, I will explore the last case: the resource theory of asymmetry. In the most general form, it can be interpreted as a strategy to understand the structure of quantum states and allowed operations when some of the information about a quantum system is inaccessible.

Consider the following scenario: a quantum state $\rho$ is prepared in the laboratory $A$. The state is subsequently sent to the laboratory $B$ with the task of transforming it into the target state $\sigma$, with the necessary information like description of the initial and final states. However, the state $\rho$ is prepared with reference to the system $A$, which $B$ may not have access to. In the case of optical systems, the phase reference of $A$ may be different than in $B$, leading to diverging views on the state in question [58]. The quantization axis of spin states is defined with respect to some coordinate system, and relation between the two frames used in laboratories may be unknown. In this context, not every quantum operation $\mathcal{E}$ can be implemented in the laboratory $B$. As a result, not every *interconversion task* of turning $\rho$ into $\sigma$ can be performed, even if $B$ has complete control over its part of the quantum system.

[58]: Bartlett et al. (2006), 'Dialogue concerning two views on quantum coherence: factist and fictionist'

Here, *interconversion* is synonymous with *transformation*, as is customary in the literature. Outside of this context, *interconversion* may mean *mutual conversion* – two objects turning into each other at the same time – but this is not the case here.

If the relation between $A$ and $B$ is described with the language of group $G$ – e.g., $G = U(1)$ for phases of optical systems, $G = SU(2)$ for rotations of spins – then the implementable channels are exactly those which commute with $G$. Such operations are called group-covariant [59, 60]:

[59]: Marvian (2012), 'Symmetry, Asymmetry and Quantum Information'
[60]: Gour et al. (2008), 'The resource theory of quantum reference frames: manipulations and monotones'

**Definition 5.1** *A channel $\mathcal{E}$ is group-covariant with an unitary representation $\mathcal{U}$ of a group G acting on the set of quantum states if and only if it commutes with every group element: for every $g \in G$ and quantum state $\rho$,*

$$\mathcal{U}_g \mathcal{E}(\rho) \mathcal{U}_g^\dagger = \mathcal{E}(\mathcal{U}_g \rho \mathcal{U}_g^\dagger). \tag{5.1}$$

Here, I will analyze the simplest example of a nontrivial asymmetry theory: the theory of coherence of pure states of an energy ladder system. This is related to the problem of phase estimation in linear optics as well as quantum thermodynamics. The system in question is Hilbert space spanned by infinitely many kets $|n\rangle$ for each $n \in \mathbb{Z}$. Resource important for this kind of questions is coherence, related to the symmetry under time evolution generated by the following Hamiltonian:

Thermodynamical work that can be performed with the aid of a particular state can be measured by its coherence.

$$\hat{N} = \sum_{n\in\mathbb{Z}} n\, |n\rangle \langle n| \,. \tag{5.2}$$

The unitary evolution operator $\mathcal{U}_t = \exp(-it\hat{N})$ is periodic in time, denoted by the symbol $t$ – thefore, the set of all such unitaries $\{\mathcal{U}_t\}$ forms a representation of the group $U(1)$.

The same questions can be raised for arbitrary scenarios.

In the context of this particular resource theory, one could ask the most basic question: given two states,

**Definition 5.2** *The state $\rho$ can be transformed G-covariantly to $\sigma$ if and only if there exists a G-covariant channel $\mathcal{E}$ such that $\mathcal{E}(\rho) = \sigma$. This is also called state interconversion or reachability and denoted by $\rho \xrightarrow{G-\text{cov.}} \sigma$.*

$$\begin{aligned} |\psi\rangle &= \sum_{n\in\mathbb{Z}} e^{i\alpha_n} \sqrt{p_n}\, |n\rangle \,, \\ |\phi\rangle &= \sum_{n\in\mathbb{Z}} e^{i\beta_n} \sqrt{q_n}\, |n\rangle \,, \end{aligned} \tag{5.3}$$

is it possible to turn one into another using a quantum channel $\mathcal{E}$ which is *covariant* with respect to $U(1)$?

In this chapter, I will mostly restrict the analysis to the case of pure states.

This question – when can $|\psi\rangle$ be covariantly transformed into $|\phi\rangle$ – has been in part answered in previous works, presenting various methods than can be used to analyze the simple case of $U(1)$. Direct algebraic examination of the quantum channels compatible with the group structure is the basis of [60]. More abstract approach utilizing functions defined over groups and their relation to quantum states is discussed in [59].

**Definition 5.3** *The convolution of vectors $\vec{w}$ and $\vec{q} \in \mathbb{R}^{\mathbb{Z}}$ is the bilinear operation $* : \mathbb{R}^{\mathbb{Z}} \times \mathbb{R}^{\mathbb{Z}} \to \mathbb{R}^{\mathbb{Z}}$ defined by*

$$(\vec{w} * \vec{q})_n = \sum_{k\in\mathbb{Z}} w_{n-k} q_k. \tag{5.4}$$

All of them eventually reach the same condition, which can be expressed as a question of whether the probability vector $\vec{p}$ defining the initial state $|\psi\rangle$ is a *convolution* of $\vec{q}$ (corresponding to the target state $|\phi\rangle$) with a nonnegative weight vector $\vec{w}$.

Here, $\mathbb{R}^{\mathbb{Z}}$ denotes the infinite-dimensional vector space of functions $\mathbb{Z} \to \mathbb{R}$ – that is, the coordinates are indexed with integers.

Exact deconvolution of infinite-dimensional vectors is a nontrivial operation, and the previously developed convertibility criteria are often unwieldy to use as a result. In this chapter, I provide a simplier and easily implementable test of convertibility with respect to $U(1)$-covariant channels. In the test, the algebraic structure of *circulant matrices* is used. Such operators are defined by finite-dimensional vectors – and the multiplication of them implements convolution with periodic boundary conditions. Therefore, while the convolutional structure is still there, its complexity is greatly reduced due to the finite number of dimensions.

Additional remarks about the set of accessible states can be noted: while there is a continuum of states which lead to a particular state $|\psi\rangle$, only a discrete set of states can be accessed from it. This observation is proved here with the aid of another convolution-implementing algebraic structure: the ring of polynomials of one variable.

## 5.2 Theoretical background

The problem posed in the introduction can be expressed in the following way: given normalized states $|\psi\rangle$ and $|\phi\rangle$ expressed by Eq. (5.3), decide whether a transformation $|\psi\rangle \xrightarrow{U(1)-\text{cov.}} |\phi\rangle$ between the two pure states is possible, where the representation of $U(1)$ is the time evolution:

$$\mathcal{U}_t = \exp(-it\hat{N}) = \sum_{n\in\mathbb{Z}} \exp(-int)\,|n\rangle\,\langle n|\,. \tag{5.5}$$

At least two approaches to answer this problem have been proposed. I will briefly sketch them, both providing exhaustive answers to the problem of $U(1)$ state reachability (i.e., necessary and sufficient conditions). Hence, it comes as no surprise that the two approaches – despite starting from different places – eventually converge to yield the same statement. However, as mentioned previously, the provided condition is inconvenient to use – I will provide a solution to this problem in the section following the outlines of the two methods. The simplification allows for a more in-depth study of the accessibility structure.

### 5.2.1 Approach 1: decomposition of $U(1)$-covariant channels

Quantum channels commuting with elements of a given group $G$ are formed from Kraus operators belonging to irreducible representations of $G$ – the general form of admissible Kraus operators is given the the Lemma 1 of [60], stating the following:

[60]: Gour et al. (2008), 'The resource theory of quantum reference frames: manipulations and monotones'

**Lemma 5.1** *Any G-covariant channel admits a decomposition into Kraus operators,*

$$\mathcal{E}(\rho) = \sum_{j,m,\alpha} K_{j,m,\alpha}\rho K^{\dagger}_{j,m,\alpha}, \tag{5.6}$$

*where $j$ denotes index of the irreducible representation, $m$ is an internal index of the representation, and $\alpha$ labels potential copies. The Kraus operators can only mix under conjugation with $\mathcal{U}_g$ with the unitary representation $u^{(j)}$ of $G$ inside the single $j$ and $\alpha$:*

$$\mathcal{U}_g K_{j,m,\alpha}\mathcal{U}_g^{\dagger} = \sum_{m,n} u^{(j)}_{mn}(g) K_{j,n,\alpha}. \tag{5.7}$$

The most important case here is the group $G = U(1)$ with an unitary representation defined by Eq. (5.5). Since the irreducible representations of $U(1)$ are numbered by $k \in \mathbb{Z}$ only,[1] the lemma amounts to the following statement:

1: Irreps of $U(1)$ have no internal structure, thus the matrices $u_j(g)$ appearing in Lemma 5.1 are $1 \times 1$, and I assumed that only single instances of representations are present.

$$\mathcal{U}_\theta K_k \mathcal{U}_\theta^\dagger = \exp(-ik\theta) K_k. \tag{5.8}$$

Substitution by the explicit, most general form of Kraus operators[2] leads to

2: The full matrix: $K_k = \sum_{m,n \in \mathbb{Z}} c_{m,n}^{(k)} |m\rangle \langle n|$.

$$\begin{aligned} \sum_{m,n \in \mathbb{Z}} c_{m,n}^{(k)} \exp(-i\theta(m-n)) |m\rangle \langle n| = \\ \sum_{m,n \in \mathbb{Z}} c_{m,n}^{(k)} \exp(-ik\theta) |m\rangle \langle n| . \end{aligned} \tag{5.9}$$

This equation needs to hold for an arbitrary phase $\theta$, thus the exponent arguments need to be equal. Consequently, $m - n = k$ and the Kraus operator reads

$$K_k = \sum_{n \in \mathbb{Z}} \kappa_n^{(k)} |n + k\rangle \langle n| , \tag{5.10}$$

where $\kappa_n^{(k)} = c_{n+k,n}^{(k)}$. Therefore, $U(1)$-invariant channels $\mathcal{E}$ are defined by a set of Kraus operators $\{K_k\}$ such that $K_k$ is a Kraus operator shifting the state along the infinite ladder by $k$ energy levels with possible changes in relative amplitude.

The objects of interest here are pure states only. However, the state resulting from the application of a quantum channel composed from such operators to a pure state $|\psi\rangle$,

$$\mathcal{E}(|\psi\rangle \langle\psi|) = \sum_{k \in \mathbb{Z}} K_k |\psi\rangle \langle\psi| K_k^\dagger, \tag{5.11}$$

in general might not be pure. The requirement of $\mathcal{E}(|\psi\rangle \langle\psi|)$ being pure restricts the set of admissible quantum channels: for this to be true, every $K_k |\psi\rangle$ must be proportional to the same ket – the target state, $|\phi\rangle$:

The minus sign of index in Eq. 5.12 is to ensure consistency with later reasoning.

$$K_k |\psi\rangle = \sqrt{w_{-k}} |\phi\rangle . \tag{5.12}$$

By inserting the explicit decomposition of states $|\psi\rangle$ and $|\phi\rangle$ in the $\{|n\rangle\}_{n \in \mathbb{Z}}$ basis (Eq. (5.3)) and the action of the Kraus operator $K_k$ (Eq. (5.10)), the sought $U(1)$-covariant accessibility criterion can be derived. As the form of Kraus operators – shifting and mixing different parts of the quantum state – suggests, one state is reachable from another if the states are related via a properly defined convolution. This observation is one of the main results of [60].

[60]: Gour et al. (2008), 'The resource theory of quantum reference frames: manipulations and monotones'

**Theorem 5.2** (Gour-Spekkens) *For the states*

$$|\psi\rangle = \sum_{n\in\mathbb{Z}} e^{i\alpha_n}\sqrt{p_n}\,|n\rangle\,,\qquad |\phi\rangle = \sum_{n\in\mathbb{Z}} e^{i\beta_n}\sqrt{q_n}\,|n\rangle\,, \tag{5.13}$$

*there exists an $U(1)$-covariant channel $\mathscr{E}$ such that $\mathscr{E}(|\psi\rangle\langle\psi|) = |\phi\rangle\langle\phi|$ if and only if the the probability vectors $\vec{p},\vec{q}\in\mathbb{R}^{\mathbb{Z}}$ are connected through a discrete convolution:*

$$\vec{p} = \vec{w} * \vec{q}, \tag{5.14}$$

*such that $\vec{w}$ is a probability vector.*
*The quantum channel realizing the transformation is then formed by Kraus operators: $\mathscr{E}(\rho) = \sum_{k\in\mathbb{Z}} K_k\rho K_k^\dagger$, where*

$$K_k = \sum_{\substack{n\in\mathbb{Z}\\ p_n\neq 0}} \underbrace{\sqrt{\frac{w_{-k}q_{n+k}}{p_n}}}_{\kappa_n^{(k)}} |n+k\rangle\langle n|\,. \tag{5.15}$$

Note that the phases $\alpha_n$ and $\beta_n$ do not have any effect on the accessibility structure. This is because any channel which only action is to change the phase of basis kets ($|n\rangle \mapsto \exp(i\gamma_n)\,|n\rangle$) is $U(1)$-covariant. In the general case of $G$-covariance for an arbitrary group $G$ this applies as well: the relative phases of substates belonging to different representations of $G$ can be arbitrarily changed. Here, every $|n\rangle$ forms a one-dimensional representation of $U(1)$.
The phases can take arbitrary values, but they *must be definite*. A mixed state of this system with the same probabilities (expectation values of $|n\rangle\langle n|$ for every $n$) has less resource.
Formally, the channel defined by Kraus operators of form appearing in Eq. (5.15) is trace decreasing, due to the fact that the only potential nonzero elements are for $n\in\mathbb{Z}$ such that $p_n \neq 0$. This can be fixed by adding auxillary elements to this sum for the omitted indices, such that $\sum_k K_k^\dagger K_k = \mathbb{1}$. In further considerations I omit this extension, since it does not change the accessibility structure.

By reformulating the problem in such terms, most of the complexity vanished: *a priori*, there is no obvious way to constraint $\kappa_n^{(k)}$ appearing in Eq. (5.10). This theorem allows us to search only for the relative weights between shift operators; the finer structure of amplitude multipliers $\kappa_n^{(k)}$ is then completely determined.

Note the reverse order: a state vector defined by $\vec{p}$ can be turned into the one defined by $\vec{q}$ if and only if $\vec{p}$ is *a convolution of* $\vec{q}$ with another probability vector, even though the intuition based on usual probability theory would suggest otherwise.

Experience from the theory of stochastic operations could suggest that a *peaked state* ($p_i = \delta_{i,k}$ for some $k$) is the most powerful one – here, the converse is true.

This is because in this theory, the coherence across a range of eigenstates is a resource. Therefore, states with singleton support ($|n\rangle$), having no meaningful coherence, can always be obtained from *any* state. Note that the *constant quantum channel* always returning one of the basis states $|n\rangle$ is $U(1)$-covariant![3]

3: E.g., a channel formed by $K_{-k} = |0\rangle\langle k|$ for every $k\in\mathbb{Z}$ always yields $|0\rangle$ regardless of input.

### 5.2.2 Approach 2: characteristic functions of states

There exists another approach to the theory of state transformation, allowing for a different, more abstract view. The feasibility of a state interconversion can be formulated as a statement about the positive definiteness of a function defined over group elements, as described by the following theorem [59]:

[59]: Marvian (2012), 'Symmetry, Asymmetry and Quantum Information'

**Definition 5.4** *The expression $\langle\psi| U_g |\psi\rangle$ is called the characteristic function of $|\psi\rangle$ and is denoted by $\chi_\psi(g)$.*

In the case of a discrete group $G$, the integral is changed to sum over pairs of group elements.

**Theorem 5.3** (Marvian) *There exists a G-covariant (see Definition 5.1) quantum channel $\mathcal{E}$ transforming the pure state $|\psi\rangle$ into $|\phi\rangle$,*

$$\mathcal{E}(|\psi\rangle\langle\psi|) = |\phi\rangle\langle\phi|, \tag{5.16}$$

*if and only if there exists a positive semidefinite function $f$ defined over elements of group G such that*

$$\langle\psi| U_g |\psi\rangle = f(g) \langle\phi| U_g |\phi\rangle. \tag{5.17}$$

*Here, $f$ is positive semidefinite in the sense of positivity of the following double integral over the group G for all functions $c : G \longrightarrow \mathbb{C}$:*

$$\int c_g^* f(gh^{-1}) c_h dg dh \geq 0. \tag{5.18}$$

The analysis of the group $G = U(1)$ based on this result is substantially different to the reasoning presented in the previous section, but leads to the same conclusion. Since it is a simple nontrivial application of this theorem and illustrates its main points well, I demonstrate it below.

[61]: Michael Reed (1975), *Methods of Modern Mathematical Physics, vol. 2*

$p_n = |\langle n|\psi\rangle|^2$ and $q_n = |\langle n|\phi\rangle|^2$.

**Example 5.1** In the case of $U(1)$, the characteristic function of a state $|\psi\rangle$,

$$\chi_\psi(\theta) = \langle\psi| \exp(-i\hat{N}\theta) |\psi\rangle, \tag{5.19}$$

is the Fourier transform of the discrete probability distribution $|\langle n|\psi\rangle|^2$,

$$\chi_\psi(\theta) = \int \exp(-i\theta x) \left( \sum_{n\in\mathbb{Z}} \delta(x-n) \left|\langle n|\psi\rangle\right|^2 \right) dx. \tag{5.20}$$

Positive semidefiniteness of $f = \chi_\psi \chi_\phi^{-1}$ is equivalent to the positivity of its Fourier transform $\tilde{f}$ – a result known as Bochner's theorem [61]. The characteristic functions are explicitly expressed as Fourier transforms as well, therefore the condition for the existence of a covariant quantum channel described in Theorem 5.3 can be written out as follows:

If $\vec{p}$ and $\vec{q}$ are defined as in Eq. (5.3), then $|\psi\rangle \xrightarrow{U(1)\text{-cov.}} |\phi\rangle$ if and only if there exists a vector $\vec{w}$ with nonnegative entries such that

$$\vec{p} = \vec{w} * \vec{q}, \tag{5.21}$$

where $*$ denotes the discrete convolution. Normalization of $\vec{w}$ to the unit sum is enforced by the properties of the convolution – thus, $\vec{w}$ acts as a probability distribution, or vector of weights, for shifting $\vec{q}$ to arrive at $\vec{p}$ – the central problem becomes to invert Eq. (5.21) and check if $\vec{w}$ has only nonnegative elements.

In the case of more general groups, the results of [59] , while powerful, do not present explicit and easily applicable criteria for interconvertibility;[4] little is known about the properties of accessible states in these cases. Limited results arising from this theory in the case of $SU(2)$ and Weyl-Heisenberg groups are presented at the end of this chapter.

[59]: Marvian (2012), 'Symmetry, Asymmetry and Quantum Information'

4: Anything beyond numerical analysis is challenging.

## 5.3 Finite-dimensional relaxations

Two complementary approaches to arrive at the same condition of accessibility of states via $U(1)$-covariant channels have been presented. The condition does not offer much insight into the structure of accessible states on its own – mainly because of the infinite-dimensional nature of the system. To aid in understanding, two mathematically similar – and easier to interpret – structures are used: the circulant matrices and the ring of polynomials.

In the case of $U(1)$-covariant interconversion, the convolution of probability distributions plays an important role. Therefore, algebraic structures which implement convolution naturally can be used to tackle this problem in a simplified way. There are two possible natural choices: circulant matrices and polynomials.

**Circulant matrices** are parameterized by finite-dimensional vectors. Multiplication of two such matrices is equivalent to the convolution of the defining vectors – however, the convolution is cyclic (with periodic boundary conditions).

**Polynomials of one variable** have the same property: coefficients of the product $fg$ of two polynomials $f$ and $g$ are formed by the convolution of the coefficients of $f$ and $g$. The convolution is not cyclic, but still finite-dimensional due to the fact that polynomials always have finite degrees.

**Lemma 5.4** *For polynomials $f = \sum_{i=0}^{m} f_i x^i$ and $g = \sum_{i=0}^{n} g_i x^i$, their coefficients of the product $fg = \sum_{i=0}^{m+n} h_i x^i$ are formed by the discrete convolution of the coefficients of $f$ and $g$. Loosely speaking, $\vec{h} = \vec{f} * \vec{g}$.*

*Proof.* Direct multiplication yields $fg = \sum_{i=0}^{m} \sum_{j=0}^{n} f_i g_j x^{i+j}$. Therefore, the coefficient $h_k$ of $x^k$ is formed by the sum of $f_i g_j$ such that $i+j=k$. □

Throughout this chapter, circulant matrices are used for applications of the theory – linear algebra is appropriate to actually perform the computations – polynomials are suited well for proving the existence of accessible states due to the rich theory of polynomial equations.

Circulant matrices may be viewed as arising from considering the convolution of a *fixed* vector $\vec{w}$ and a variable $\vec{q}$,

$$\vec{q} \mapsto \vec{w} * \vec{q} \tag{5.22}$$

as a linear map in a restricted, finite-dimensional basis – this map is represented exactly[5] by a circulant matrix $C(\vec{w})$. As a result of

5: In the natural basis consisting of peaked vectors, $(\vec{e}_k)_i = \delta_{ik}$.

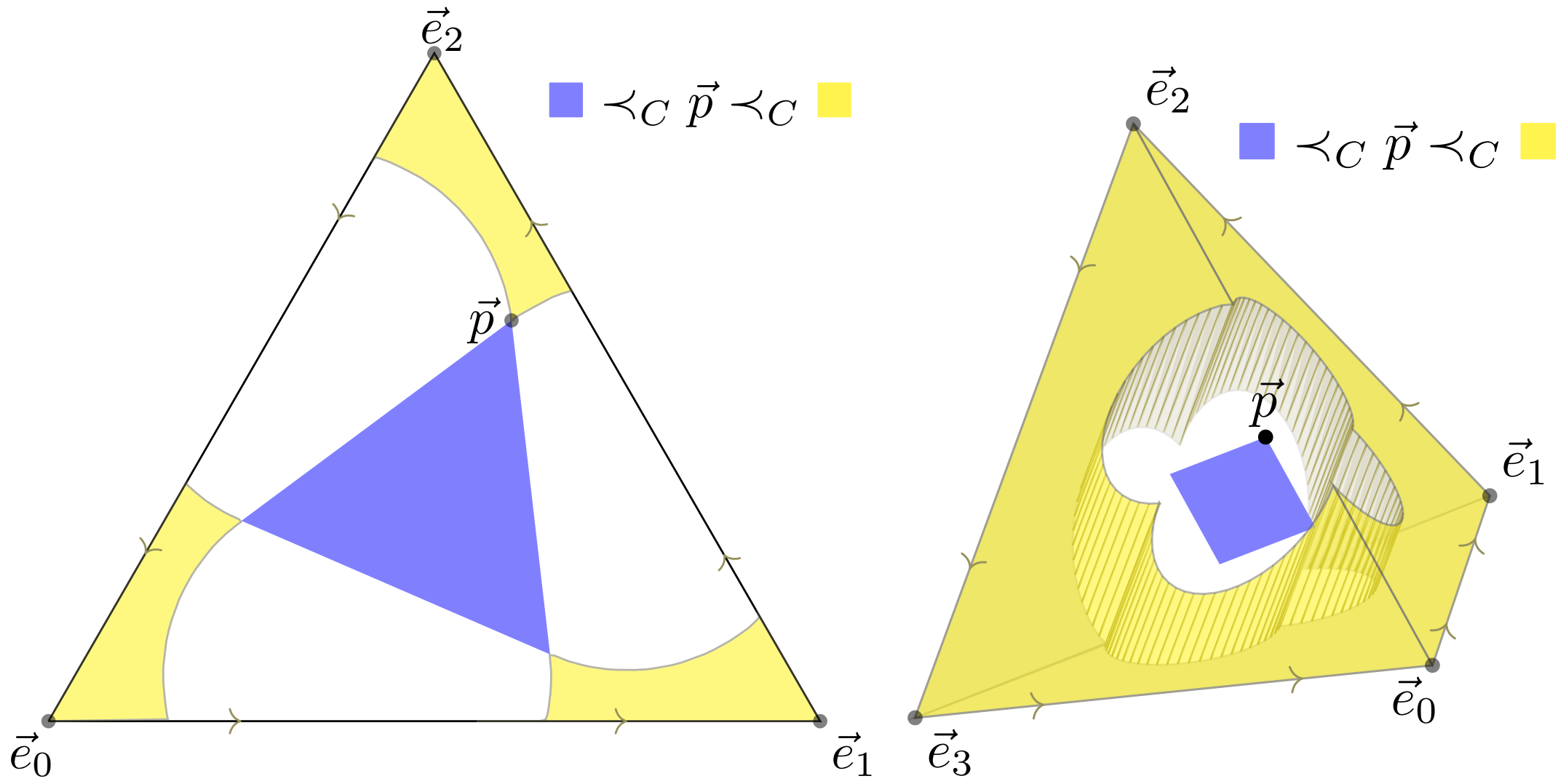

**Figure 5.1:** *Left*: Visualization of the set of precedence of the relation $\prec_C$ in the case of probability space of dimension three, in which case the probability simplex is a regular triangle, with the vertices $\vec{e}_i$ corresponding to the pure states. Here, the set of points accessible from $\vec{p}$ is always a rotated and scaled down probability simplex for every $\vec{p}$, and the set of points preceding $\vec{p}$ is connected to $\vec{p}$.
*Right*: Sets of precedence for probability space of dimension four. In this case, the probability simplex is a tetrahedron. The precedence structure is more complex, with degenerated (flat) sets preceded by $\vec{p}$ and disconnected sets preceding $\vec{p}$ becoming possible. The stereographic (3D) version of this picture is available in Figure A.3 (Appendix A, page 103).

being constructed to mimic convolution, this matrix shares many properties with it[6] . Crucially, if a circulant matrix is nonsingular, it is also easily invertible, which helps with providing an explicit interconversion criterion.

6: The convolutional and multiplication structures are compatible:

**Remark 5.1** For all $\vec{w}$ and $\vec{q}$, $C(\vec{w})C(\vec{q}) = C(\vec{w} * \vec{q})$.

**Definition 5.5** (Circulant matrices) *Let $\vec{p} \in \mathbb{R}^N$. The set of matrices defined by*

$$C(\vec{p}) = \begin{pmatrix} p_1 & p_2 & \dots & p_n \\ p_n & p_1 & \dots & p_{n-1} \\ \vdots & & \ddots & \vdots \\ p_2 & p_3 & \dots & p_1 \end{pmatrix}, \tag{5.23}$$

*forms an abelian semigroup. The elements of this set are called circulant matrices.*

To put the result of Theorem 5.2 in a simpler form, a special kind of majorization over vectors in $\mathbb{R}^{\mathbb{Z}}$ can be introduced. This is a partial order relation based on reachability through convolution. Convolution can be translated into the language of circulant matrices – hence, also the partial order relation has a *circulant* (or *cyclic*) counterpart, which replicates many of its properties. This will come in handy later.

**Definition 5.6** ($\mathbb{Z}$-majorization) *The vector $\vec{q} \in \mathbb{R}^{\mathbb{Z}}$ is said to $\mathbb{Z}$-majorize $\vec{p} \in \mathbb{R}^{\mathbb{Z}}$ (written as $\vec{p} \prec_{\mathbb{Z}} \vec{q}$) if and only if*

$$\vec{p} \in \operatorname{conv}\left\{\Delta^k \vec{q} : k \in \mathbb{Z}\right\}, \tag{5.24}$$

*where $\Delta$ is an operator shifting the whole probability distribution up by a single step:*

$$\left(\Delta \vec{q}\right)_n = q_{n-1}. \tag{5.25}$$

*In other words: $\vec{p} \prec_{\mathbb{Z}} \vec{q}$ if and only if $\vec{p} = \vec{q} * \vec{w}$ for some $\vec{w}$ with nonnegative entries summing up to unity.*

Note that this relation exactly captures the reachability structure corresponding to the $U(1)$-covariant quantum channel. It might at this point sound just like a rephrasing of the same criterion – however, it has a circulant (cyclic) counterpart.

**Definition 5.7** (Cyclic majorization) *The vector $\vec{q} \in \mathbb{R}^N$ is said to cyclically majorize $\vec{p} \in \mathbb{R}^N$, $\vec{p} \prec_C \vec{q}$, if and only if $\vec{p}$ is contained in the convex hull of all circulant shifts of $\vec{q}$ [62]:*

$$\vec{p} \prec_C \vec{q} \Leftrightarrow \vec{p} \in \operatorname{conv}\left\{P^k \vec{q} : k \in \mathbb{Z}_N\right\}, \tag{5.26}$$

*where $P$ is the elementary circulant permutation matrix of size $N$:*

$$P = \begin{pmatrix} 0 & 1 & 0 & \dots & 0 \\ 0 & 0 & 1 & \dots & 0 \\ 0 & 0 & 0 & & \vdots \\ \vdots & & & \ddots & 1 \\ 1 & 0 & 0 & \dots & 0 \end{pmatrix}. \tag{5.27}$$

Note the similarities between cyclic majorization and $\mathbb{Z}$-majorization in Definition 5.6.

[62]: Giovagnoli et al. (1996), 'Cyclic majorization and smoothing operators'

The shift operator $P$ can be viewed as the restriction of the operator $\Delta$ (Eq. (5.25)) to $N$ levels with additionally imposed cyclic boundary conditions ($p_i = p_{i \bmod N}$).

Another view on the circulant matrices is that they form an affine subspace spanned by all $P^n$.

With the notation provided by this definition, the relation defined by Eq. (5.26) admits an explicit suffient (but not necessary) condition. The following Lemma is a simplification of the reasoning presented in [62].

[62]: Giovagnoli et al. (1996), 'Cyclic majorization and smoothing operators'

**Lemma 5.5** *The vector $\vec{p} \in \mathbb{R}^N$ is cyclically majorized by $\vec{q} \in \mathbb{R}^N$ ($\vec{p} \prec_C \vec{q}$) if both following criteria are met:*

*a) $C(\vec{q})$ is nonsingular.*
*b) The entries of the matrix $C(\vec{w})$, defined by*

$$C(\vec{w}) = C(\vec{p})C(\vec{q})^{-1}, \tag{5.28}$$

*are nonnegative.*

*Proof.* If $\vec{p}$ is contained in the convex hull of $\{P^k\vec{q}\}$, there exists a convex combination of the extremal points resulting in $\vec{p}$:

$$\vec{p} = \sum_{k\in\mathbb{Z}_N} w_k P^k \vec{q}. \tag{5.29}$$

Conveniently, the total combination of shift operators, also known as *smoothing operator*, is expressible by $C(\cdot)$:

$$C(\vec{w}) = \sum_{k\in\mathbb{Z}_N} w_k P^k. \tag{5.30}$$

Therefore, $\vec{p} \prec_C \vec{q}$ if and only if there exists $\vec{w}$ (with positive entries and $\sum_k w_k = 1$) such that

$$\vec{p} = C(\vec{w})\vec{q}. \tag{5.31}$$

This equation essentially ensures that $\vec{p}$ is a discrete convolution of $\vec{q}$ with weights of $\vec{w}$. Since circulant matrices naturally implement convolution as part of their algebraic structure, this can be rewritten as

$$C(\vec{p}) = C(\vec{w})C(\vec{q}). \tag{5.32}$$

This, in turn, implies that (provided $C(\vec{q})$ is invertible):

$$C(\vec{w}) = C(\vec{p})C(\vec{q})^{-1}. \tag{5.33}$$

Entries of $C(\vec{w})$ are exactly the coordinates of $\vec{w}$ – therefore, if all elements of $C(\vec{p})C(\vec{q})^{-1}$ are nonnegative, this implies that $\vec{p} \prec_C \vec{q}$. □

It may happen that $C(\vec{q})$ is noninvertible – in such case the authors of [62] suggest to use the Moore-Penrose pseudoinverses of proper full-rank submatrices of $C(\vec{q})$. They claim that if the procedure works for one of these matrices, it follows that $\vec{p} \prec_C \vec{q}$ – therefore, the substitution of $C(\vec{q})^{-1}$ by one of the pseudoinverses of $C(\vec{q})$ leads to the theorem working in both directions. However, in our application it is possible to ensure the invertibility of the matrices appearing during the calculations.

[62]: Giovagnoli et al. (1996), 'Cyclic majorization and smoothing operators'

## 5.4 Cyclic majorization in interconversion

The original interconversion criterion (Theorem 5.2) makes use of the $\mathbb{Z}$-majorization,[7] with all its interpretation problems. There is a way to use the finite-dimensional cyclic majorization $\prec_C$ instead. This is feasible if the support of both vectors $\vec{p}$ and $\vec{q}$ is bounded.

7: The state $|\phi\rangle$ is reachable from $|\psi\rangle$ if and only if for the defining vectors $\vec{p}$ and $\vec{q}$ in $\mathbb{R}^{\mathbb{Z}}$, one precedes the other: $\vec{p} \prec_{\mathbb{Z}} \vec{q}$.

The main idea is to excise the nonzero part of $\vec{p}$ and $\vec{q}$ and embed it in a probability space with cyclic structure. That is, the embedded vectors are contained in finite-dimensional space $\mathbb{R}^N$:

$$\mathbb{R}^{\mathbb{Z}} \ni \vec{p} \mapsto \vec{p}^\circ \in \mathbb{R}^N, \quad \mathbb{R}^{\mathbb{Z}} \ni \vec{q} \mapsto \vec{q}^\circ \in \mathbb{R}^N. \tag{5.34}$$

Then the cyclic majorization is applied to check whether $\vec{p}^\circ \prec_C \vec{q}^\circ$. If this condition holds and the embedding is designed in a proper way, then one can ascertain that $\vec{p} \prec_{\mathbb{Z}} \vec{q}$.

### 5.4.1 Vector embedding

The reasoning presented below will use the following notion:

**Definition 5.8** *The support of a probability vector* supp *is the set of its nonzero elements:*

$$\operatorname{supp} \vec{x} = \{k \in \mathbb{Z} : x_k \neq 0\}. \tag{5.35}$$

*Additionally, the diameter of support is*

$$\operatorname{diam} \vec{x} = \max \operatorname{supp} \vec{x} - \min \operatorname{supp} \vec{x}. \tag{5.36}$$

Without loss of generality[8] it can be assumed that the supports of $\vec{p}$ and $\vec{q} \in \mathbb{R}^{\mathbb{Z}}$ are bounded:

$$0 \leq \operatorname{supp} \vec{p}, \operatorname{supp} \vec{q} \leq n. \tag{5.37}$$

8: The initial problem of $U(1)$ interconversion is translationally invariant.

If this condition is met, the vectors $\vec{p}$ and $\vec{q}$ can be embedded into $\mathbb{R}^N$ yielding $\vec{p}^\circ$ and $\vec{q}^\circ$:

$$\begin{aligned} \vec{p} \mapsto \vec{p}^\circ &= (p_0, p_1, \ldots, p_n, \overbrace{0, 0, \ldots, 0}^{N-n-1 \text{ times}}), \\ \vec{q} \mapsto \vec{q}^\circ &= (q_0, q_1, \ldots, q_n, \underbrace{0, 0, \ldots, 0}_{N-n-1 \text{ times}}). \end{aligned} \tag{5.38}$$

With the embedding now defined, let us formulate the following theorem.

**Theorem 5.6** *Let the probability vectors $\vec{p}$ and $\vec{q} \in \mathbb{R}^{\mathbb{Z}}$ have finite support satisfying Eq.* (5.37) *and* $0 \in \operatorname{supp} \vec{q}$. *If the embedding (according to Eq.* (5.38)*) is of high enough dimension: $N > 2n+1$, the cyclic majorization of embedded vectors is equivalent to $\mathbb{Z}$-majorization of the original vectors:*

$$\vec{p} \prec_{\mathbb{Z}} \vec{q} \iff \vec{p}^\circ \prec_C \vec{q}^\circ. \tag{5.39}$$

*Proof.* If the majorization criterion $\vec{p} \prec_{\mathbb{Z}} \vec{q}$ holds, then there exists $\vec{w}$ which realizes the transformation:[9]

$$\vec{p} = \sum_{k \in \mathbb{Z}} w_k \Delta^k \vec{q} = \sum_{k=0}^{n-m} w_k \Delta^k \vec{q}. \tag{5.40}$$

9: Such that $\forall_k w_k \geq 0$ and $\sum w_k = 1$.

The last equality comes from the fact that the support of $\vec{w}$ is a subset of $\{0, \ldots, n-m\}$: any $w_k \neq 0$ for $k$ outside this range would be inconsistent with the assumptions placed on the supports of the vectors $\vec{p}$ and $\vec{q}$ – see Eq. (5.37).

Now, let $N > 2n+1$ – a requirement that ensures the proper size of the embedding of $\vec{p}$ and $\vec{q}$ performed according to Eq. (5.38). The embedded vectors are denoted by $\vec{p}^{\circ}$ and $\vec{q}^{\circ} \in \mathbb{R}^N$, respectively.

Then the expression analogous to Eq. (5.40),

$$\left(\sum_{k=0}^{n-m} w_k P^k \vec{q}^{\circ}\right)_i = \sum_{k=0}^{n-m} w_k q^{\circ}_{i-k} \tag{5.41}$$

depends on $q^{\circ}_i, \ldots, q^{\circ}_{i-(n-m)}$. However, it is easy to observe that since $n - m \leq n$, then for all relevant $k$ and $i$, the equation $q^{\circ}_{i-k} = q_{i-k}$ holds and consequently

$$\begin{aligned}\left(\sum_{k=0}^{n-m} w_k P^k \vec{q}^{\circ}\right)_i &= \sum_{k=0}^{n-m} w_k q_{i-k} \\ &= \left(\sum_{k=0}^{n-m} w_k \Delta^k \vec{q}\right)_i = p_i.\end{aligned} \tag{5.42}$$

In other words, $\vec{q}^{\circ}$ is transformed into $\vec{p}^{\circ}$ by the circulant matrix $C(\vec{w})$ the same way as $\vec{q}$ is transformed into $\vec{p}$ – therefore, $\vec{p} \prec_{\mathbb{Z}} \vec{q}$ implies $\vec{p}^{\circ} \prec_C \vec{q}^{\circ}$. Since all transformations here are equivalent, the converse is also true: if $\vec{p}^{\circ} \prec_C \vec{q}^{\circ}$ for $\vec{p}^{\circ}, \vec{q}^{\circ}$ padded with enough zeros, then $\vec{p} \prec_{\mathbb{Z}} \vec{q}$. □

The assumptions on $\operatorname{supp} \vec{p}$ and $\operatorname{supp} \vec{q}$ can be relaxed: since powers of the shift operator $\Delta$ are invertible, $U(1)$-covariant, and can shift the supports of $\vec{p}, \vec{q}$ to any position, the vectors may be translated anywhere. Binding the supports to 0 only helps with calculations and the statement of the theorem.

In practice, some choices of embedding dimension $N$ may lead to the matrix $C(\vec{q})$ appearing in the cyclic majorization criterion to be singular. Treatment of this case is difficult and can be easily avoided by increasing the embedding dimension $N$. This is guaranteed to eventually lead to a nonsingular case, as is guaranteed by the following theorem.

**Theorem 5.7** *For every vector* $\vec{q} \in \mathbb{R}^{\mathbb{Z}}$ *such that* $0 \leq \operatorname{supp} \vec{q} \leq n$, *there exists* $N > 2n+1$ *such that the circulant matrix* $C(\vec{q}^{\circ})$ *of the embedded vector* $\vec{q}^{\circ} \in \mathbb{R}^N$ *is invertible.*

*Proof.* $C(\vec{q}^\circ)$ has full rank if and only if its *associated polynomial*,

$$f_{\vec{q}^\circ}(x) = \sum_{i=0}^{N-1} x^i q_i^\circ, \tag{5.43}$$

has no common divisors with $1 - x^N$ other than 1 – see [63]. The roots $f_{\vec{q}^\circ}$ are independent from the embedding dimension $N$, but the roots $1 - x^N$ are not: the only common root of $1 - x^N$ and $1 - x^M$ when $N$ and $M$ are prime is $x = 1$. Since 1 is not a root of the associated polynomial,[10] it can not share roots with an infinite family of polynomials corresponding to prime embedding dimensions. As a consequence, it can not share divisors with an infinite number of polynomials corresponding to different embedding dimensions. Thus, it is possible to find an embedding dimension $N$ for which $C(\vec{q}^\circ)$ is invertible. □

[63]: Ingleton (1956), 'The Rank of Circulant Matrices'

10: As proved by direct calculation: $f_{\vec{q}^\circ}(1) = \sum_{i=0}^{N-1} q_i^\circ = 1$.

**Example 5.2** Is the transformation $\mathscr{E}(|\psi\rangle) = |\phi\rangle$ possible with $U(1)$-covariant channel $\mathscr{E}$? The states $|\psi\rangle$ and $|\phi\rangle$ are defined on an infinite ladder as:

$$\begin{aligned} |\psi\rangle &= \frac{1}{\sqrt{6}}\left(|1\rangle + \sqrt{2}\,|2\rangle + \sqrt{2}\,|3\rangle + |4\rangle\right). \\ |\phi\rangle &= \frac{1}{\sqrt{2}}\left(|0\rangle + |1\rangle\right). \end{aligned} \tag{5.44}$$

As a result of Theorem 5.2, the problem is equivalent to determining whether $\vec{p} \prec_{\mathbb{Z}} \vec{q}$, where

$$\begin{aligned} \vec{p} &= \left(\ldots, 0, \frac{1}{6}, \frac{1}{3}, \frac{1}{3}, \frac{1}{6}, 0, 0, \ldots\right) \\ \vec{q} &= \left(\ldots, \frac{1}{2}, \frac{1}{2}, 0, 0, \ldots\right). \end{aligned} \tag{5.45}$$

This is done by considering whether the vectors $\vec{p}^\circ, \vec{q}^\circ \in \mathbb{R}^N$ embedded according to Theorem 5.6 (see Figure 5.2) are cyclically majorized: $\vec{p}^\circ \prec_C \vec{q}^\circ$. The answer to this question is affirmative, as proved by Lemma 5.5. In this case, the circulant matrix $C(\vec{q}^\circ)$ for $N = 12$ is singular – however, the simple change to $N = 13$ leads to an invertible $C(\vec{p}^\circ)$. The corresponding shift weight vector $\vec{w}$ reads

$$\vec{w} = \left(0, \frac{1}{3}, \frac{1}{3}, \frac{1}{3}, 0, \ldots\right), \tag{5.46}$$

which leads to the realization of $\vec{p} \prec_{\mathbb{Z}} \vec{q}$ with transformation

$$\vec{p} = \left(\frac{1}{3}\Delta^1 + \frac{1}{3}\Delta^2 + \frac{1}{3}\Delta^3\right)\vec{q}. \tag{5.47}$$

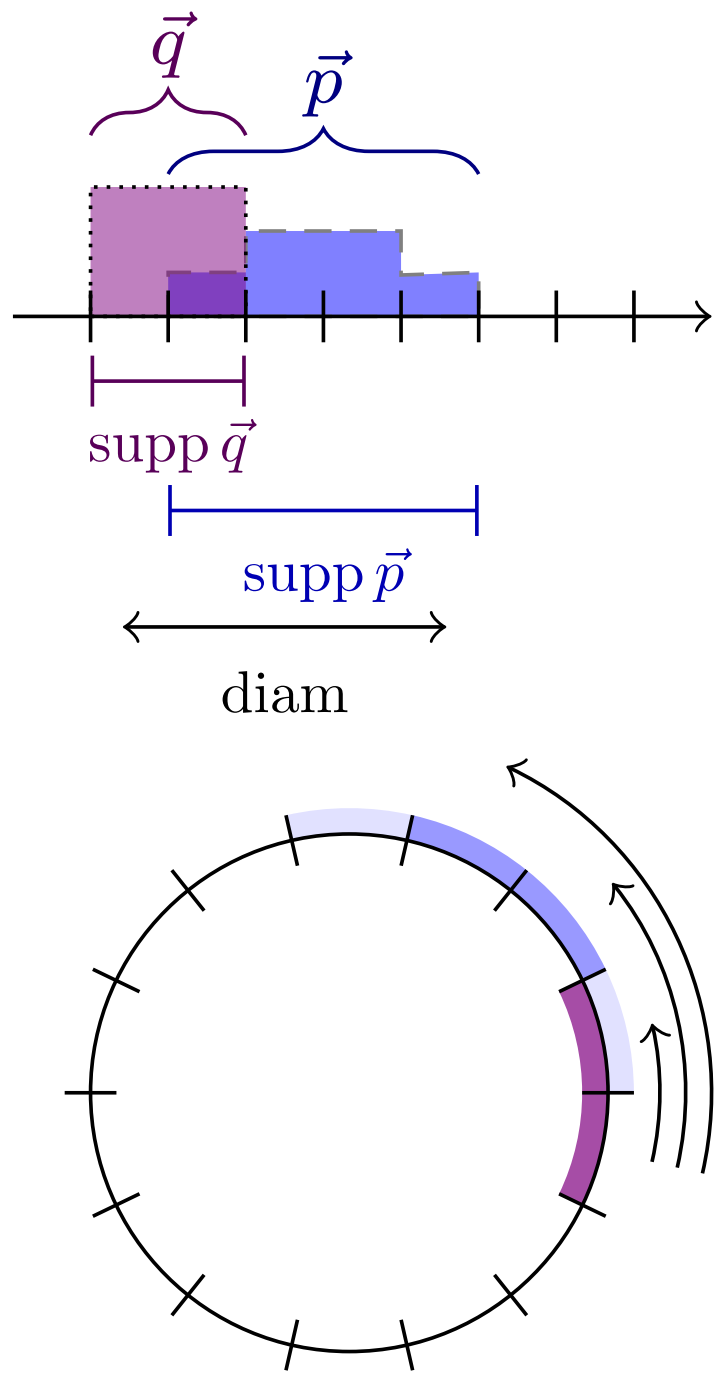

**Figure 5.2:** *Top*: Visualization of probability vectors defined in Example 5.2 by Eq. (5.45).
*Bottom*: Embedding of the same vectors in a finite-dimensional space (shaded arcs) according to Eq. (5.38) along with the shifts used in convolution (arrows).

This, in turn, defines Kraus operators transforming $|\psi\rangle$ to $|\phi\rangle$:

$$\begin{aligned} K_{-1} &= |0\rangle\langle 1| + \frac{1}{\sqrt{2}}|1\rangle\langle 2|, \\ K_{-2} &= \frac{1}{\sqrt{2}}|0\rangle\langle 2| + \frac{1}{\sqrt{2}}|1\rangle\langle 3|, \\ K_{-3} &= \frac{1}{\sqrt{2}}|0\rangle\langle 3| + |1\rangle\langle 4|. \end{aligned} \tag{5.48}$$

The resulting channel,

$$\mathscr{E}(\rho) = K_{-1}\rho K_{-1}^{\dagger} + K_{-2}\rho K_{-2}^{\dagger} + K_{-3}\rho K_{-3}^{\dagger}, \tag{5.49}$$

is $U(1)$-covariant by construction.

The channel $\mathscr{E}$ defined by Eq. (5.49) can be made explicitly trace preserving by adding elements $|n-1\rangle\langle n|$ to $K_{-1}$ for $n \in \mathbb{Z}$ such that $p_n = 0$. This is omitted in Eq. (5.48) for clarity, since it does not change the action of $\mathscr{E}$ on $|\psi\rangle$.
Note that a properly scaled single $U(1)$-covariant Kraus operator applied to $|\psi\rangle$ could yield $|\phi\rangle$, but would not form a quantum channel on its own.

## 5.5 Accessible states

A natural question arises: what states can be accessed with $U(1)$-covariant operations? The most general statement of this question, with states spread over an infinite number of levels of the ladder, is difficult to analyze. However, if the support of the initial state is finite, a simple reasoning ensures that the set of target states is discrete. It is infinite, but only trivially: all accessible states are translations of a finite set of states.

**Theorem 5.8** *The set of states accessible by U(1)-covariant operations from a state $|\psi\rangle = \sum_n \sqrt{p_n}\,|n\rangle$ with finite support is finite up to the overall translation. If the diameter of the support is denoted by $\operatorname{diam}\vec{p} = m$, then there are at most $2^m$ substantially distinct target states.*

*Proof.* Action of smoothing with weighted shifts translates to the language of associated polynomials given by Eq. (5.43) naturally:[11] if $\vec{p} = \sum w_k \Delta^k \vec{q}$, then $f_{\vec{p}}(x) = f_{\vec{w}}(x) f_{\vec{q}}(x)$. This can be explicitly verified by expansion of the product of two polynomials: the sequence of coefficients of the result is a convolution of the coefficients of both terms in the product. When viewed this way, the number of accessible states is equal to the number of decompositions of $f_{\vec{p}}(x)$ into a product of two polynomials in such a way that both terms define vectors in the probability simplex (the coefficients must be nonnegative, their sum must be equal to 1).

11: See Lemma 5.4 on page 75.

The number of such decompositions is finite: if the polynomial $f_{\vec{p}}(x)$ is factored into $m$ linear terms defined by its roots $x_1, \ldots, x_n$

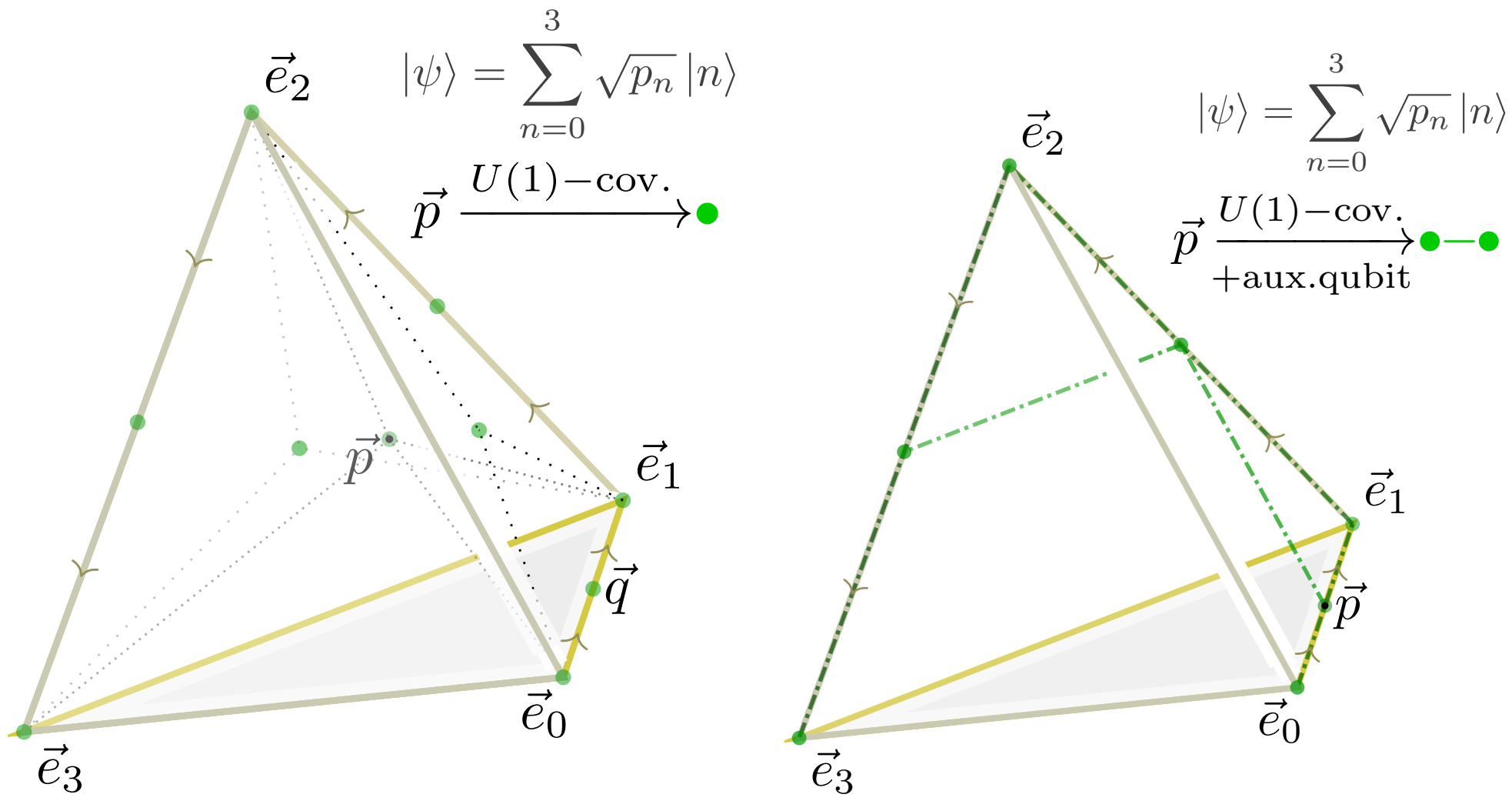

**Figure 5.3:** Visualization of states $|\phi\rangle = \sum_{n=0}^{3} \sqrt{q_n}\,|n\rangle$ defined by $\vec{q}$ accessible from a state corresponding to $\vec{p}$ (Eq. (5.45), black dot inside the tetrahedron). Two different accessibility structures arising from different scenarios are presented.
*Left*: $U(1)$-interconversion without any external resource, accessible states (green dots) are described by Theorem 5.2, and form a discrete set as proved by Theorem 5.8.
*Right*: $U(1)$-interconversion with a controllable qubit resource. The set of accessible states is an union of segments (green dashdotted lines) – see Theorem 5.9 (and discussion below its proof) for details.

and a constant $c$,

$$f_{\vec{p}}(x) = c(x - x_1) \cdot \ldots \cdot (x - x_m), \tag{5.50}$$

then *any* decomposition of $f_{\vec{p}}(x)$ into a product of two polynomials $g(x)$ and $h(x)$ takes the form of

$$f_{\vec{p}}(x) = \overbrace{\left(\frac{c}{t}\prod_{i\in S}(x - x_i)\right)}^{g(x)}\overbrace{\left(t\prod_{i\notin S}(x - x_i)\right)}^{h(x)}, \tag{5.51}$$

where $S$ is a subset of $\{1, \ldots, m\}$ and $t$ is an arbitrary number. For every choice of $S$, there is only a single choice of $t$ which may correspond to an associated polynomial of a vector contained in the probability simplex: it must be such that $g(1) = 1$. Since there are $2^m$ subsets of $\{1, \ldots, m\}$ and a state which support has diameter $m$ defines a polynomial of order $m$ (after a proper translation), there are at most $2^m$ untranslated target states. □

## 5.6 Effect of an auxiliary system

A different structure of accessible states is present in the case when controllable auxiliary system is present: in this case, it is possible to reach states which probability vectors are majorized by the one of the initial state (i.e. in the opposite direction to the 'pure' case).

**Theorem 5.9** *Consider a bipartite system: one part is an infinite ladder $\{|n\rangle\}_{n\in\mathbb{Z}}$, while the other is a $(2d+1)$-dimensional qudit. The total Hamiltonian defining $U(1)$ evolution is*

$$H = \overbrace{\sum_{n\in\mathbb{Z}} n\,|n\rangle\langle n|}^{\text{infinite ladder}} + \overbrace{\sum_{m=-d}^{d} m\,|m'\rangle\langle m'|}^{\text{qudit part}}. \tag{5.52}$$

*The initial state of a ladder subsystem is $|\psi\rangle$, and the goal is to reach $|\phi\rangle$, where*

Due to the normalization of states, $\sum_{n\in\mathbb{Z}} p_n = \sum_{n\in\mathbb{Z}} q_n = 1$.

$$\begin{aligned} |\psi\rangle &= \sum_{n\in\mathbb{Z}} \sqrt{p_n}\,|n\rangle\,, \\ |\phi\rangle &= \sum_{n\in\mathbb{Z}} \sqrt{q_n}\,|n\rangle\,. \end{aligned} \tag{5.53}$$

*If $\vec{q} \in \mathrm{conv}\{\Delta^{-d}\vec{p},\ldots,\vec{p}\ \ldots,\Delta^{d}\vec{p}\}$, then there exists a qudit vector $|\beta\rangle$ such that*

$$|\psi\rangle\otimes|\beta\rangle \xrightarrow{U(1)-\text{cov.}} |\phi\rangle\otimes|0'\rangle\,. \tag{5.54}$$

*Proof.* Initially, the ladder state is $|\psi\rangle = \sum_{n\in\mathbb{Z}} \sqrt{p_n}\,|n\rangle$ and the $(2d+1)$-dimensional qudit system can be in arbitrary state:

$$|\beta\rangle = \sum_{m=-d}^{d} \sqrt{w_m}\,|m'\rangle\,, \tag{5.55}$$

such that the total initial state reads $|\psi\rangle\otimes|\beta\rangle$. Any unitary operation performed inside the energy eigenspace is $U(1)$-invariant in this case [60] – and the energy eigenspace (to energy $n$) is spanned by $\{|n-m\rangle\otimes|m'\rangle\}_{m=-d}^{d}$. One of the $U(1)$-covariant operations restricted to this subspace is the rotation of the part of the state vector parallel to it onto $|n\rangle\otimes|0'\rangle$:

[60]: Gour et al. (2008), 'The resource theory of quantum reference frames: manipulations and monotones'

$$\Pi_n\left(\sum_{m=-d}^{d} c_m\,|n-m\rangle\otimes|m'\rangle\right) = \sqrt{\sum_{m=-d}^{d} |c_m|^2}\,|n\rangle\otimes|0'\rangle\,. \tag{5.56}$$

Let me denote the complete operation performed for all $n$ by $\bar{\Pi}$:

$$\bar{\Pi} = \prod_{n\in\mathbb{Z}} \Pi_n\,. \tag{5.57}$$

If $\bar{\Pi}$ is applied to any product state $|\psi\rangle \otimes |\beta\rangle$, the result reads

$$\bar{\Pi}\Big(\sum_{n\in\mathbb{Z}}\sum_{m=-d}^{d}\sqrt{p_{n-m}w_m}\,|n-m\rangle\otimes|m'\rangle\Big) = \underbrace{\left(\sum_{n\in\mathbb{Z}}\sqrt{q_n}\,|n\rangle\right)}_{|\phi\rangle}\otimes|0'\rangle\,. \tag{5.58}$$

Elementary calculation based on the unitarity of the above operation shows that $q_n = \sum_{m=-d}^{d} p_{n-m}b_m$. In other words, the probability vector $\vec{q}$ is a convolution of $\vec{p}$ and $\vec{w}$ with $\vec{w}$ having only nonnegative components. □

The statement of Theorem 5.9 is simplified to exemplify the direct application of the additional qudit resource. The resource can be defined differently,[12] and the controlled qubit state can also be used indirectly, in a sequence of $U(1)$-covariant operations – the example shown in Figure 5.3 takes the entire set of accessible states into account.

12: For example, it can be qubit with energies of 0 and 1. The accessible states are then defined similarly, with only the summation limits changed.

## 5.7 Weyl-Heisenberg group

Due to the simple description of the action of quantum channels in the Weyl-Heisenberg formalism (see Eq. (2.77)), one can formulate the following statement, which uses the language introduced in [24] to describe the main result of [64].

[24]: Koukoulekidis et al. (2021), 'Constraints on magic state protocols from the statistical mechanics of Wigner negativity'

[64]: Datta et al. (2006), 'Complementarity and additivity for covariant channels'

**Theorem 5.10** *Weyl-Heisenberg-covariant interconversion of the state* $\rho \in \mathcal{M}_d$ *to* $\sigma \in \mathcal{M}_d$ *is possible if and only if the Wigner distribution* $W_\rho$ *is a two-dimensional cyclic convolution of* $W_\sigma$ *with a nonnegative, normalized kernel.*

Weyl-Heisenberg group and Wigner functions are defined in section 2.3, pages 21-23.

*Proof.* A quantum channel $\mathcal{E}$ is covariant with respect to the Weyl-Heisenberg group if and only if it commutes with every displacement operator $D_{\vec{x},\vec{q}}$:

$$D_{\vec{x},\vec{q}}\mathcal{E}(\rho)D^{\dagger}_{\vec{x},\vec{q}} = \mathcal{E}(D_{\vec{x},\vec{q}}\,\rho\,D^{\dagger}_{\vec{x},\vec{q}}). \tag{5.59}$$

For this to happen, the channel $\mathcal{E}$ must be a convex combination of displacement operators by itself [64]:

$$\mathcal{E}(\rho) = \sum_{\vec{x},\vec{q}} p_{\vec{x},\vec{q}}\left(D_{\vec{x},\vec{q}}\,\rho\,D^{\dagger}_{\vec{x},\vec{q}}\right). \tag{5.60}$$

Here, $p_{\vec{x},\vec{q}} \geq 0$ and $\sum p_{\vec{x},\vec{q}} = 1$.

Since the displacement operators shift the Wigner distribution and the operation is linear, the action of a channel $\mathcal{E}$ on a Wigner distribution is a convolution with a nonnegative kernel. □

Therefore, the case of Weyl-Heisenberg-covariant interconversion can be analyzed with almost the same methods as for the group $U(1)$.

## 5.8 $SU(2)$ case

The case of $SU(2)$ group is substantially more difficult than the two other mentioned previously. While the approach presented in [60] does yield a characterization of $SU(2)$-covariant channels by their Kraus operators, it does not immediately offer clear-cut answer to the question whether the two given states are interconvertible.[13] The methods of [59] *do* work in this regard, and the interconvertibility test is numerically feasible. However, it is nontrivial to express this result in a direct analytical form and to characterize the set of accessible states.

[60]: Gour et al. (2008), 'The resource theory of quantum reference frames: manipulations and monotones'

13: The authors of [60] do, however, consider a restricted family of states for which their procedure offers a solution.

[59]: Marvian (2012), 'Symmetry, Asymmetry and Quantum Information'

In this section, I explore the reasons behind this complexity and provide nontrivial examples of interconversion. The results of [59] are most important in understanding the problem on an abstract level: interconvertibility can be reinterpreted as a statement about a particular decomposition of the states in question:[14]

14: The following lemma is a direct application of Theorem 63 of [59] to the $SU(2)$ case. The initial $SU(2)$-invariant auxiliary state can be proved to be pure by Theorem 25 of the same work – and the only invariant pure state here is the singlet.

Throughout this section the ket $|j, m\rangle$ denotes the spin state of total spin $j$ and the spin projection quantum number $m$. Potential indices beyond the two indicate different copies of the same representation $j$.

**Lemma 5.11** *Consider two arbitrary spin states $|\phi\rangle$ and $|\psi\rangle$:*

$$|\psi\rangle = \sum_{j,m} \psi_{j,m} |j, m\rangle , \qquad |\phi\rangle = \sum_{j,m} \phi_{j,m} |j, m\rangle . \tag{5.61}$$

*The interconversion is possible – $|\psi\rangle \xrightarrow{SU(2)-\text{cov.}} |\phi\rangle$ – if there exists a third state $|\omega\rangle = \sum_{j,m} \omega_{j,m} |j, m\rangle$ such that*

$$|\psi\rangle \approx |\phi\rangle \otimes |\omega\rangle . \tag{5.62}$$

*Here, $\approx$ means the following: the state $|\phi\rangle \otimes |\omega\rangle$ expressed in the irreducible representations of $SU(2)$ is reducible to $|\psi\rangle$ by means of $SU(2)$-invariant unitary $U$. Every such operator can only mix degeneracy indices within one representation: $U = \sum_j \mathbb{1}_j \otimes u^{(j)}$, where $u^{(j)}$ are unitaries.*

Operators $\mathbb{1}_j \otimes u^{(j)}$ act only on the degeneration indices:

$$\mathbb{1}_j \otimes u^{(j)} |j, m, \mu\rangle = \sum_\nu u^{(j)}_{\mu,\nu} |j, m, \nu\rangle .$$

The state $|\omega\rangle$ is traced over in the end and the target state $|\phi\rangle$ is recovered. While the simple picture suffices for $U(1)$, the description gets more complicated for $SU(2)$. To see why, let me reiterate the spin composition properties [65]:

[65]: Griffiths (2008), *Introduction to Elementary Particles*

**Theorem 5.12** *Consider arbitrary two states $|\phi\rangle$ and $|\omega\rangle$ defined as in Eq.* (5.61)*. The state $|\psi\rangle = |\phi\rangle \otimes |\omega\rangle$ expressed in the basis of the total spin $j$ is:*

$$|\psi\rangle = \sum_{j,m;j_1,m_1;j_2,m_2} C^{j_1,j_2;j}_{m_1,m_2;m} \omega_{j_1,m_1} \phi_{j_2,m_2} |j, m, (j_1, j_2)\rangle , \tag{5.63}$$

*where $C^{j_1,j_2;j}_{m_1,m_2;m}$ is the $SU(2)$ Clebsch-Gordan coefficient [66].*

**Definition 5.9** *The $SU(2)$ characteristic function of $|\alpha\rangle$ is $\chi_\alpha(g) = \langle\alpha|U_g|\alpha\rangle$, where $g \in SU(2)$.*

The state $|\psi\rangle$ has the same SU(2) characteristic function as $|\phi\rangle \otimes |\omega\rangle$.

Note the additional index $(j_1, j_2)$ tracking the degeneracies – it is important here. Since the original state $|\psi\rangle$ is pure, the states $|\omega\rangle$ and $|\phi\rangle$ must combine in such a way that $|\psi\rangle$ is reachable by means of unitaries $u^{(j)}$ within a single representation $j$:

$$u^{(j)} \sum_{(j_1,j_2),m} C^{j_1,j_2;j}_{m_1,m_2;m} \omega_{j_1,m_1} \phi_{j_2,m_2} |j, m, (j_1, j_2)\rangle = \sum_m \psi_m |j, m, (j, 0)\rangle . \tag{5.64}$$

This is clearly not always possible.[15] Consider the state $|\alpha\rangle = 1/\sqrt{2}\,(|0, 0\rangle + |1/2, \uparrow\rangle)$. The state $|\alpha\rangle \otimes |\alpha\rangle$ expressed in the basis of the total spin is the following:

$$|\alpha\rangle \otimes |\alpha\rangle = \frac{1}{2} (|0, 0\rangle + |1/2, \uparrow, (1/2, 0)\rangle + |1/2, \uparrow, (0, 1/2)\rangle + |1, 1\rangle) . \tag{5.65}$$

If the substates $|1/2, \uparrow, (1/2, 0)\rangle + |1/2, \uparrow, (0, 1/2)\rangle$ were to be combined into the pure state $2\,|1/2, \uparrow\rangle$, the resulting state $|\alpha\rangle \otimes |\alpha\rangle$ would not be normalized.[16] This is one sign that this is not the right way of thinking about the problem.

This is the reason why the problem of $SU(2)$-covariant interconversion is much more involved than $U(1)$. For $U(1)$, the irreducible representations are always one-dimensional, and two copies of the same representation can be effortlessly combined.[17] Given this problem, what can be said about the reachability structure?

The first simplification is to consider a restricted family of states which belong only to a single irreducible representation. All such maps are expressible as a combination $\mathcal{E}(\rho) = \sum_{n=0}^{2j} x_n \zeta^{(n)}(\rho)$, where

$$\zeta(\rho) = \frac{1}{j(j+1)} \sum_{\sigma \in \{X,Y,Z\}} J_\sigma \rho J_\sigma . \tag{5.66}$$

Here, $\zeta^{(n)}$ denotes $n$-times application of $\zeta$.

The numbers $x_n$ are real and sum up to unity, but are not necessarily positive. However, since $\zeta^{(n)}$ do commute, the allowed set of values forms a simplex.

The values outside the simplex are maps which are not completely positive, but still are $SU(2)$-covariant. Among those, there exist nontrivial covariant *antiunitary* maps, so the pure state is mapped

[66]: Khersonskii et al. (1988), *Quantum Theory of Angular Momentum*

15: For example: as far as spin observables are considered, a maximally mixed state of the irreducible $j = 1/2$ representation is indistinguishable from $1/\sqrt{2}\,(|1/2, \uparrow, A\rangle + |1/2, \downarrow, B\rangle)$. Hence, it can not be $SU(2)$-invariantly turned into a pure state of $j = 1/2$ representation. Here, for readability $|1/2, \uparrow\rangle$ denotes the state $|1/2, 1/2\rangle$, similarly for $|1/2, \downarrow\rangle$, and unimportant degeneracy indices are omitted.

16: The substates are, however, unitarily reducible to $\sqrt{2}\,|1/2, \uparrow\rangle$.

17: In $U(1)$, by an abuse of notation:

$$\bigoplus_{k\in\mathbb{Z}} \sqrt{p_{n-k}}\sqrt{q_k} \approx \sqrt{\sum_{k\in\mathbb{Z}} p_{n-k} q_k}.$$

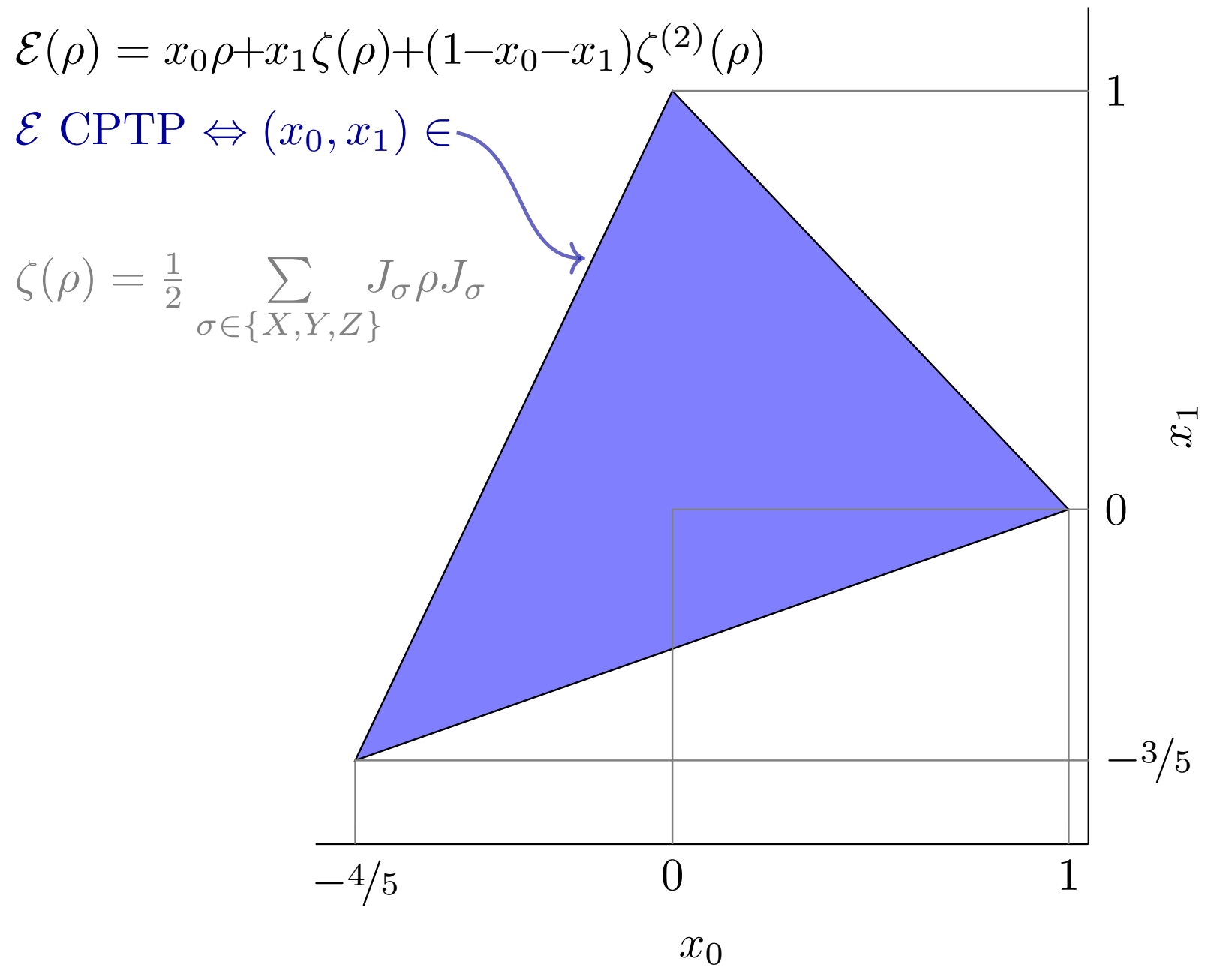

**Figure 5.4:** Visualisation of the simplex of $SU(2)$-covariant quantum channels constrained to the $j = 1$ representation. All the possible covariant trace-preserving maps are parameterized with two real numbers $x_0$ and $x_1$. A map is a channel if the two numbers belong to the simplex spanned by $\{(1, 0), (0, 1), (-4/5, -3/5)\}$. Identity map ($x_0 = 1, x_1 = 0$) is the only channel that maps pure states to pure states.
Outside the CPTP simplex (at $(x_0, x_1) = (-1, -2)$) lies the antiunitary map defined in Eq. (5.67).

to a pure state. For $j = 1$, the following antiunitary map is $SU(2)$-covariant:

$$\rho \mapsto \left(R\rho R^\dagger\right)^T, \tag{5.67}$$

This map can also be expressed as $\rho \mapsto -\rho - 2\zeta(\rho) + 4\zeta^{(2)}(\rho)$.

where $R = \exp(i\pi J_Y)$ and the transpose is defined in the $J_Z$ eigenbasis.

As it turns out, the set of pure states reachable from an initial state $|\psi\rangle$ is trivial: it only contains the original state $|\psi\rangle$.[18]

18: There are, however, many reachable *mixed* states – for instance, the maximally mixed state is always accessible due to its invariance under arbitrary unitaries.

**Lemma 5.13** *Let the state $|\psi\rangle = \sum_{m=-j}^{j} \psi_m |j, m\rangle$ be contained completely within a single representation $j$ of $SU(2)$. If only the $SU(2)$-covariant operations restricted to the representation $j$ are considered, the only pure state covariantly reachable from $|\psi\rangle$ is $|\psi\rangle$ itself.*

*Proof.* If $|\psi\rangle \xrightarrow{SU(2)-\text{cov.}} |\phi\rangle$, then a state $|\omega\rangle$ exists such that $|\phi\rangle \otimes |\omega\rangle$ is $SU(2)$-invariantly unitarily reducible within one representation to $|\psi\rangle$. However, any $|\omega\rangle$ different from $|0, 0\rangle$ does make the state $|\phi\rangle \otimes |\omega\rangle$ leave the original representation $j$. Hence, only $|\omega\rangle = |0, 0\rangle$ is possible, and in such case $|\psi\rangle \propto |\phi\rangle$. □

Other restrictions on the set of spin states aimed at reducing the complexity of the problem are possible. In [60] the set of states spanned by maximal angular momentum projections (coherent states) is considered. Every state in this set admits the following form:

[60]: Gour et al. (2008), 'The resource theory of quantum reference frames: manipulations and monotones'

$$|\psi\rangle = \sum_{j=0,1/2,\ldots} \psi_j |j, j\rangle . \tag{5.68}$$

This greatly simplifies the reasoning, as such states combine trivially with Clebsch-Gordan coefficients[19] in a manner similar to the $U(1)$ case. [20]

19: To form another coherent state: $C^{j_1,j_2;j}_{j_1,j_2;j} = \delta_{j-(j_1+j_2)}$.

20: Additionally, *every* state $SU(2)$-covariantly reachable from such a $|\psi\rangle$ lies in this set.
The degeneracy of representations still arises, but all states corresponding to a given $j$ are proportional (lie on a one-dimensional subspace). The same happens for $U(1)$.

Similar simplification can be done for other families of states. For instance, if the $SU(2)$ inverconversion problem $|\psi\rangle \xrightarrow{SU(2)-\text{cov.}} |\phi\rangle$ is restricted to the eigenvectors of $J_Z$:

$$J_Z |\psi\rangle \propto |\psi\rangle , \;\; J_Z |\phi\rangle \propto |\phi\rangle , \tag{5.69}$$

then the problem once again becomes tractable:

**Theorem 5.14** *Consider two states $|\phi\rangle$ and $|\omega\rangle$ defined as in Eq.* (5.61) *with an additional constraint: the states must be eigenvectors of $J_Z$, such that $J_Z|\phi\rangle = M|\phi\rangle$ and $J_Z|\omega\rangle = M'|\omega\rangle$. The pure state with the same $SU(2)$ characteristic function as $|\phi\rangle \otimes |\omega\rangle$ is the following:*

$$|\psi\rangle = \sum_j \sqrt{\sum_{j_1,j_2} \left| C^{j_1,j_2;j}_{M,M',M+M'} \phi_{j_1,M} \omega_{j_2,M'} \right|^2} \, |j, M+M'\rangle . \tag{5.70}$$

If a state $|\alpha\rangle = \sum_{j,m} \alpha_{j,m} |j,m\rangle$ is an eigenvector of $J_Z$ to the eigenvalue $M$, then $\alpha_{j,m} \neq 0$ only if $m = M$.

*Proof.* The states combine according to Theorem 5.12. Then, the state $|\psi\rangle$ is accessible due to Lemma 5.11: different copies of a representation $j$ all are eigenstates of $J_Z$, and therefore proportional. □

This means that the state $|\phi\rangle$ is $SU(2)$-covariantly accessible from $|\psi\rangle$.[21] As a direct result, the following nontrivial and example of $SU(2)$ interconversion can be stated:

21: The states are defined by their probabilities similarly to the $U(1)$ case – therefore, a structure similar to the convolution appears:

**Remark 5.2** Let $M, M' \in \mathbb{Z}$ be fixed and $|\phi\rangle = \sum_j \sqrt{q_j} |j,M\rangle$, $|\omega\rangle = \sum_j \sqrt{w_j} |j,M'\rangle$. Then the state discussed in Theorem 5.14 can be written as $|\psi\rangle = \sum_j \sqrt{p_j} |j, M+M'\rangle$ such that

$$p_j = \sum_{j_1,j_2} G_{j,j_1,j_2} q_{j_1} w_{j_2},$$

where $G_{j,j_1,j_2} = \left( C^{j_1,j_2;j}_{M,M',M+M'} \right)^2$.

**Example 5.3** Interconvertiblity occurs ($|\psi\rangle \xrightarrow{SU(2)-\text{cov.}} |\phi\rangle$) for the following states:

$$|\psi\rangle = \sqrt{3/10}\,|1,-1\rangle + \sqrt{43/126}\,|2,-1\rangle + \sqrt{97/360}\,|3,-1\rangle + \sqrt{5/56}\,|4,-1\rangle ,$$
$$|\phi\rangle = 1/\sqrt{3}\,|1,-1\rangle + 1/\sqrt{3}\,|2,-1\rangle + 1/\sqrt{3}\,|3,-1\rangle . \tag{5.71}$$

This is ascertained by Theorem 5.14: the state $|\psi\rangle$ is constructed as $|\phi\rangle \otimes |\omega\rangle$ with

$$|\omega\rangle = 1/\sqrt{2}\,|0,0\rangle + 1/\sqrt{2}\,|1,0\rangle . \tag{5.72}$$

The case of $SU(2)$-covariant interconversion, while difficult, admits a restricted analysis. The main novel result here is the Theorem 5.14 and its application in Example 5.3.

## 5.9 Concluding remarks

The problem of state interconversion using a restricted set of quantum channels appears naturally in various quantum mechanical scenarios. In this chapter, I presented the prototypical scenario of $U(1)$-covariant channels. The preexisting theoretical background has been shown, as well as novel results, providing a geometrical interpretation of the state accessibility structure. This perspective arises from a simplification of the theory, allowing for a finite-dimensional treatment. I also briefly sketched two other important cases – the Weyl-Heisenberg and $SU(2)$ groups – and provided results regarding these cases.

# Conclusion 6

Geometry plays a special role in quantum mechanics. Many of the tools and objects appearing in quantum theory admit a geometric perspective, often greatly simplifying interpretation and reasoning. The goal of this thesis was to present some of them and to demonstrate certain exemplary applications.

## 6.1 Numerical ranges

Major component of this thesis is the discussion of images of the sets of quantum states – the entire set $\mathcal{M}_d$ of density operators of size $d$ and its useful subsets – under linear maps. Such images are called *numerical ranges*, and they capture the relevant information concerning quantum states while often greatly reducing the dimensionality of the problem. Thus, such a technique allows for tractable treatment and straightforward visualization of multiple questions in quantum information science. The topic discussed in chapter 4 include the following:

**Geometry of numerical ranges** and its connection with commutativity properties of the defining operators. In certain cases, a full explicit description of the numerical ranges is known, shedding light on the algebraic structure of the defining observables. In this thesis, I presented the classification regarding qutrit numerical ranges of three observables (Theorem 4.10) – the proof of the classification theorem appearing here is a significant simplification of the original arguments of [1]. Every classification applies to the spectrahedra as well as a result of the intersection-projection duality (Theorem 3.4).

[1]: Szymański et al. (2018), 'Classification of joint numerical ranges of three hermitian matrices of size three'

**Phase transitions and numerical ranges** are closely connected, since the boundary of the latter is formed by the ground states of the combinations of the defining operators. As a result, properties of the operators and their ground states are determined by appropriately constructed numerical ranges. In this thesis, I formulated an effective criterion (Theorem 4.18) confirming the vanishing value of the energy gap $\delta(H)$ of a Hamiltonian $H$ by studying the geometry of the numerical range $W(V, H)$ with a suitably selected interaction operator $V$. This result has been recently published [4].

[4]: Szymański et al. (2021), 'Universal witnesses of vanishing energy gap'

**State-independent uncertainty relations** involving sums of variances can be stated in the form of a question about the geometric properties of a three-dimensional numerical range, regardless of the dimension of the original physical system. This allows for reformulation as an efficiently solvable polynomial equation system (Theorems 4.11 and 4.13). This approach has been successfully applied in the case of spin uncertainty relations to yield tight and analytical bounds regarding spin squeezing (see Example 4.3 and [3]) – sum of two variances of two spin components is a limiting factor of the experimental accuracy. Parameterized uncertainty relations are related to the *uncertainty range* – the range of values of variances of a set of operators. In [3] and in this thesis, I constructed a method of variance approximation by a family of observables constructed from the original operators. Using this, I provided a method of approximation of the uncertainty range and presented an application of the theory to the case of spin observables.

[3]: Szymański et al. (2019), 'Geometric and algebraic origins of additive uncertainty relations'

**Entanglement detection** has multiple approaches, one of them is the usage of witnesses – observables which negative expectation value indicates entanglement. Separable numerical ranges – analogues of numerical ranges with the set of states restricted to the set of separable states – are closely connected to this idea (and its derivatives), effectlvely providing a parameterized family of entanglement witnesses which encapsulates the entire information one may infer from a set of expectation values. Approximation of separable numerical ranges is an NP-hard problem in general, but there are significant simplifications in some cases: bipartite qubit-qubit and qubit-qutrit systems can be efficiently solved using semidefinite optimization. In this thesis, I described a novel method based on Theorem 4.19 capable of controlled approximation of entanglement witnesses of multipartite systems consisting of multiple qubits and a single qudit of arbitrary size. The algorithm has found use in research regarding multipartite entanglement [2, 5]. The notion of separable numerical range is used in Theorem 4.22, in which I presented a way to determine the range of expectation values of an observable when the set of states is restricted to the states of Schmidt rank at most 2.

[2]: Czartowski et al. (2019), 'Separability gap and large-deviation entanglement criterion'
[5]: Simnacher et al. (2021), 'Confident entanglement detection via the separable numerical range'

Additionally, the set of positive partial transpose states admits an interpretation as a spectrahedron dual to a certain *numerical* range – a novel observation formulated in Theorem 4.21. This provides an alternative description of the geometry of PPT numerical ranges, which in some cases coincides with the separable counterparts.

## 6.2 State interconversion

The second part of the thesis covers the topic of state interconversion using group-covariant quantum channels. It is a subset of quantum resource theory, with the emphasis on *asymmetry with respect to the group operations* being the object of interest. In this context, symmetric (invariant) states do not convey much information – and assymetry may be used to determine some properties of the quantum system of interest. One may ask the question: given an initial state $|\psi\rangle$, which states can be reached using some quantum channel $\mathscr{E}$ with the only restriction being that $\mathscr{E}$ commutes with every group element? Different groups correspond to different quantum systems:

**U(1) group** models the free evolution of a doubly infinite, equally spaced harmonic ladder with eigenstates of the Hamiltonian numbered by integers: $H|n\rangle = n|n\rangle$ for $n \in \mathbb{Z}$. In this case, the problem of state interconversion can be mapped to a question whether two probability distributions over $\mathbb{Z}$ are connected through a convolution with a nonnegative kernel. In this thesis I explore the consequences of this observation: a simplification of the original criterion is proposed, allowing for a direct algebraic test (Theorem 5.6). The states accessible via $U(1)$-covariant operations are characterized using the theory of polynomials – a novel result (Theorem 5.8) implies that, under certain assumptions, there are only finitely many substantially distinct accessible states. The effect of an auxiliary system is shown to invert the direction of the accessibility structure (Theorem 5.9).

**Weyl-Heisenberg group** is the language of a large part of the theory of quantum error correcting codes and quantum computing. The group is discrete, but the structure of accessible states turns out to be similar to the $U(1)$ case – a state $\rho$ can be transformed to $\sigma$ if the Wigner distribution $W_\rho$ is a two-dimensional cyclic convolution of $W_\sigma$ with a nonnegative kernel. I provide a proof of this result, based on previous research in this direction.

**SU(2) group** naturally arises as the description of spin and angular momentum. The asymmetry in this case corresponds to the usefulness of a state as a direction reference – the completely symmetric states are rotationally invariant. Because $SU(2)$ is nonabelian, the structure of accessible states is much more difficult to analyze than $U(1)$ or Weyl-Heisenberg group – however, some patterns still can be determined. I present the origins of the complexity of this problem and provide exemplary instances of a tractable, yet nontrivial $SU(2)$-covariant interconversion problem (Theorem 5.14 and Example 5.3).

## 6.3 Further research

Quantum information is still riddled with a plethora of open problems – this thesis sheds a new light on some of them. Below I propose the directions of further investigations:

**Classification of numerical ranges**

The set of quantum states $\mathcal{M}_d$ has an intricate structure depending on the dimension $d$. In the generic case – when the Hermitian operators $X_1, \ldots, X_k$ are identically and independently sampled from Gaussian Unitary Ensemble and $k \ll d^2$ – the numerical range $W(X_1, \ldots, X_k)$ typically forms an oval, so a lot of higher-dimensional information is lost during such a projection. However, if the operators $\{X_i\}_{i=1}^k$ are chosen carefully, some of the information is preserved in the form of the nonanalytical parts of the boundary.

Therefore, from a mathematician point of view, it is interesting to determine the possible shapes a numerical range may take. The first results in the theory of numerical ranges, dating back to Hausdorff [35] and Toeplitz [34], show that the numerical range of qubit operators is an affine image of the Bloch ball. The qutrit case proved to be much harder, with the first classification for the $k = 2$ operators case dating to the late 20th century [47], followed by the recent $k = 3$ analysis [1]. Limited classification is known about $d = 4$ [67], and beyond this not much is known about the structure of numerical ranges.

[35]: Hausdorff (1919), 'Der Wertvorrat einer Bilinearform'

[34]: Toeplitz (1918), 'Das algebraische Analogon zu einem Satze von Fejér'

[47]: Keeler et al. (1997), 'The numerical range of 3×3 matrices'

[1]: Szymański et al. (2018), 'Classification of joint numerical ranges of three hermitian matrices of size three'

[67]: Chien et al. (2012), 'Singular points of the ternary polynomials associated with 4-by-4 matrices'

Perhaps it is feasible to analyze the higher-dimensional cases in a systematic way, by combining the lower and higher dimensional information. The flat parts of the boundary are affine images of numerical ranges corresponding to the lower dimensional subspaces of partially commuting operators – a fact which proved to be useful in the $k = d = 3$ case. Hypothetically, this could be extended to more complicated cases, with constraints on the boundary features implied by the combinatorial properties of the subspaces involved.

Other way to look at this is through the progressive projection of high-dimensional objects [1]. A 'full' numerical range of $k = d^2 - 1$ operators of size $d$ is, under general assumptions, affinely equivalent to the set $\mathcal{M}_d$ itself, thus the numerical ranges of a high number of operators are all alike. What changes if the number of operators is dropped by one? And further: maybe it is possible to determine the possible shapes of the numerical range of $k$ operators using the $k + 1$ case. Certainly, the number of flat parts of the boundary can not increase under this operation.[1]

1: Flat parts correspond to normal vectors, and omission of one coordinate selects a subset of them.

**Applications of numerical ranges in phase transitions**

The formalism of numerical ranges already offers a fresh look on the physical problems related to quantum phase transitions. Further research in this direction may lead to new developments. In particular, the approach to the estimation of the energy gap value presented in this thesis could be extended to other kinds of gap (e.g., band gaps) by the usage of auxiliary operators selecting the proper subspace of interest. Possibly a similar approach could be used to estimate the *lower* bound for the value of the energy gap.

This idea can be illustrated with a simple extension of the Theorem 4.16. The original statement implies the unreachability of the derivative of the ground state of $H + \lambda V$ as another ground state. By adding operators selecting certain eigenspaces into the reasoning, a stronger theorem can be derived: the derivative of any eigenstate generically can not be the eigenstate for any value of the parameter $\lambda$.

Suppose it is not the case, and let the $n$-th eigenstate of $H + \lambda V$ to energy $E_n^{(\lambda)}$ be denoted by $|n_\lambda\rangle$. Then, the derivative state (normalized to unity),

$$|\bar{n}_\lambda\rangle \propto \frac{\mathrm{d}\,|n_\lambda\rangle}{\mathrm{d}\lambda}, \tag{6.1}$$

can be always set to be orthogonal to $|n_\lambda\rangle$ by a proper gauge choice – an elementary result of the perturbation theory. Now, if there exist operators $H'$ and $V'$ such that

$$\begin{aligned} &|n_0\rangle \text{ is the ground state of } H', \\ &|\bar{n}_0\rangle \text{ is the ground state of } H' + V', \end{aligned} \tag{6.2}$$

then $H'$ and $V'$ must commute on the subspace spanned by $|n_0\rangle$ and $|\bar{n}_0\rangle$, by the same reasoning as presented in the proof of the Theorem 4.16 – the two states are ortogonal, and therefore antipodal on the ellipse in $W(V', H')$, and this can only happen if the ellipse is degenerated.

However, such operators are fairly easy to construct: if $\Pi_n^{(\lambda)}$ denotes the projection operator onto the lowest $n-1$ eigenstates of $H + \lambda V$, then for

$$\begin{aligned} H' &= H + C\Pi_n^{(0)}, \\ V' &= \lambda V - C\Pi_n^{(0)} + C'\Pi_m^{(\lambda)}, \end{aligned} \tag{6.3}$$

$\Pi_n^{(\lambda)} = \sum_{m<n} |m_\lambda\rangle\langle m_\lambda|$

and finite $C > E_n^{(0)}$, $C' > E_m^{(\lambda)}$, the operator $H'$ has $|n_0\rangle$ as its ground state and the ground state of $H' + V'$ is $|m_\lambda\rangle$. The derivative state $|\bar{n}_0\rangle$ can only be the ground state of $H' + V'$ if the ellipse is degenerated, which generically does not happen as is further ascertained by freedom of choice of $C$ and $C'$.

This implies that the eigenstates of Hamiltonians have a certain structure. This topic has obtained attention before [68], however, not much is known about the structure of eigenstates of a restricted family of Hamiltonians.

[68]: Haselgrove et al. (2003), 'Quantum states far from the energy eigenstates of any local Hamiltonian'

**Interconvertibility of $SU(2)$ states**

A spin state $|\psi\rangle$ can be transformed to $|\phi\rangle$ using $SU(2)$-covariant quantum channel if and only if the ratio of the characteristic functions $\chi_\psi \chi_\phi^{-1}$ is positive semidefinite. This criterion can be efficiently implemented numerically, but such an approach does not offer much insight into the structure of states accessible by group covariant operations. Full analysis similar to the $U(1)$ case remains elusive due to the difficulty of working with a nonabelian group.

Limited results are available, and possibly by going in this direction the full structure of $SU(2)$ accessibility can be uncovered. Perhaps the *only* pure accessible states have some special structure similar to the assumptions of Theorem 5.14 or results presented in [60] – the incoherent mixing of different representations does suggest this might be the case.

[60]: Gour et al. (2008), 'The resource theory of quantum reference frames: manipulations and monotones'

**Separable and restricted Schmidt rank numerical ranges**

Entanglement is an essential feature of various quantum systems: it manifests itself in the eigenstates of important models, it is required for efficient quantum computation [69], and is a sign of nonclassical phenomena. Methods of entanglement detection are therefore needed, and miscellaneous approaches are widely used [70, 71].

[69]: Linden et al. (2001), 'Good Dynamics versus Bad Kinematics: Is Entanglement Needed for Quantum Computation?'

[70]: Horodecki et al. (2001), 'Separability of n-particle mixed states: necessary and sufficient conditions in terms of linear maps'
[71]: Gühne et al. (2009), 'Entanglement detection'

The most straightforward one is the concept of *entanglement witness*: an observable $W$ such that the expectation value $\langle W\rangle_\rho$ is nonnegative for separable $\rho$ – hence, if a negative expectation value is observed, the state $\rho$ is entangled. Natural generalizations follows: given arbitrary observable $A$, what is the range $[\lambda^{\otimes}_{\min}(A), \lambda^{\otimes}_{\max}(A)]$ of expectation values for separable states? What if the expectation value $\langle B\rangle_\rho$ of other observable $B$ is known? All these approaches can be unified in the formalism of *separable numerical ranges* – sets of simultaneously attainable expectation values of separable states:

$$W_{\mathrm{SEP}}(X_1, \ldots, X_k) = \left\{(\langle X_1\rangle_\rho, \ldots, \langle X_k\rangle_\rho) : \rho \in \mathcal{M}^{\mathrm{SEP}}\right\}. \quad (6.4)$$

Some of the properties of separable numerical ranges are already presented in Chapter 4 of this thesis. The most significant obstacle to further research is the fact that the calculation of $W_{\mathrm{SEP}}$ is NP-hard in the general case – a common problem with anything related to entanglement.[2] This does not apply to bipartite qubit-qubit and qubit-qutrit systems, for which semidefinite programming can be applied.[3] I presented a numerical algorithm capable of determining the separable numerical range for multipartite qubit systems – the problem is still difficult, but the approximation error is known and easily controlled.

2: This is an obstacle only in the case of large system sizes.

3: This is due to the fact that every trace-preserving map $F : \mathcal{A}_2 \to \mathcal{A}_2$ between two spaces of Hermitian matrices of size 2 is decomposable by two channels:

$$F(X) = p\mathcal{E}_1(X) + (1-p)\mathcal{E}_2(X^T).$$

Better methods for calculating separable numerical ranges would greatly help in understanding of the properties of entangled states. This is also of purely mathematical interest – classification of the structure of the set $W_{\text{SEP}}$ for the simple case of two bipartite qubit operators would be a novel result. In this case, the semidefinite structure of the positive partial transpose criterion could be applied.

Separable numerical ranges could also be used in the analysis of 2-entangled states, as suggested by Theorem 4.22. Since 2-entangled states are related to entanglement distillation, this could help in a wide array of related problems.

**Entropic numerical ranges**

The von Neumann entropy is a direct analogue of the classical Gibbs entropy: it is the entropy of the probability distribution of observing eigenstates of the density operator.

**Definition 6.1** *The von Neumann entropy of a quantum state $\rho \in \mathcal{M}_d$ is defined as*

$$S(\rho) = -\operatorname{Tr} \rho \log \rho, \tag{6.5}$$

*where $\log \cdot$ is a matrix logarithm, properly defined for strictly positive definite matrices. If the rank of a state $\rho$ is not full, the logarithm does not exist, but the expression $\rho \log \rho$ is still well defined in the set of positive semidefinite operators by any sequence of full rank density operators converging to $\rho$.*

The quantum entropy is a concave function of $\rho$ with the global maximum (over $\mathcal{M}_d$) of $\log d$ for the maximally mixed state and a minimum of 0 for any pure state. Due to its concavity, its superlevel sets are convex and question about the shape of their projections[4] comes naturally.

4: By analogy with other projections of restricted subsets of the set of density operators, one could call them *entropic numerical ranges*:

**Definition 6.2**
*The entropic numerical range of Hermitian operators $X_1, \ldots, X_k$ of size $d$ with the minimal entropy of $\gamma$ is the set*

$$W_{S\geq\gamma}(X_1,\ldots,X_k) = \{(\langle X_i\rangle_\rho)_{i=1}^k : \rho \in \mathcal{M}_d, S(\rho) \geq \gamma\}.$$

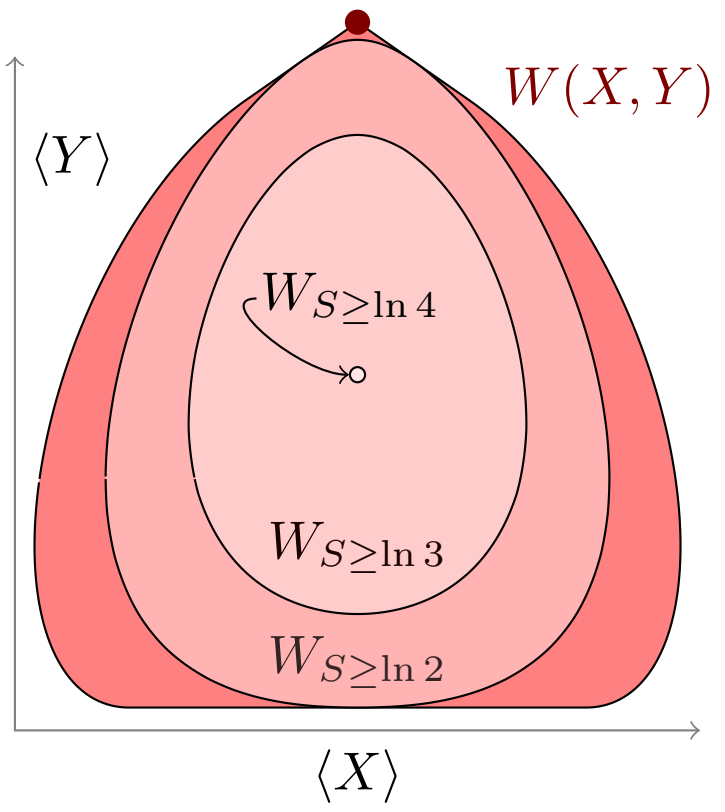

**Figure 6.1:** Entropic numerical ranges superimposed on the joint numerical range $W(X,Y)$. The operators $X$ and $Y$ used here are the same as in Figure 4.3 appearing on page 42.

Similar approaches using the notion of *maximum entropy inference map* were studied previously due to the connection with phase transitions [72, 73]. All of these ideas are closely connected to the properties of nonzero temperature phase transitions: the nonanalytical behavior of entropy and expectation values should be visible in the geometry of entropic numerical ranges. Such approach linking the geometry of subsets of quantum states with physical phenomena would certainly help in further understanding of quantum phase transitions.

[72]: Weis (2014), 'Continuity of the maximum-entropy inference'
[73]: Weis (2016), 'Maximum-entropy inference and inverse continuity of the numerical range'

# Appendix

# A

# Stereograms

The pictures in this sections are *cross-eyed stereograms*, each consisting of a *stereoscopic pair*. When viewed in the right way, the images become combined and give rise to the illusion of depth. To perceive the 3D nature of images, they need to be superimposed on each other, with the right image being seen by the left eye and *vice versa*.[1]

1: This is different from parallel or *side-by-side stereograms*, for which the right image is meant for the right eye. This kind of stereograms are, however, require more training to view and are limited in size due to physiologic constraints.

To aid in viewing the stereograms, the following method may be used:

1. Place a pen between the two images (on the paper or screen) forming a stereoscopic pair.
2. While keeping focus on the pen, slowly bring it closer to the eyes. The two images will appear to split and move in the background.
3. At a certain distance, three copies of the image will be visible: the central one with two others on the side.[2] The images will appear blurry at this point.
4. While keeping the eyes converged, shift the focus to the central, background image.[3] A seemingly three-dimensional image will appear.

2: The central image is a combination formed from the two peripheral ones.

3: Bringing the pen slightly closer to eyes just before switching focus may help.

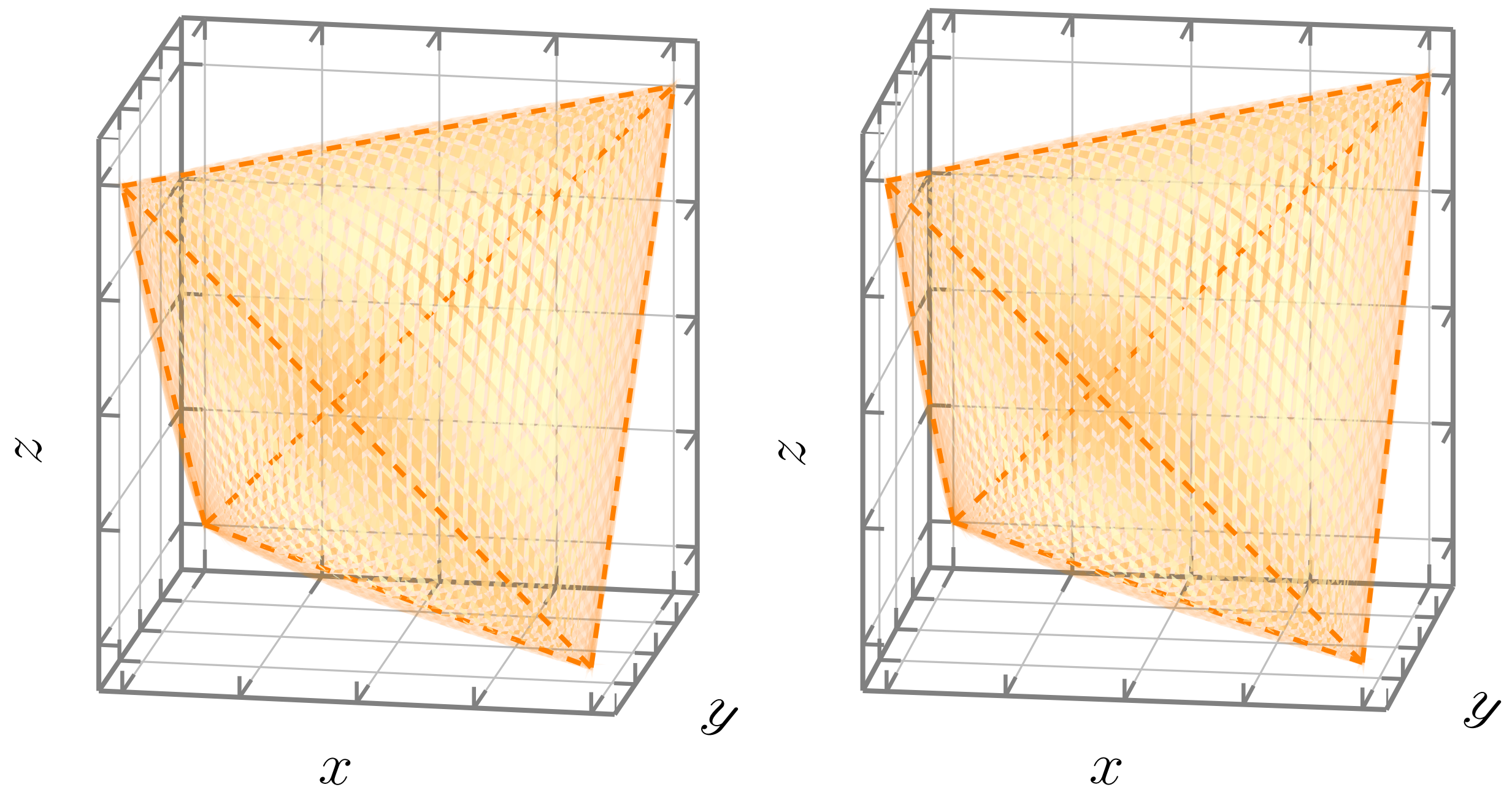

**Figure A.1:** Stereographic version of the Figure 4.4 appearing on page 44.

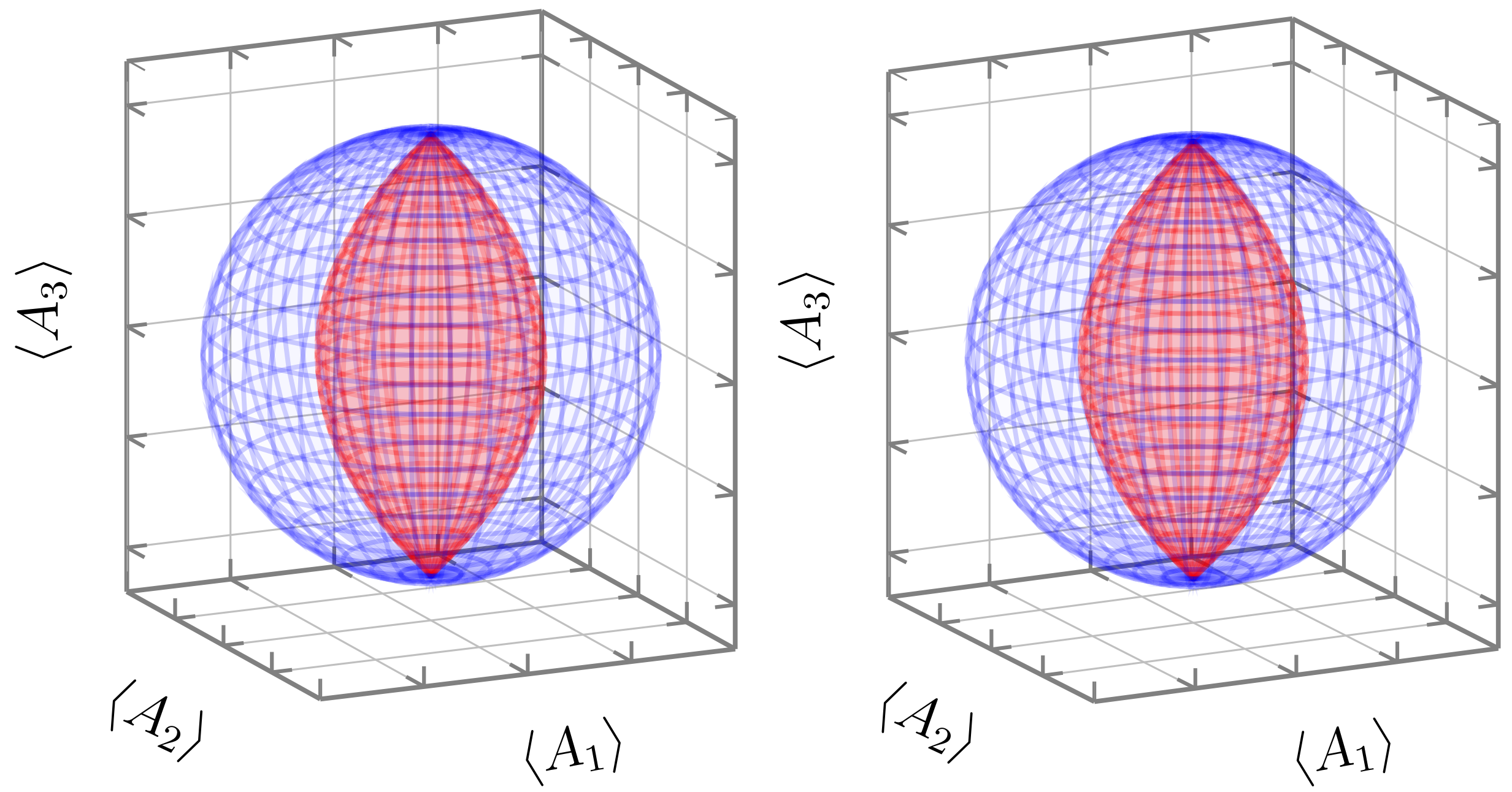

**Figure A.2:** Stereographic version of the Figure 4.14 appearing on page 61.

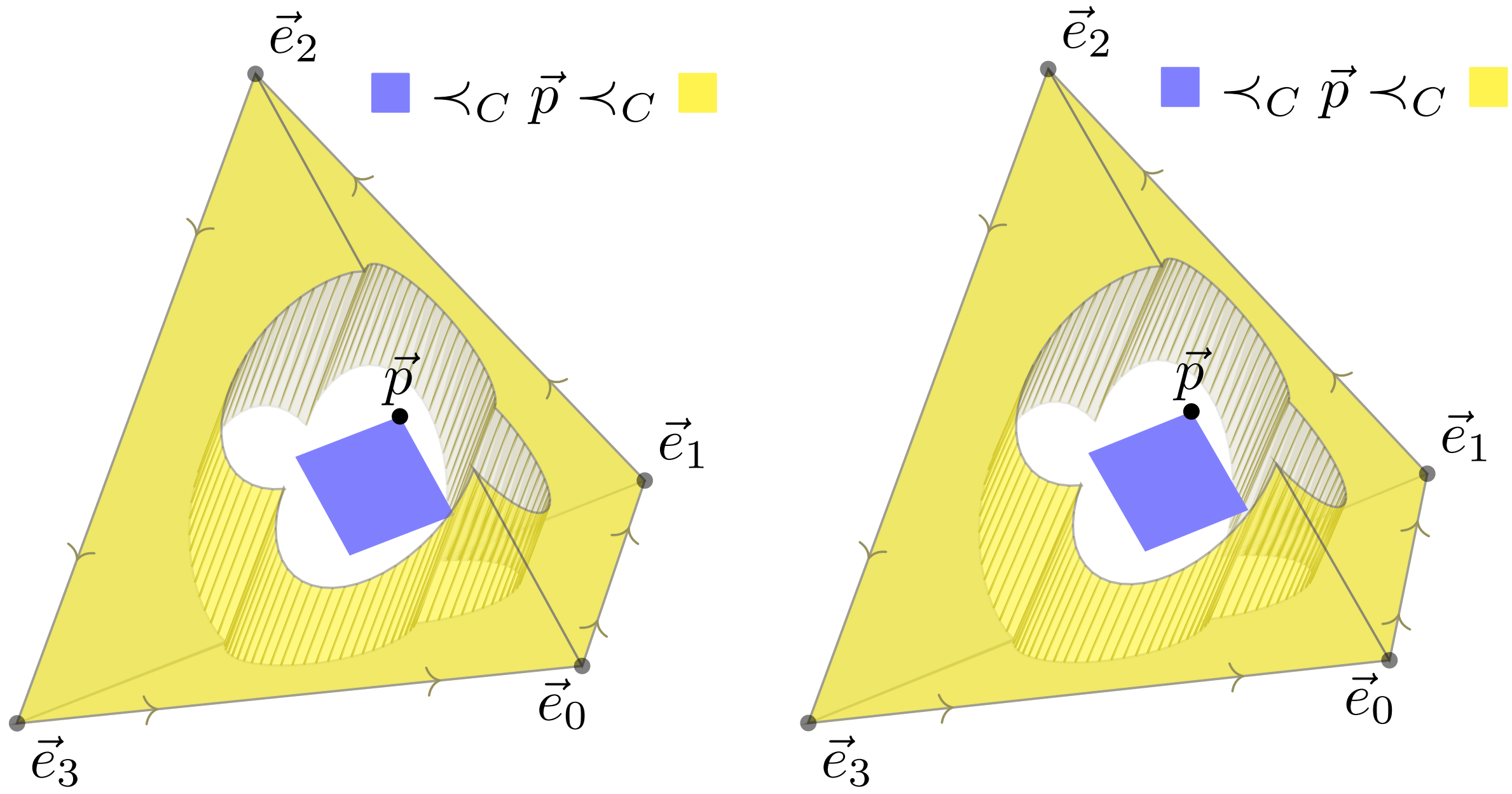

**Figure A.3:** Stereographic version of the Figure 5.1 appearing on page 76.

# B

# Code

Useful snippets of Mathematica code are contained here. They are short, untested against adversarial input, and cover all the different topics of this thesis – hence, it is not worthwhile to release them as a full fledged code package. Still, they may be useful for a person wanting to implement some of the functions defined here. The code is listed here as `InputForm` (mostly: `Association` objects are written with the `<| |>` notation and anonymous functions use `|->`), without any special formatting – it is less readable as a result, but does work when pasted directly from the PDF file into Mathematica frontend with only minimal changes: line breaks must be manually removed in some places.

## B.1 Joint numerical range

```
SupportVec[Xs_, n_] := (Normalize @* Last @* First @* MaximalBy[First] @*
                              Transpose @* Eigensystem @* N)[n . Xs]
SupportPoint[Xs_, n_] := Function[v, (Chop[Conjugate[v] . #1 . v] & ) /@ Xs]
                      [SupportVec[Xs, n]]
JNR[{X_, Y_}, npts_:1000] := ConvexHullMesh[(SupportPoint[{X, Y}, #1] & )
                                               /@ CirclePoints[npts]]
JNR[{X_, Y_, Z_}, npts_:1000] := ConvexHullMesh[(SupportPoint[{X, Y, Z}, #1] & )
                                               /@ SpherePoints[npts]]
```

The function `JNR` takes a list of two or three matrices (and optional number of boundary points to be sampled, by default $10^3$) and returns a `BoundaryMeshRegion` which can be later plotted with `RegionPlot`, used for geometric calculations, or saved to a `STL` file suitable for 3D printing with the builtin `Export` function.

The polynomial description of the numerical range boundary can be found with the following functions, separate for the 2D and 3D cases. Both take matrices as an input and return a Gröbner basis of the system of polynomials similar to the one described in Theorem 4.4 with the auxillary variables eliminated, such that the numerical range is the convex hull of the set of real roots of the resulting polynomials. In some cases, spurious parts not contained in the numerical range may appear and must be manually eliminated.

```
GroebnerNR[X_,Y_]:=GroebnerBasis[{Det[t IdentityMatrix[3] + u X+v Y]}~Join~
      (D[Det[t IdentityMatrix[3] + u X+v Y],#2]-#1&@@@{{a,t},{x,u},{y,v}}),
                                                  {x,y,a},{u,v,t}]/.a->1

GroebnerNR[X_,Y_,Z_]:=GroebnerBasis[{Det[t IdentityMatrix[3]+u X+v Y+w Z]}~Join~
   (D[Det[t IdentityMatrix[3]+u X+v Y+w Z],#2]-#1&@@@{{a,t},{x,u},{y,v},{z,w}}),
                                                  {x,y,z,a},{u,v,w,t}]/.a->1
```

## B.2 Multipartite systems: partial transpose, partial trace and PPT numerical range

```
IndexOf[indices_, subdims_] := Sequence @@ (1 + Reverse[#1 - 1] . Most[FoldList
           [Times, 1, Reverse[subdims]]] & ) /@ Transpose[Partition[indices, 2]]
MatrixReshape[(matrix_)?SquareMatrixQ, subdims_] := Array[matrix[[IndexOf[{##1},
        subdims]]] & , Riffle[subdims, subdims]]/;Times@@subdims==Length[matrix]
PartialTrace[(matrix_)?SquareMatrixQ, subdims_, (which_)?IntegerQ] := NestWhile
          [ArrayFlatten, Map[Tr[#1, Plus, 2] & , MatrixReshape[matrix, subdims],
          {2*which - 2}], Length[Dimensions[#1]] > 2 & ]
          /; Times @@ subdims == Length[matrix] && 1 <= which <= Length[subdims]
PartialTranspose[(matrix_)?SquareMatrixQ,subdims_, (which_)?IntegerQ]:=NestWhile[
          ArrayFlatten, Transpose[MatrixReshape[matrix, subdims], 2*which - 1 <->
          2*which], Length[Dimensions[#1]] > 2 & ]
          /; Times @@ subdims == Length[matrix] && 1 <= which <= Length[subdims]
```

The functions `PartialTrace` and `PartialTranspose` both take a matrix (a two-dimensional square array), description of subdimensions `subdims`, and the subdimension index `which` for which the appropriate action should take place. For instance, a matrix of size 8 can describe the operator acting on three qubit space – in this case, the `subdims` argument should be `{2,2,2}`. `MatrixReshape` is a helper function creating an array of nested arrays (nested block matrix) given a matrix and subdimension description.

```
MinimizeBipartitePT[matrix_, subdims_] := Module[{rho, vars, rhoPT, tooptim, sol,
 total, S, A},
  total = Times @@ subdims;
  rho=Array[S@@Sort[{#1, #2}]+I*Sign[#2-#1]*A@@Sort[{#1, #2}]&, {total, total}];
  rhoPT = PartialTranspose[rho, subdims, 2];
  vars = Union[Cases[rho, S[_, _], 4], Cases[rho, A[_, _], 4]];
  tooptim = Chop[Tr[matrix . rho]]//ComplexExpand//Chop;
  sol = SemidefiniteOptimization[tooptim,
     VectorGreaterEqual[{rho,0}, {"SemidefiniteCone", total}] &&
     VectorGreaterEqual[{rhoPT,0},{"SemidefiniteCone",total}]&&Tr[rho]==1,vars];
  rho /. (sol//Chop)]                           /; Times @@ subdims == Length[matrix]
```

The function `MinimizeBipartitePT` minimizes the expectation value of a given observable `matrix` defined on bipartite system with local dimensions `subdims=={`$d_1$`,`$d_2$`}` over the states with positive partial transpose $\mathcal{M}^{\mathrm{PPT}}_{d_1 \times d_2}$ and returns the minimizing state. With this, the functions calculating PPT numerical range are defined, analogously to the case of joint numerical range:

```
SupportPointPPTNR[Xs_, ns_, subdims_] := Function[rho,
            (Chop[Tr[rho . #1]] & ) /@ Xs][MinimizeBipartitePT[ns . Xs, subdims]]
PPTNR[{X_, Y_}, subdims_, npts_:1000] := ConvexHullMesh[
            (SupportPointSNR[{X, Y}, #1, subdims] & ) /@ CirclePoints[npts]     ]
PPTNR[{X_, Y_, Z_}, subdims_, npts_:1000] := ConvexHullMesh[
         (SupportPointSNR[{X, Y, Z}, #1, subdims] & ) /@ SpherePoints[npts]     ]
```

## B.3 Weyl-Heisenberg operators and discrete Wigner function

```
X[dim_] := DiagonalMatrix[ConstantArray[1, dim - 1], -1] +
                                  DiagonalMatrix[{1}, dim - 1]
Z[dim_] := DiagonalMatrix[Table[Exp[(2*Pi*I*k)/dim], {k, 0, dim - 1}]]
Dis[x_, p_, (dim_)?PrimeQ] := Dis[x, p, dim] =
        (-Exp[(I*Pi)/dim])^(x*p)*MatrixPower[X[dim], x] . MatrixPower[Z[dim], p]
Dis[xs_, ps_, (dims_)?ListQ] := (Dis[xs, ps, dims] =
              KroneckerProduct @@ Apply[Dis, Transpose[{xs, ps, dims}], {1}])
                                   /; Length[dims] >= 2
Dis[{x_}, {p_}, {(dim_)?PrimeQ}] := Dis[{x}, {p}, {dim}] = Dis[x, p, dim]
A0[(dim_)?PrimeQ] := A0[dim] =
     (1/dim)*FullSimplify[Sum[Dis[x, p, dim], {x, 0, dim - 1}, {p, 0, dim - 1}]]
A0[(dims_)?ListQ] := (A0[dims] =
     KroneckerProduct @@ A0 /@ dims) /; Length[dims] >= 2
A0[{(dim_)?PrimeQ}] := A0[{dim}] = A0[dim]
A[xs_, ps_, (dims_)?ListQ] := A[xs, ps, dims] =
     Dis[xs, ps, dims] . A0[dims] . ConjugateTranspose[Dis[xs, ps, dims]]
```

The displacement operators are defined by function `Dis` taking three arguments `x`, `p` and `dim(s)`. The first two define the displacement parameters, while `dim(s)` describes the quantum system size. If the dimension is prime, all argument can be input as numbers, in the other case they have to be lists (vectors) describing each of the subsystems of prime dimensions.

```
Wigner[state_, dims_] := Module[{dim = Length[state], ZZ},
 ZZ = Tuples[(Range[0, #1 - 1] & ) /@ dims];
 Chop@Table[(1/dim)*Tr[A[xs,ps,dims].ConjugateTranspose[state]],{xs,ZZ},{ps,ZZ}]]
FromWigner[wigner_, dims_] := Module[{dim = Length[wigner], ZZ},
 ZZ = Tuples[(Range[0, #1 - 1] & ) /@ dims];
 Chop@Sum[wigner[[xidx,pidx]]*A[ZZ[[xidx]],ZZ[[pidx]],dims],
                      {xidx, Length[ZZ]}, {pidx, Length[ZZ]}] ]
```

The function `Wigner` takes a state – a density operator – along with the subsystem dimensions description and returns its Wigner function as a real square matrix. `FromWigner` does the opposite – given a Wigner function, it returns a density operator. In either case the argument `dims` must be a list of prime numbers – the dimensions of subsystems.

## B.4 Composition of spin states and $SU(2)$ characteristic functions

```
Off[ClebschGordan::phy];
Splus[0] = {{0}};
Splus[j_] := DiagonalMatrix[Table[Sqrt[j*(j+1) - m*(m+1)], {m,-j,j-1}], 1];
Sminus[j_] := Transpose[Splus[j]]
Sx[j_] := Sx[j] = (Splus[j] + Sminus[j])/2
Sy[j_] := Sy[j] = (Splus[j] - Sminus[j])/(2*I)
Sz[j_] := Sz[j] = DiagonalMatrix[Table[m, {m, j, -j, -1}]]
SpinRotation[j_, {x_, y_, z_}] := MatrixExp[I*(x*Sx[j] + y*Sy[j] + z*Sz[j])]
```

The functions above define spin operators for given total spin $j$, as well as the rotation matrix $\exp\left(i(xJ_X + yJ_Y + zJ_Z)\right)$ inside a single representation. To avoid variable name collision, the matrices $J_X, J_Y, J_Z$ are denoted in the code by `Sx, Sy, Sz`.

```
ElementarySpinCombine[substate1_,substate2_]:=Module[{j1,m1,j2,m2},
  {j1,m1}={"J","M"}/.substate1;
  {j2,m2}={"J","M"}/.substate2;
  Select[<|Table[<|"J"->j,"M"->m1+m2,"other"->{j1,j2}|>->
      ClebschGordan[{j1,m1},{j2,m2},{j,m1+m2}],{j,Abs[j1-j2],j1+j2}]|>,#!=0&] ]
SpinCombine[state1_,state2_]:=Merge[
  ElementarySpinCombine[#1,#2]state1[#1]state2[#2]&@@@
      Tuples[{Keys[state1],Keys[state2]}],Total ]
```

The function `SpinCombine` represent a tensor product of the two input spin states in the total spin basis in the sense of Theorem 5.12. The input and output states are represented as `Association` objects – see below for an example of notation.

```
SpinGather[state_]:=<|#->Lookup[KeyMap[key|->"M"/.key,
  KeySelect[state,key|->KeyDrop[key,{"M"}]==#]],
    Range["J"/.#,-"J"/.#,-1],0]&/@ (Union@(KeyDrop[{"M"}]/@Keys[state]))|>
SpinCharacteristicFunction[state_,param_]:=Total@KeyValueMap[
  {desc,substate}|->Conjugate[substate].SpinRotation["J"/.desc,param].substate,
    SpinGather[state]]
SpinEigenstatesCombine[state_]:=Merge[
  KeyValueMap[KeyTake[#1,{"J","M"}]->#2&,state], Norm ]
```

The helper function `SpinGather` takes the output of `SpinCombine` and returns an `Association` of the kets corresponding to the same representation and multiplicity indices (`"J"` and `"other"`, respectively). This is used by the self-explanatory `SpinCharacteristicFunction`. If the combined spin states corresponding to the same index $j$ are proportional, they can be combined in the sense of Theorem 5.14 by `SpinEigenstatesCombine`. Below is the application of this code in the randomized verification of Example 5.3 (characteristic function of a combined spin state is compared with a product of characteristic functions of two compound states at a random point parameterized by `n`):

```
phi=<|<|"J"->1,"M"->-1|>->√1/3,<|"J"->2,"M"->-1|>->√1/3,<|"J"->3,"M"->-1|>->
√1/3|>;
omega=<|<|"J"->0,"M"->0|>->-√1/2,<|"J"->1,"M"->0|>->√1/2|>;
psi=SpinCombine[phi,omega]//SpinEigenstatesCombine;
n=RandomReal[{-10,10},3];
Print@Chop[SpinCharacteristicFunction[phi,n]SpinCharacteristicFunction[omega,n]];
Print@SpinCharacteristicFunction[psi,n];
```

# Bibliography

Here are the references in citation order.

[1] Konrad Szymański, Stephan Weis, and Karol Życzkowski. 'Classification of joint numerical ranges of three hermitian matrices of size three'. In: *Linear Algebra and its Applications* 545 (2018), pp. 148–173 (cited on pages 1, 46, 48, 50, 91, 94).

[2] Jakub Czartowski, Konrad Szymański, Bartłomiej Gardas, Yan Fyodorov, and Karol Życzkowski. 'Separability gap and large-deviation entanglement criterion'. In: *Physical Review A* 100.4 (2019), p. 042326 (cited on pages 1, 63, 92).

[3] Konrad Szymański and Karol Życzkowski. 'Geometric and algebraic origins of additive uncertainty relations'. In: *Journal of Physics A: Mathematical and Theoretical* 53.1 (2019), p. 015302 (cited on pages 1, 51–53, 92).

[4] Konrad Szymański and Karol Życzkowski. 'Universal witnesses of vanishing energy gap'. In: *Europhysics Letters* 136 (2021), p. 30003 (cited on pages 1, 57, 59, 60, 91).

[5] Timo Simnacher, Jakub Czartowski, Konrad Szymański, and Karol Życzkowski. 'Confident entanglement detection via the separable numerical range'. In: *Physical Review A* 104 (2021), p. 042420 (cited on pages 1, 63, 92).

[6] Muhammad ibn Musa al-Khwarizmi. *Al-Kitab al-mukhtasar fi hisab al-gabr wa'l-muqabala*. (translation: https://openlibrary.org/works/OL18101578W/ and original text: https://w.wiki/4pN7). Baghdad, Abbasid Caliphate, circa 820 (cited on page 3).

[7] Jie Xie, Aonan Zhang, Ningping Cao, Huichao Xu, Kaimin Zheng, Yiu-Tung Poon, Nung-Sing Sze, Ping Xu, Bei Zeng, and Lijian Zhang. 'Observing geometry of quantum states in a three-level system'. In: *Physical Review Letters* 125.15 (2020), p. 150401 (cited on pages 3, 50).

[8] Jun John Sakurai and Eugene Commins. *Modern Quantum Mechanics*. 1995 (cited on page 5).

[9] Alexander Holevo. *Statistical Structure of Quantum Theory*. Vol. 67. Springer Science & Business Media, 2003 (cited on pages 9, 17).

[10] Ingemar Bengtsson, Stephan Weis, and Karol Życzkowski. 'Geometry of the set of mixed quantum states: An apophatic approach'. In: *Geometric Methods in Physics*. Springer, 2013, pp. 175–197 (cited on page 12).

[11] Christopher Eltschka, Marcus Huber, Simon Morelli, and Jens Siewert. 'The shape of higher-dimensional state space: Bloch-ball analog for a qutrit'. In: *Quantum* 5 (2021), p. 485 (cited on page 12).

[12] Donald Bures. 'An extension of Kakutani's theorem on infinite product measures to the tensor product of semifinite $w^*$-algebras'. In: *Transactions of the American Mathematical Society* 135 (1969), pp. 199–212 (cited on page 12).

[13] Armin Uhlmann. 'The "transition probability" in the state space of $a^*$-algebra'. In: *Reports on Mathematical Physics* 9.2 (1976), pp. 273–279 (cited on page 12).

[14] Armin Uhlmann. 'Density operators as an arena for differential geometry'. In: *Reports on Mathematical Physics* 33.1/2 (1993), pp. 253–263 (cited on page 12).

[15] Angelo Carollo, Davide Valenti, and Bernardo Spagnolo. 'Geometry of quantum phase transitions'. In: *Physics Reports* 838 (2020), pp. 1–72 (cited on page 13).

[16] Michael Victor Berry. 'Quantal phase factors accompanying adiabatic changes'. In: *Proceedings of the Royal Society of London. A. Mathematical and Physical Sciences* 392.1802 (1984), pp. 45–57 (cited on page 15).

[17] Edward Davies. *Quantum Theory of Open Systems*. Academic Press, 1976 (cited on page 17).

[18] Marcus Appleby, Ingemar Bengtsson, Markus Grassl, Michael Harrison, and Gary McConnell. 'SIC-POVMs from Stark Units'. In: *arXiv preprint arXiv:2112.05552* (2021) (cited on page 17).

[19] Carlton Caves, Christopher Fuchs, and Rüdiger Schack. 'Unknown quantum states: the quantum de Finetti representation'. In: *Journal of Mathematical Physics* 43.9 (2002), pp. 4537–4559 (cited on page 18).

[20] Man-Duen Choi. 'Completely positive linear maps on complex matrices'. In: *Linear Algebra and its Applications* 10.3 (1975), pp. 285–290 (cited on pages 19, 20).

[21] Andrzej Jamiołkowski. 'Linear transformations which preserve trace and positive semidefiniteness of operators'. In: *Reports on Mathematical Physics* 3.4 (1972), pp. 275–278 (cited on pages 19, 20).

[22] Wolfgang Dür, Marc Hein, Ignacio Cirac, and Hans Briegel. 'Standard forms of noisy quantum operations via depolarization'. In: *Physical Review A* 72.5 (2005), p. 052326 (cited on pages 20, 21).

[23] Min Jiang, Shunlong Luo, and Shuangshuang Fu. 'Channel-state duality'. In: *Physical Review A* 87.2 (2013), p. 022310 (cited on page 21).

[24] Nikolaos Koukoulekidis and David Jennings. 'Constraints on magic state protocols from the statistical mechanics of Wigner negativity'. In: *arXiv preprint arXiv:2106.15527* (2021) (cited on pages 22, 23, 85).

[25] Robert Raussendorf, Juani Bermejo-Vega, Emily Tyhurst, Cihan Okay, and Michael Zurel. 'Phase-space-simulation method for quantum computation with magic states on qubits'. In: *Physical Review A* 101 (1 2020), p. 012350 (cited on page 23).

[26] Reinhard Werner. 'Quantum states with Einstein-Podolsky-Rosen correlations admitting a hidden-variable model'. In: *Physical Review A* 40.8 (1989), p. 4277 (cited on page 25).

[27] Lin Chen and Dragomir Đoković. 'Distillability of non-positive-partial-transpose bipartite quantum states of rank four'. In: *Physical Review A* 94.5 (2016), p. 052318 (cited on page 26).

[28] Rainer Verch and Reinhard Werner. 'Distillability and positivity of partial transposes in general quantum field systems'. In: *Reviews in Mathematical Physics* 17.05 (2005), pp. 545–576 (cited on page 26).

[29] Frank Verstraete and Ignacio Cirac. 'Matrix product states represent ground states faithfully'. In: *Physical Review B* 73.9 (2006), p. 094423 (cited on page 27).

[30] Robin Hudson and Graham Moody. 'Locally normal symmetric states and an analogue of de Finetti's theorem'. In: *Zeitschrift für Wahrscheinlichkeitstheorie und verwandte Gebiete* 33.4 (1976), pp. 343–351 (cited on page 28).

[31] Jianxin Chen, Cheng Guo, Zhengfeng Ji, Yiu-Tung Poon, Nengkun Yu, Bei Zeng, and Jie Zhou. 'Joint product numerical range and geometry of reduced density matrices'. In: *Science China: Physics, Mechanics & Astronomy* 60.2 (2017), pp. 1–9 (cited on page 28).

[32] Steven Lay. *Convex Sets and Their Applications*. Courier Corporation, 2007 (cited on pages 29, 36).

[33] David Cox, John Little, and Donal O'Shea. *Ideals, varieties, and algorithms: an introduction to computational algebraic geometry and commutative algebra*. Springer Science & Business Media, 2013 (cited on pages 32, 36).

[34] Otto Toeplitz. 'Das algebraische Analogon zu einem Satze von Fejér'. In: *Mathematische Zeitschrift* 2.1 (1918), pp. 187–197 (cited on pages 37, 39, 94).

[35] Felix Hausdorff. 'Der Wertvorrat einer Bilinearform'. In: *Mathematische Zeitschrift* 3.1 (1919), pp. 314–316 (cited on pages 37, 39, 94).

[36] Panayiotis Psarrakos and Michael Tsatsomeros. *Numerical range: (in) a Matrix Nutshell*. Vol. 45. Mathematical Notes from Washington State University. 2002 (cited on page 39).

[37] Piotr Gawron, Zbigniew Puchała, Jarosław Adam Miszczak, Lukasz Skowronek, and Karol Życzkowski. 'Restricted numerical range: a versatile tool in the theory of quantum information'. In: *Journal of Mathematical Physics* 51.10 (2010), p. 102204 (cited on page 40).

[38] Motakuri Ramana. 'An exact duality theory for semidefinite programming and its complexity implications'. In: *Mathematical Programming* 77.1 (1997), pp. 129–162 (cited on pages 41, 45).

[39] Rudolf Kippenhahn. 'Über den Wertevorrat einer Matrix'. In: *Mathematische Nachrichten* 6.3-4 (1951), pp. 193–228 (cited on page 43).

[40] Miroslav Fiedler. 'Geometry of the Numerical Range of Matrices'. In: *Linear Algebra and its Applications* 37 (1981), pp. 81–96 (cited on page 43).

[41] Mao-Ting Chien and Hiroshi Nakazato. 'Joint numerical range and its generating hypersurface'. In: *Linear Algebra and its Applications* 432.1 (2010), pp. 173–179 (cited on page 43).

[42] René Schwonnek and Reinhard Werner. 'The Wigner distribution of n arbitrary observables'. In: *Journal of Mathematical Physics* 61.8 (2020), p. 082103 (cited on page 43).

[43] Daniel Plaumann, Rainer Sinn, and Stephan Weis. 'Kippenhahn's Theorem for joint numerical ranges and quantum states'. In: *SIAM Journal on Applied Algebra and Geometry* 5.1 (2021), pp. 86–113 (cited on page 43).

[44] Mateusz Michałek and Bernd Sturmfels. *Invitation to Nonlinear Algebra*. Vol. 211. American Mathematical Society, 2021 (cited on page 44).

[45] Fernando Brandao and Reinaldo Vianna. 'Robust semidefinite programming approach to the separability problem'. In: *Physical Review A* 70.6 (2004), p. 062309 (cited on page 45).

[46] Kazimierz Kuratowski. 'Sur le problème des courbes gauches en topologie'. In: *Fundamenta Mathematicae* 15.1 (1930), pp. 271–283 (cited on page 49).

[47] Dennis Keeler, Leiba Rodman, and Ilya Spitkovsky. 'The numerical range of 3×3 matrices'. In: *Linear Algebra and its Applications* 252.1-3 (1997), pp. 115–139 (cited on pages 49, 94).

[48] Qiongyi He, Shi-Guo Peng, Peter Drummond, and Margaret Reid. 'Planar quantum squeezing and atom interferometry'. In: *Physical Review A* 84.2 (2011), p. 022107 (cited on pages 50, 53).

[49] Paolo Giorda, Lorenzo Maccone, and Alberto Riccardi. 'State-independent uncertainty relations from eigenvalue minimization'. In: *Physical Review A* 99 (5 2019), p. 052121 (cited on page 51).

[50] Lorenzo Maccone and Arun Pati. 'Stronger Uncertainty Relations for All Incompatible Observables'. In: *Physical Review Letters* 113 (26 Dec. 2014), p. 260401 (cited on page 51).

[51] Lars Dammeier, René Schwonnek, and Reinhard Werner. 'Uncertainty relations for angular momentum'. In: *New Journal of Physics* 17.9 (2015), p. 093046 (cited on pages 51, 54).

[52] René Schwonnek, Lars Dammeier, and Reinhard Werner. 'State-independent uncertainty relations and entanglement detection in noisy systems'. In: *Physical Review Letters* 119.17 (2017), p. 170404 (cited on page 51).

[53] Toby Cubitt, David Perez-Garcia, and Michael Wolf. 'Undecidability of the spectral gap'. In: *Nature* 528.7581 (2015), pp. 207–211 (cited on page 57).

[54] Elliott Lieb, Theodore Schultz, and Daniel Mattis. 'Two soluble models of an antiferromagnetic chain'. In: *Annals of Physics* 16.3 (1961), pp. 407–466 (cited on page 59).

[55] Jose Lado. *DMRGPY: Python library to spin and fermionic Hamiltonians with DMRG*. `https://github.com/joselado/dmrgpy/`. 2021 (cited on page 60).

[56] Robert Edwards. *Functional Analysis, Theory and Applications, Holt, Rinehart and Winston, New York*. 1965 (cited on page 61).

[57] Shmuel Friedland and Lek-Heng Lim. 'Nuclear norm of higher-order tensors'. In: *Mathematics of Computation* 87.311 (2018), pp. 1255–1281 (cited on page 63).

[58] Stephen Bartlett, Terry Rudolph, and Robert Spekkens. 'Dialogue concerning two views on quantum coherence: factist and fictionist'. In: *International Journal of Quantum Information* 4.01 (2006), pp. 17–43 (cited on page 69).

[59] Iman Marvian. 'Symmetry, Asymmetry and Quantum Information'. PhD thesis. 2012 (cited on pages 69, 70, 73, 75, 86).

[60] Gilad Gour and Robert Spekkens. 'The resource theory of quantum reference frames: manipulations and monotones'. In: *New Journal of Physics* 10.3 (2008), p. 033023 (cited on pages 69–72, 84, 86, 88, 96).

[61] Barry Simon Michael Reed. *Methods of Modern Mathematical Physics, vol. 2*. Methods of Modern Mathematical Physics. Academic Press, 1975 (cited on page 74).

[62] Alessandra Giovagnoli and Henry Wynn. 'Cyclic majorization and smoothing operators'. In: *Linear Algebra and its Applications* 239 (1996), pp. 215–225 (cited on pages 77, 78).

[63] Aubrey Ingleton. 'The Rank of Circulant Matrices'. In: *Journal of the London Mathematical Society* s1-31.4 (Oct. 1956), pp. 445–460 (cited on page 81).

[64] Nilanjana Datta, Motohisa Fukuda, and Alexander Holevo. 'Complementarity and additivity for covariant channels'. In: *Quantum Information Processing* 5.3 (2006), pp. 179–207 (cited on page 85).

[65] David Griffiths. *Introduction to Elementary Particles*. 2nd. Wiley-VCH, 2008 (cited on page 86).

[66] Valery Kelmanovich Khersonskii, Anatoly Nikolaevich Moskalev, and Dmitry Alexandrovich Varshalovich. *Quantum Theory of Angular Momentum*. World Scientific Publishing Company, 1988 (cited on page 87).

[67] Mao-Ting Chien and Hiroshi Nakazato. 'Singular points of the ternary polynomials associated with 4-by-4 matrices'. In: *The Electronic Journal of Linear Algebra* 23 (2012), pp. 755–769 (cited on page 94).

[68] Henry Haselgrove, Michael Nielsen, and Tobias Osborne. 'Quantum states far from the energy eigenstates of any local Hamiltonian'. In: *Physical Review Letters* 91.21 (2003), p. 210401 (cited on page 96).

[69] Noah Linden and Sandu Popescu. 'Good Dynamics versus Bad Kinematics: Is Entanglement Needed for Quantum Computation?' In: *Physical Review Letters* 87.4 (2001), p. 047901 (cited on page 96).

[70] Michał Horodecki, Paweł Horodecki, and Ryszard Horodecki. 'Separability of n-particle mixed states: necessary and sufficient conditions in terms of linear maps'. In: *Physics Letters A* 283.1-2 (2001), pp. 1–7 (cited on page 96).

[71] Otfried Gühne and Géza Tóth. 'Entanglement detection'. In: *Physics Reports* 474.1-6 (2009), pp. 1–75 (cited on page 96).

[72] Stephan Weis. 'Continuity of the maximum-entropy inference'. In: *Communications in Mathematical Physics* 330.3 (2014), pp. 1263–1292 (cited on page 97).

[73] Stephan Weis. 'Maximum-entropy inference and inverse continuity of the numerical range'. In: *Reports on Mathematical Physics* 77.2 (2016), pp. 251–263 (cited on page 97).